\documentclass[twocolumn]{aastex631}

\usepackage{natbib,aas_macros,amsmath}
\usepackage{multirow,color}

\newcommand{\rqq}{\textquotedblright}
\newcommand{\lqq}{\textquotedblleft}

\begin{document}

\shorttitle{Morphological Demographics of Galaxies at $z\sim 10$--$16$
} 
\shortauthors{Ono et al.}
\submitjournal{ApJ in press}

\title{%
Morphological Demographics of Galaxies at $z\sim 10$--$16$:\\
Log-Normal Size Distribution and Exponential Profiles Consistent with the Disk Formation Scenario}


\author[0000-0001-9011-7605]{Yoshiaki Ono}
\affiliation{Institute for Cosmic Ray Research, The University of Tokyo, 5-1-5 Kashiwanoha, Kashiwa, Chiba 277-8582, Japan}

\author[0000-0002-1049-6658]{Masami Ouchi}
\affiliation{National Astronomical Observatory of Japan, 2-21-1 Osawa, Mitaka, Tokyo 181-8588, Japan}
\affiliation{Institute for Cosmic Ray Research, The University of Tokyo, 5-1-5 Kashiwanoha, Kashiwa, Chiba 277-8582, Japan}
\affiliation{Kavli Institute for the Physics and Mathematics of the Universe (WPI), The University of Tokyo, 5-1-5 Kashiwanoha, Kashiwa-shi, Chiba, 277-8583, Japan}

\author[0000-0002-6047-430X]{Yuichi Harikane}
\affiliation{Institute for Cosmic Ray Research, The University of Tokyo, 5-1-5 Kashiwanoha, Kashiwa, Chiba 277-8582, Japan}

\author[0000-0002-1319-3433]{Hidenobu Yajima}
\affiliation{Center for Computational Sciences, University of Tsukuba, Ten-nodai, 1-1-1 Tsukuba, Ibaraki 305-8577, Japan}

\author[0000-0003-2965-5070]{Kimihiko Nakajima}
\affiliation{National Astronomical Observatory of Japan, 2-21-1 Osawa, Mitaka, Tokyo 181-8588, Japan}

\author[0000-0001-7201-5066]{Seiji Fujimoto}
\affiliation{Department of Astronomy, The University of Texas at Austin, Austin, TX 78712, USA}

\author[0009-0000-1999-5472]{Minami Nakane}
\affiliation{Institute for Cosmic Ray Research, The University of Tokyo, 5-1-5 Kashiwanoha, Kashiwa, Chiba 277-8582, Japan}
\affiliation{Department of Physics, Graduate School of Science, The University of Tokyo, 7-3-1 Hongo, Bunkyo, Tokyo 113-0033, Japan}

\author[0000-0002-5768-8235]{Yi Xu}
\affiliation{Institute for Cosmic Ray Research, The University of Tokyo, 5-1-5 Kashiwanoha, Kashiwa, Chiba 277-8582, Japan}
\affiliation{Department of Astronomy, Graduate School of Science, The University of Tokyo, 7-3-1 Hongo, Bunkyo, Tokyo 113-0033, Japan}

\begin{abstract}

We homogeneously investigate the morphological properties of $169$ galaxies at $z\sim10$--$16$
with deep JWST NIRCam images employing our established techniques of GALFIT modeling 
and uncertainty evaluation (systematics+statistics) . 
We obtain effective radii $r_{\rm e}$ ranging $20$--$500$ pc, 
with a distribution significantly broader than the scatter made by the uncertainties. 
We find that the $r_{\rm e}$ distribution is well described by a log-normal distribution 
with a mean of $r_{\rm e}=133^{+13}_{-12}$ pc
and a standard deviation of $\sigma_{{\rm ln}r_{\rm e}}=0.52\pm0.08$. 
The standard deviation is comparable to that of local galaxies, 
indicating no significant evolution over $z\sim0$--$10$. 
We also find that the axis ratio distribution is nearly uniform, 
statistically similar to that of local spirals.
We estimate the virial radius $r_{\rm vir}$ from the stellar masses 
via the star-formation main sequence and stellar-to-halo mass relation, 
obtaining a stellar-to-halo size ratio $r_{\rm e}/r_{\rm vir}=0.015^{+0.015}_{-0.005}$, 
which is comparable to those of star-forming galaxies in the local and low-$z$ Universe. 
Our results of 
1) the log-normal $r_{\rm e}$ distribution, 
2) the uniform axis ratio distribution, 
and 
3) a mean radial profile consistent with an exponential profile ($n=1.3\pm0.6$) 
suggest that galaxies at $z\sim10$--$16$ generally follow the classical galaxy disk formation scenario
with a specific disk angular momentum fraction of $j_{\rm d} / m_{\rm d} \sim 0.5$--$1$. 
Interestingly, we identify two remarkable outliers 
GN-z11 ($z_{\rm spec}=10.60$) and GHZ2 ($z_{\rm spec}=12.34$) 
with $r_{\rm e}=55^{+5}_{-6}$ pc and $39\pm11$ pc, respectively, 
that may not be explained by disk structures  
but by AGN or compact star-forming galaxies merging underway 
in short periods of time, as reproduced in numerical simulations.

\end{abstract}

\keywords{%
\href{http://astrothesaurus.org/uat/734}{High-redshift galaxies (734)}; 
\href{http://astrothesaurus.org/uat/979}{Lyman-break galaxies (979)}; 
\href{http://astrothesaurus.org/uat/594}{Galaxy evolution (594)}; 
\href{http://astrothesaurus.org/uat/595}{Galaxy formation (595)}; 
\href{http://astrothesaurus.org/uat/622}{Galaxy structure (622)}; 
\href{http://astrothesaurus.org/uat/617}{Galaxy radii (617)}; 
\href{http://astrothesaurus.org/uat/573}{Galaxies (573)}; 
\href{http://astrothesaurus.org/uat/563}{Galactic and extragalactic astronomy (563)} 
}

\section{Introduction} \label{sec:introduction}

Morphological properties of galaxies play a vital role in characterizing galaxy evolution 
(\citealt{2014ARA&A..52..291C}; \citealt{2025arXiv250117078S}). 
Thanks to the superior spatial resolution and sensitivity of 
NIRCam (\citealt{2005SPIE.5904....1R}; \citealt{2023PASP..135b8001R})
onboard the James Webb Space Telescope 
(JWST; \citealt{2006SSRv..123..485G}; \citealt{2023PASP..135f8001G}; \citealt{2023PASP..135e8001M}), 
significant progress has been made 
in investigating the morphological properties of high-$z$ star-forming galaxies 
primarily found based on their rest-frame UV spectral features 
(\citealt{2022ApJ...938L..17Y}; \citealt{2023ApJ...942L..28T}; \citealt{2023ApJ...951...72O}; 
\citealt{2024ApJ...963....9M}; \citealt{2024PASJ...76..219O}). 
Additionally, morphological studies have also been conducted for high-$z$ galaxies 
selected with other methods, including dusty star-forming galaxies 
(\citealt{2023A&A...676A..26G}; \citealt{2024A&A...688A..53L}; 
\citealt{2023arXiv230907834F}; \citealt{2024arXiv240715846H}; 
\citealt{2025arXiv250201868U}), 
strong emission-line selected star-forming galaxies 
(\citealt{2024ApJ...963L..38N}; \citealt{2024ApJ...964L..33L}; \citealt{2024ApJ...966..210L}; 
\citealt{2024arXiv240919205G}), 
quiescent galaxies 
(\citealt{2024ApJ...964..192I}; \citealt{2024ApJ...964L..10W}; 
\citealt{2024ApJ...967L..23C}; \citealt{2024arXiv240100934J}; 
\citealt{2025arXiv250104788C}) 
and little red dots
(\citealt{2024ApJ...977L..13B}; \citealt{2024arXiv241114383R}). 
Furthermore, several studies have focused on robust samples of galaxies 
limited to spectroscopically confirmed ones 
(\citealt{2024A&A...690A.290C}; 
\citealt{2023arXiv230706336L}) 
and on more general samples selected using photometric redshifts 
(\citealt{2022ApJ...937L..33S}; 
\citealt{2023ApJ...954..113Y}; 
\citealt{2023ApJ...955L..12B}; 
\citealt{2024ApJ...960...53V}; 
\citealt{2024ApJ...960..104S}; 
\citealt{2024MNRAS.527.6110O}; 
\citealt{2024ApJ...962..176W}; 
\citealt{2024ApJ...966..113L}; 
\citealt{2024ApJ...972..134M}; 
\citealt{2024MNRAS.533.3724V}; 
\citealt{2024arXiv241016354A}; \citealt{2024arXiv241107458J}; \citealt{2024arXiv241206957M}; 
\citealt{2024arXiv241214970W}). 
Comprehensive morphological classifications have also been performed 
including those based on visual inspection 
(\citealt{2022ApJ...938L...2F}; \citealt{2023ApJ...946L..15K}; \citealt{2023ApJ...955...94F}; 
\citealt{2024MNRAS.531.4857C}; \citealt{2025MNRAS.539.1359S}; \citealt{2025ApJ...987...74G}) 
and approaches utilizing machine learning techniques 
(\citealt{2023ApJ...942L..42R}; \citealt{2024ApJ...961...51V}; \citealt{2024ApJ...962..164T}; 
\citealt{2024A&A...685A..48H}; \citealt{2025arXiv250203532H}).
In addition, theoretical studies based on cosmological simulations 
have explored the morphological properties of high-$z$ galaxies 
and compared their results with observations 
(e.g., \citealt{2023ApJ...946...71C}; \citealt{2023MNRAS.525.4832Y}; \citealt{2024MNRAS.527.2835S}; 
\citealt{2024MNRAS.534.1433S}; \citealt{2024ApJ...975..238N}; \citealt{2024arXiv240116608L}).

Through deep surveys conducted by the Hubble Space Telescope (HST), 
the morphological properties of high-$z$ galaxies have been investigated 
(e.g., \citealt{2004ApJ...600L.107F}; \citealt{2004ApJ...611L...1B}; \citealt{2006ApJ...653...53B}; 
\citealt{2008ApJ...673..686H}; \citealt{2010ApJ...709L..21O}; \citealt{2012A&A...547A..51G}; 
\citealt{2013ApJ...777..155O}; \citealt{2013ApJ...765...68H}; \citealt{2013ApJ...773..153J}; 
\citealt{2015ApJ...804..103K}; \citealt{2015ApJ...808....6H}; \citealt{2015ApJS..219...15S}; 
\citealt{2016MNRAS.457..440C}; \citealt{2017ApJ...834L..11A}; \citealt{2017MNRAS.466.3612B}; 
\citealt{2017ApJ...843...41B}; \citealt{2018ApJ...855....4K}; \citealt{2019ApJ...882...42B}; 
\citealt{2020AJ....160..154H}; \citealt{2022ApJ...927...81B}). 
Nevertheless, many open questions regarding the early stages of galaxy evolution remain unresolved, 
such as when the first galaxy disks formed in the early universe, 
which is related to the strength of feedback processes in the early stages of galaxy formation.  
Using HST WFC3 data, \cite{2014ApJ...788...28V} have demonstrated that 
the slope of the galaxy size--stellar mass relation and the scatter in the galaxy size distribution 
show no significant evolution for galaxies at $z\sim0$--$3$. 
\cite{2015ApJS..219...15S} have reported that, 
based on HST extragalactic legacy survey data, 
the median S{\'e}rsic index for star-forming galaxies at $z\sim0$--$8$ is $1.5$, 
and that their size distributions follow a log-normal form with standard deviations 
that do not evolve significantly 
(see also, \citealt{2015ApJ...804..103K}; \citealt{2018ApJ...855....4K}). 
These findings are consistent with the picture that high-$z$ star-forming galaxies possess disk-like stellar components 
(e.g., \citealt{1998MNRAS.295..319M}; \citealt{2003MNRAS.343..978S}).

Observational studies using the Atacama Large Millimeter/submillimeter Array (ALMA) 
have explored the formation of disks at high redshift 
from the perspective of kinematic properties, utilizing far-infrared emission lines 
such as [{\sc Cii}]$158\mu$m and [{\sc Oiii}]$88\mu$m.  
\cite{2020Natur.584..201R} have analyzed the kinematics of 
the high-resolution [{\sc Cii}] emission of the gravitationally lensed $z=4.22$ galaxy 
SPT0418-47 (\citealt{2013ApJ...767...88W}), 
reporting that its rotation curve resembles those of typical nearby spiral galaxies 
and that the ratio of rotational velocity to velocity dispersion is as high as those obtained for nearby spiral galaxies.
\cite{2024MNRAS.535.2068R} have reported that, 
from high-resolution ALMA observations of the [{\sc Cii}] emission from the $z=7.31$ galaxy REBELS-25, 
the [{\sc Cii}] flux distribution closely resembles an exponential disk 
with a S{\'e}rsic index of $n=1.3\pm0.2$, 
and the ratio of rotational velocity to velocity dispersion is consistent with a dynamically cold disk.  
These findings suggest that galaxies with disk structures already exist at $z\sim4$--$9$  
(see also, \citealt{2018Natur.553..178S}; 
\citealt{2020Natur.581..269N}; 
\citealt{2021Sci...371..713L}; 
\citealt{2021A&A...647A.194F}; 
\citealt{2021ApJ...911..141N}; 
\citealt{2021Sci...372.1201T}; 
\citealt{2021ApJ...917...99Y}; 
\citealt{2022ApJ...933L..19T}; 
\citealt{2022A&A...665L...8H}; 
\citealt{2022A&A...667A...5R}; 
\citealt{2023A&A...669A..46P}; 
\citealt{2023A&A...673A.153P}; 
\citealt{2023MNRAS.521.1045R}; 
\citealt{2023ApJ...958..132N}; 
\citealt{2024ApJ...968....9W}).

Early results from JWST have also reported findings on the morphological and kinematic properties of high-$z$ galaxies.  
\cite{2023ApJ...942L..42R} have derived the S{\'e}rsic index for $z=2$--$5$ galaxies 
classified as disks using convolutional neural networks applied to JWST images, 
showing that the average S{\'e}rsic index is approximately $n=1$.  
\cite{2024ApJ...962..176W} have examined the size--stellar mass relation for galaxies at $z=0.5$--$5.5$ 
using both HST and JWST images 
and reported that the galaxy size scatter exhibits little to no evolution  
(see also, \citealt{2024MNRAS.533.3724V}; \citealt{2024ApJ...963....9M}; 
\citealt{2024arXiv241016354A}; \citealt{2024arXiv241206957M}).  
From a kinematic perspective, 
\cite{2024A&A...684A..87D} have carried out JWST NIRSpec spectroscopy 
for six galaxies at $z=5.5$--$7.4$ 
and found that half of them show rotation-dominated velocity structures 
based on the H$\alpha$ and [{\sc Oiii}] emission lines.  
In addition, using NIRSpec integral field spectroscopy, 
\cite{2024ApJ...976..142X} have demonstrated that 
the {\sc Ciii}]$\lambda\lambda1907,1909$ emission in the $z=10.6$ galaxy GN-z11 
exhibits a velocity gradient consistent with disk rotation 
(see also, \citealt{2024ApJ...976L..27N}; \citealt{2023arXiv231009327L}; \citealt{2024arXiv240218543F}).  
However, most of these results focus on galaxies up to $z\sim10$, 
and a statistical investigation into the presence of disks in galaxies 
at even higher redshifts is worth pursuing.

In this paper, 
we investigate the morphological properties of $z\sim10$--$16$ galaxies 
using deep JWST NIRCam images obtained from 
the Cosmic Evolution Early Release Science 
(CEERS; \citealt{2025arXiv250104085F}; 
see also, \citealt{2022ApJ...940L..55F}; \citealt{2023ApJ...946L..13F}; \citealt{2023ApJ...946L..12B}) survey 
and the JWST Advanced Deep Extragalactic Survey 
(JADES; \citealt{2023arXiv230602465E}; \citealt{2024A&A...690A.288B}; \citealt{2023ApJS..269...16R}; 
\citealt{2023arXiv231012340E}; \citealt{2024arXiv240406531D}; \citealt{2024ApJ...964...71H}). 
Our analysis primarily relies on the high-$z$ galaxy catalogs constructed 
for the CEERS fields by \cite{2024ApJ...969L...2F} 
and for the JADES fields by \cite{2024ApJ...964...71H}.
This paper is structured as follows. 
Section \ref{sec:data} describes the JADES and CEERS NIRCam data 
and provides an overview of $z\sim10$--$16$ galaxies analyzed in this study. 
Section \ref{sec:SBprofile_fitting} details the two-dimensional surface brightness profile fitting method that we employ, 
along with an investigation of its systematic and statistical uncertainties based on Monte Carlo simulations. 
In Section \ref{sec:results}, we present the results of 
profile fittings such as sizes and total magnitudes for 
$z\sim10$--$16$ galaxies in the rest-frame UV continuum 
and derive their size and axis ratio 
distributions. 
In addition, we explore the relation between galaxy UV size and luminosity, 
as well as the redshift evolution of galaxy sizes. 
In Section \ref{sec:discussion}, 
we estimate the stellar-to-halo size ratios of $z\sim10$--$16$ galaxies 
and discuss whether the morphological properties of $z\sim10$--$16$ galaxies 
are consistent with the disk formation scenario. 
Furthermore, we compare our observational results with cosmological simulation results 
to discuss the origin of the diversity in galaxy sizes ranging from compact to extended ones.
Section \ref{sec:summary} provides a summary of this study.

Throughout this paper, 
we use magnitudes in the AB system (\citealt{1983ApJ...266..713O}) 
and assume a flat universe (e.g., \citealt{2020A&A...641A...6P}) with 
$\Omega_{\rm m}=0.3$, $\Omega_\Lambda=0.7$, and $H_0=70$ km s$^{-1}$ Mpc$^{-1}$. 
In this cosmological model, an angular dimension of $1\farcs0$ 
corresponds to a physical dimension of 
$4.159$ kpc at $z=10$, $3.659$ kpc at $z=12$, and $3.268$ kpc at $z=14$ 
(Equation 18 of \citealt{1999astro.ph..5116H}).\footnote{The cosmic ages 
at these redshifts are $464$ Myr, $361$ Myr, and $291$ Myr, respectively.} 
These values are provided as illustrative examples; 
in our analysis, we calculate the physical dimensions 
using the exact redshifts of individual galaxies, 
adopting spectroscopic redshifts when available and photometric redshifts otherwise.
Following previous studies, we describe the UV luminosities of galaxies 
in units of the characteristic luminosity of $z\sim3$ galaxies, 
$L_{z=3}^\ast$, which corresponds to $M_{\rm UV} = -21.0$ mag
(\citealt{1999ApJ...519....1S}).
In this case, $0.048L_{z=3}^\ast$, $0.12L_{z=3}^\ast$, and $0.3L_{z=3}^\ast$ 
correspond to $M_{\rm UV} = -17.7$, $-18.7$, and $-19.7$ mag, respectively.

\begin{deluxetable*}{lcccc} 
\tablecolumns{5} 
\tablewidth{0pt} 
\tabletypesize{\small}
\tablecaption{Numbers of Objects Analyzed in This Study 
\label{tab:sample_numbers}}
\tablehead{
    \colhead{ } 
    &  \colhead{CEERS}
    &  \colhead{JADES Deep}
    &  \colhead{JADES Medium}
    &  \colhead{JADES GOODS-N}
}
\startdata 
Number of galaxies at $z=9.5$--$13$ 		& 26 (5) 	& 56 (5) 	& 34 (0) 	& 41 (1) \\
\hspace{1em}Number of galaxies at $z=9.5$--$13$ 
				with S/N$>10$ in F200W	& \hspace{2em}7 (4) 	& \hspace{2em}10 (4) 	& \hspace{2em}8 (0) 	& \hspace{2em}7 (1) \\
Number of galaxies at $z=13$--$16$			& 3 (0) 	& 9 (3) 	& 0 (0) 	& 0 (0) \\ 
\hspace{1em}Number of galaxies at $z=13$--$16$ 
				with S/N$>10$ in F277W	& \hspace{2em}0 (0) 	& \hspace{2em}3 (3) 	& \hspace{2em}0 (0) 	& \hspace{2em}0 (0) \\
\hline
Total									& 29 (5)	& 65 (8)	& 34 (0)	& 41 (1) \\
\enddata 
\tablecomments{
The numbers in parentheses indicate the number of objects with spectroscopic redshifts.
The total number of $z=9.5$--$16$ galaxies analyzed in this study is $29+65+34+41=169$, 
and the number of $z=9.5$--$16$ galaxies with S/N $>10$ is $7 + 10 + 8 + 7 + 3 = 35$.  
}
\end{deluxetable*} 

\begin{figure}[h]
\begin{center}
   \includegraphics[width=0.5\textwidth]{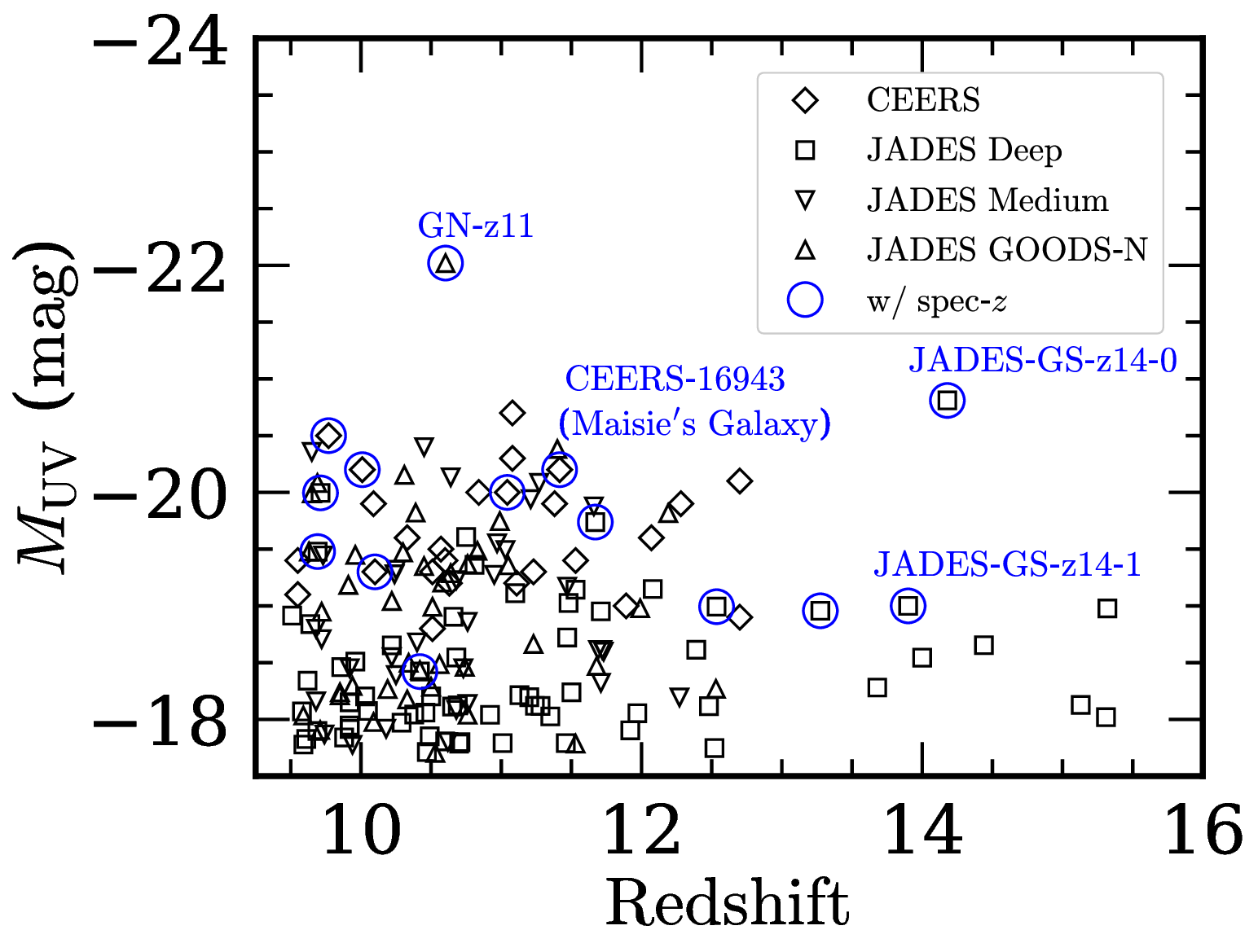}
\caption{
$M_{\rm UV}$ vs. redshift for the $z\sim10$--$16$ galaxies analyzed in this study. 
The black diamonds represent CEERS galaxies (\citealt{2024ApJ...969L...2F}), 
the black squares correspond to JADES Deep, 
the black downward triangles indicate JADES Medium, 
and the black upward triangles denote JADES GOODS-N galaxies 
(\citealt{2024ApJ...964...71H}; \citealt{2024Natur.633..318C}; 
\citealt{2024arXiv240920549S}; \citealt{2024arXiv240920533C}). 
Objects outlined with large blue circles are spectroscopically confirmed.
The names of well-studied galaxies are labeled in blue near their respective data points.
}
\label{fig:Muv_redshift}
\end{center}
\end{figure}

\begin{figure}[ht]
\begin{center}
   \includegraphics[width=0.5\textwidth]{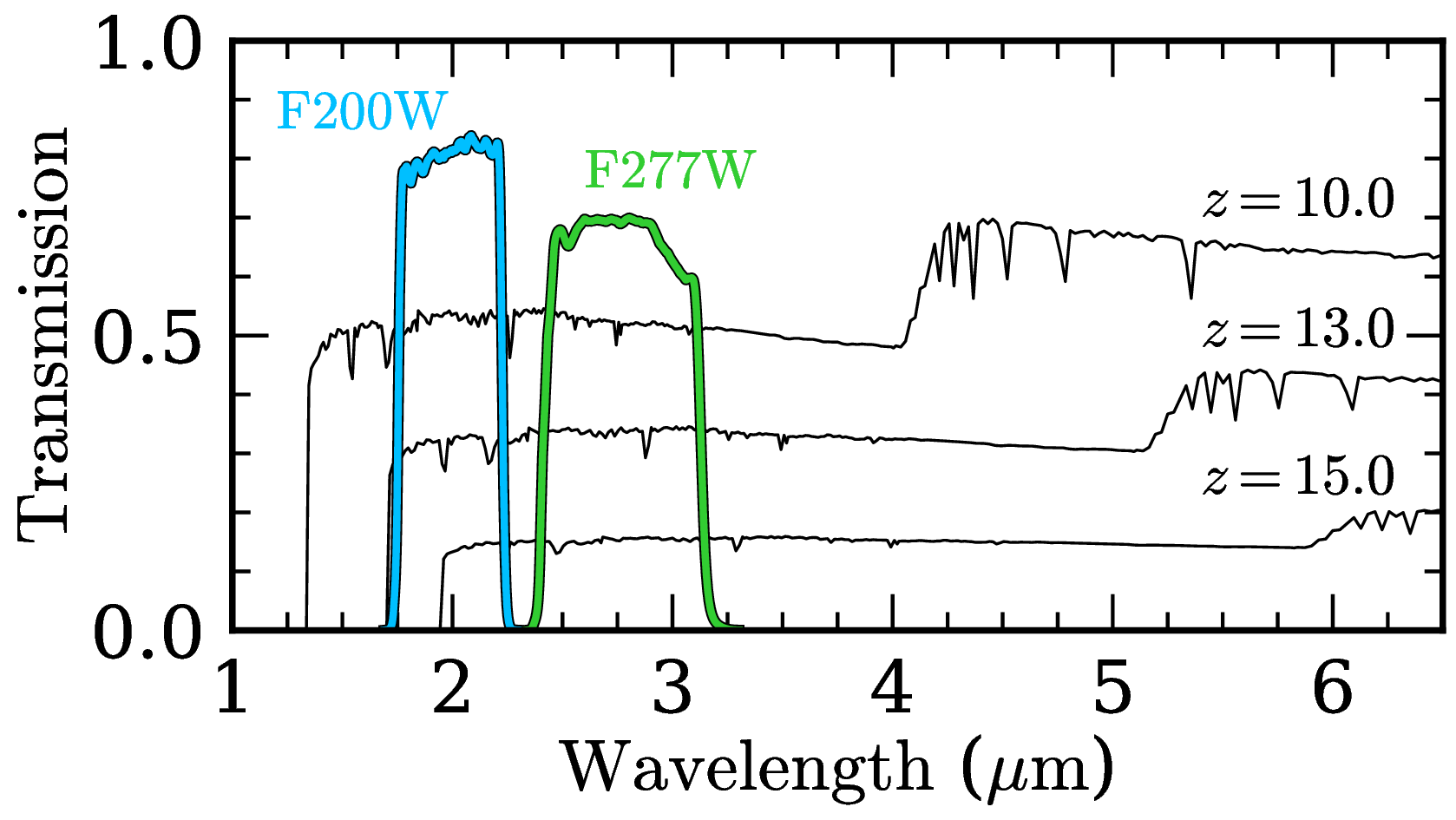}
\caption{
Transmission curves of the NIRCam broadband filters F200W (cyan) and F277W (green), 
along with the spectra of a star-forming galaxy at $z=10$, $z=13$, and $z=15$  
based on the \cite{2003MNRAS.344.1000B} library (black).
}
\label{fig:SED}
\end{center}
\end{figure}

\begin{deluxetable}{lcc} 
\tablecolumns{3} 
\tablewidth{0pt} 
\tabletypesize{\small}
\tablecaption{Numbers of Stacked Objects in Each Field and Luminosity Bin
\label{tab:sample_stacked_numbers}}
\tablehead{
    \colhead{ID} 
    &  \colhead{luminosity range}
    &  \colhead{$N$} \\
    \colhead{ } 
    &  \colhead{$L / L_{z=3}^\ast$}
    &  \colhead{ } 
}
\startdata 
\multicolumn{3}{c}{$z=9.5$--$13$} \\ 
z11-stack-CEERS		& $0.3$--$1$ 		& $5$ 	\\ 
z11-stack-CEERS-f		& $0.12$--$0.3$ 	& $14$ 	\\ 
z11-stack-JADES-f		& $0.12$--$0.3$ 	& $5$ 	\\ 
z11-stack-JADES-f2		& $0.048$--$0.12$ 	& $41$ 	\\ 
z11-stack-JADESM-f		& $0.12$--$0.3$ 	& $8$ 	\\ 
z11-stack-JADESM-f2	& $0.048$--$0.12$ 	& $17$ 	\\ 
z11-stack-JADESGN		& $0.3$--$1$ 		& $3$ 	\\ 
z11-stack-JADESGN-f	& $0.12$--$0.3$ 	& $12$ 	\\ 
z11-stack-JADESGN-f2	& $0.048$--$0.12$ 	& $19$ 	\\ 
\multicolumn{3}{c}{$z=13$--$16$} \\ 
z15-stack-JADES-f		& $0.048$--$0.3$ 	& $6$ 	\\ 
\enddata 
\tablecomments{
The third part of the ID from the left, separated by hyphens, 
indicates the field where the stacked objects are located. 
{\lqq}CEERS{\rqq} corresponds to the CEERS field, 
{\lqq}JADES{\rqq} to the JADES Deep field, 
{\lqq}JADESM{\rqq} to the JADES Medium field, 
and {\lqq}JADESGN{\rqq} to the JADES GOODS-N field.
}
\end{deluxetable} 

\section{Data and Samples} \label{sec:data}

We measure the sizes of $z\sim10$--$16$ galaxies found in the CEERS and JADES programs. 
For CEERS, we use the high-$z$ galaxy sample from \cite{2024ApJ...969L...2F}, 
and for JADES, we use the sample from \cite{2024ApJ...964...71H}. 
From these high-$z$ galaxy samples, which are primarily selected based on photometric redshifts, 
we include those with $z>9.5$. 
The GOODS-S field in JADES is divided into deeper regions (JADES Deep and 1210 Parallel) 
and moderately deep regions (JADES Medium). 
For simplicity, we refer to the former as JADES Deep and the latter as JADES Medium. 
For GOODS-N in JADES, we basically focus on the SE region, 
which has a depth almost comparable to JADES Medium.
GN-z11 is an exception among the objects in the NW region of JADES GOODS-N, 
because it is a spectroscopically confirmed luminous high-$z$ source that has been studied in detail 
(e.g., \citealt{2016ApJ...819..129O}; \citealt{2023ApJ...952...74T}; \citealt{2023A&A...677A..88B}; 
\citealt{2024ApJ...966...92S}; \citealt{2024Natur.627...59M}; \citealt{2024ApJ...976..142X}; 
\citealt{2024ApJ...976..122N}; \citealt{2024arXiv241212826A}). 
We thus include it in our sample as a special case.
Additionally, for JADES, we incorporate two spectroscopically confirmed galaxies 
at $z=14.18$ and $z=13.90$, namely JADES-GS-z14-0 and JADES-GS-z14-1 
(\citealt{2024Natur.633..318C}; \citealt{2024arXiv240920549S}; \citealt{2024arXiv240920533C}). 
On the other hand, JADES-GS-53.12692-27.79102, 
included in the \cite{2024ApJ...964...71H} catalog,
has been identified as a lower-$z$ transient interloper 
by the JADES Transient Survey (\citealt{2024arXiv240605060D}),
and is therefore excluded from this study.
The total number of $z\sim10$--$16$ galaxies analyzed in this study is 
$169$, of which $14$ are spectroscopically confirmed. 
The number of $z\sim10$--$16$ galaxies in each region is summarized in Table \ref{tab:sample_numbers}. 
Figure \ref{fig:Muv_redshift} illustrates the distribution of their redshifts and rest-frame UV magnitudes, $M_{\rm UV}$. 
More detailed lists are provided in Appendix \ref{sec:galaxy_catalogs}.
Among these, two objects in CEERS and one object in JADES are included in the catalog of little red dots 
compiled by \cite{2024arXiv240403576K}; we mark them with a dagger symbol 
in Table \ref{tab:sample_CEERS} and Table \ref{tab:sample_JADES}.

\begin{deluxetable}{cccc} 
\tablecolumns{4} 
\tablewidth{0pt} 
\tabletypesize{\small}
\tablecaption{Limiting Magnitudes and PSF FWHMs of the JWST NIRCam Images for Size Analysis 
\label{tab:limitmag}}
\tablehead{
\multicolumn{3}{c}{$10 \sigma$ Depth / PSF FWHM} \\ 
    \colhead{Field} 
    &  \colhead{F200W}
    &  \colhead{F277W}
}
\startdata 
CEERS1 			& $28.5$ / $0\farcs0716$ & --- / --- \\
CEERS2 			& $28.9$ / $0\farcs0792$ & --- / --- \\
CEERS3 			& $28.6$ / $0\farcs0711$ & --- / --- \\
CEERS4 			& $28.5$ / $0\farcs0711$ & --- / --- \\
CEERS5 			& $28.5$ / $0\farcs0725$ & --- / --- \\
CEERS6 			& $28.5$ / $0\farcs0744$ & --- / --- \\
CEERS7 			& $28.5$ / $0\farcs0729$ & --- / --- \\
CEERS8 			& $28.5$ / $0\farcs0720$ & --- / --- \\
CEERS9 			& $28.5$ / $0\farcs0762$ & --- / --- \\
CEERS10			& $28.5$ / $0\farcs0764$ & --- / --- \\
JADES Deep 		& $29.3$ / $0\farcs0771$ & $29.8$ / $0\farcs1215$ \\
JADES Medium 	& $28.8$ / $0\farcs0771$ & $29.1$ / $0\farcs1215$ \\
JADES GOODS-N 	& $29.0$ / $0\farcs0789$ & $29.2$ / $0\farcs1161$ \\
\enddata 
\tablecomments{
Limiting magnitudes are measured with randomly distributed $0\farcs2$ diameter circular apertures 
(\citealt{2023ApJS..265....5H}; \textcolor{blue}{Y. Harikane et al. in preparation}; \citealt{2024ApJ...964...71H}; 
see also, \citealt{2024ApJ...969L...2F}).  
}
\end{deluxetable} 

To investigate the sizes of $z\sim10$--$16$ galaxies in their rest-frame UV continuum, 
we use JWST NIRCam F200W images for $z=9.5$--$13$ galaxies 
and F277W images for $z=13$--$16$ galaxies. 
Figure \ref{fig:SED} shows the transmission curves of these filters 
plotted together with example SEDs of high-$z$ star-forming galaxies.
For CEERS, we use NIRCam images reduced by 
\cite{2023ApJS..265....5H} and \textcolor{blue}{Y. Harikane et al. in preparation}, 
with a pixel scale of $0\farcs015$ pix$^{-1}$. 
For JADES, we utilize the official JADES images available on MAST,\footnote{\url{https://archive.stsci.edu/hlsp/jades}} 
which have a pixel scale of $0\farcs03$ pix$^{-1}$.\footnote{Initially, we considered 
using the images from the DAWN JWST Archive, which have a finer pixel scale of $0\farcs02$ pix$^{-1}$. 
However, the latest versions available at that time, i.e., the v7.0 to v7.2 images of JADES GOODS-S, were 
unexpectedly cropped in the southwest direction; we thus decided not to use them in this paper.}
The $10\sigma$ limiting magnitudes of these images are summarized in Table \ref{tab:limitmag}.
Following the previous studies, we select sources with a signal-to-noise ratio (S/N) 
greater than 10 for individual analyses, 
based on apparent magnitudes measured within $0\farcs2$ diameter circular apertures. 
The numbers of sources meeting this S/N threshold is listed in Table \ref{tab:sample_numbers}.

\begin{figure}
\begin{center}
   \includegraphics[width=0.4\textwidth]{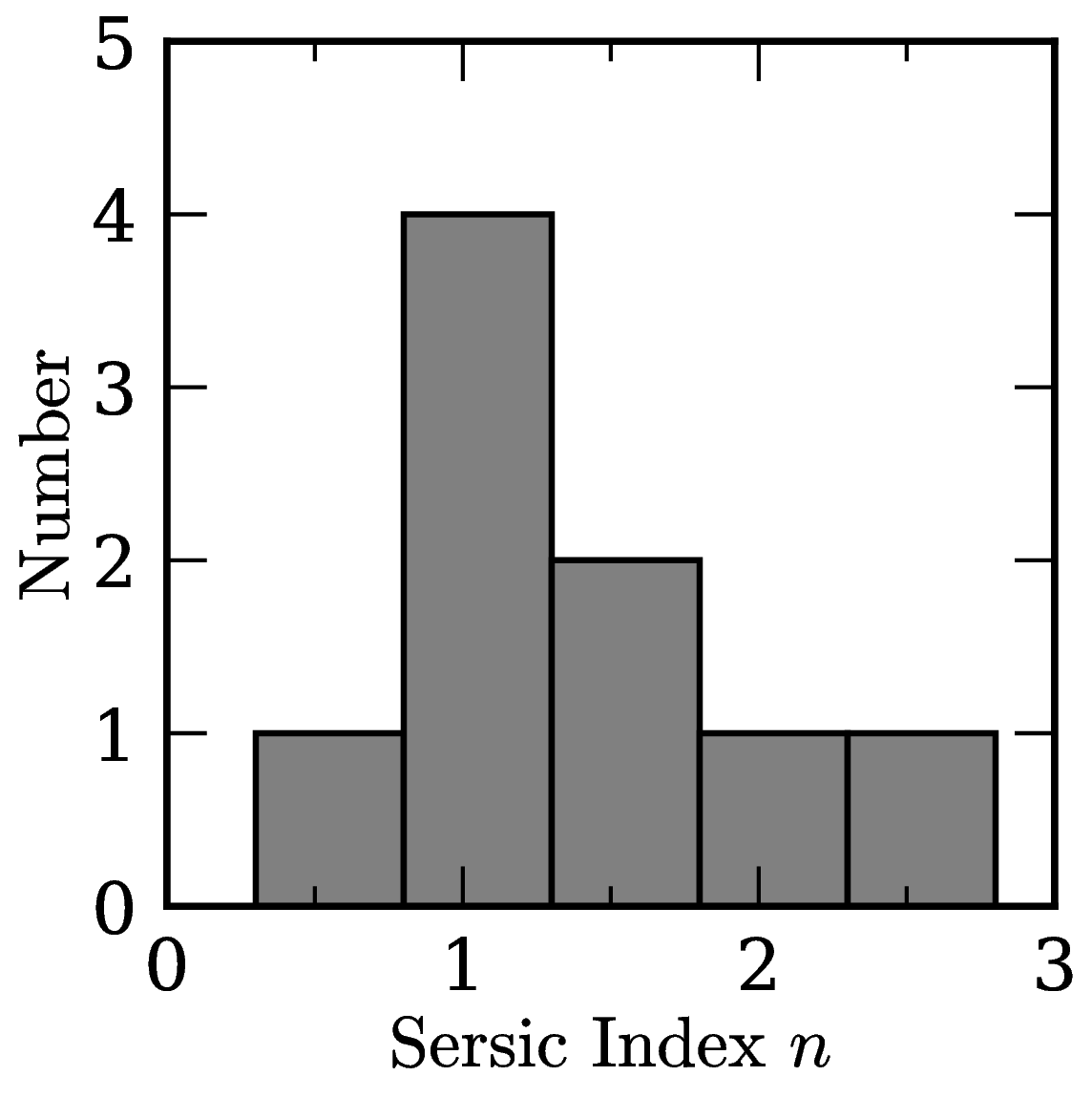}
\caption{Histogram of the best-fit S{\'e}rsic index values 
for spectroscopically confirmed galaxies in our sample with $\mathrm{S/N} > 10$. 
The mean value is $n = 1.3 \pm 0.6$.
}
\label{fig:histo_Sersic}
\end{center}
\end{figure}

To extend our size measurements to fainter objects, 
we divide the sample in each field basically into three luminosity bins, 
$L/L_{z=3}^\ast = 0.3$--$1$, $0.12$--$0.3$, and $0.048$--$0.12$, 
based on their $M_{\rm UV}$ magnitudes reported in the original catalogs 
(\citealt{2024ApJ...969L...2F}; \citealt{2024ApJ...964...71H}), 
and make median stacked images separately for these luminosity bins. 
The number of objects stacked for each luminosity bin in each field 
is summarized in Table \ref{tab:sample_stacked_numbers}. 
For $z=9.5$--$13$, 
since there are no objects in the $L/L_{z=3}^\ast = 0.3$--$1$ luminosity bin 
in JADES Deep, and only one object in this bin in JADES Medium, 
stacked images are not created for these cases. 
Additionally, no objects in CEERS fall into the $L/L_{z=3}^\ast < 0.12$ luminosity bins. 
For $z=13$--$16$, no stacked images are created for CEERS due to the small number of objects. 
In JADES Deep, due to the limited number of objects, 
we combine the objects 
in the $L/L_{z=3}^\ast = 0.048$--$0.12$ and $L/L_{z=3}^\ast = 0.12$--$0.3$ luminosity bins 
into a single stacked image. 
We confirm that the S/Ns of the aperture magnitudes of these stacked objects are $\gtrsim 10$.

Measurements of galaxy sizes require accounting for image smearing 
characterized with the point spread function (PSF). 
To correct for this effect, we use empirical PSFs constructed by 
stacking bright point sources from the NIRCam images. 
In each field, we select four to eleven unsaturated bright point sources 
with magnitudes of $\sim 22$--$24$ mag. 
More specifically, we run SExtractor (\citealt{1996A&AS..117..393B}) 
on each image to detect sources and select bright unresolved objects 
that lie in the stellar locus in the half-light radius vs. magnitude plane 
as point sources (e.g., Figure 2 of \citealt{2024MNRAS.529.1067H}).
Among these selected point sources, we further choose 
those that are isolated from nearby objects and not located around the edges of the image.
We then stack these sources to construct the empirical PSFs.
In the CEERS field, 
\cite{2024MNRAS.529.1067H} have similarly identified unresolved sources 
in the stellar locus and classified brown dwarf types using the $k$-means nearest neighbor algorithm.
We confirm that the vast majority of our selected point sources in the CEERS field 
($55$ out of $57$) match those listed in the catalog of \cite{2024MNRAS.529.1067H}.
The FWHM values of the resulting PSFs are summarized in Table \ref{tab:limitmag}.

\begin{figure*}[ht]
\begin{center}
   \includegraphics[width=0.24\textwidth]{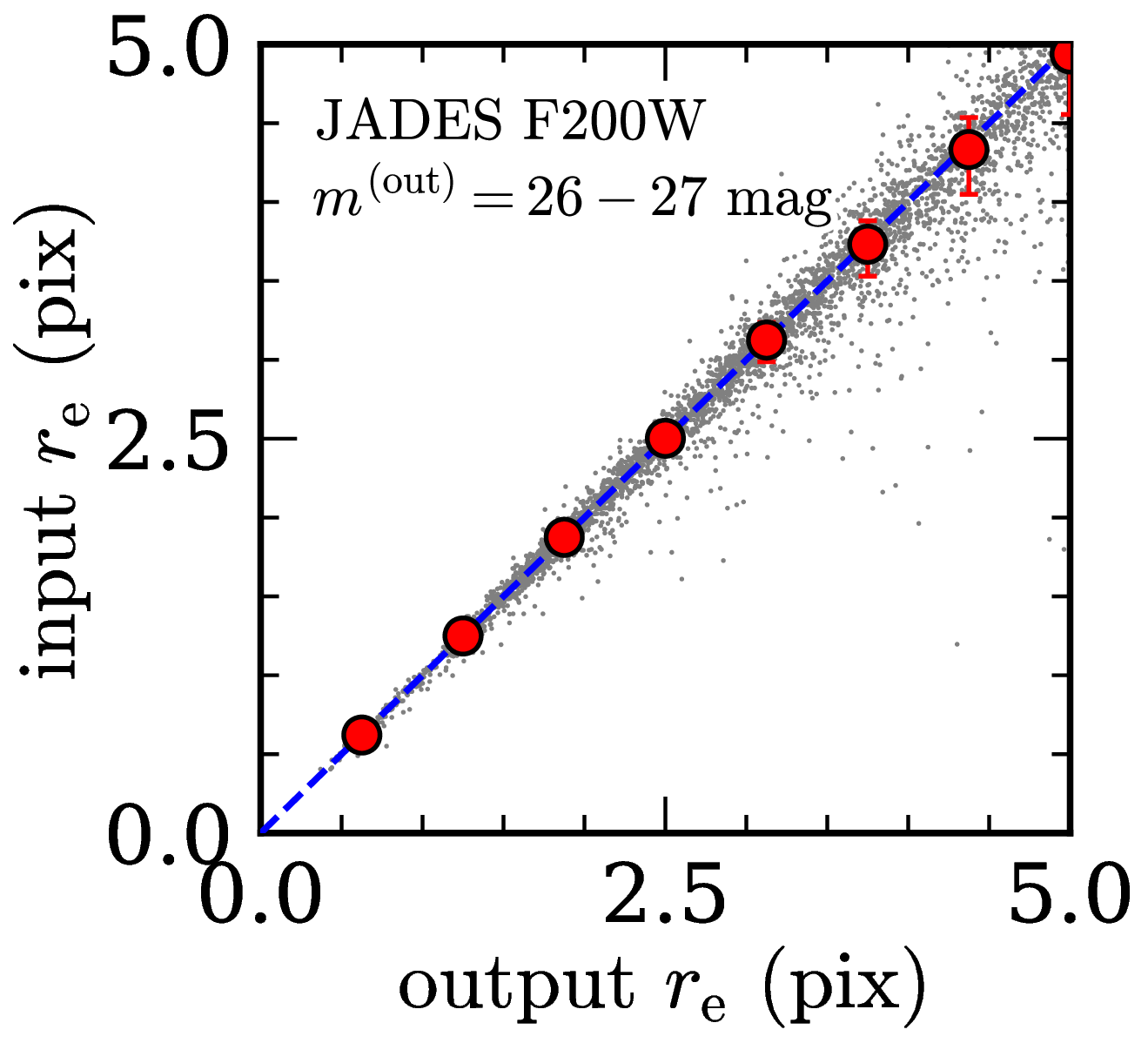}
   \includegraphics[width=0.24\textwidth]{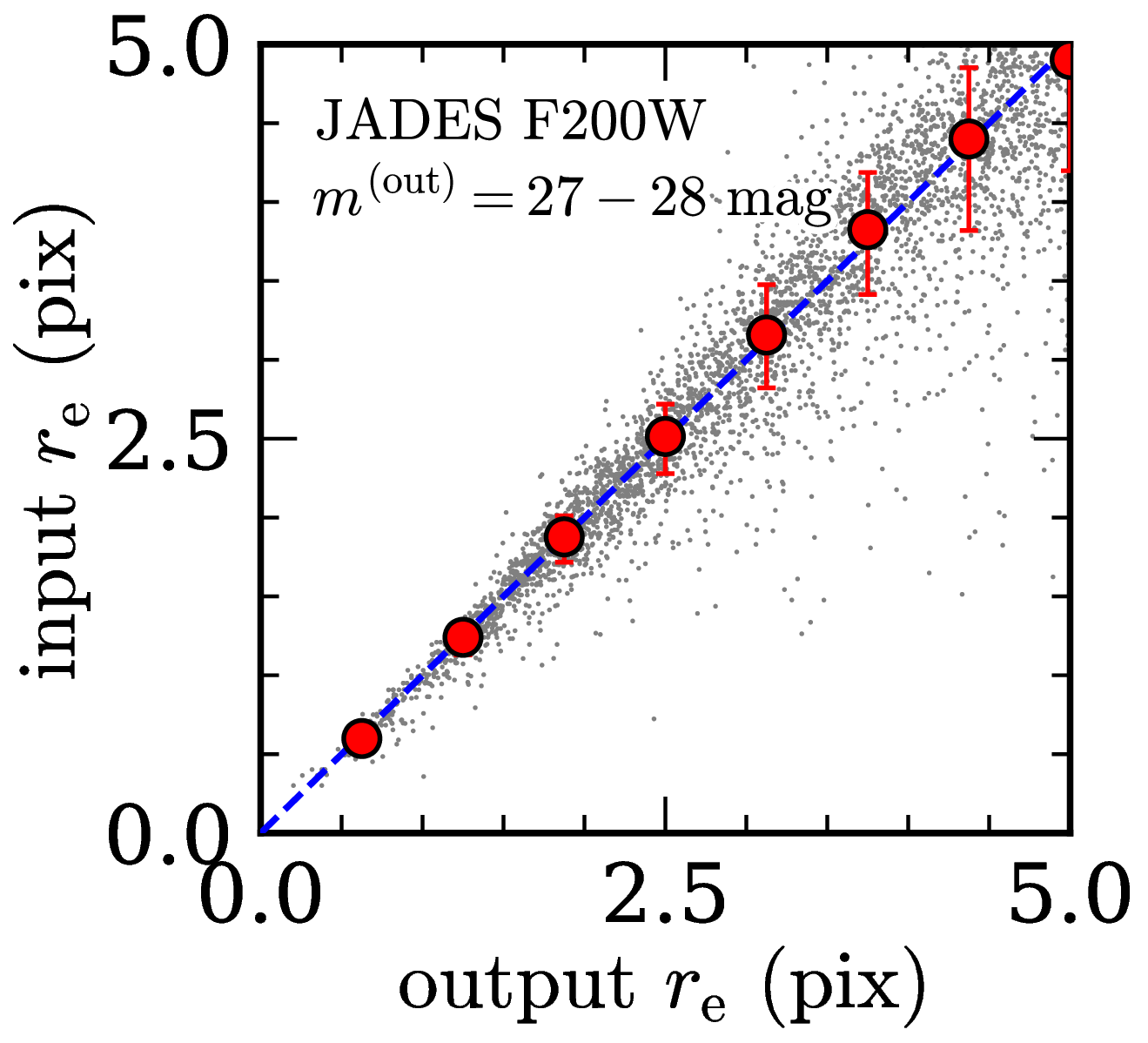}
   \includegraphics[width=0.24\textwidth]{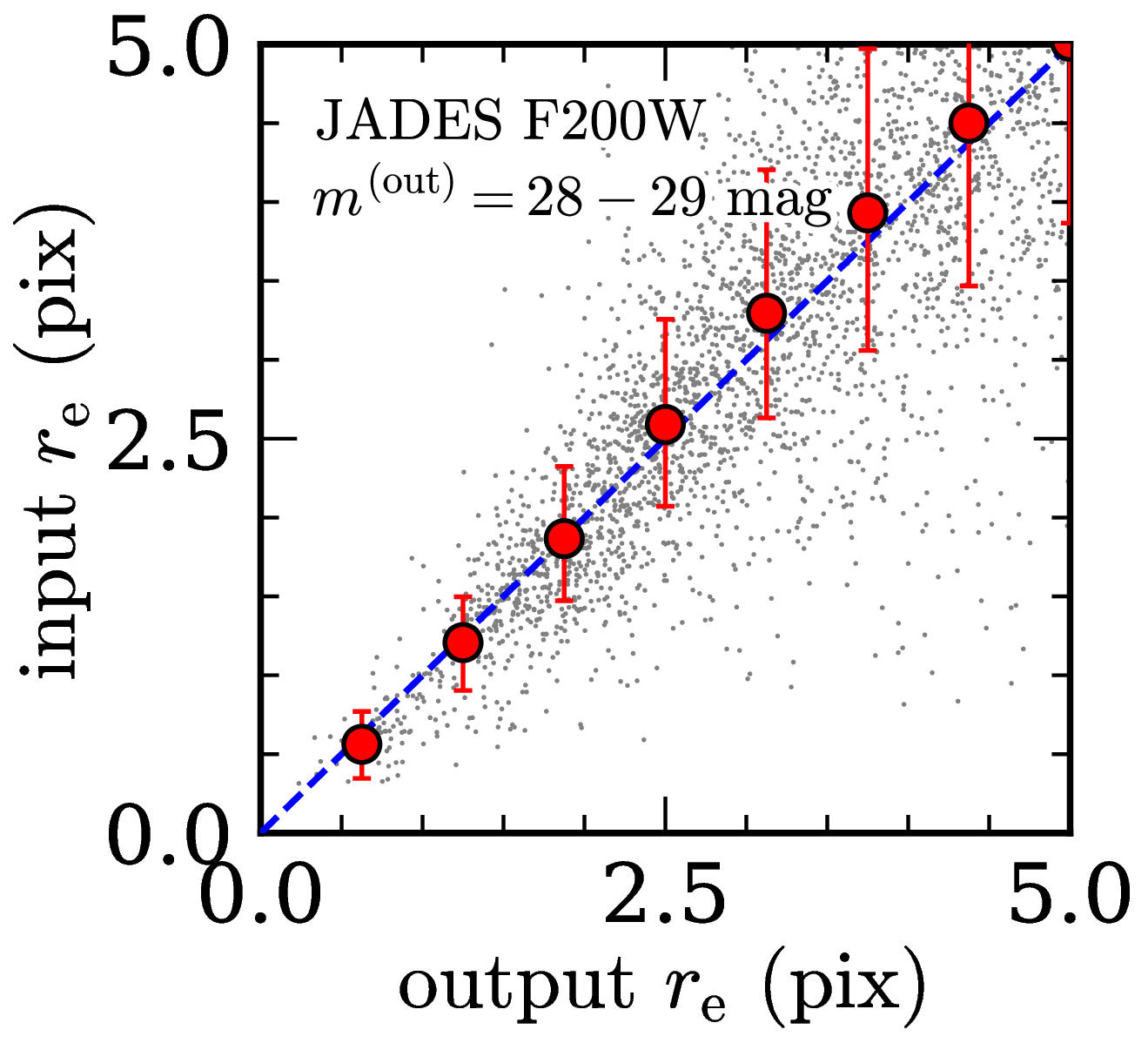}
   \includegraphics[width=0.24\textwidth]{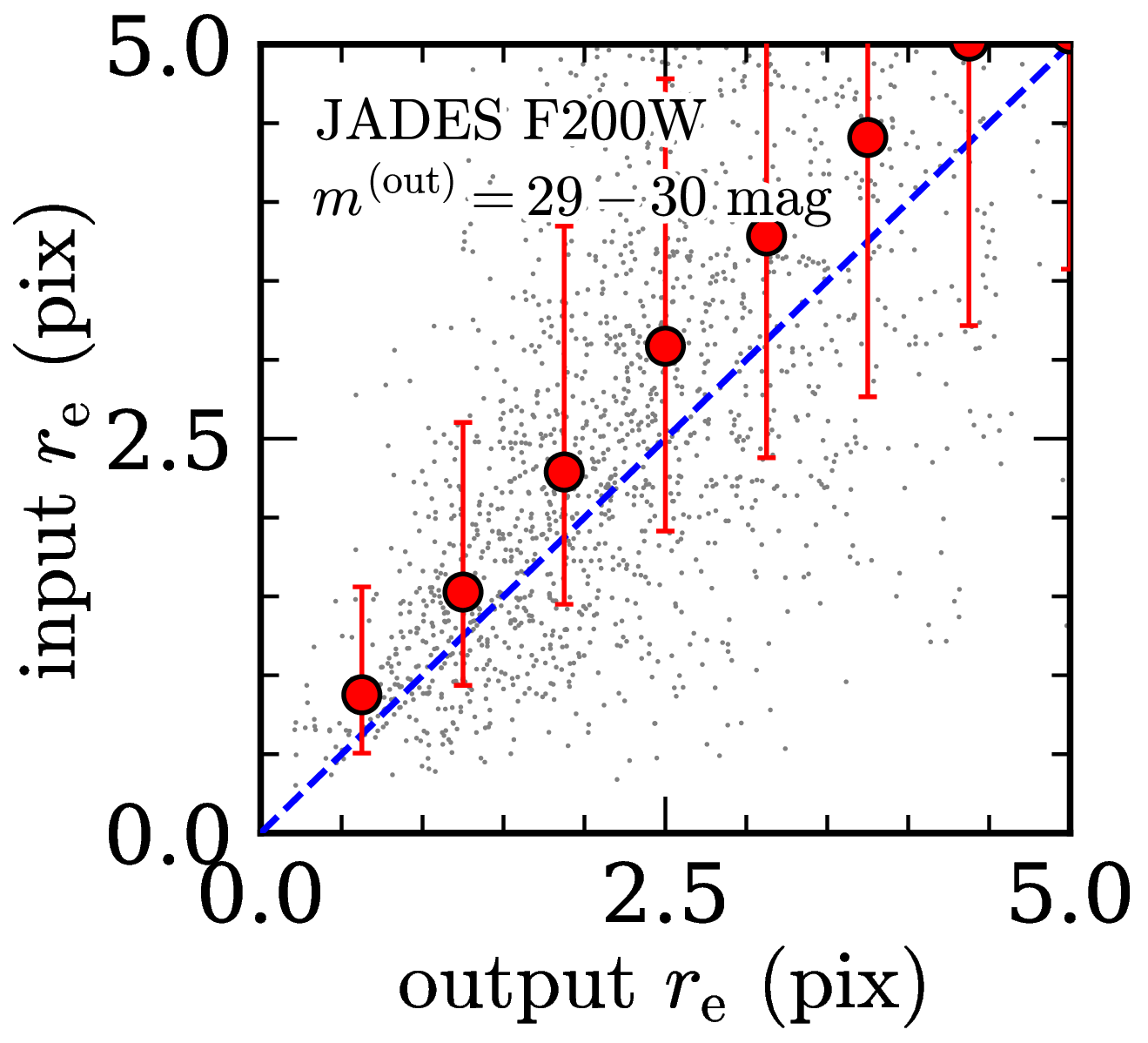}
   \includegraphics[width=0.24\textwidth]{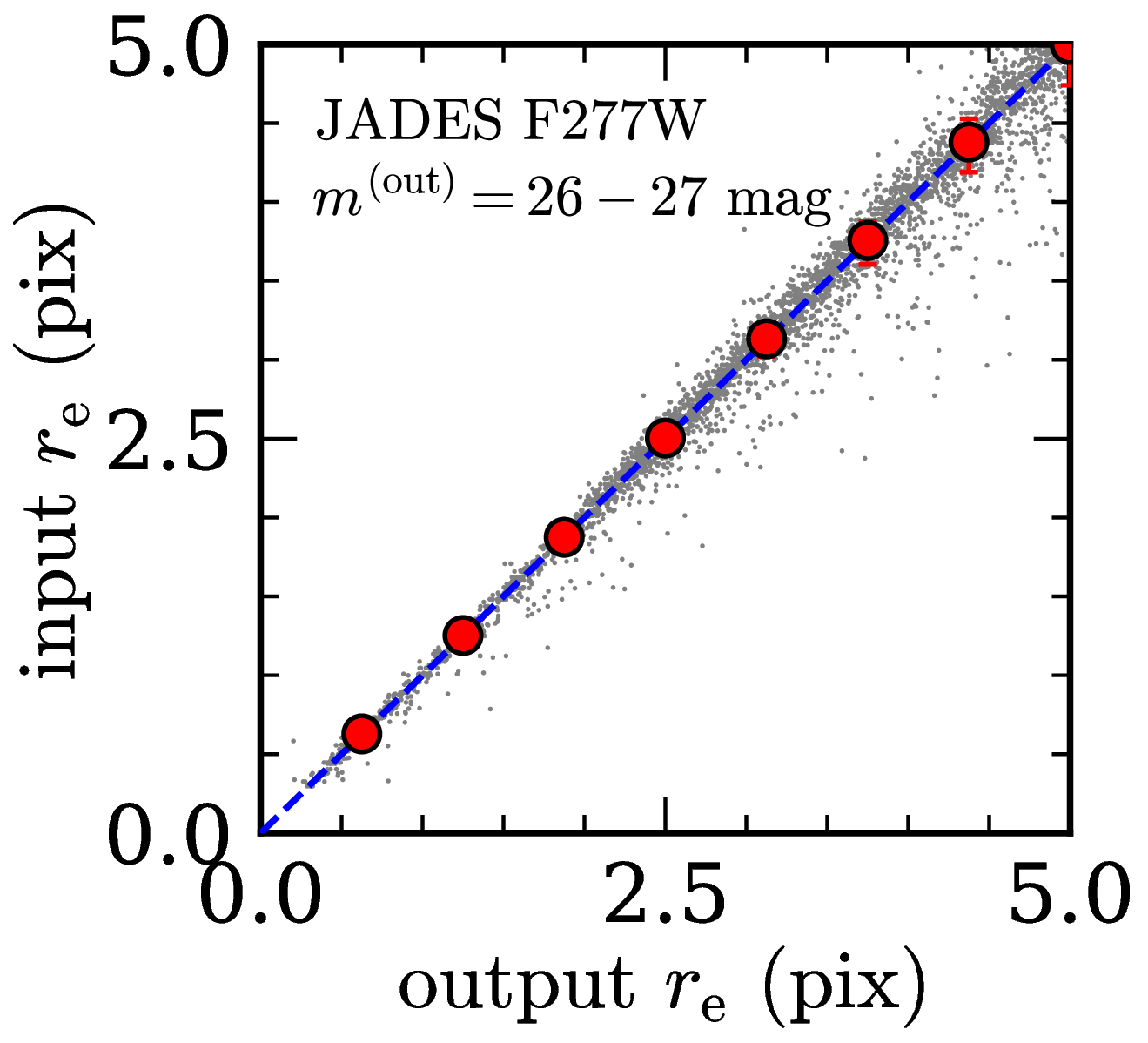}
   \includegraphics[width=0.24\textwidth]{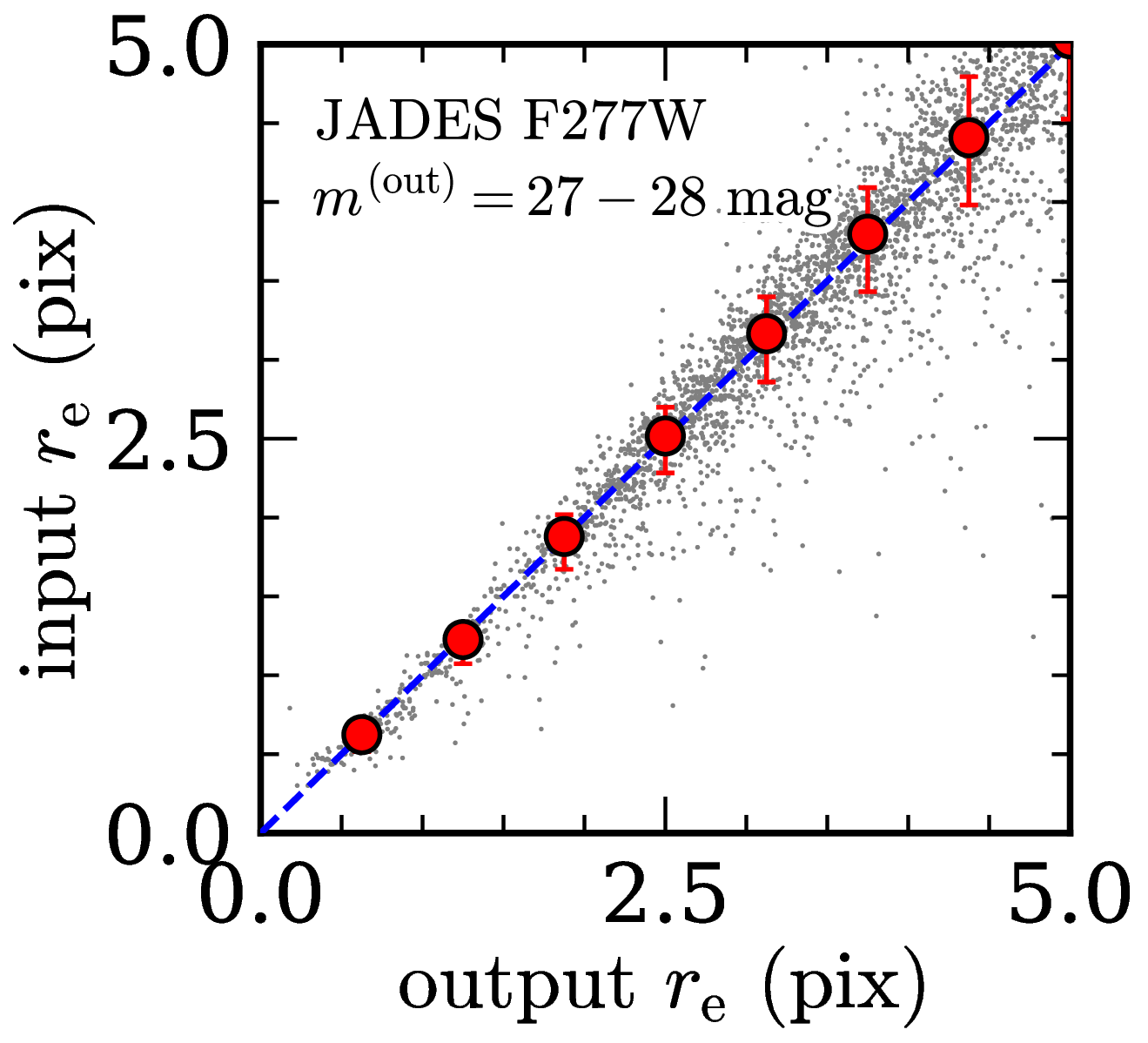}
   \includegraphics[width=0.24\textwidth]{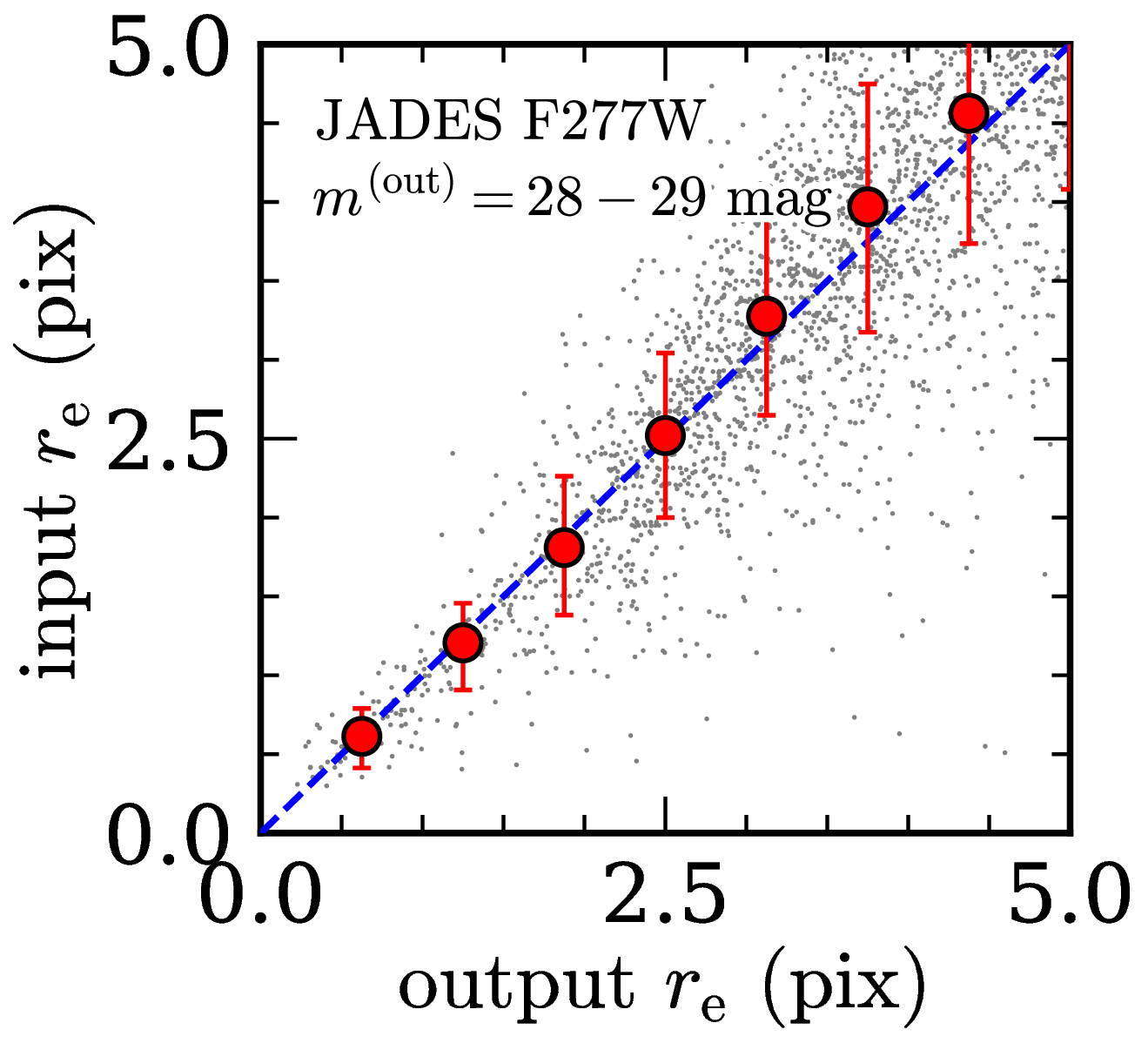}
   \includegraphics[width=0.24\textwidth]{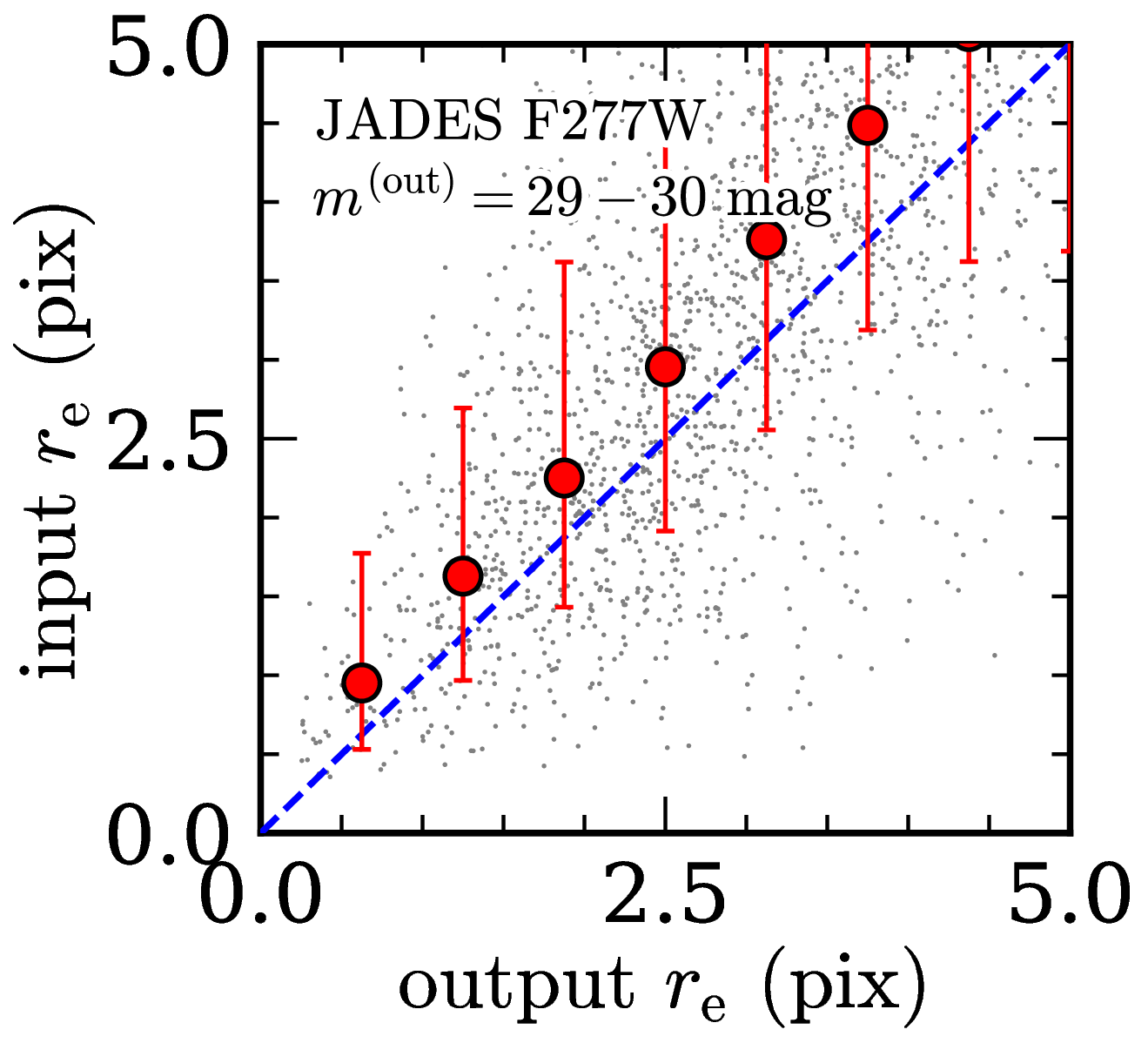}
   \includegraphics[width=0.24\textwidth]{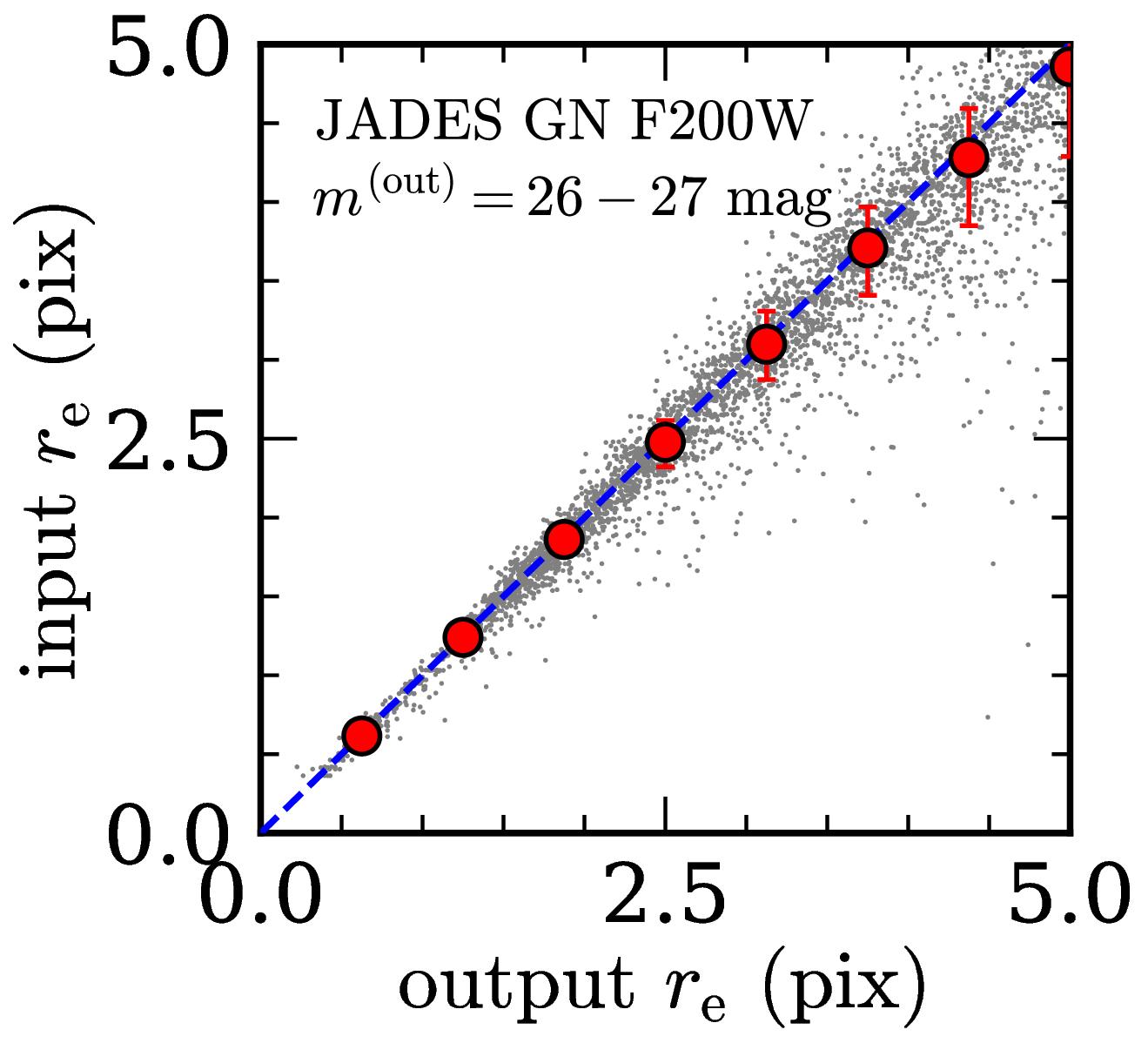}
   \includegraphics[width=0.24\textwidth]{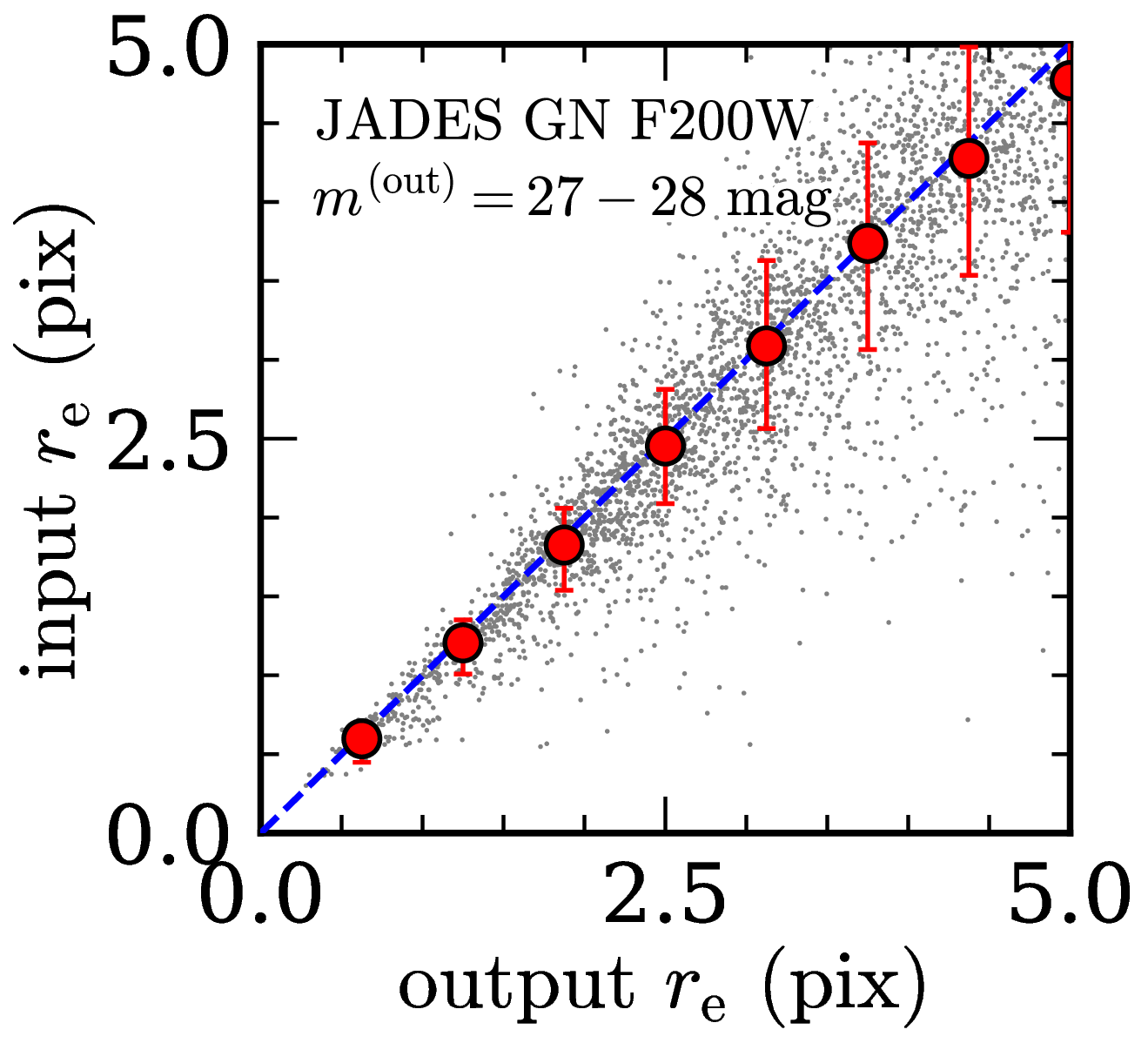}
   \includegraphics[width=0.24\textwidth]{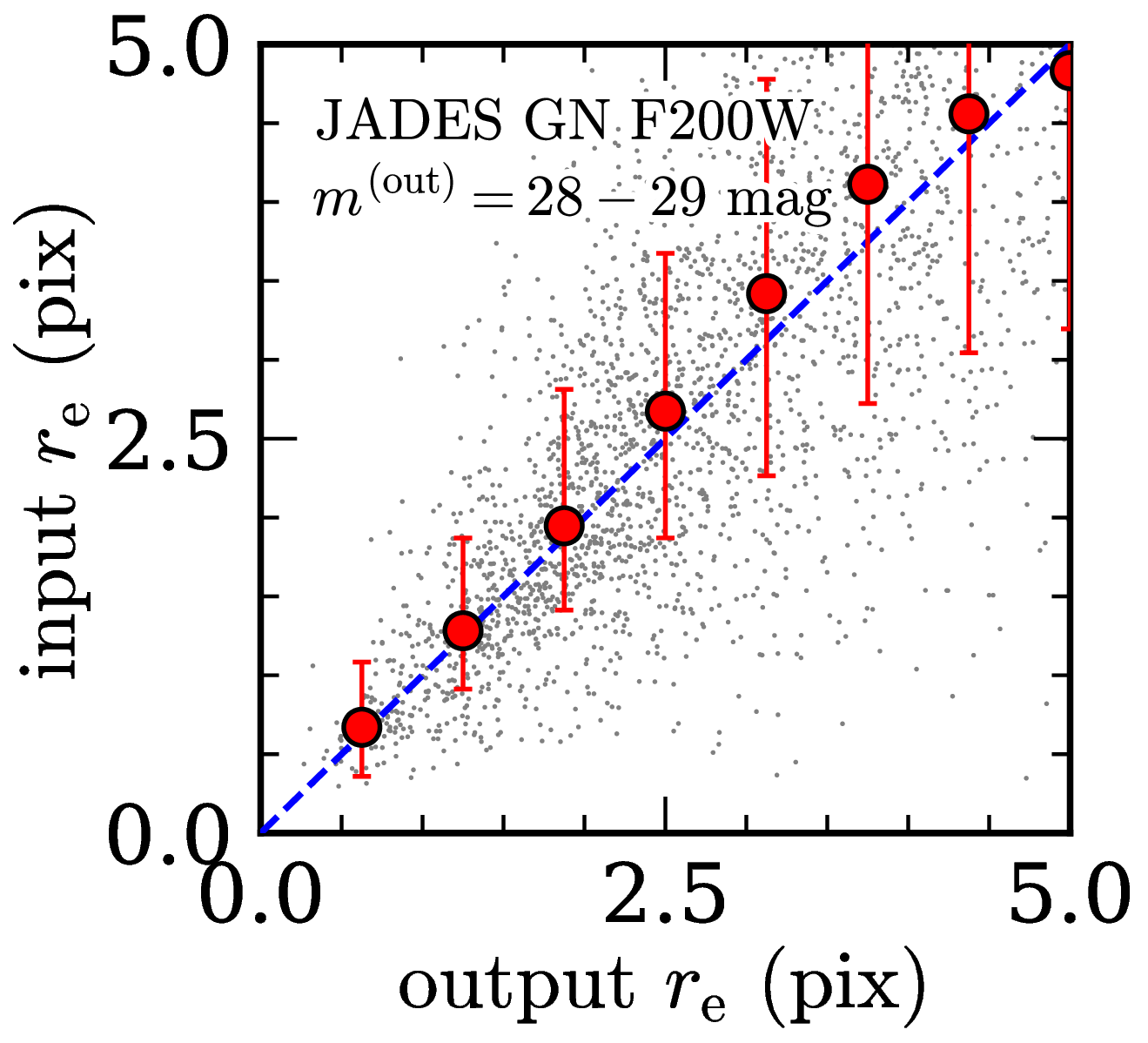}
   \includegraphics[width=0.24\textwidth]{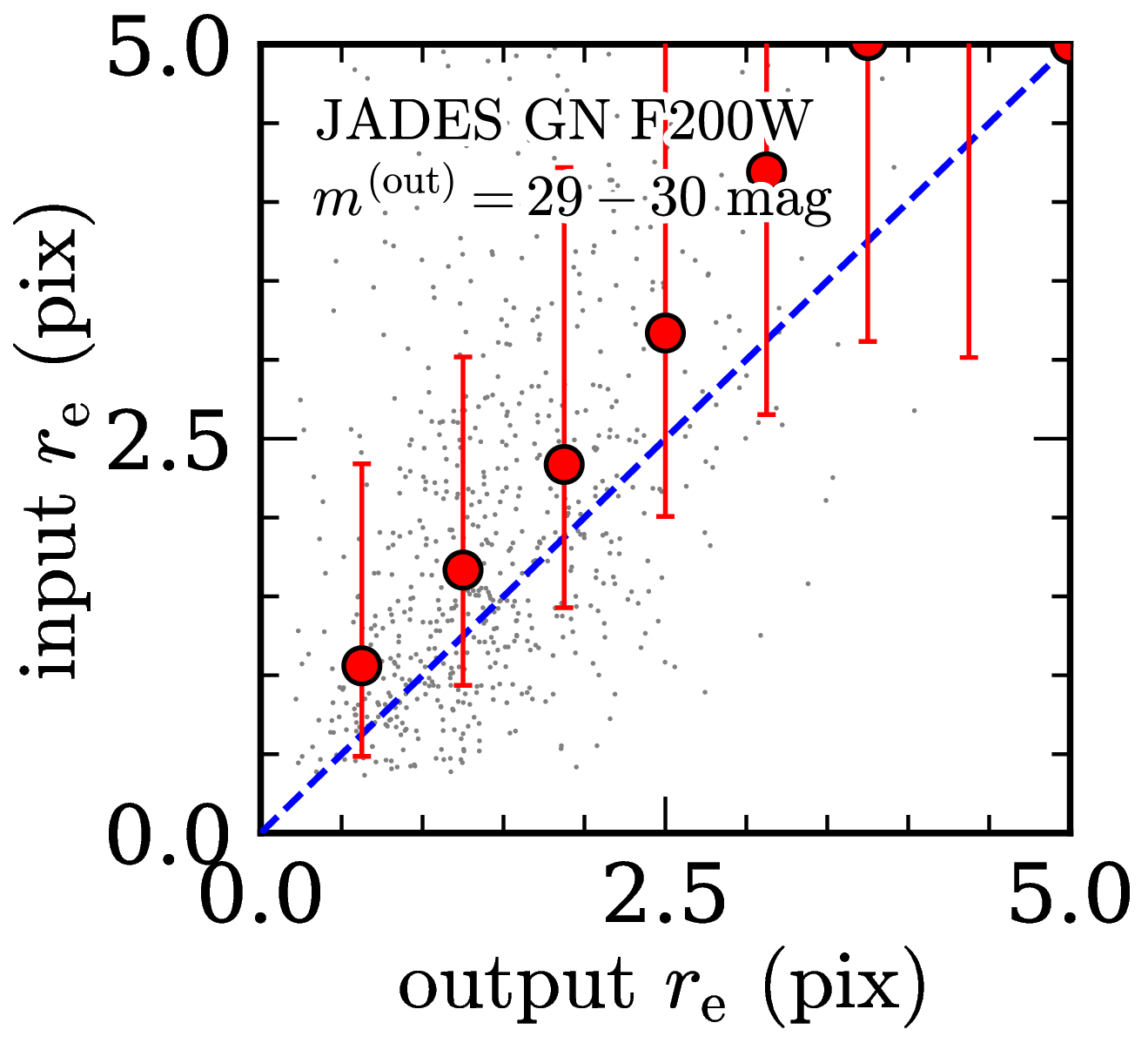}
   \includegraphics[width=0.24\textwidth]{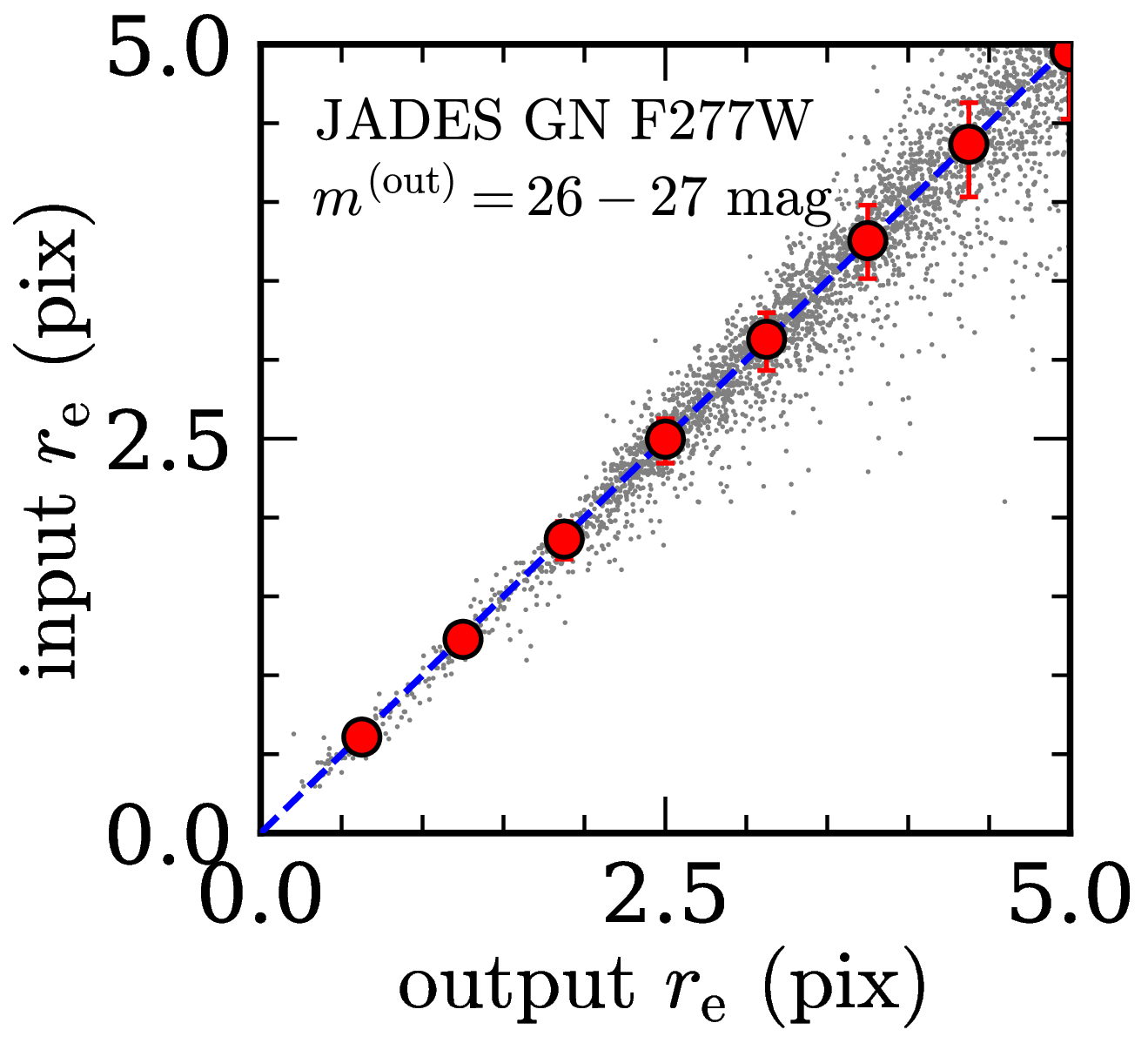}
   \includegraphics[width=0.24\textwidth]{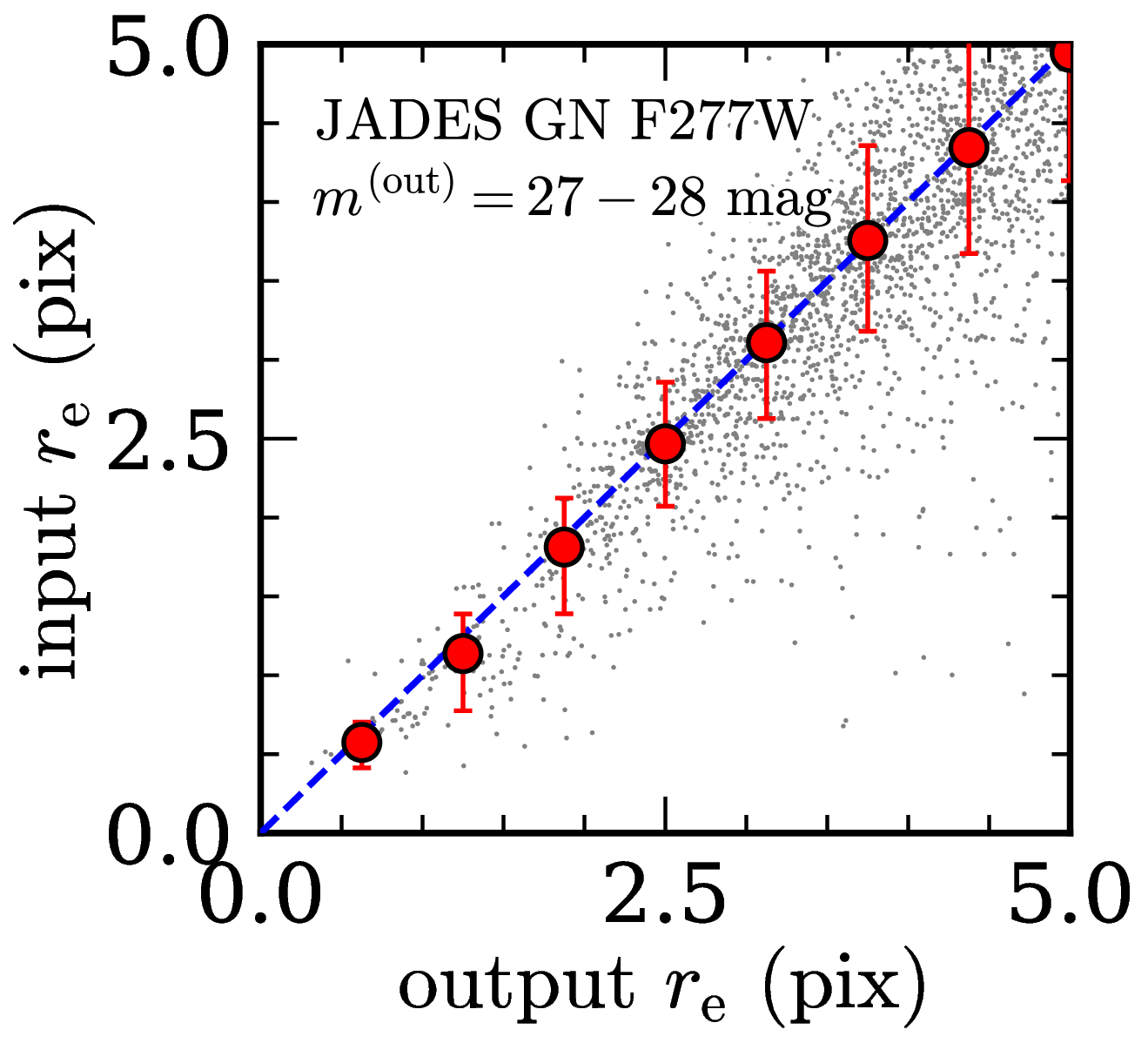}
   \includegraphics[width=0.24\textwidth]{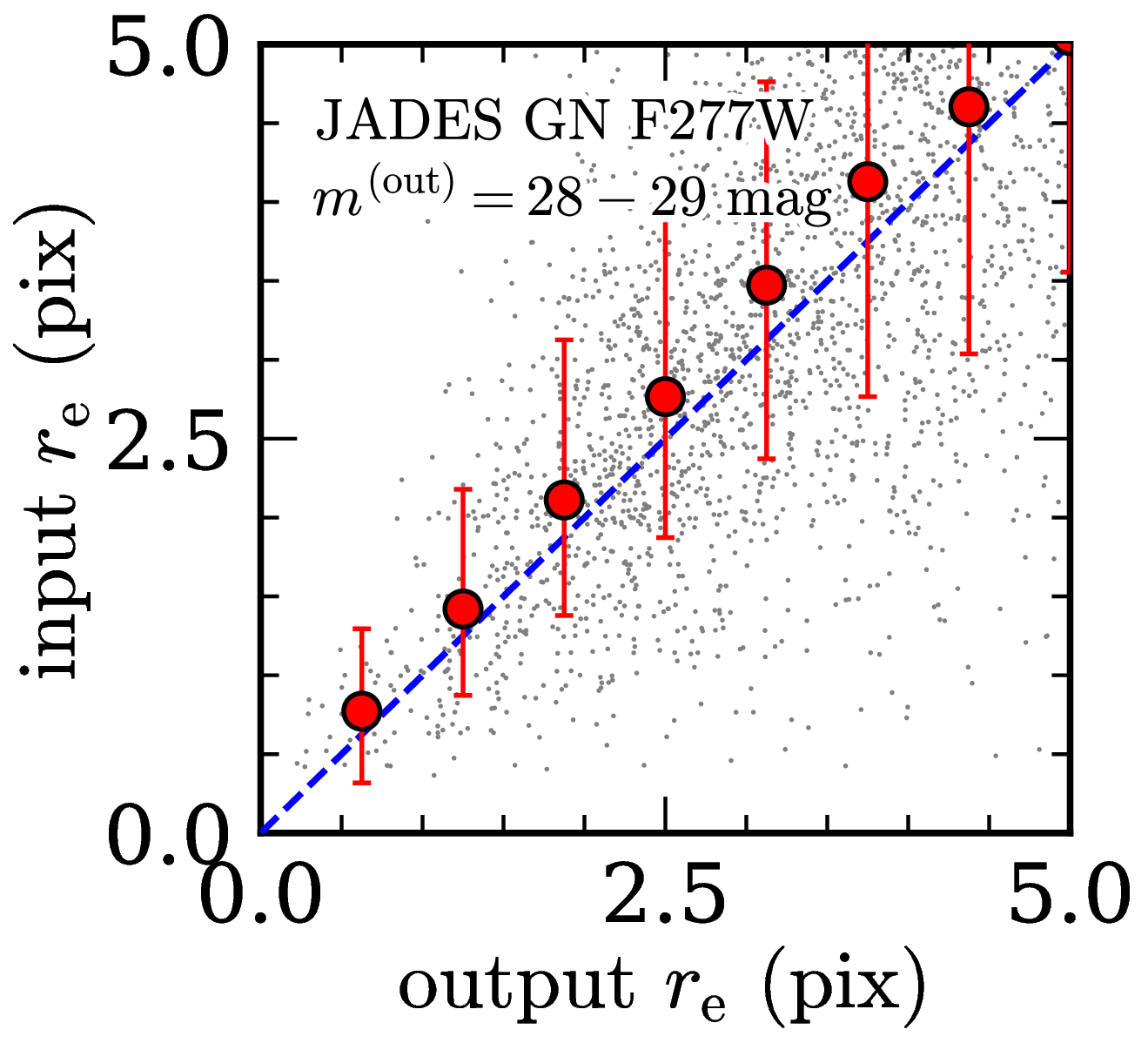}
   \includegraphics[width=0.24\textwidth]{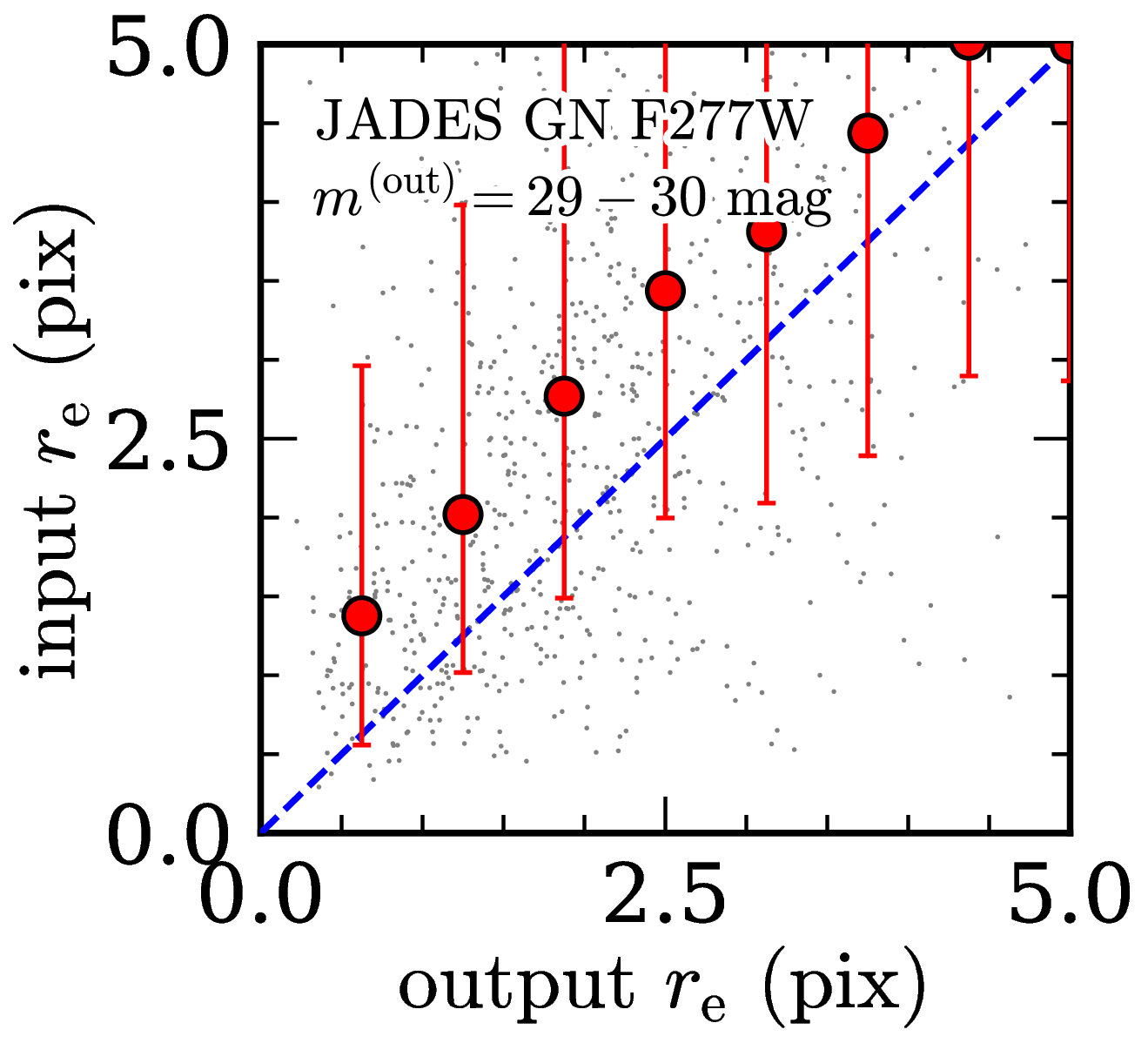}
\caption{
Input circularized radius  vs. output circularized radius 
for a range of output total magnitude 
$m^{\rm (out)} = 26$--$27$ mag, $27$--$28$ mag, $28$--$29$ mag, and $29$--$30$ mag from left to right, 
based on our GALFIT Monte Carlo simulations. 
From top to bottom, the results for 
the JADES Deep field in F200W and F277W, 
and  the JADES GOODS-N field in F200W and F277W  
are presented. 
The red filled circles and the red error bars correspond to 
the median values of the difference between the input and output circularized radii 
and the 68 percentile ranges, respectively. 
The gray dots are the results for individual simulated objects. 
The blue dashed line represents the relation that 
the input and output circularized radius are equal.  
The measured sizes are generally consistent with the input sizes on average, 
even for galaxies with sizes smaller than the PSF FWHMs, 
i.e., about $2.5$ and $4.0$ pixels in F200W and F277W, respectively.
}
\label{fig:input_output_re}
\end{center}
\end{figure*}

\begin{figure}[h]
\begin{center}
   \includegraphics[width=0.23\textwidth]{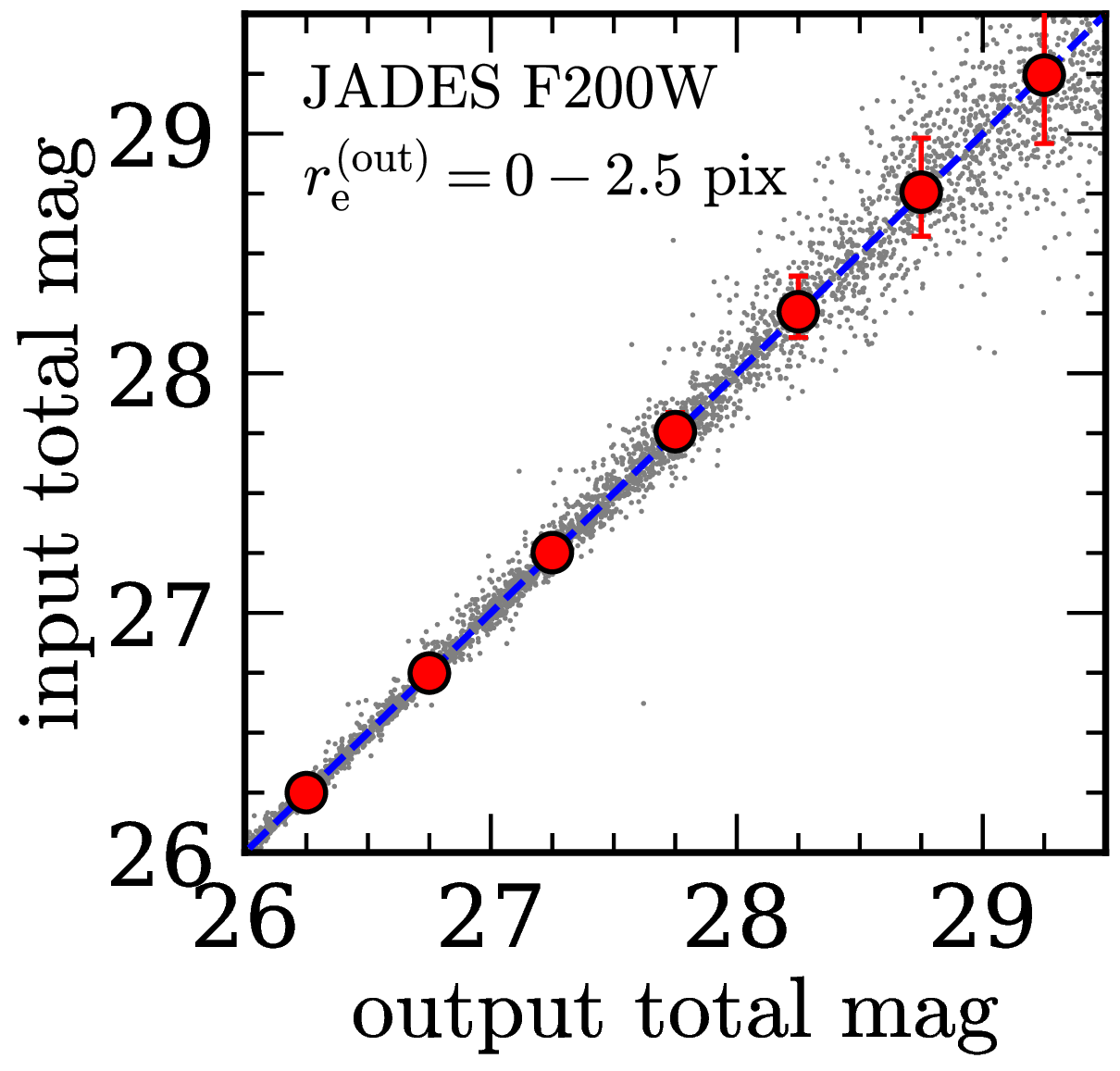}
   \includegraphics[width=0.23\textwidth]{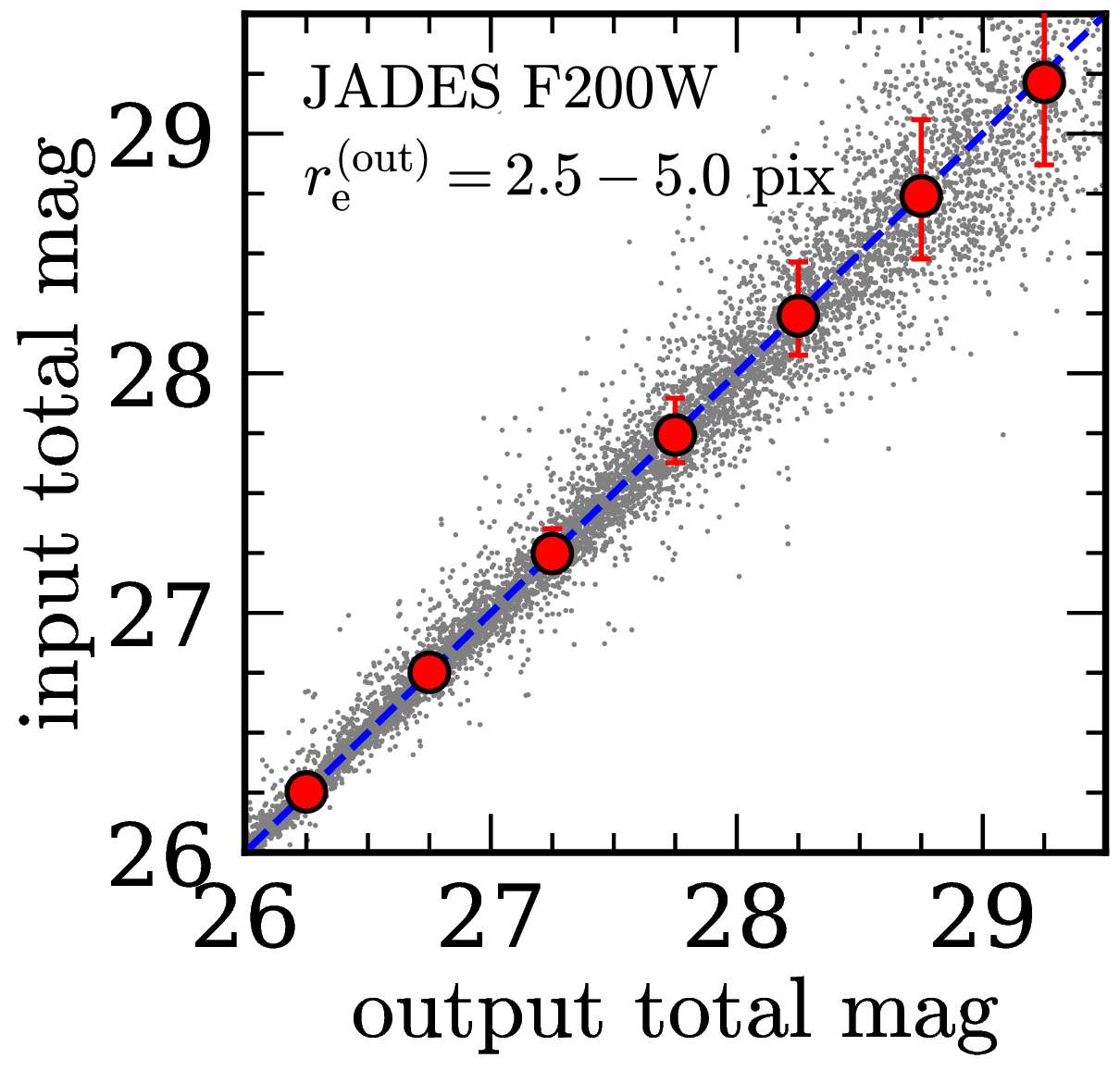}
   \includegraphics[width=0.23\textwidth]{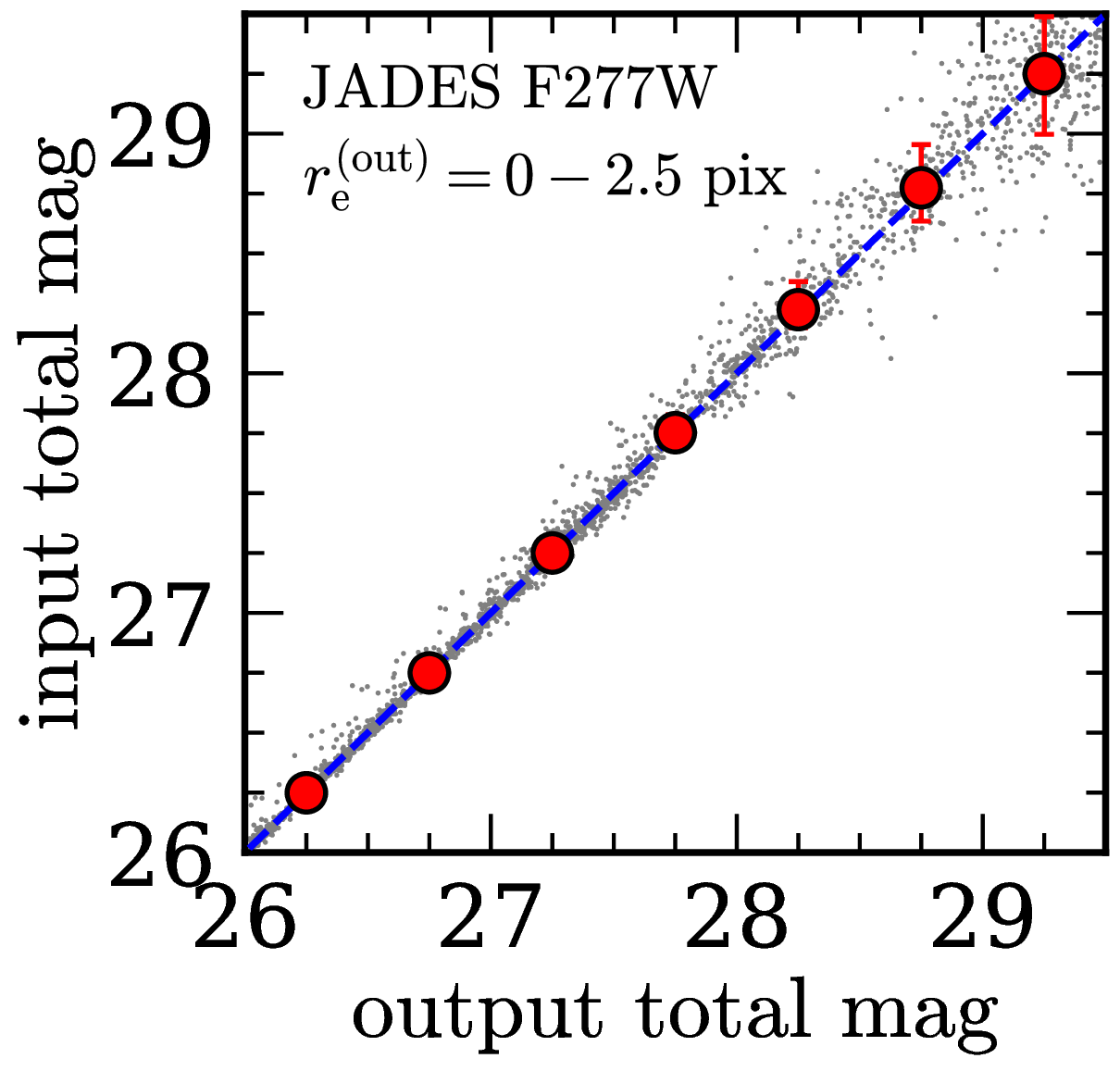}
   \includegraphics[width=0.23\textwidth]{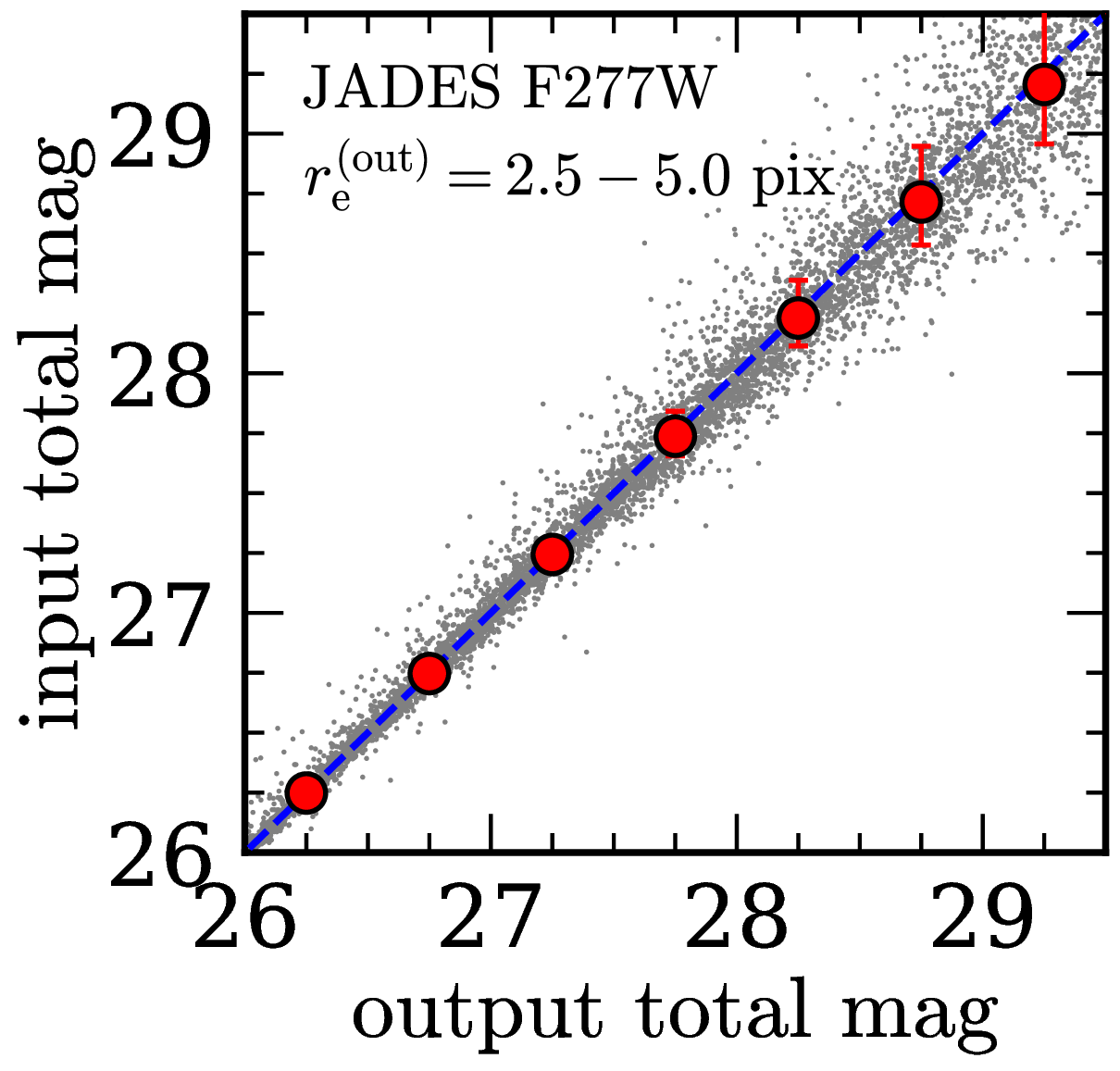}
   \includegraphics[width=0.23\textwidth]{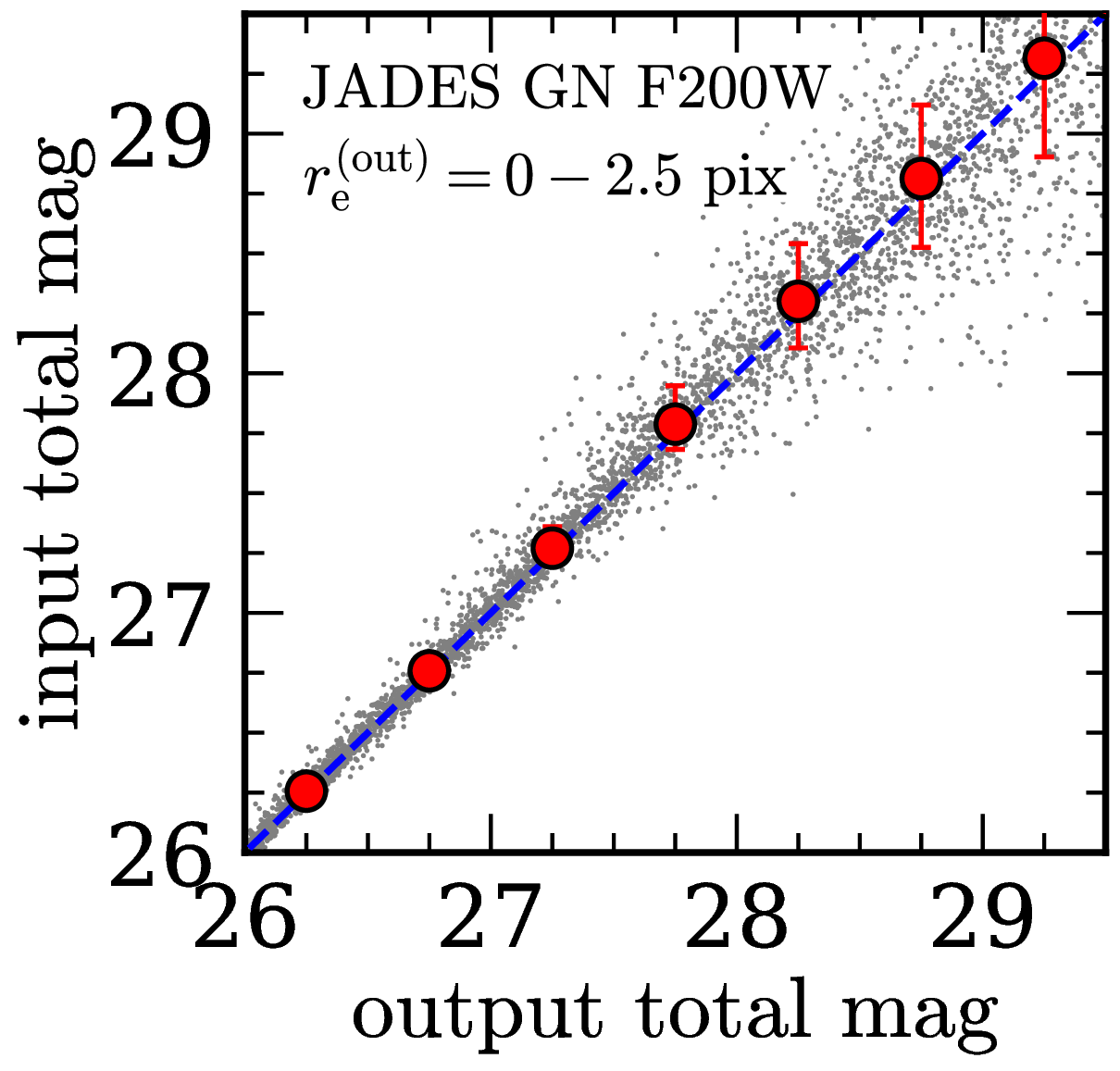}
   \includegraphics[width=0.23\textwidth]{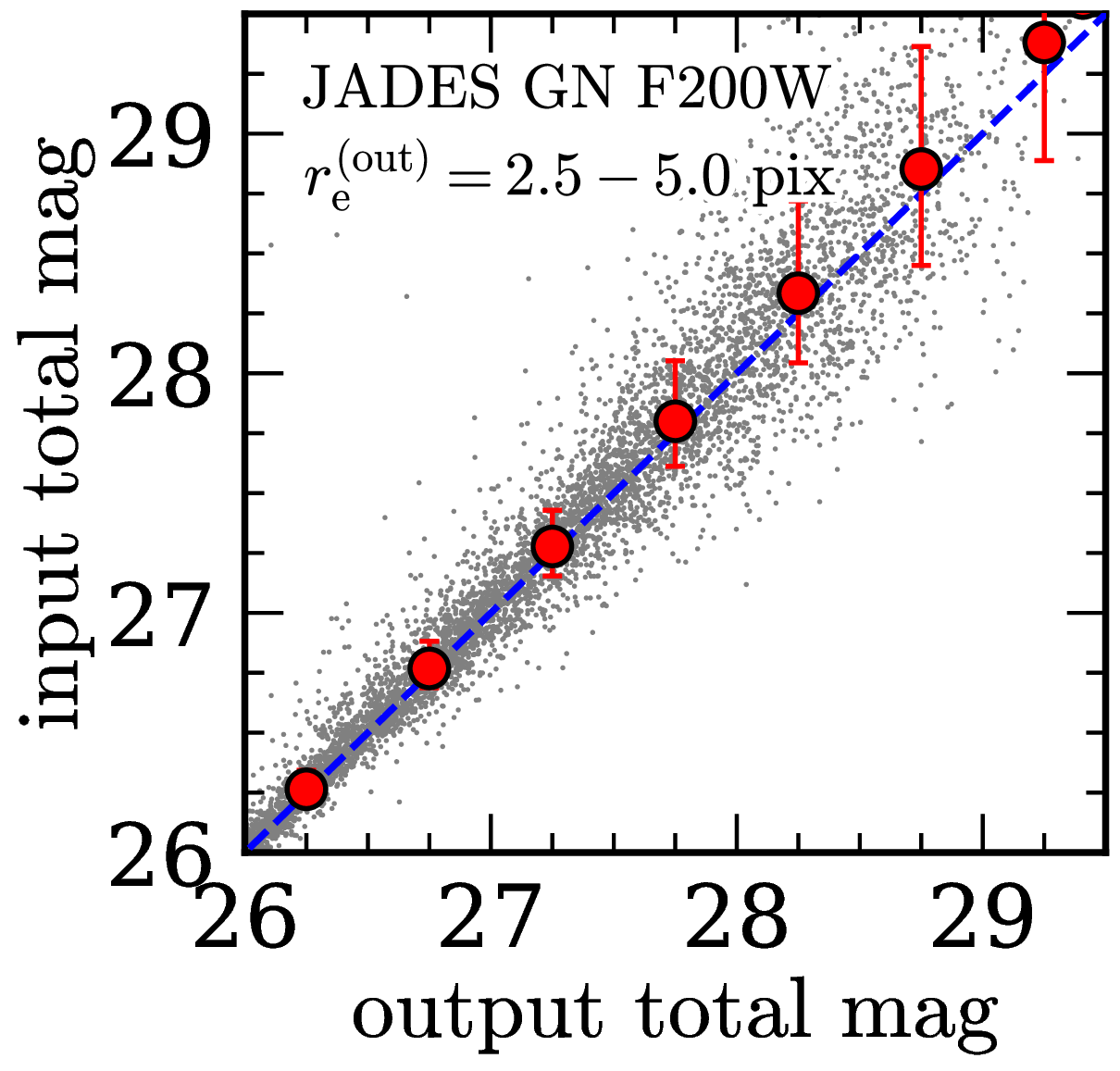}
   \includegraphics[width=0.23\textwidth]{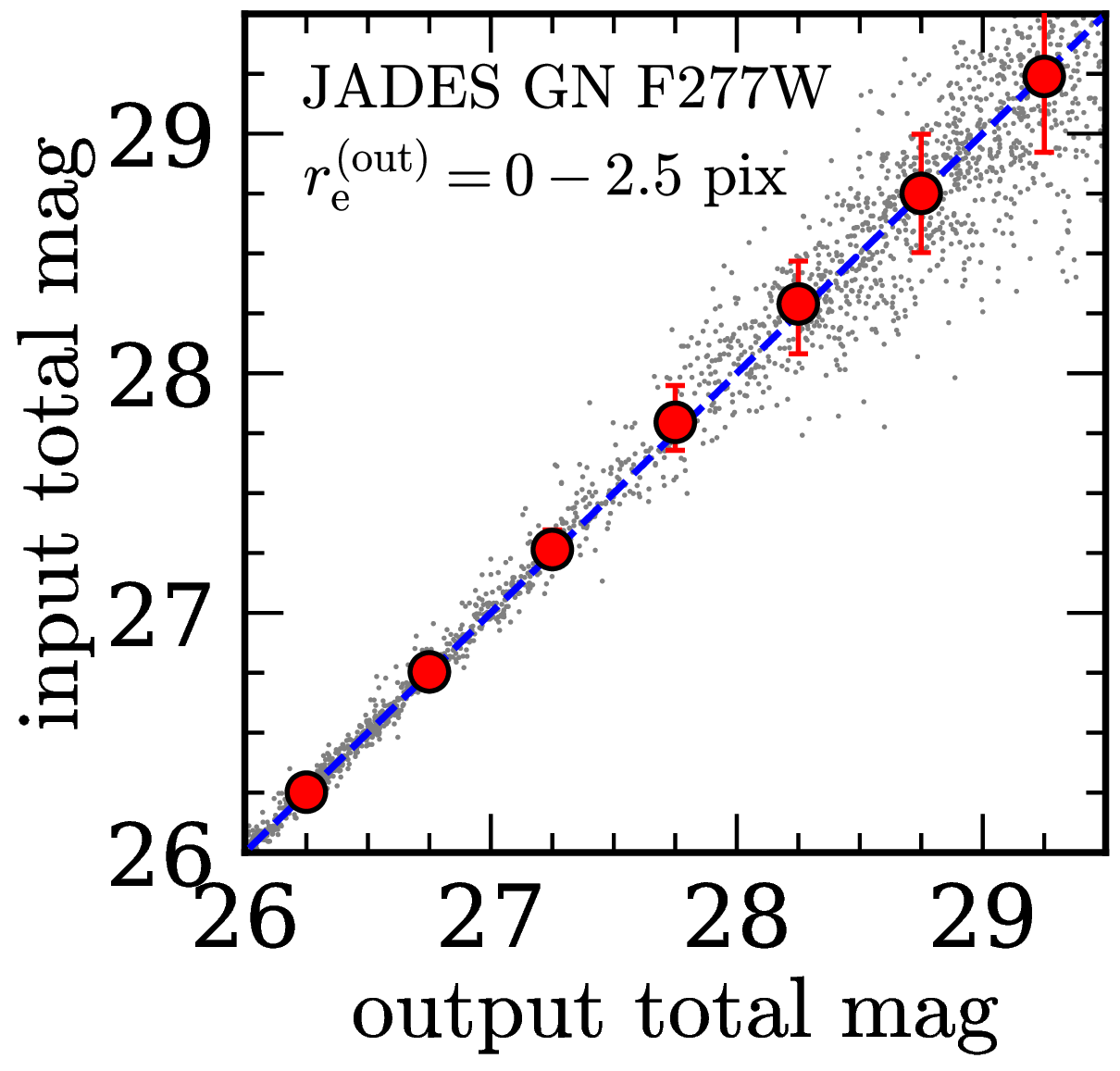}
   \includegraphics[width=0.23\textwidth]{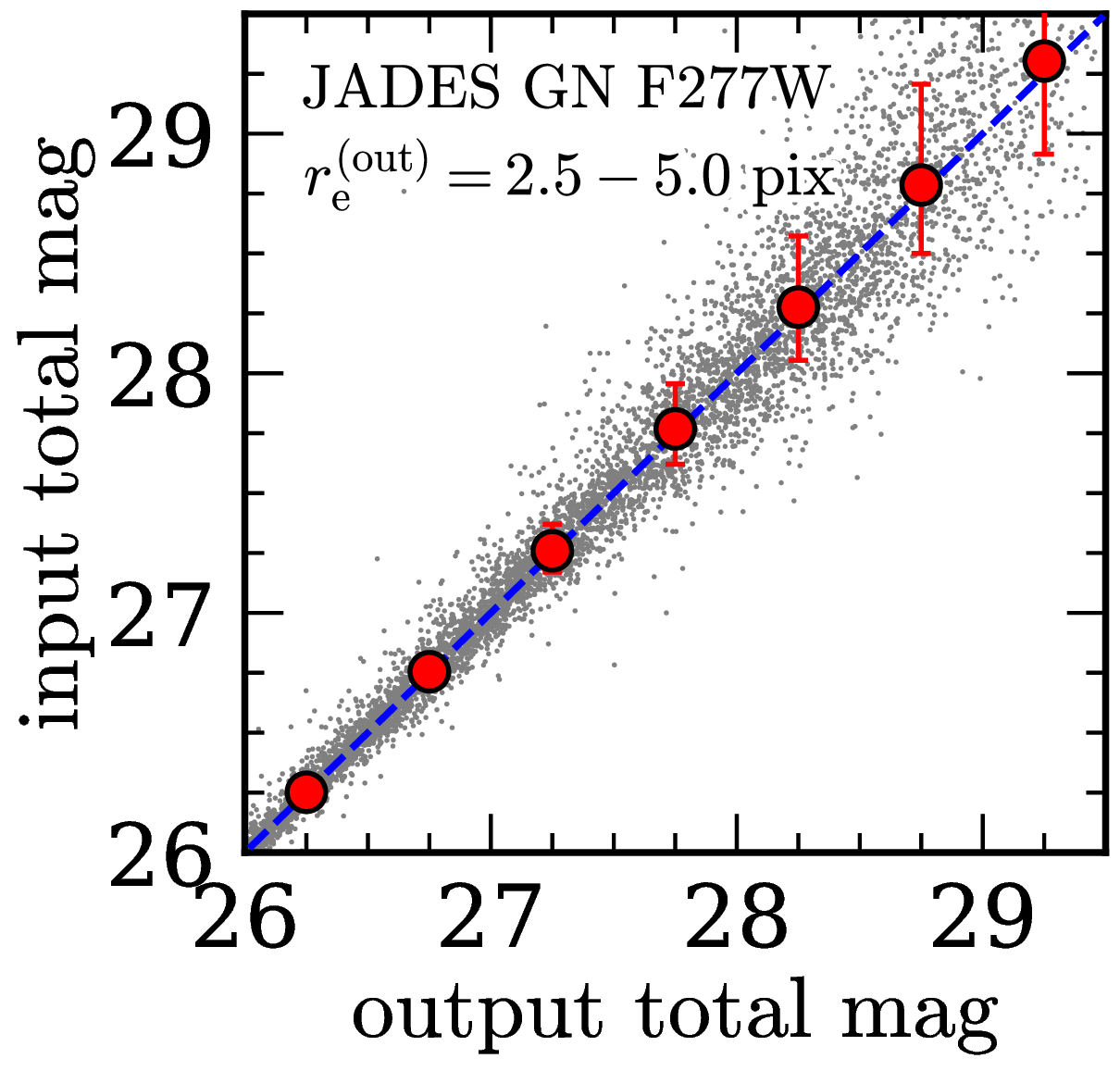}
\caption{
Input total magnitude vs. output total magnitude 
for a range of output half-light radii 
$r_{\rm e}^{\rm (out)} = 0$--$2.5$ pixels (left) and $2.5$--$5.0$ pixels (right) 
based on our  GALFIT Monte Carlo simulations. 
From top to bottom, the results for 
the JADES Deep field in F200W and F277W, 
and the JADES GOODS-N field in F200W and F277W 
are presented. 
The red filled circles and the red error bars correspond to 
the median values of the difference between the input and output magnitudes 
and the 68 percentile ranges, respectively. 
The gray dots are the results for individual simulated objects. 
The blue dashed line represents the relation that 
the input and output magnitudes are equal.  
}
\label{fig:input_output_mag}
\end{center}
\end{figure}

\begin{figure*}[ht]
\begin{center}
   \includegraphics[width=0.24\textwidth]{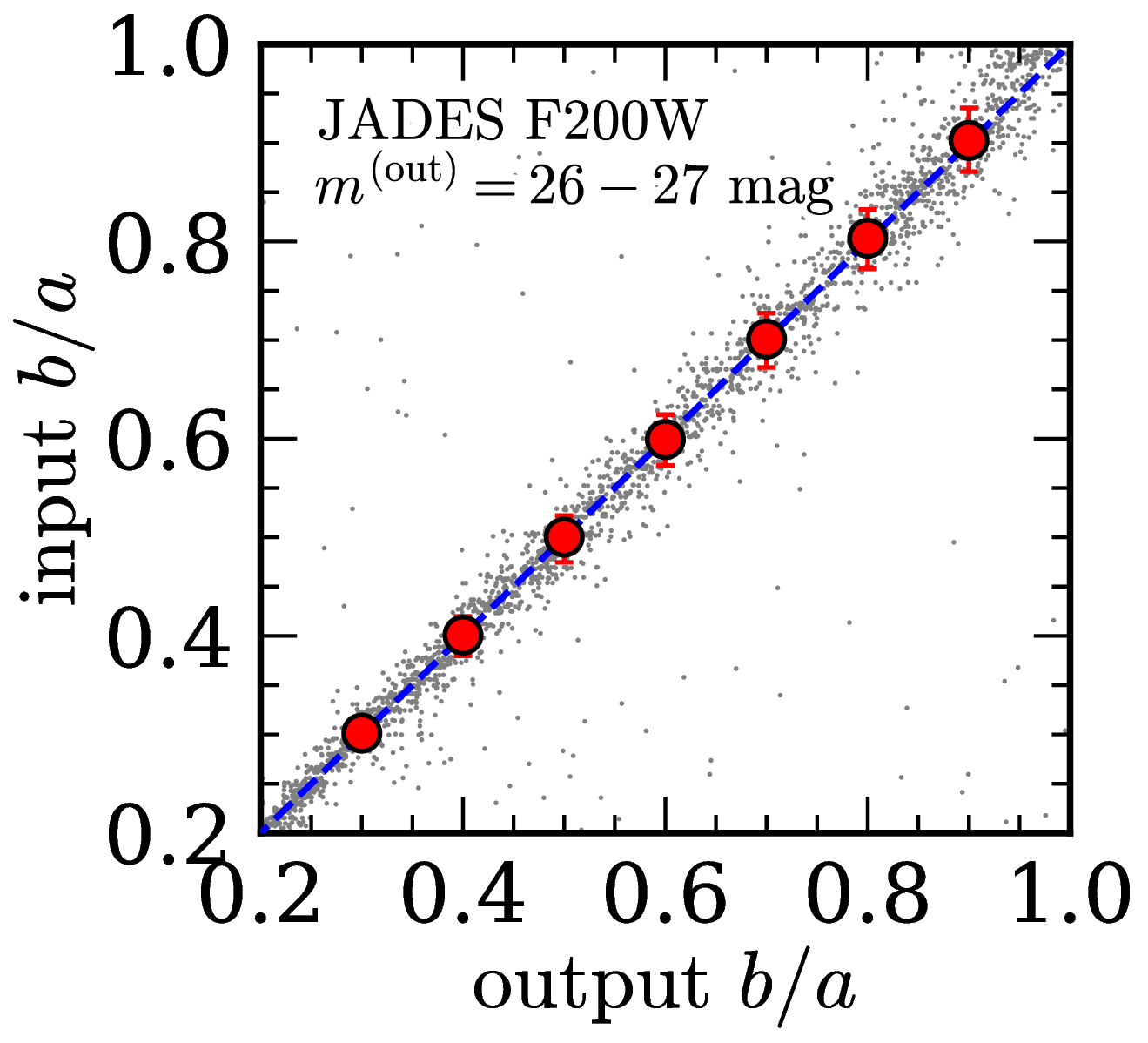}
   \includegraphics[width=0.24\textwidth]{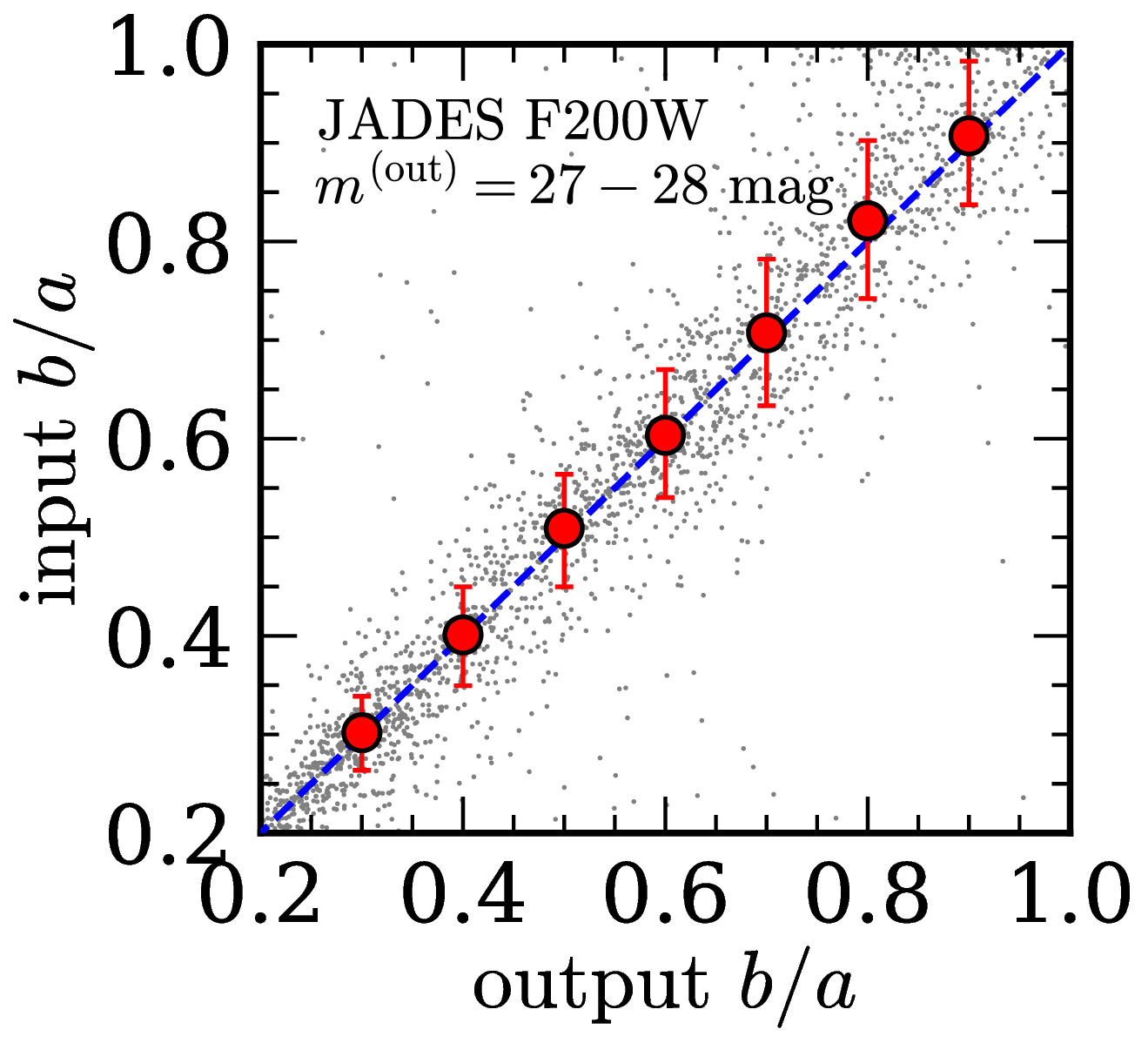}
   \includegraphics[width=0.24\textwidth]{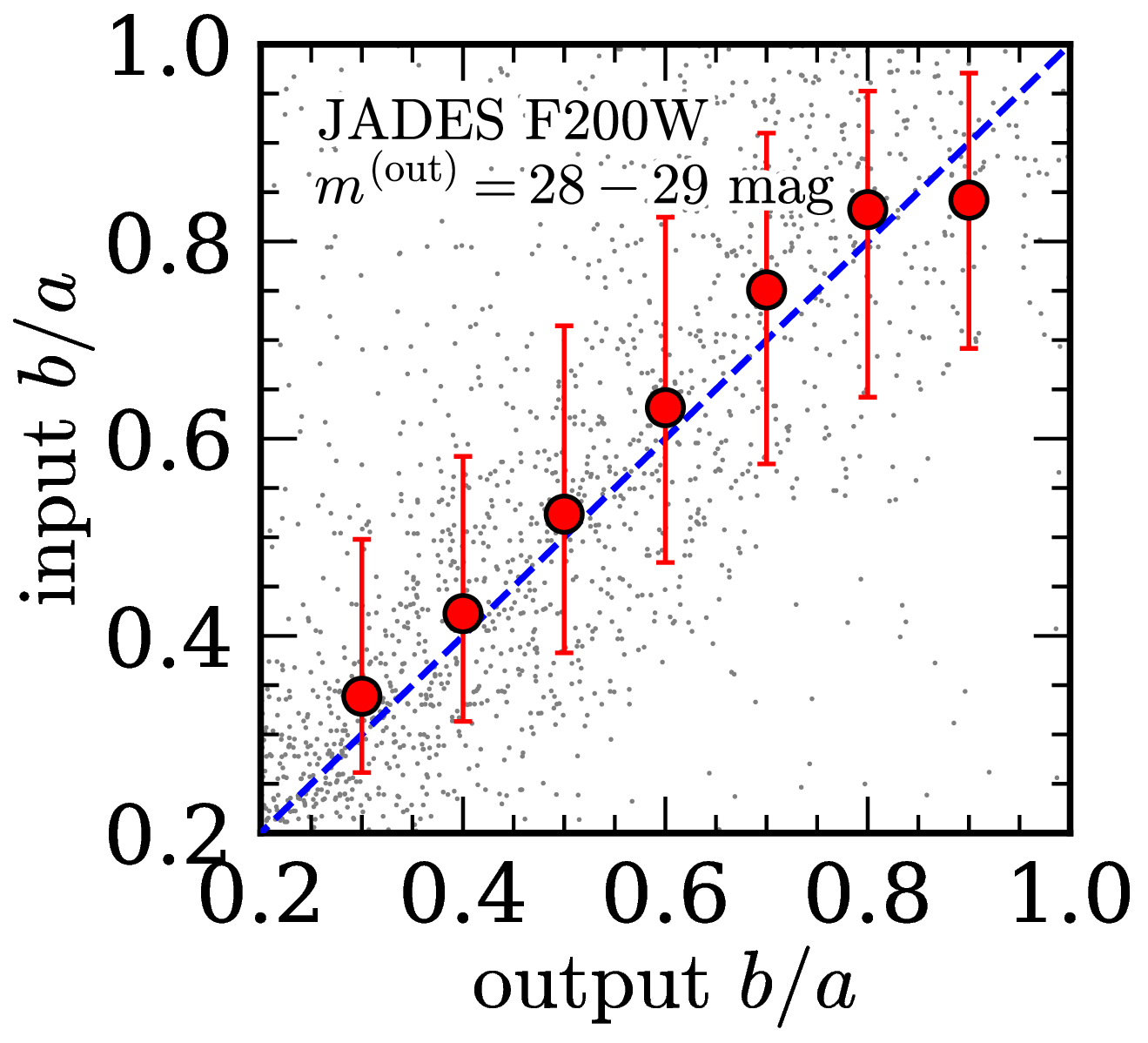}
   \includegraphics[width=0.24\textwidth]{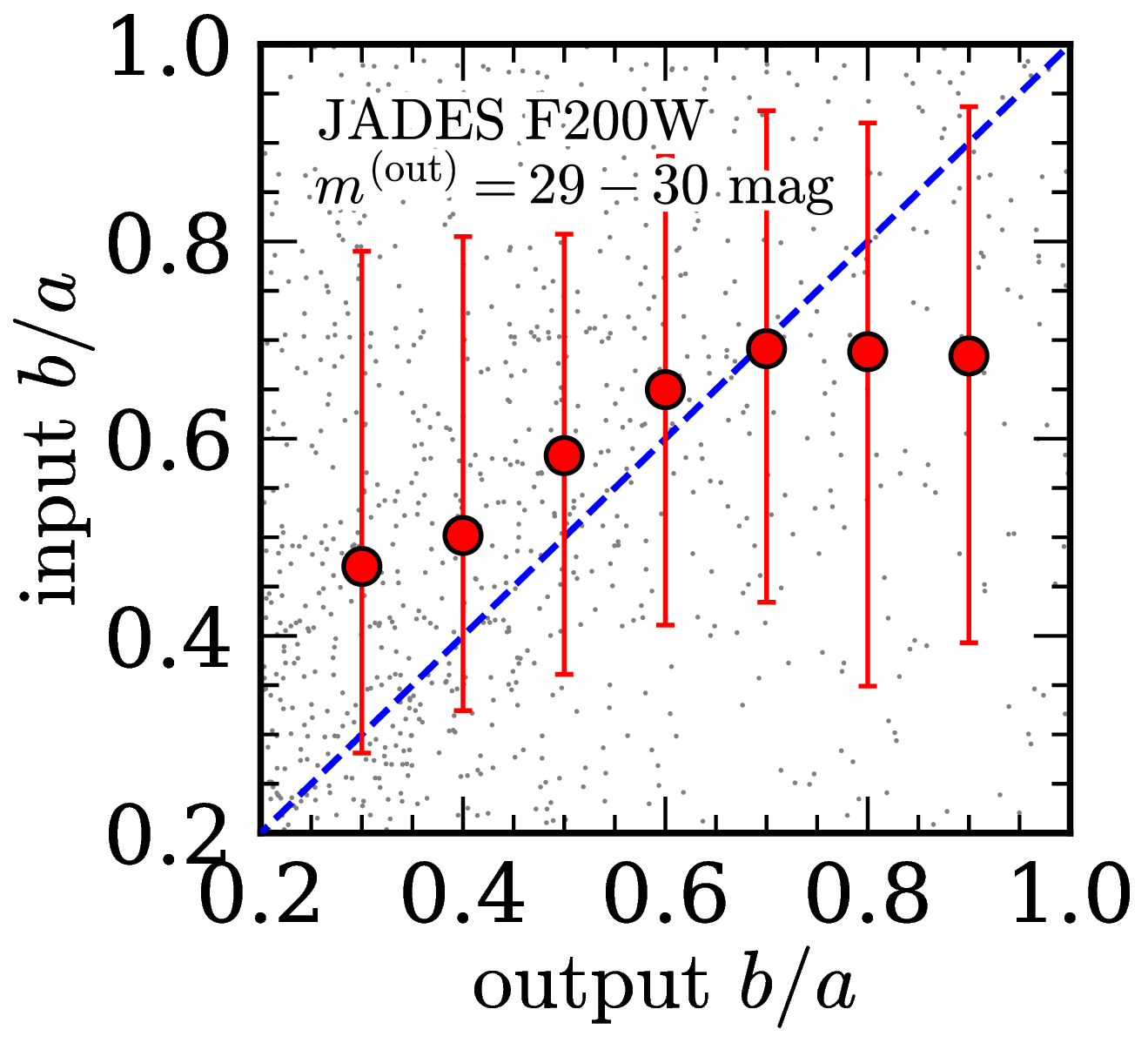}
   \includegraphics[width=0.24\textwidth]{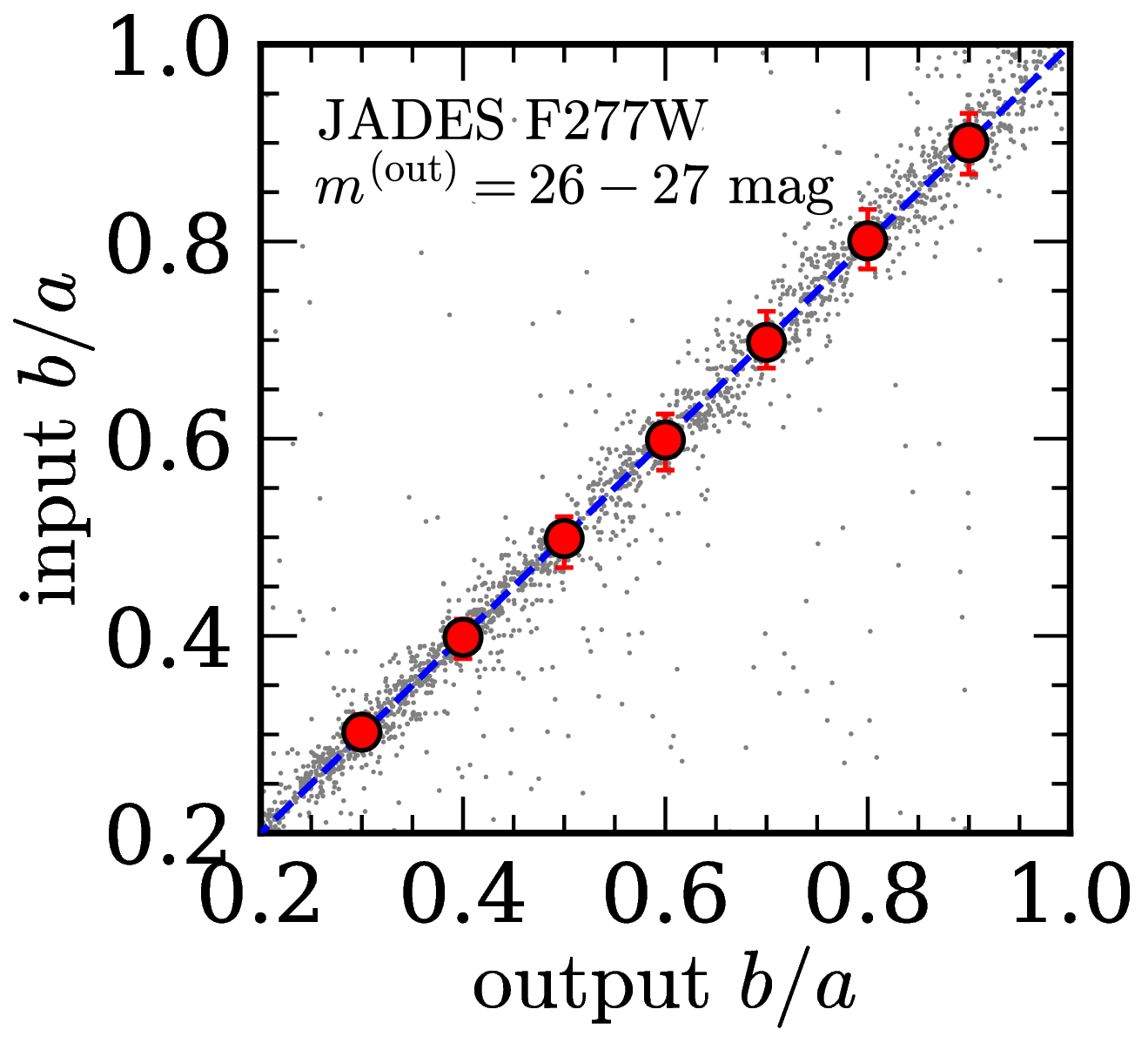}
   \includegraphics[width=0.24\textwidth]{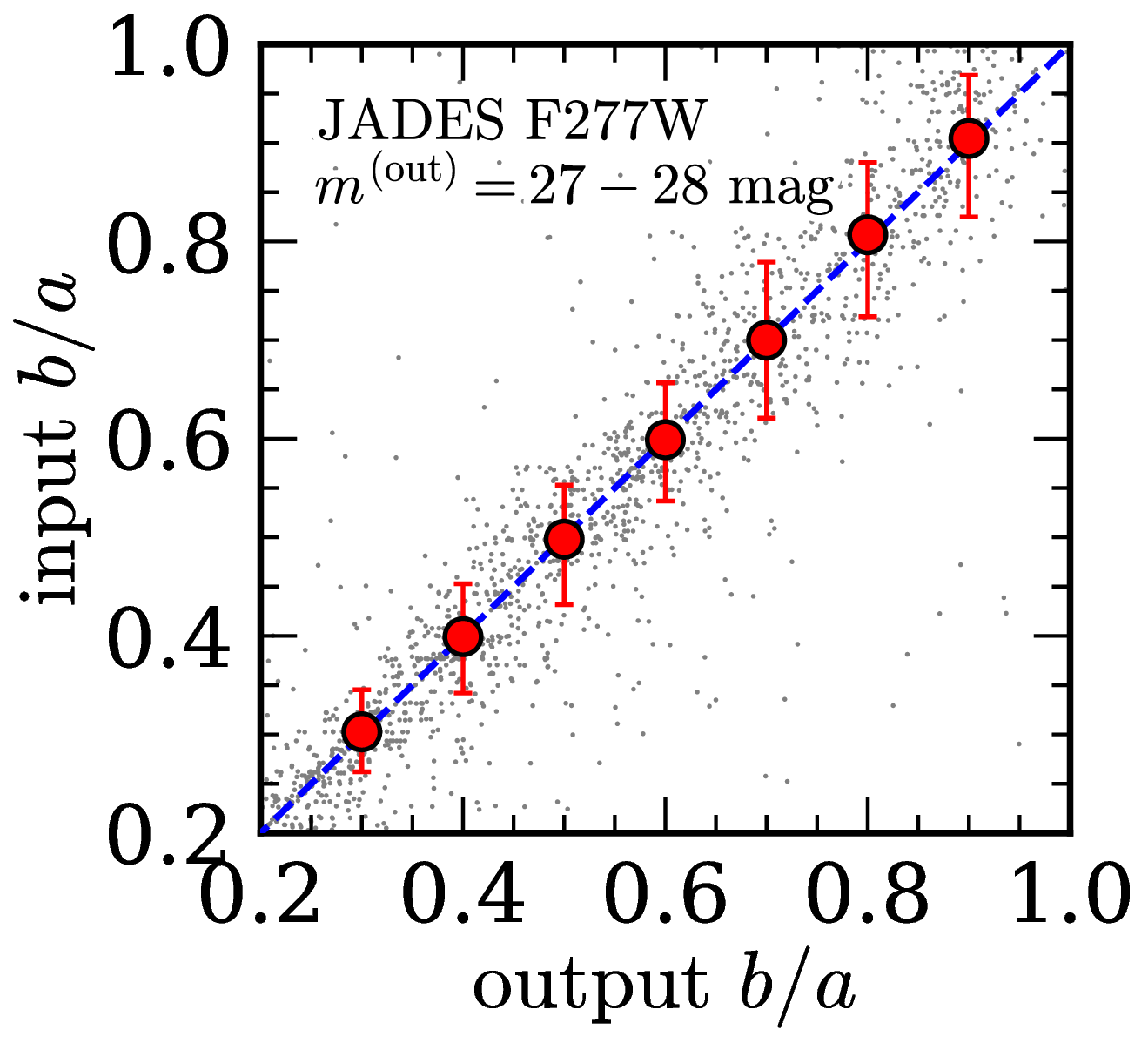}
   \includegraphics[width=0.24\textwidth]{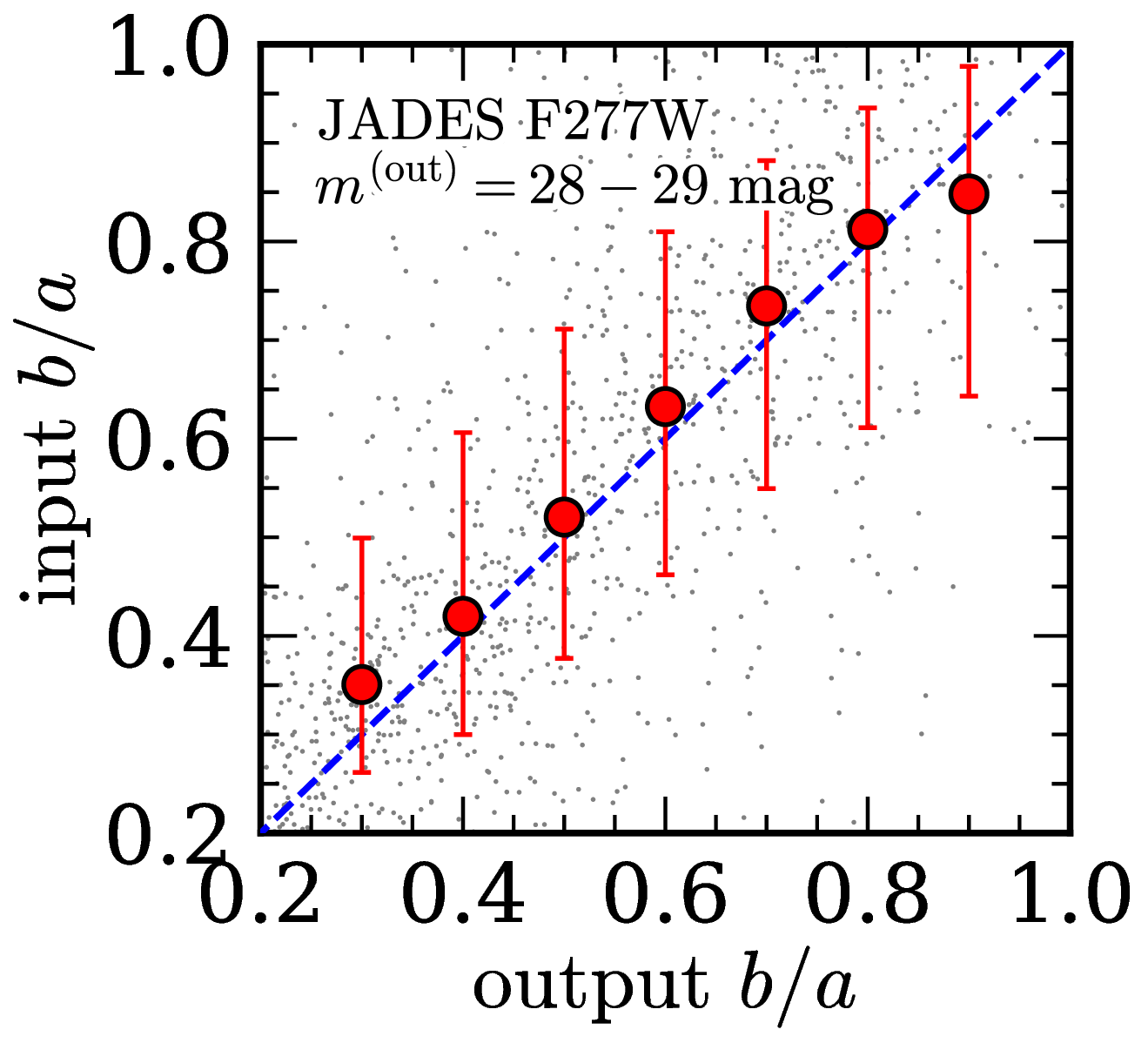}
   \includegraphics[width=0.24\textwidth]{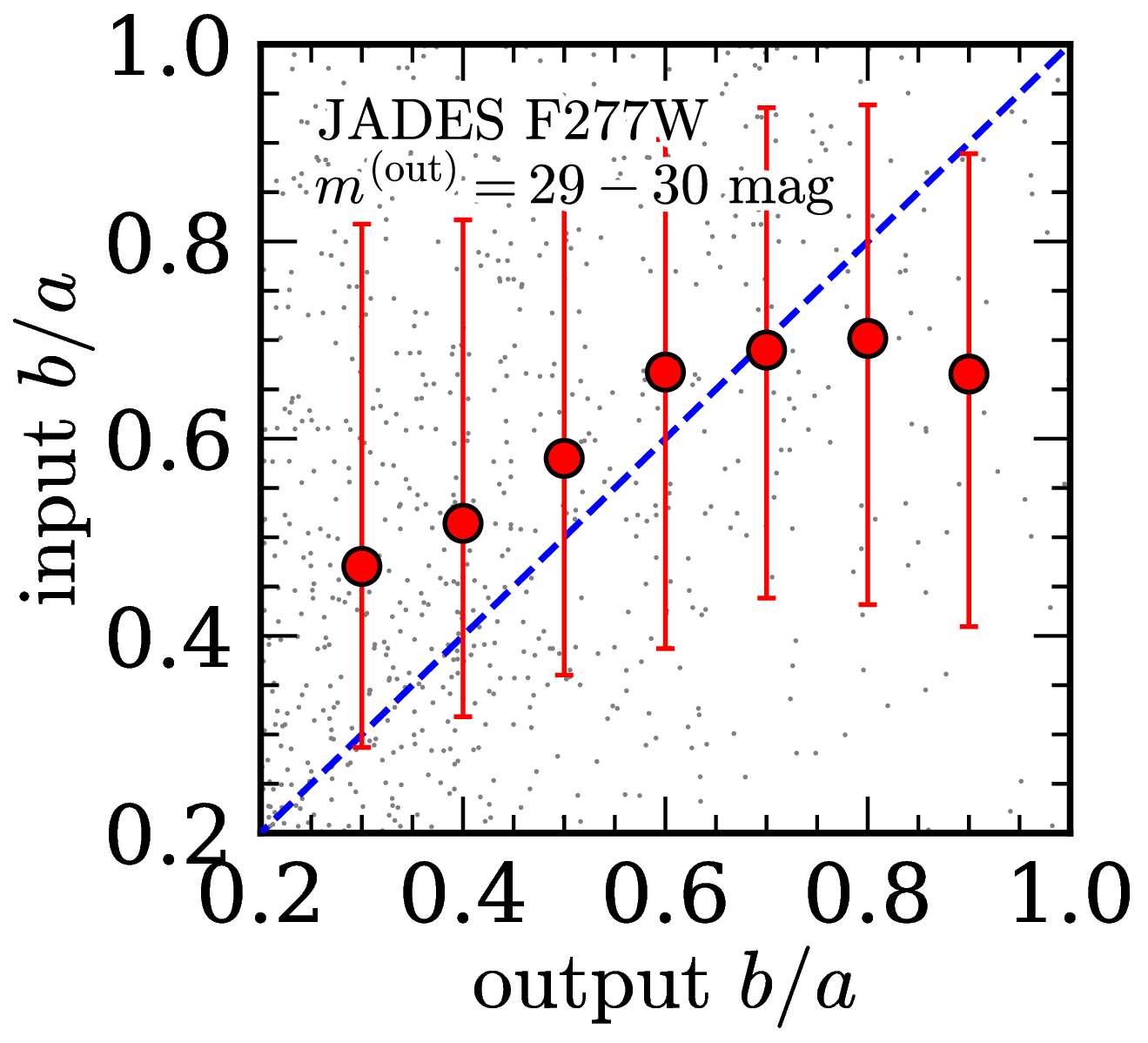}
   \includegraphics[width=0.24\textwidth]{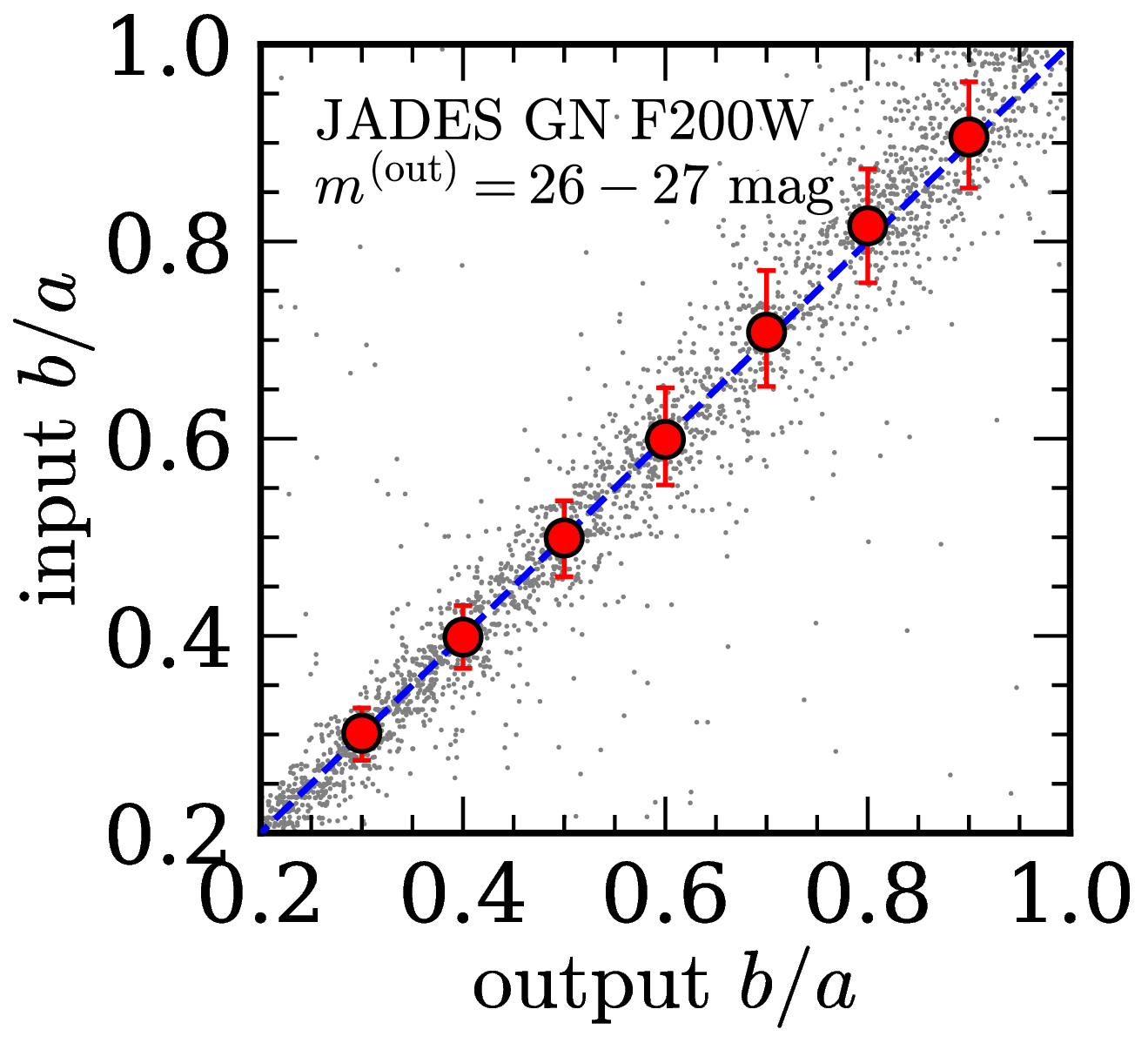}
   \includegraphics[width=0.24\textwidth]{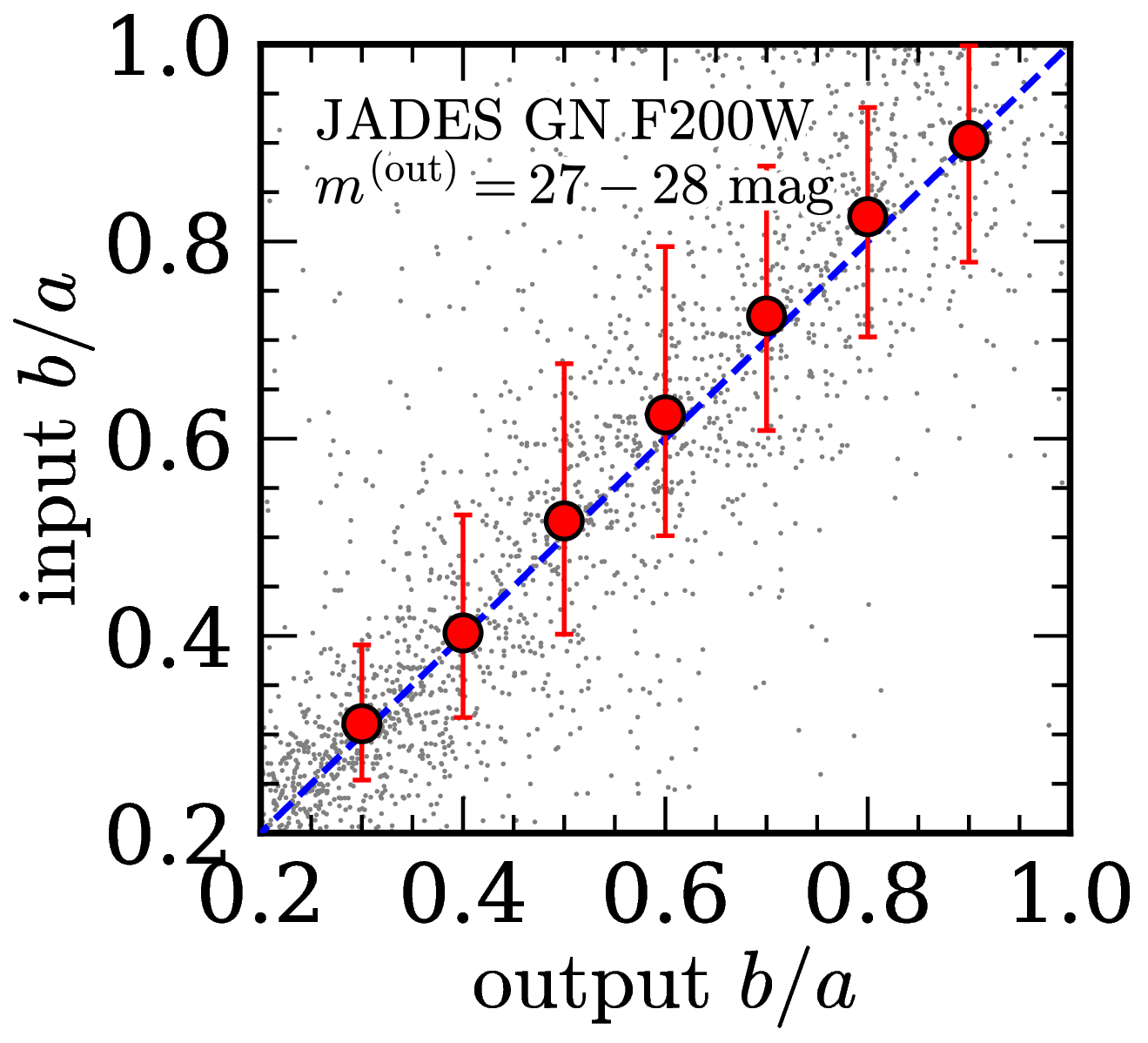}
   \includegraphics[width=0.24\textwidth]{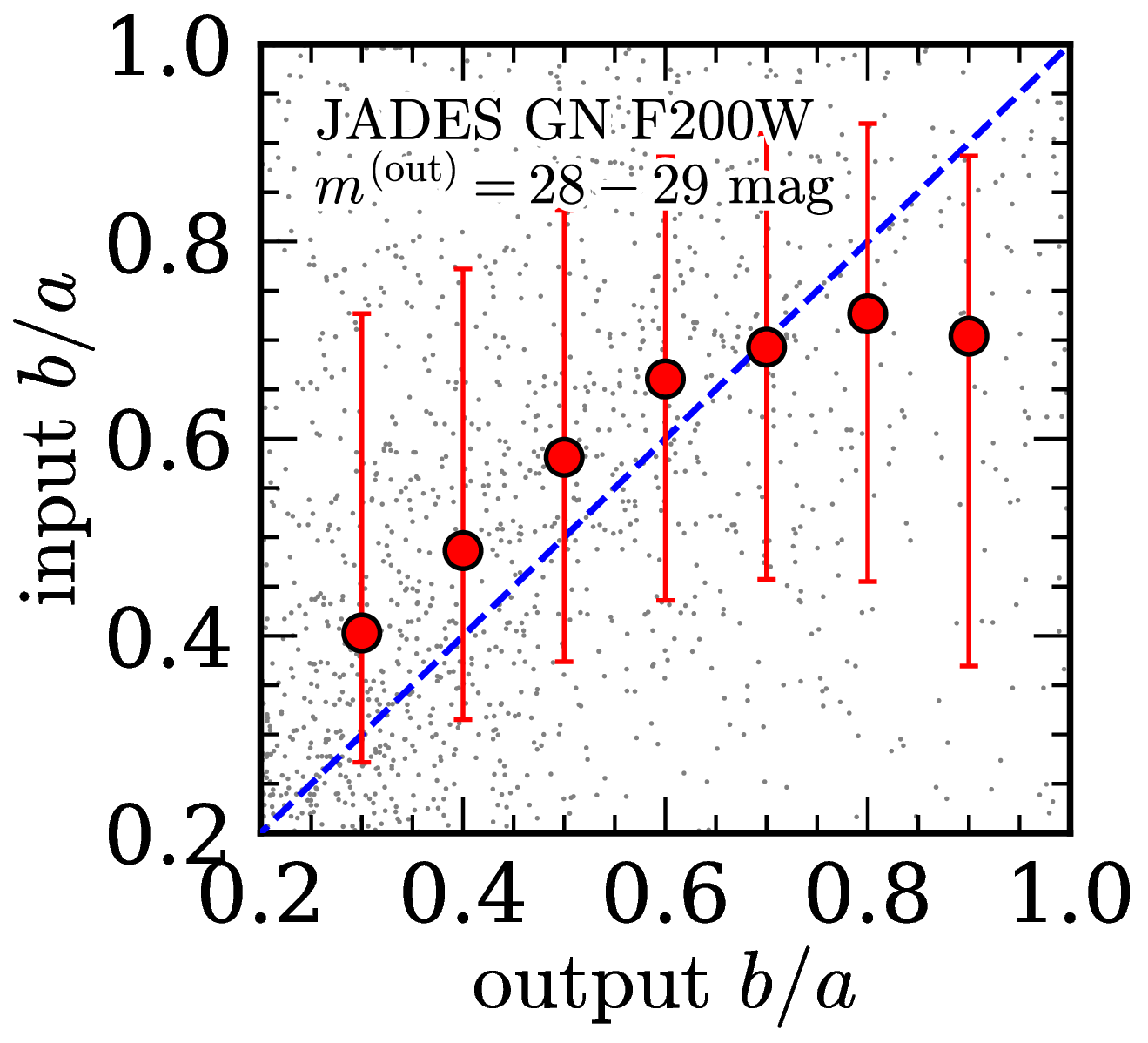}
   \includegraphics[width=0.24\textwidth]{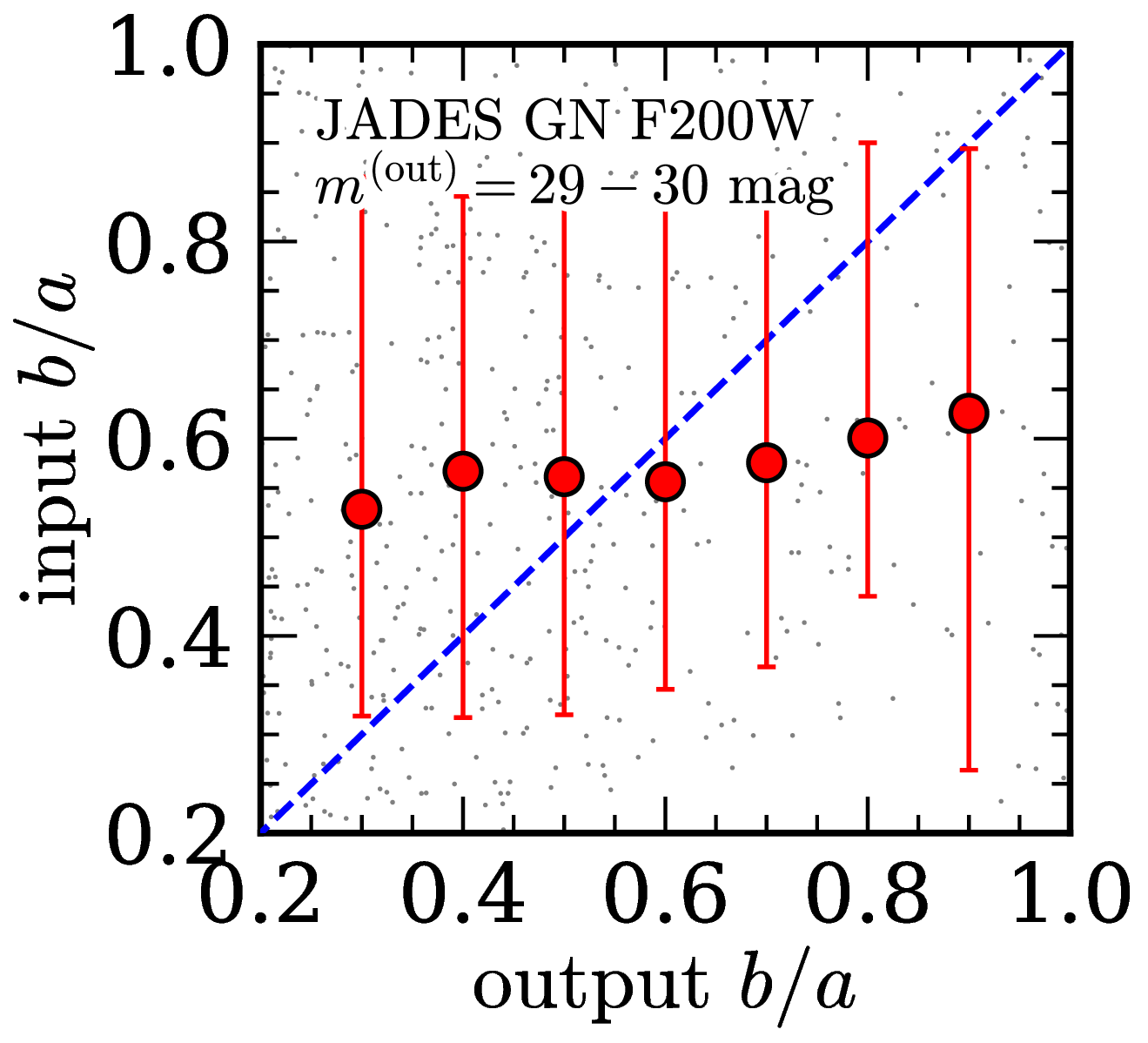}
   \includegraphics[width=0.24\textwidth]{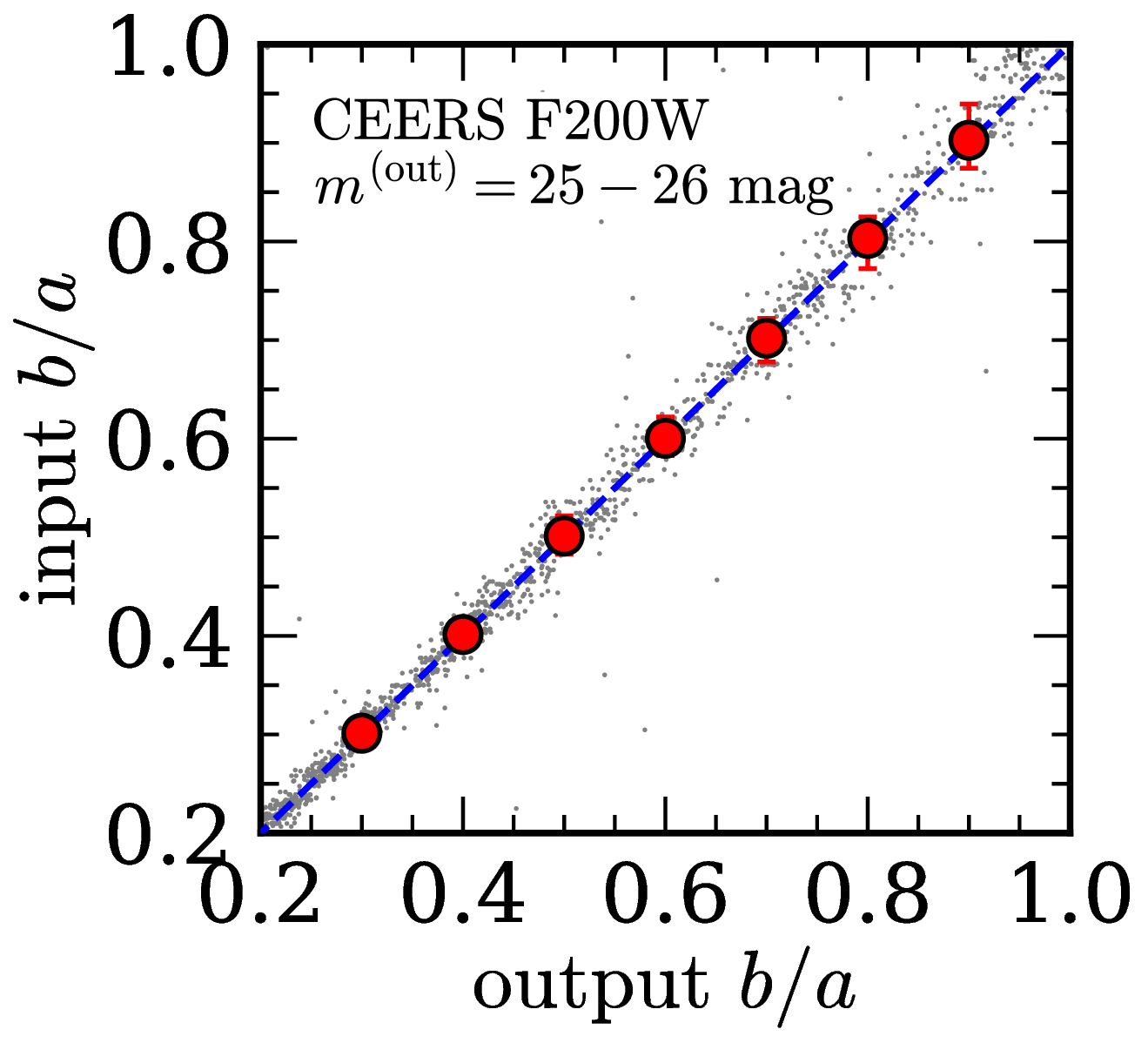}
   \includegraphics[width=0.24\textwidth]{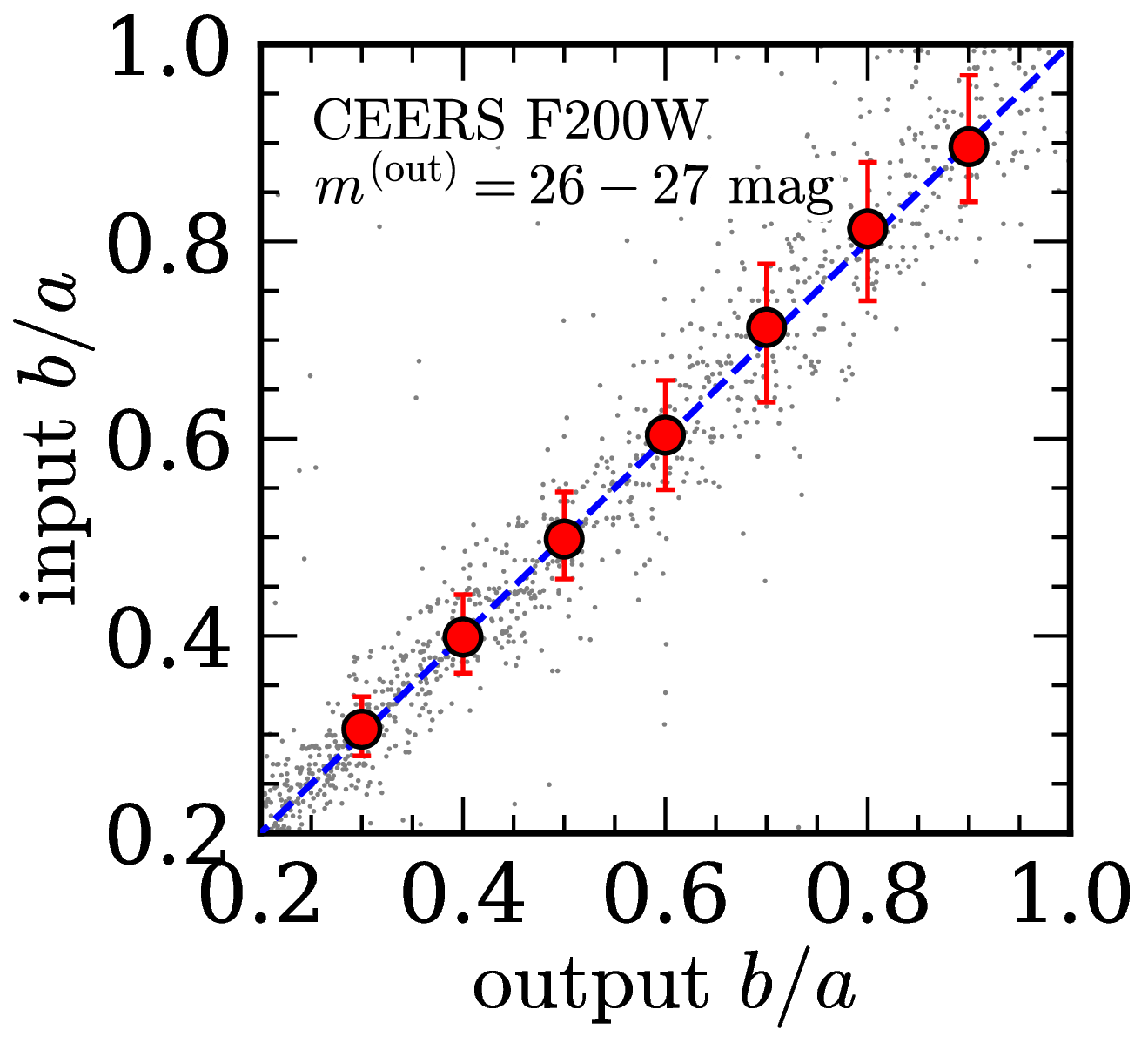}
   \includegraphics[width=0.24\textwidth]{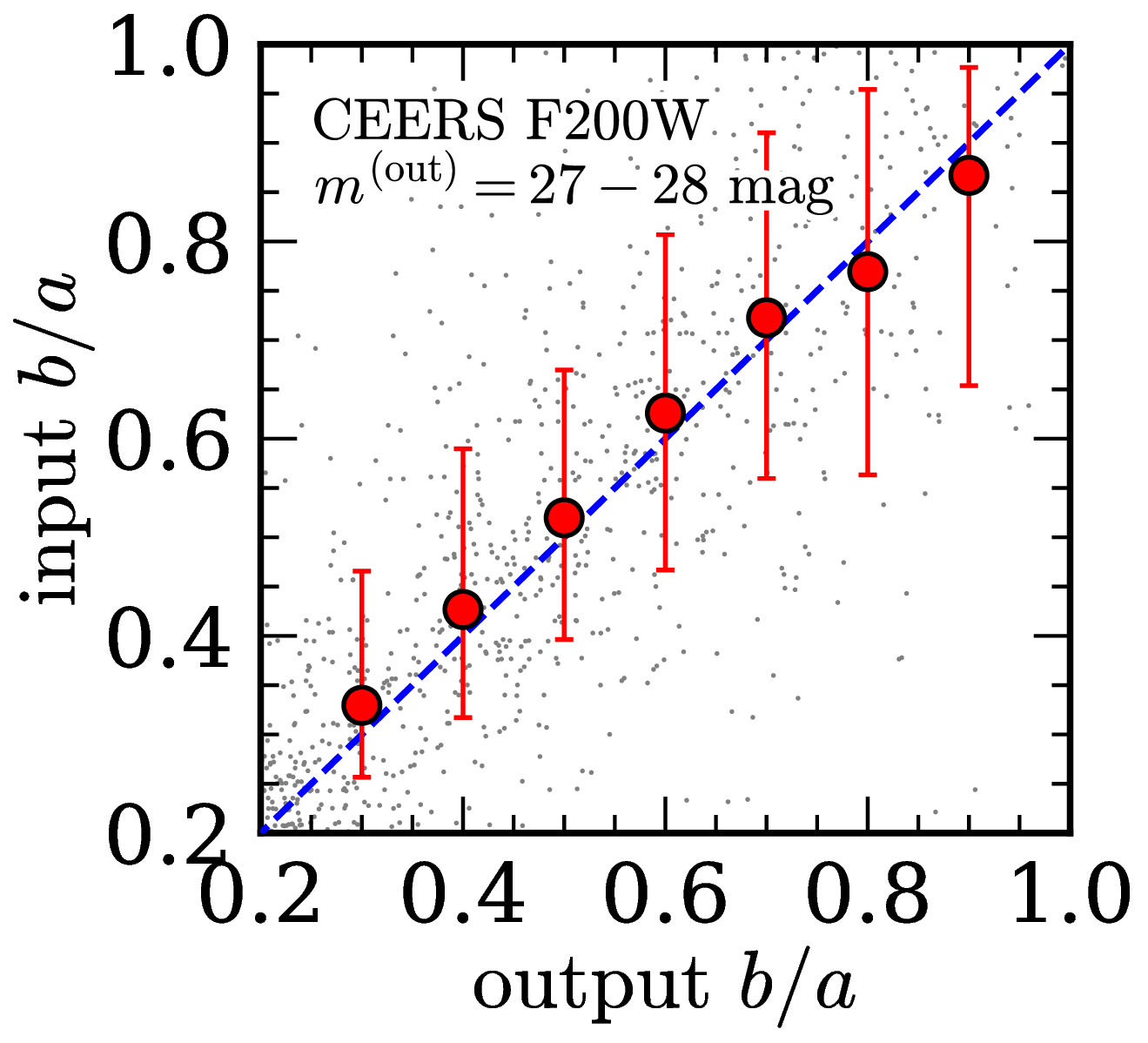}
   \includegraphics[width=0.24\textwidth]{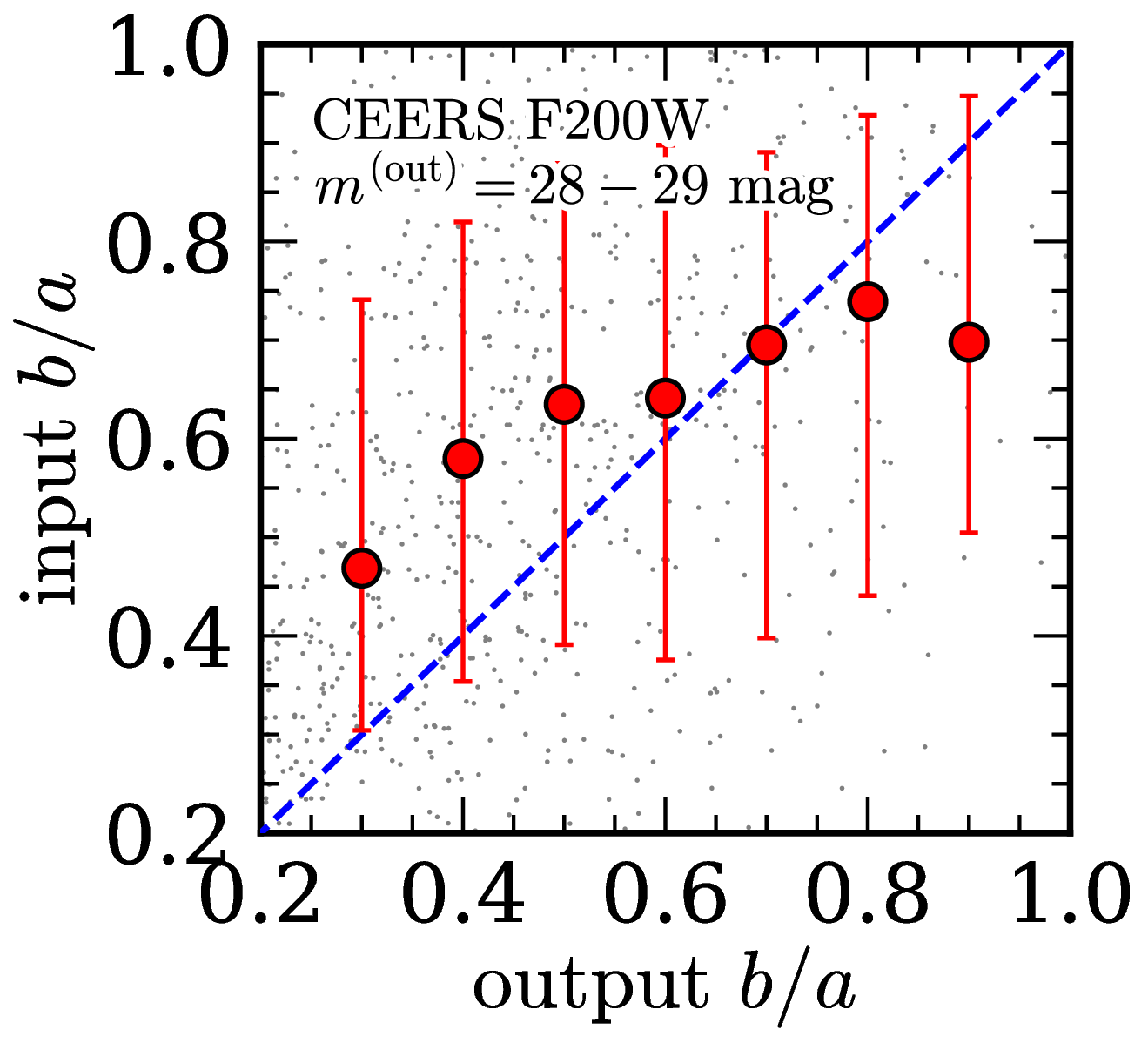}
\caption{Input axis ratio  vs. output axis ratio 
based on our GALFIT Monte Carlo simulations. 
From top to bottom, the results for 
the JADES Deep field in F200W and F277W, 
the JADES GOODS-N field in F200W 
and the CEERS field in F200W 
are presented. 
From left to right, the results for fainter magnitude ranges are shown.
The red filled circles and associated error bars are 
the median difference between the input and output axis ratios 
and their 68 percentile ranges, respectively. 
The gray dots depict the results for individual simulated objects. 
The blue dashed line marks 
the one-to-one correspondence between input and output axis ratios. 
On average, the output axis ratios demonstrate general agreement with their input values.
}
\label{fig:input_output_ba}
\end{center}
\end{figure*}

\section{Surface Brightness Profile Fitting} \label{sec:SBprofile_fitting}

To measure the half-light radii of the $z\sim10$--$16$ galaxies in the CEERS and JADES fields 
summarized in Section \ref{sec:data}, 
we fit their two-dimensional surface brightness profiles with a S{\'e}rsic profile (\citealt{1968adga.book.....S}).
The S{\'e}rsic profile is defined as 
\begin{equation}
\Sigma (r)
	= \Sigma_e \exp \left( - b_n \left[ \left( \frac{r}{r_e} \right)^{1/n} -1 \right] \right), 
\end{equation}
where $\Sigma_e$ is the surface brightness at the half-light radius $r_{\rm e}$, 
$n$ is the S{\'e}rsic index, 
and $b_n$ is a constant determined such that the flux within $r_{\rm e}$ equals half of the total flux.
We perform surface brightness profile fitting using GALFIT version 3 
(\citealt{2002AJ....124..266P}; \citealt{2010AJ....139.2097P}), 
a software package that optimizes the fit 
by convolving the surface brightness distribution with the PSF profile 
and minimizing $\chi^2$ using the Levenberg-Marquardt algorithm. 
The output parameters from GALFIT include 
the central coordinates, total magnitude, semi-major axis radius $a$, 
S{\'e}rsic index $n$, axis ratio $b/a$, and position angle. 
For each object, we calculate the circularized half-light radius as $r_{\rm e} = a \sqrt{b/a}$, 
which is commonly used as the galaxy size in previous studies 
(e.g., \citealt{2012ApJ...756L..12M}; \citealt{2012ApJ...746..162N}; \citealt{2013ApJ...777..155O}; 
\citealt{2015ApJS..219...15S}; \citealt{2018ApJ...855....4K}).
The initial parameters for GALFIT are derived from running SExtractor on the images. 
All the parameters except the S{\'e}rsic index are treated as free parameters during the fitting process. 
To weight individual pixels during the profile fitting, 
we use noise images that are calculated as the inverse square root of the weight maps. 
Additionally, segmentation images generated by SExtractor are used 
to mask nearby objects during profile fitting to focus on the target objects of interest.

We fix the S{\'e}rsic index at $n=1.5$. 
This value corresponds to the median S{\'e}rsic index 
determined for star-forming galaxies with UV luminosities 
comparable to our objects reported in the literature (\citealt{2015ApJS..219...15S}). 
Indeed, when we allow the S{\'e}rsic index to vary as a free parameter in the GALFIT profile fitting 
for $12$ spectroscopically confirmed galaxies with $\mathrm{S/N} > 10$ 
in our sample,\footnote{Our analysis 
of the varying S{\'e}rsic index is limited to spectroscopically confirmed galaxies 
to ensure high reliability and minimize contamination from, e.g., low-$z$ interlopers. 
While these galaxies tend to be brighter within our overall sample, 
it would be important to note that 
reliably determining the S{\'e}rsic index requires a sufficiently high S/N, 
meaning that such measurements are inherently biased toward brighter objects, 
even when galaxies without spectroscopic confirmation are included.}
the fitting converges successfully for 9 galaxies and the resulting S{\'e}rsic index values yield 
a mean value of $n=1.3 \pm 0.6$ 
(see also, \citealt{2023ApJ...957...46M}; \citealt{2024ApJ...960..104S}; 
\citealt{2025arXiv250102956M}).\footnote{The median value is also $n=1.3$.}
Figure \ref{fig:histo_Sersic} shows the histogram of the obtained S{\'e}rsic index values. 
However, as noted in our previous work, 
the currently available data are not sufficient to obtain reliable measurements 
on the S{\'e}rsic index and the other morphological parameters simultaneously 
for high-$z$ galaxies (see Section 3.1 of \citealt{2024PASJ...76..219O}). 
Therefore, following the approach of previous high-$z$ galaxy size studies, 
we perform surface brightness profile fitting with a fixed S{\'e}rsic index.
Our choice to fix the S{\'e}rsic index at $n=1.5$ 
represents a pragmatic approach. 
We acknowledge that while S{\'e}rsic index values for star-forming galaxies 
are generally found around $n=1$ (corresponding to an exponential disk) 
with a somewhat broad distribution 
(e.g., \citealt{2023ApJ...957...46M}; \citealt{2024ApJ...960..104S}; 
\citealt{2025arXiv250102956M}), 
the primary goal of this study 
is to robustly constrain $r_{\rm e}$ and total magnitude values for our sample. 
Fixing the S{\'e}rsic index helps to mitigate degeneracies between fitting parameters, 
thereby enhancing the reliability of the derived values, 
particularly for faint sources where the S/N is limited.

As in previous studies (\citealt{2013ApJ...777..155O}; \citealt{2015ApJS..219...15S}), 
we perform Monte Carlo simulations to quantify both statistical and systematic uncertainties 
in the GALFIT profile fitting. 
For CEERS, we adopt the results of the Monte Carlo simulations conducted in \cite{2023ApJ...951...72O}. 
For JADES, we run Monte Carlo simulations for JADES Deep and JADES GOODS-N. 
For JADES Medium, we shift the results of JADES Deep 
according to the difference in the limiting magnitudes.

First, we generate galaxy images using GALFIT, 
fixing the S{\'e}rsic index at $n=1.5$. 
The half-light radius is randomly selected from a range of 
$0.3$ pix to $10.0$ pix, 
while the axis ratio is randomly chosen from values between $0.2$ and $1.0$, 
and the total magnitude is chosen between $25.5$ mag and $31.0$ mag. 
These simulated galaxy images are convolved with the PSF images 
created in Section \ref{sec:data} and inserted into blank regions of the actual NIRCam images. 
We then perform surface brightness profile fitting with GALFIT in the same way 
as for real high-$z$ galaxies to measure the sizes and total magnitudes.

The comparisons between the measured sizes and the input sizes are shown in Figure \ref{fig:input_output_re}, 
while the comparisons between the measured total magnitudes and the input total magnitudes 
are presented in Figure \ref{fig:input_output_mag}. 
Figure \ref{fig:input_output_re} demonstrates that 
even when the galaxy sizes are smaller than the PSF FWHMs 
(about $2.5$ pixels in F200W and $4.0$ pixels in F277W), 
the measured sizes are, on average, reasonably consistent with the input sizes. 
In agreement with the results of \cite{2023ApJ...951...72O}, 
our Monte Carlo simulations confirm that the half-light radius is systematically underestimated for faint objects. 
For instance, examining the results for the F200W image in the JADES Deep field, 
the median difference between input and output sizes is nearly zero 
for the brightest magnitude range of $26$--$27$ mag, 
whereas it is approximately $15$--$25${\%} for the faintest range of $29$--$30$ mag.
Furthermore, statistical uncertainties increase for fainter and larger objects.
For example, for the F200W image in the JADES Deep field, 
the statistical uncertainties are several percent in the brightest magnitude range, 
but increase to about $40$--$100${\%} in the faintest magnitude range.
We account for these systematic and statistical uncertainties in the measurements of real high-$z$ galaxies. 
Briefly, for each high-$z$ galaxy in our sample, 
we correct the measured size by the median difference between the input and output sizes in the magnitude bin 
corresponding to its output total magnitude.
Additionally, we take the 68th percentile difference between the input and output sizes in each bin 
as the statistical uncertainty. 
A similar approach is applied to the total magnitudes 
(see Section 3 of \citealt{2023ApJ...951...72O} for details).

In addition to the circularized half-light radius and total magnitude, 
we also compare the input and output values for the axis ratio $b/a$. 
As shown in Figure \ref{fig:input_output_ba}, the measured $b/a$ values are, on average, 
consistent with the input $b/a$ across all input values. 
Similar to the sizes and total magnitudes, the statistical uncertainties for the axis ratio 
increase with decreasing brightness. 
For example, considering the results for the F200W image in the JADES Deep field, 
the statistical uncertainties are about several percent in the brightest magnitude range, 
but increase to about $10$--$50${\%} for the $28$--$29$ mag range. 
For the axis ratio, as with sizes and total magnitudes, 
we account for these systematic and statistical uncertainties in the measurements of real high-$z$ galaxies.

\begin{deluxetable*}{lccccccc} 
\tablecolumns{8} 
\tablewidth{0pt} 
\tablecaption{Surface Brightness Profile Fitting Results 
\label{tab:GALFITresults}}
\tablehead{
    \colhead{ID} 
    &  \colhead{$z_{\rm photo}$}
    &  \colhead{$z_{\rm spec}$}
    &  \colhead{S/N}
    &  \colhead{$m_{\rm UV}$}
    &  \colhead{$M_{\rm UV}$}
    &  \colhead{$r_{\rm e}$}
    &  \colhead{$b/a$}
\\
    \colhead{ } 
    &  \colhead{ }
    &  \colhead{ }
    &  \colhead{ }
    &  \colhead{(mag)}
    &  \colhead{(mag)}
    &  \colhead{(kpc)}
    &  \colhead{ }
\\
    \colhead{(1)} 
    &  \colhead{(2)}
    &  \colhead{(3)}
    &  \colhead{(4)}
    &  \colhead{(5)}
    &  \colhead{(6)}
    &  \colhead{(7)}
    &  \colhead{(8)}
}
\startdata 
\multicolumn{8}{c}{CEERS} \\  
                 CEERS-11384 &   11.53 &              11.043$^{+0.003}_{-0.003}$ & $13$ &  27.53$^{+0.20}_{-0.21}$ &  $-20.09^{+0.20}_{-0.21}$ &  0.336$^{+0.079}_{-0.059}$ &  $0.35^{+0.14}_{-0.07}$ \\
                 CEERS-16943$^{\dagger 1}$ &   11.08 &  11.416$^{+0.005}_{-0.005}$ & $16$ &   27.16$^{+0.13}_{-0.13}$ &  $-20.51^{+0.13}_{-0.13}$ &  0.453$^{+0.120}_{-0.082}$ &  $0.67^{+0.18}_{-0.16}$ \\
                 CEERS-35590 &   10.15 &                 10.01$^{+0.14}_{-0.19}$ & $15$ &   27.61$^{+0.12}_{-0.14}$ &  $-19.87^{+0.12}_{-0.14}$ &  0.167$^{+0.056}_{-0.049}$ &  $0.71^{+0.18}_{-0.16}$ \\
                 CEERS-54306 &   11.23 &                                     --- & $11$ &   27.93$^{+0.21}_{-0.21}$ &  $-19.72^{+0.21}_{-0.21}$ &  0.089$^{+0.052}_{-0.045}$ &  $0.75^{+0.19}_{-0.16}$ \\
                 CEERS-85546 &   11.08 &                                     --- & $11$ &   27.98$^{+0.21}_{-0.21}$ &  $-19.65^{+0.21}_{-0.21}$ &  0.106$^{+0.129}_{-0.068}$ &  $0.76^{+0.19}_{-0.30}$ \\
                 CEERS-98518$^{\dagger 2}$ &   10.09 &                                     --- & $19$ &   27.46$^{+0.12}_{-0.14}$ &  $-20.03^{+0.12}_{-0.14}$ &  0.121$^{+0.056}_{-0.048}$ &  $0.64^{+0.18}_{-0.16}$ \\
                 CEERS-99715 &    9.76 &                  9.77$^{+0.37}_{-0.29}$ & $13$ &   28.15$^{+0.27}_{-0.29}$ &  $-19.30^{+0.27}_{-0.29}$ &  0.159$^{+0.141}_{-0.074}$ &  $0.57^{+0.27}_{-0.16}$ \\
                CEERS-99715b &    9.76 &                  9.77$^{+0.37}_{-0.29}$ & $13$ &   27.47$^{+0.12}_{-0.14}$ &  $-19.98^{+0.12}_{-0.14}$ &  0.340$^{+0.086}_{-0.064}$ &  $0.78^{+0.18}_{-0.21}$ \\
\multicolumn{8}{c}{JADES Deep} \\  
              JADES-GS-z14-0 &     --- &                  $14.1793^{+0.0007}_{-0.0007}$ & $46$ &   26.91$^{+0.02}_{-0.02}$ &  $-21.08^{+0.02}_{-0.02}$ &  0.288$^{+0.005}_{-0.007}$ &  $0.48^{+0.35}_{-0.18}$ \\
              JADES-GS-z14-1 &     --- &                  13.9$^{+0.17}_{-0.17}$ & $10$ &   28.85$^{+0.18}_{-0.14}$ &  $-19.11^{+0.18}_{-0.14}$ &  0.158$^{+0.017}_{-0.020}$ &  $0.20^{+0.02}_{-0.02}$ \\
   JADES-GS-53.14988-27.7765$^{\dagger 3}$ &   13.41 &                                  13.274 & $10$ &   29.27$^{+0.18}_{-0.14}$ &  $-18.62^{+0.18}_{-0.14}$ &  0.061$^{+0.084}_{-0.043}$ &  $0.29^{+0.35}_{-0.18}$ \\
  JADES-GS-53.16476-27.77463$^{\dagger 4}$ &   12.31 &                                  11.671 & $18$ &   28.20$^{+0.08}_{-0.06}$ &  $-19.50^{+0.08}_{-0.06}$ &  0.108$^{+0.023}_{-0.024}$ &  $0.56^{+0.19}_{-0.14}$ \\
  JADES-GS-53.16635-27.82156$^{\dagger 5}$ &   12.46 &                                  12.535 & $10$ &   28.79$^{+0.23}_{-0.18}$ &  $-19.01^{+0.23}_{-0.18}$ &  0.148$^{+0.022}_{-0.023}$ &  $0.76^{+0.16}_{-0.18}$ \\
  JADES-GS-53.16736-27.80751 &    9.54 &                                   9.691 & $23$ &   28.10$^{+0.08}_{-0.06}$ &  $-19.33^{+0.08}_{-0.06}$ &  0.105$^{+0.026}_{-0.028}$ &  $0.75^{+0.16}_{-0.18}$ \\
  JADES-GS-53.17551-27.78064 &    9.66 &                                   9.712 & $29$ &   27.78$^{+0.08}_{-0.06}$ &  $-19.66^{+0.08}_{-0.06}$ &  0.122$^{+0.009}_{-0.011}$ &  $0.65^{+0.07}_{-0.06}$ \\
  JADES-GS-53.05177-27.88727 &    9.51 &                                     --- & $11$ &   28.79$^{+0.23}_{-0.18}$ &  $-18.61^{+0.23}_{-0.18}$ &  0.138$^{+0.027}_{-0.028}$ &  $0.81^{+0.16}_{-0.18}$ \\
  JADES-GS-53.16594-27.83424 &   10.75 &                                     --- & $25$ &   27.63$^{+0.06}_{-0.04}$ &  $-19.96^{+0.06}_{-0.04}$ &  0.194$^{+0.009}_{-0.010}$ &  $0.36^{+0.04}_{-0.04}$ \\
  JADES-GS-53.19051-27.74982 &   12.08 &                                     --- & $17$ &   28.47$^{+0.15}_{-0.11}$ &  $-19.29^{+0.15}_{-0.11}$ &  0.023$^{+0.023}_{-0.023}$ &  $0.20^{+0.16}_{-0.08}$ \\
\multicolumn{8}{c}{JADES Medium} \\  
  JADES-GS-53.05511-27.84555 &    9.65 &                                     --- & $15$ &   27.88$^{+0.15}_{-0.11}$ &  $-19.54^{+0.15}_{-0.11}$ &  0.150$^{+0.027}_{-0.028}$ &  $0.53^{+0.19}_{-0.14}$ \\
  JADES-GS-53.07076-27.86544 &   11.03 &                                     --- & $11$ &   28.32$^{+0.23}_{-0.18}$ &  $-19.31^{+0.23}_{-0.18}$ &  0.077$^{+0.024}_{-0.025}$ &  $0.49^{+0.16}_{-0.11}$ \\
   JADES-GS-53.10798-27.8776 &   10.97 &                                     --- & $15$ &   27.97$^{+0.15}_{-0.11}$ &  $-19.65^{+0.15}_{-0.11}$ &  0.077$^{+0.024}_{-0.025}$ &  $0.57^{+0.19}_{-0.14}$ \\
   JADES-GS-53.1216-27.90813 &   11.66 &                                     --- & $13$ &   27.92$^{+0.15}_{-0.11}$ &  $-19.78^{+0.15}_{-0.11}$ &  0.153$^{+0.023}_{-0.024}$ &  $0.36^{+0.16}_{-0.08}$ \\
  JADES-GS-53.13918-27.84849 &   10.45 &                                     --- & $12$ &   27.87$^{+0.15}_{-0.11}$ &  $-19.68^{+0.15}_{-0.11}$ &  0.192$^{+0.025}_{-0.026}$ &  $0.36^{+0.16}_{-0.08}$ \\
  JADES-GS-53.07597-27.80654 &   11.27 &                                     --- & $11$ &   28.55$^{+0.23}_{-0.18}$ &  $-19.10^{+0.23}_{-0.18}$ &  0.118$^{+0.079}_{-0.043}$ &  $0.78^{+0.25}_{-0.29}$ \\
 JADES-GS-53.07597-27.80654b &   11.27 &                                     --- & $11$ &   28.32$^{+0.23}_{-0.18}$ &  $-19.34^{+0.23}_{-0.18}$ &  0.132$^{+0.024}_{-0.025}$ &  $0.43^{+0.16}_{-0.11}$ \\
\multicolumn{8}{c}{JADES GOODS-N} \\  
 JADES-GN-189.10604+62.24204$^{\dagger 6}$ &   11.00 &                                  10.603 & $102$ &   25.97$^{+0.03}_{-0.03}$ &  $-21.59^{+0.03}_{-0.03}$ &  0.055$^{+0.005}_{-0.006}$ &  $0.83^{+0.06}_{-0.06}$ \\
 JADES-GN-189.21769+62.19949 &   10.31 &                                     --- & $22$ &   27.54$^{+0.09}_{-0.06}$ &  $-19.98^{+0.09}_{-0.06}$ &  0.122$^{+0.010}_{-0.018}$ &  $0.44^{+0.12}_{-0.09}$ \\
 JADES-GN-189.25354+62.19416 &    9.63 &                                     --- & $14$ &   28.07$^{+0.16}_{-0.11}$ &  $-19.35^{+0.16}_{-0.11}$ &  0.137$^{+0.053}_{-0.040}$ &  $0.46^{+0.32}_{-0.13}$ \\
 JADES-GN-189.26202+62.20109 &    9.69 &                                     --- & $12$ &   27.52$^{+0.09}_{-0.06}$ &  $-19.91^{+0.09}_{-0.06}$ &  0.369$^{+0.030}_{-0.037}$ &  $0.31^{+0.08}_{-0.06}$ \\
 JADES-GN-189.26804+62.15158 &   10.99 &                                     --- & $12$ &   28.19$^{+0.16}_{-0.11}$ &  $-19.43^{+0.16}_{-0.11}$ &  0.118$^{+0.048}_{-0.036}$ &  $0.40^{+0.32}_{-0.13}$ \\
 JADES-GN-189.28903+62.22905 &   10.39 &                                     --- & $15$ &   27.96$^{+0.16}_{-0.11}$ &  $-19.58^{+0.16}_{-0.11}$ &  0.137$^{+0.010}_{-0.018}$ &  $0.69^{+0.17}_{-0.12}$ \\
 JADES-GN-189.30296+62.21075 &   10.30 &                                     --- & $15$ &   28.14$^{+0.16}_{-0.11}$ &  $-19.38^{+0.16}_{-0.11}$ &  0.029$^{+0.051}_{-0.029}$ &  $0.25^{+0.32}_{-0.13}$ \\
\enddata 
\tablecomments{The systematic effects and statistical uncertainties 
in these obtained size and total magnitude measurements are considered 
based on our MC simulation results.  
(1) Object ID. 
(2) Photometric redshift (\citealt{2024ApJ...969L...2F}; \citealt{2024ApJ...964...71H}). 
(3) Spectroscopic redshift (\citealt{2024ApJ...969L...2F}; \citealt{2024ApJ...964...71H}; 
\citealt{2024Natur.633..318C}; \citealt{2024arXiv240920549S}; \citealt{2024arXiv240920533C}). 
(4) S/N values based on apparent magnitudes measured within $0\farcs2$ diameter circular apertures. 
For $z < 13$, F200W images are used, and for $z > 13$, F277W images are used. 
(5) Total apparent UV magnitude measured by GALFIT. 
(6) Total absolute UV magnitude, calculated using $z_{\rm spec}$ if available, otherwise using $z_{\rm photo}$. 
(7) Circularized half-light radius $r_{\rm e} = a \sqrt{b/a}$. 
(8) Axis ratio. 
}
\vspace{-0.5em}
\tablenotetext{$^{\dagger 1}$}{
Maisie's Galaxy (\citealt{2022ApJ...940L..55F}).}
\vspace{-0.5em}
\tablenotetext{$^{\dagger 2}$}{
CEERS 23931 selected as a little red dot (\citealt{2024arXiv240403576K}).}
\vspace{-0.5em}
\tablenotetext{$^{\dagger 3}$}{
GS-z13-0 (\citealt{2023NatAs...7..622C}).}
\vspace{-0.5em}
\tablenotetext{$^{\dagger 4}$}{
UDFj-39546284 (\citealt{2011Natur.469..504B}); 
UDF12-3954-6284 (\citealt{2013ApJ...763L...7E}); 
GS-z11-0 (\citealt{2023NatAs...7..622C}).}
\vspace{-0.5em}
\tablenotetext{$^{\dagger 5}$}{
GS-z12-0 (\citealt{2023NatAs...7..622C}).}
\vspace{-0.5em}
\tablenotetext{$^{\dagger 6}$}{
GN-z11 (\citealt{2016ApJ...819..129O}).}
\end{deluxetable*} 

\section{Results} \label{sec:results}

\begin{figure}
\begin{center}
	\includegraphics[width=0.5\textwidth]{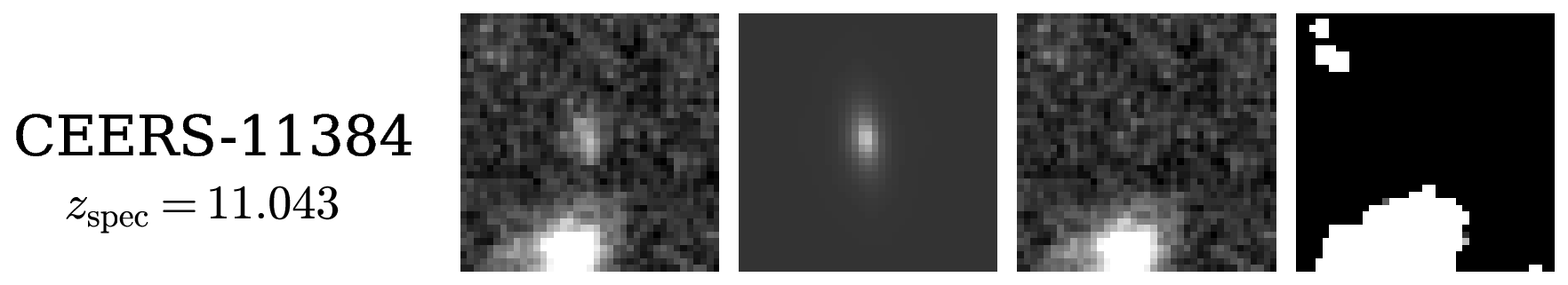}
	\includegraphics[width=0.5\textwidth]{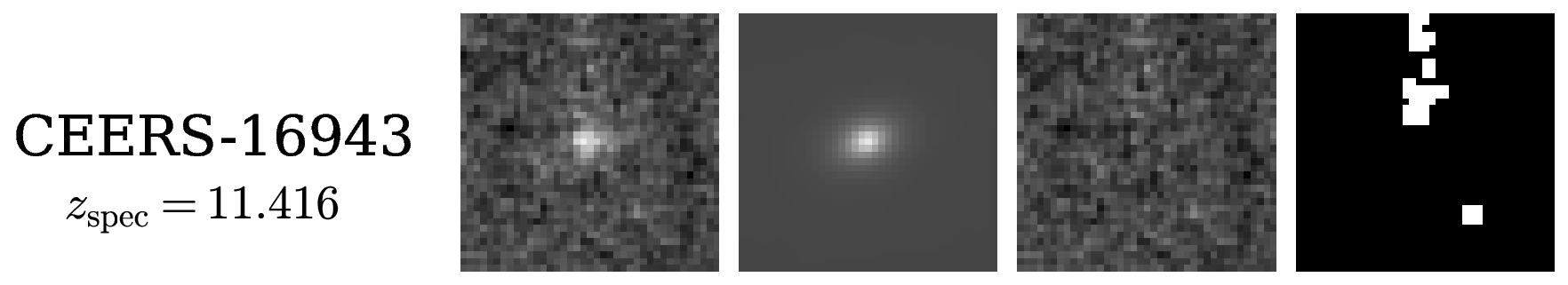}
	\includegraphics[width=0.5\textwidth]{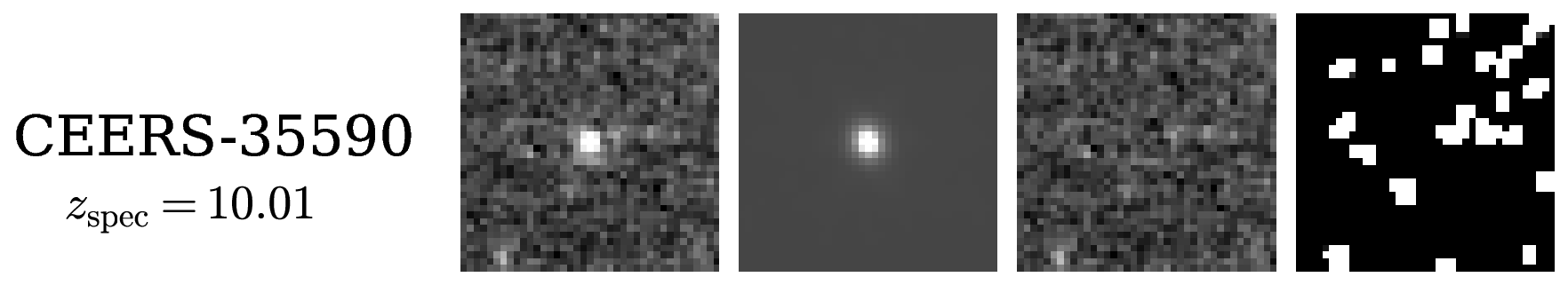}
	\includegraphics[width=0.5\textwidth]{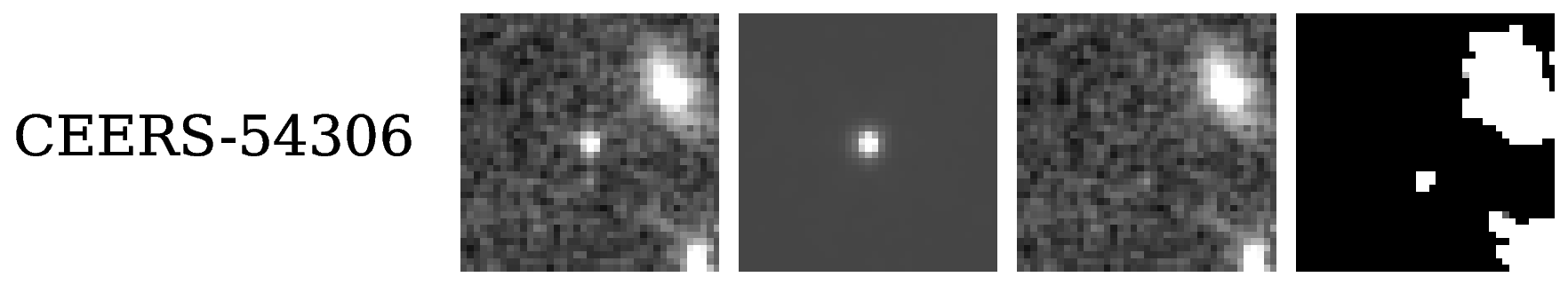}
	\includegraphics[width=0.5\textwidth]{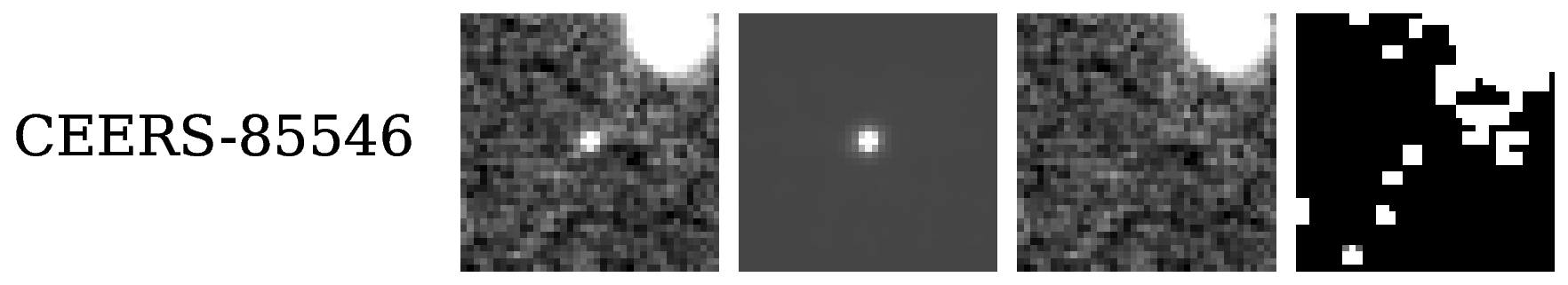}
	\includegraphics[width=0.5\textwidth]{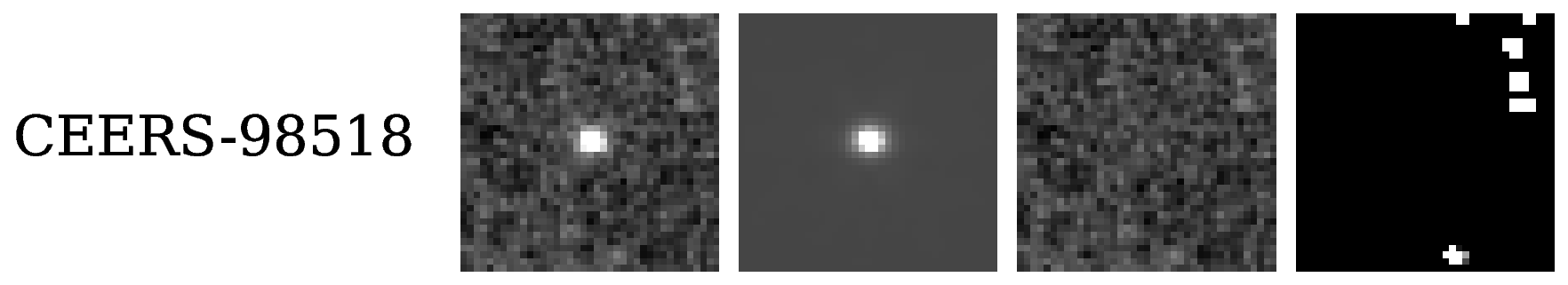}
	\includegraphics[width=0.5\textwidth]{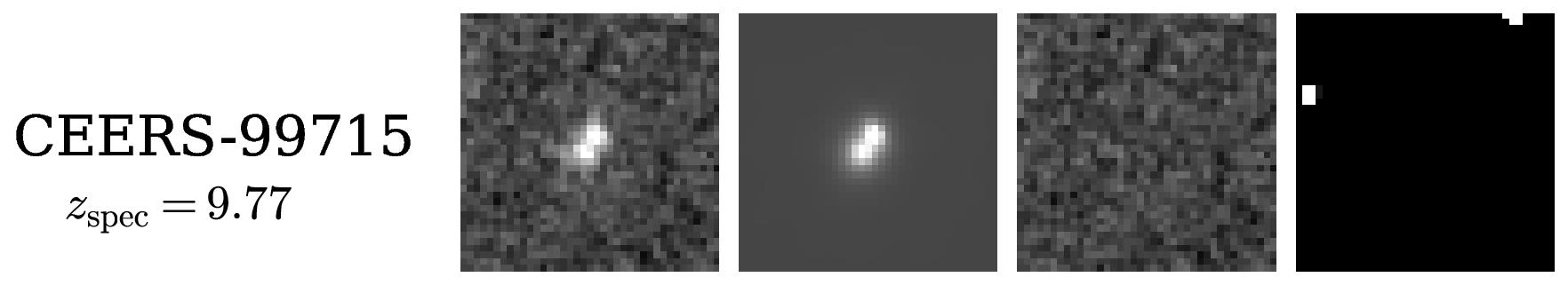}
\caption{
S{\'e}rsic profile fitting results for high-$z$ galaxies found in the CEERS field. 
Spectroscopically confirmed galaxies are labeled with their $z_{\rm spec}$ values. 
From left to right, the $1\farcs5 \times 1\farcs5$ cutouts of the original image, 
the best-fit S{\'e}rsic model profile images, 
the residual images that are made by subtracting the best-fit images from the original ones, 
and the segmentation maps used for masking all the neighboring objects during the profile fitting are presented. 
}
\label{fig:fit_results_CEERS}
\end{center}
\end{figure}

\begin{figure}[h]
\begin{center}
	\includegraphics[width=0.5\textwidth]{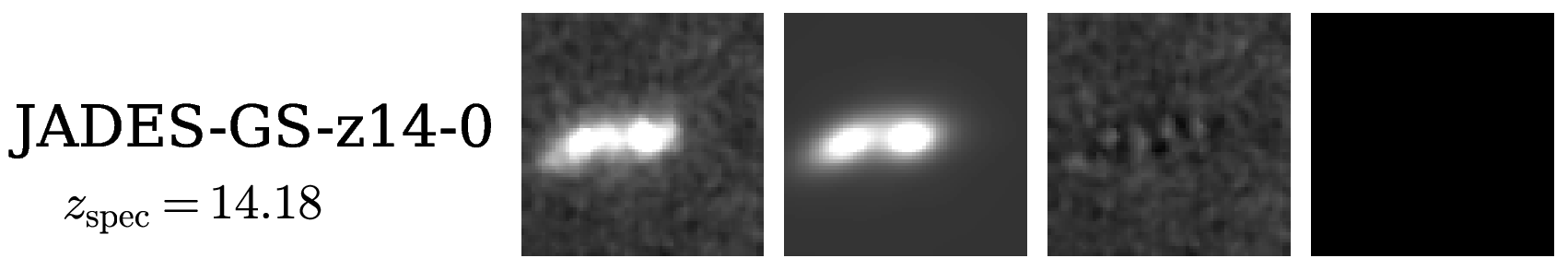}
	\includegraphics[width=0.5\textwidth]{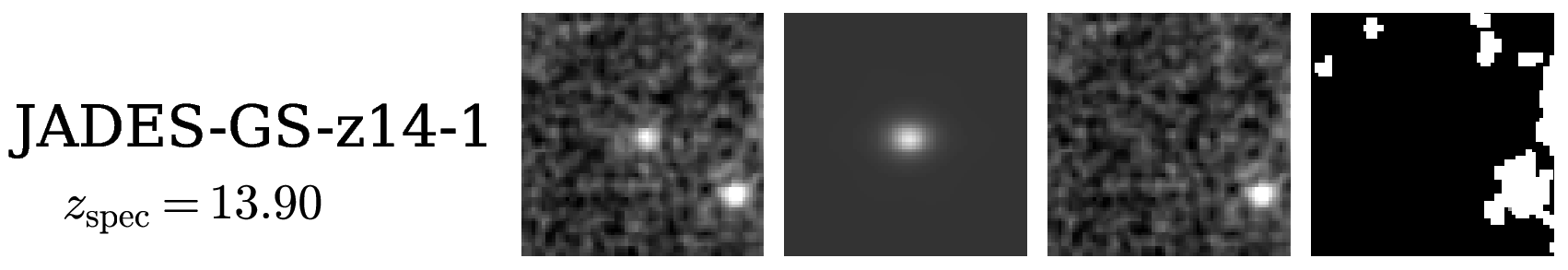}
	\includegraphics[width=0.5\textwidth]{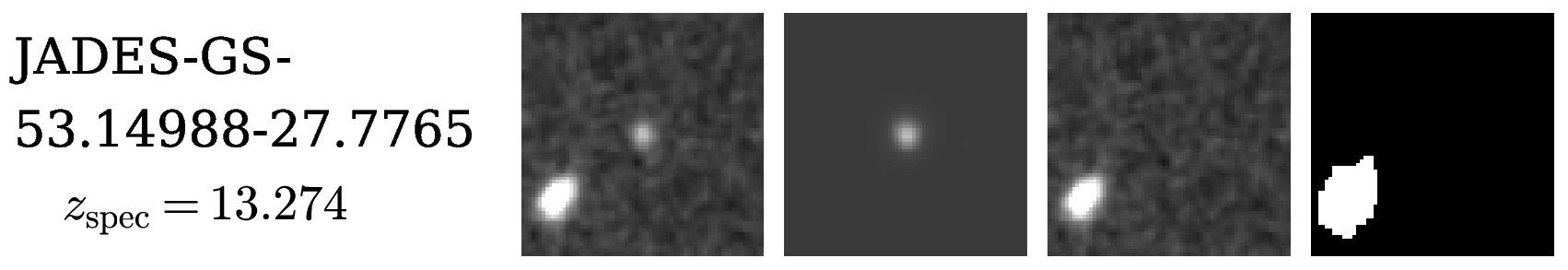}
	\includegraphics[width=0.5\textwidth]{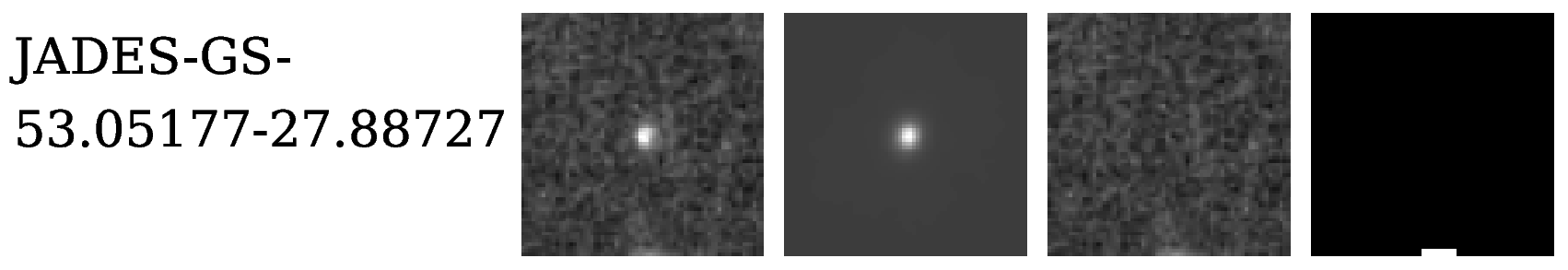}
	\includegraphics[width=0.5\textwidth]{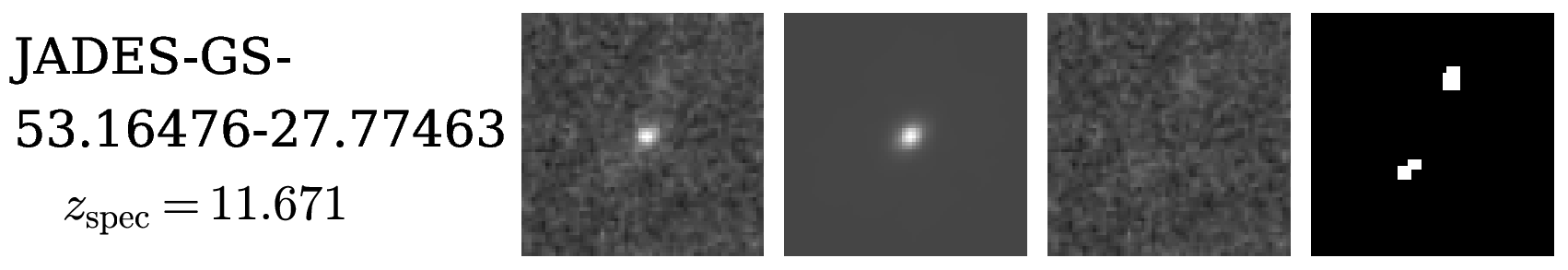}
	\includegraphics[width=0.5\textwidth]{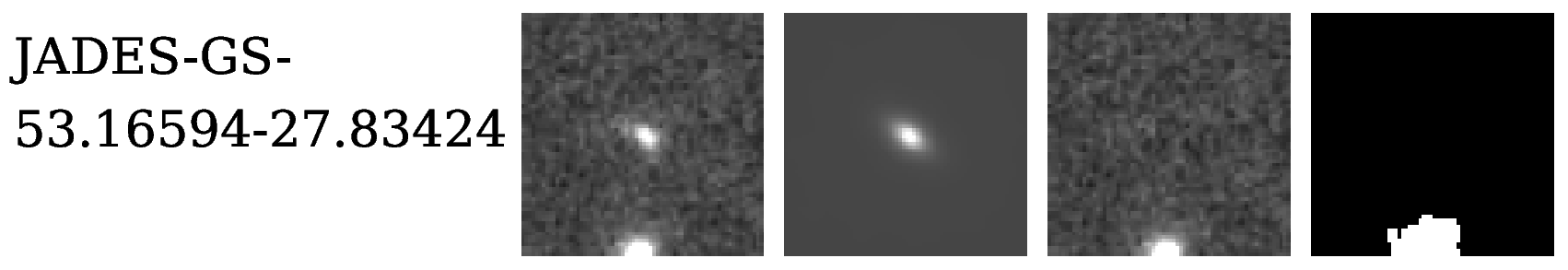}
	\includegraphics[width=0.5\textwidth]{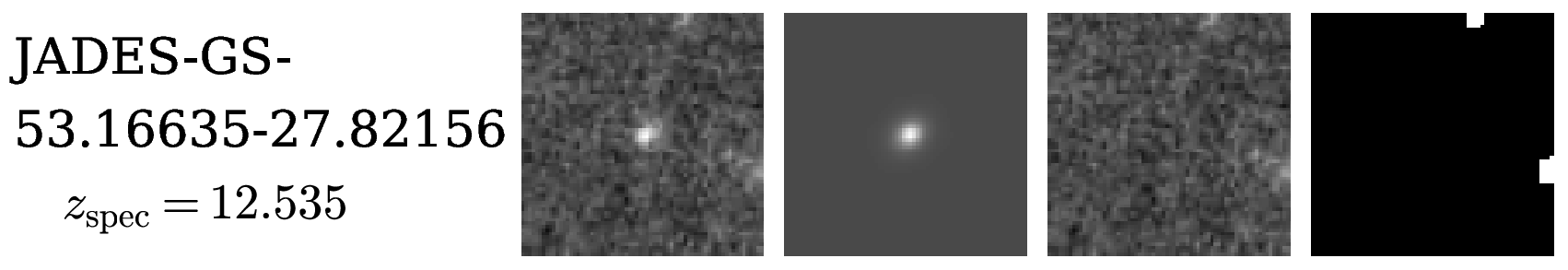}
	\includegraphics[width=0.5\textwidth]{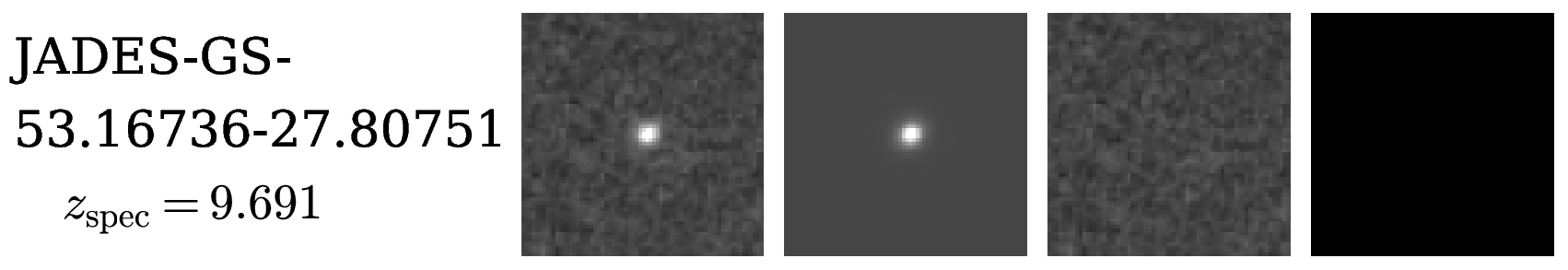}
	\includegraphics[width=0.5\textwidth]{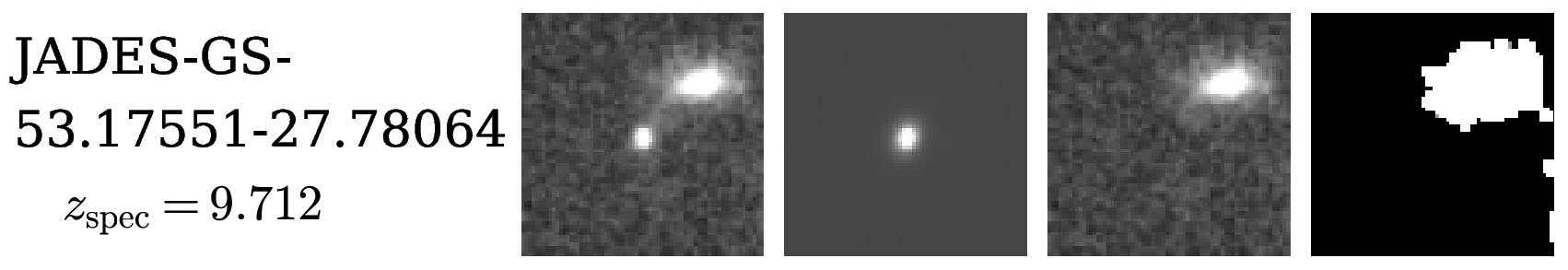}
	\includegraphics[width=0.5\textwidth]{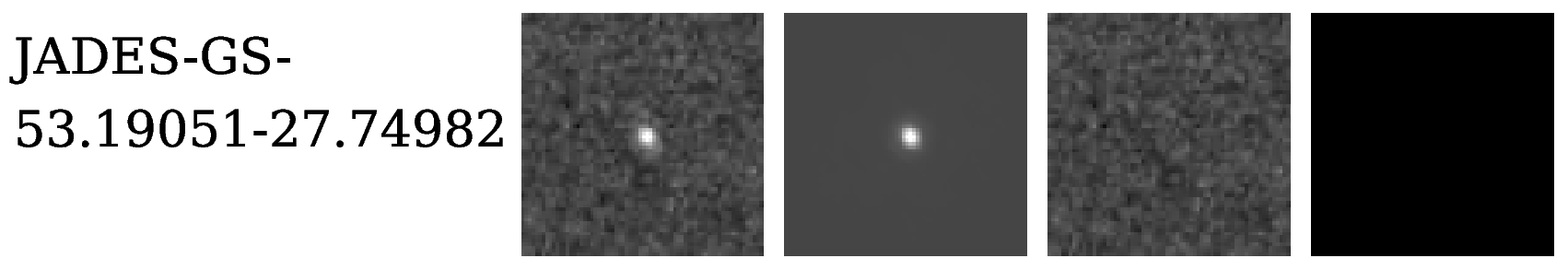}
\caption{
Same as Figure \ref{fig:fit_results_CEERS}, but for JADES Deep. 
}
\label{fig:fit_results_JADES}
\end{center}
\end{figure}

\begin{figure}[h]
\begin{center}
	\includegraphics[width=0.5\textwidth]{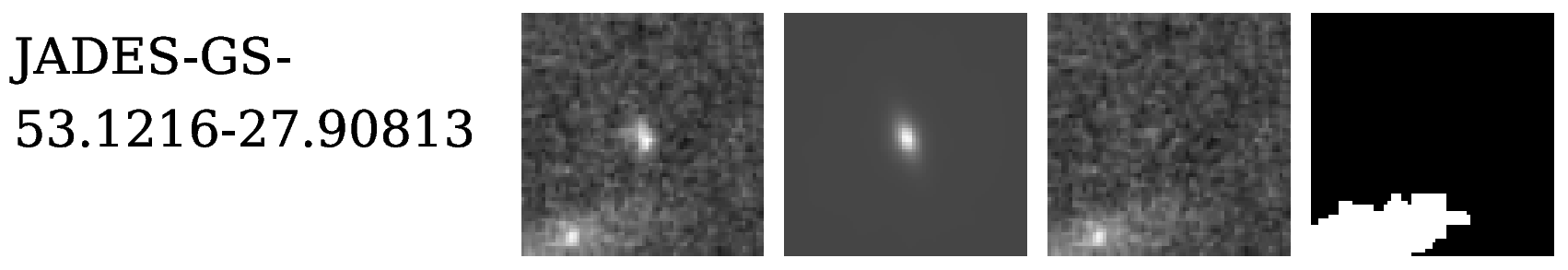}
	\includegraphics[width=0.5\textwidth]{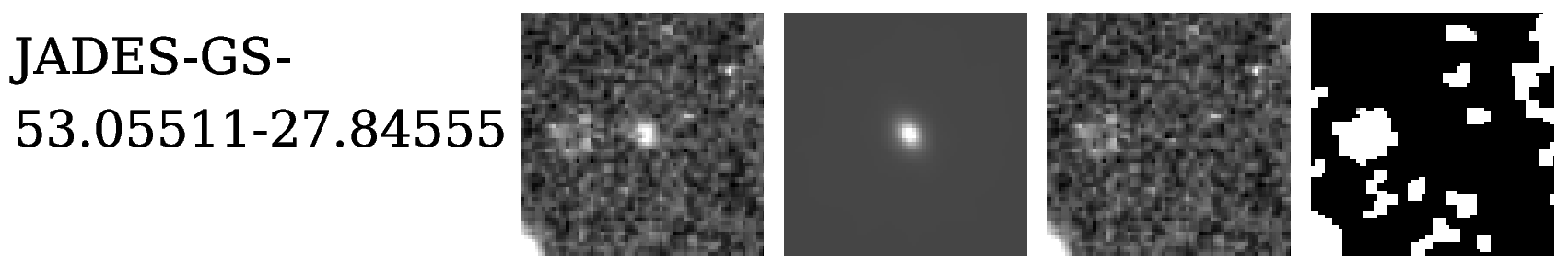}
	\includegraphics[width=0.5\textwidth]{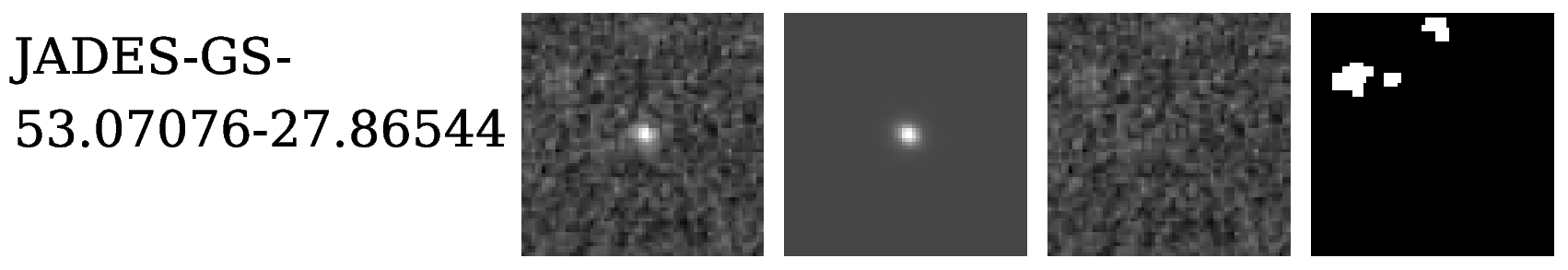}
	\includegraphics[width=0.5\textwidth]{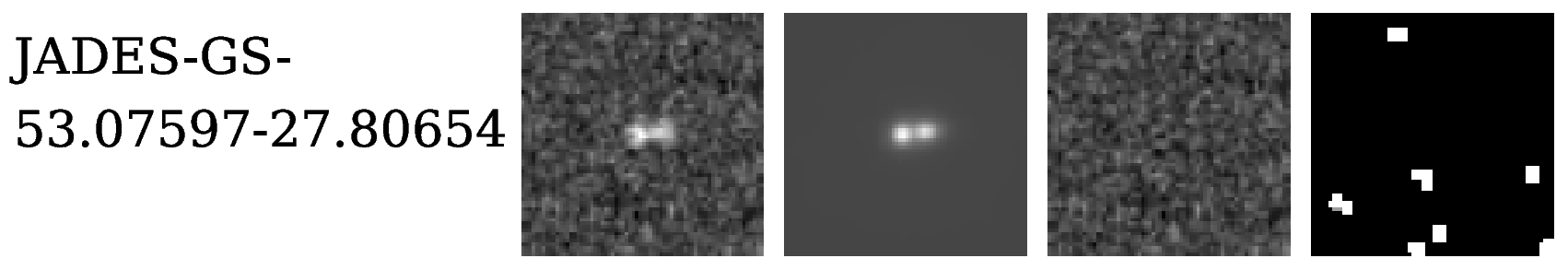}
	\includegraphics[width=0.5\textwidth]{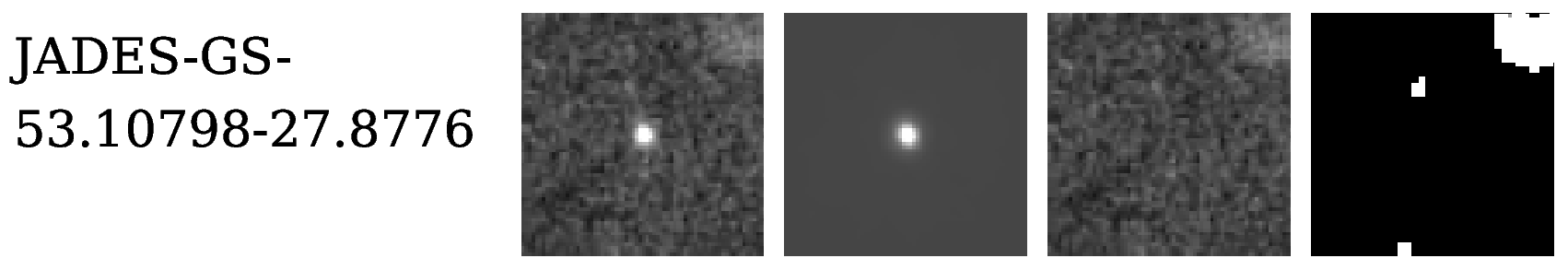}
	\includegraphics[width=0.5\textwidth]{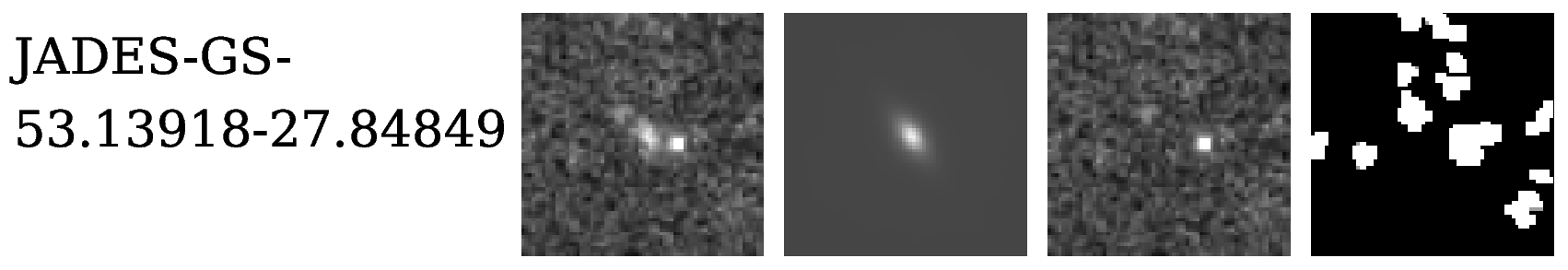}
\caption{
Same as Figure \ref{fig:fit_results_CEERS}, but for JADES Medium. 
}
\label{fig:fit_results_JADES_medium}
\end{center}
\end{figure}

\begin{figure}[h]
\begin{center}
	\includegraphics[width=0.5\textwidth]{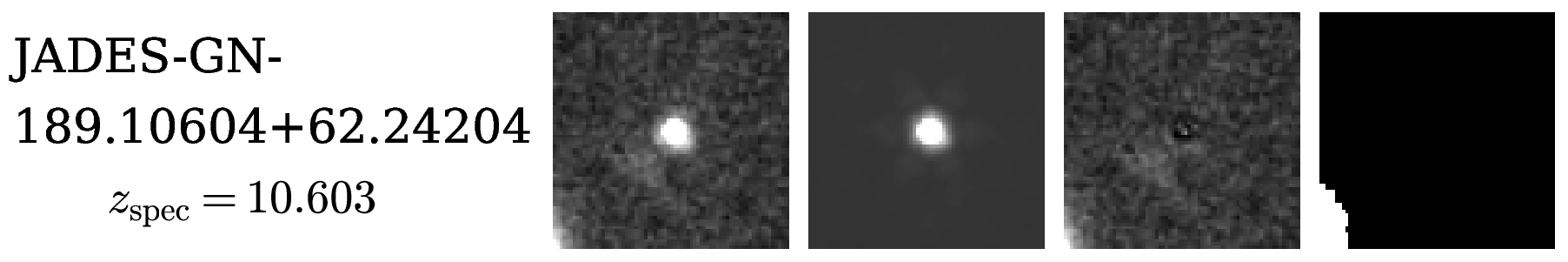}
	\includegraphics[width=0.5\textwidth]{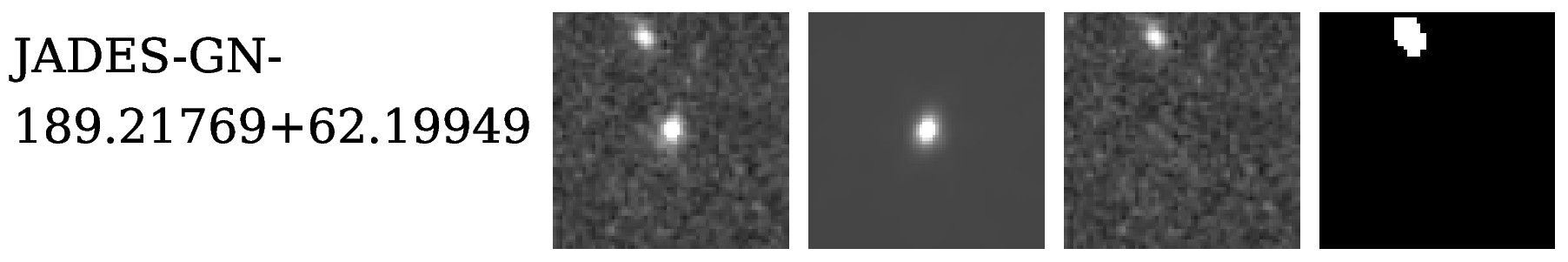}
	\includegraphics[width=0.5\textwidth]{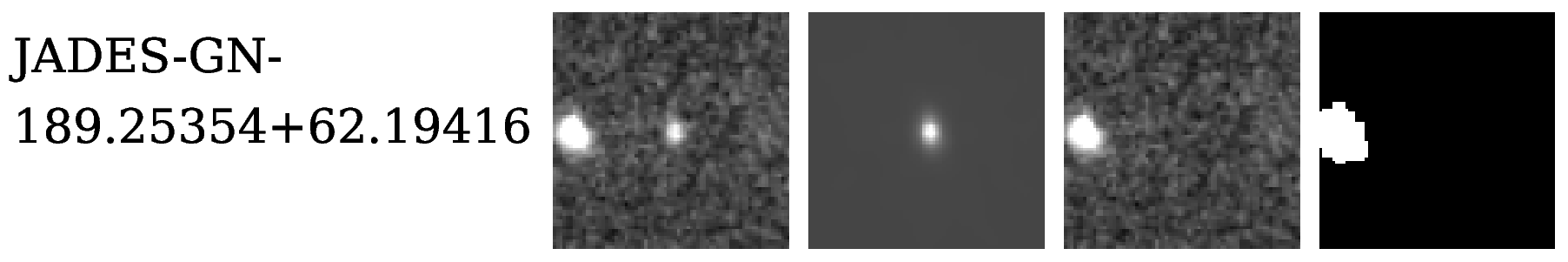}
	\includegraphics[width=0.5\textwidth]{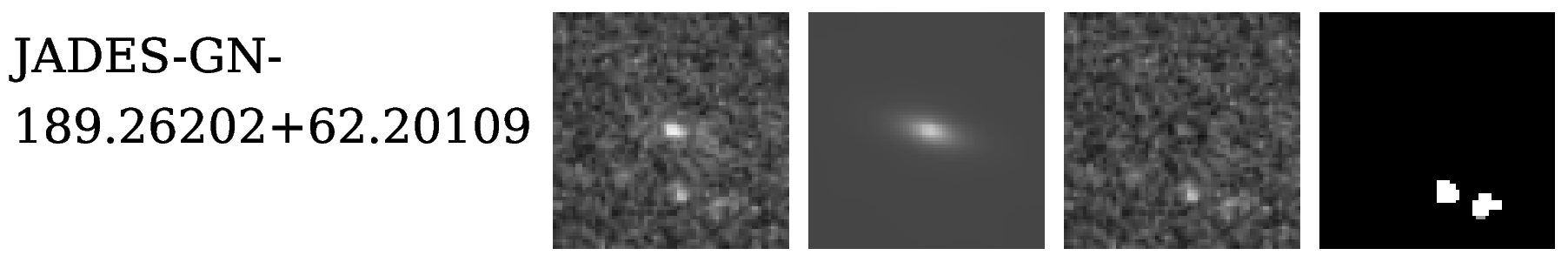}
	\includegraphics[width=0.5\textwidth]{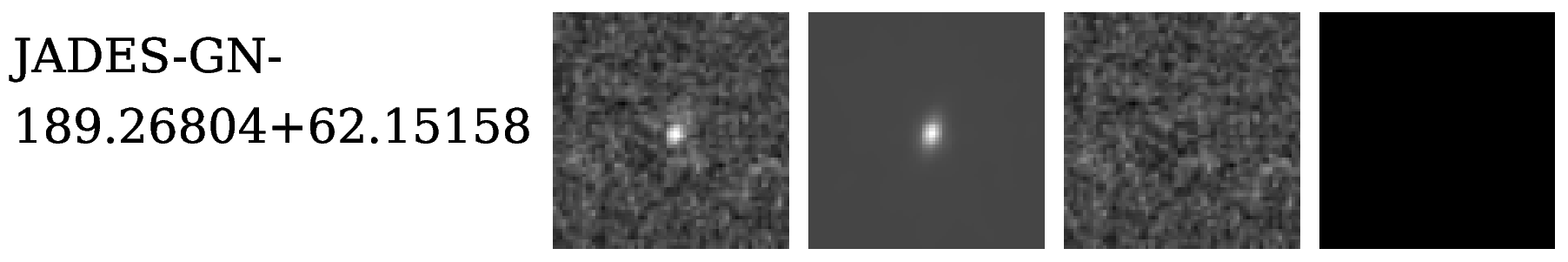}
	\includegraphics[width=0.5\textwidth]{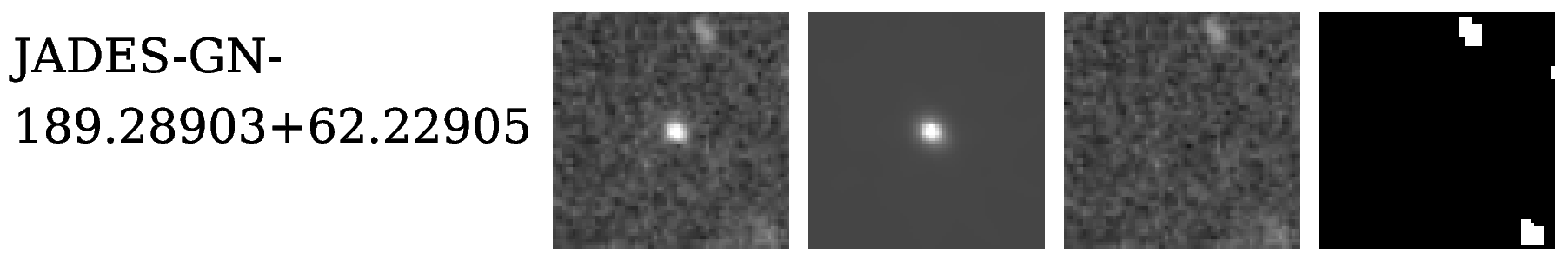}
	\includegraphics[width=0.5\textwidth]{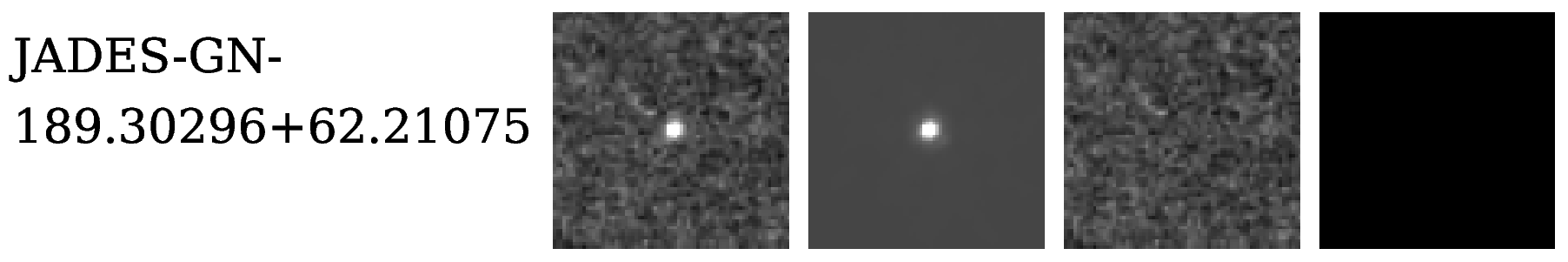}
\caption{
Same as Figure \ref{fig:fit_results_CEERS}, but for JADES GOODS-N. 
}
\label{fig:fit_results_JADES_GN}
\end{center}
\end{figure}

\begin{figure*}
\begin{center}
   \includegraphics[width=0.35\textwidth]{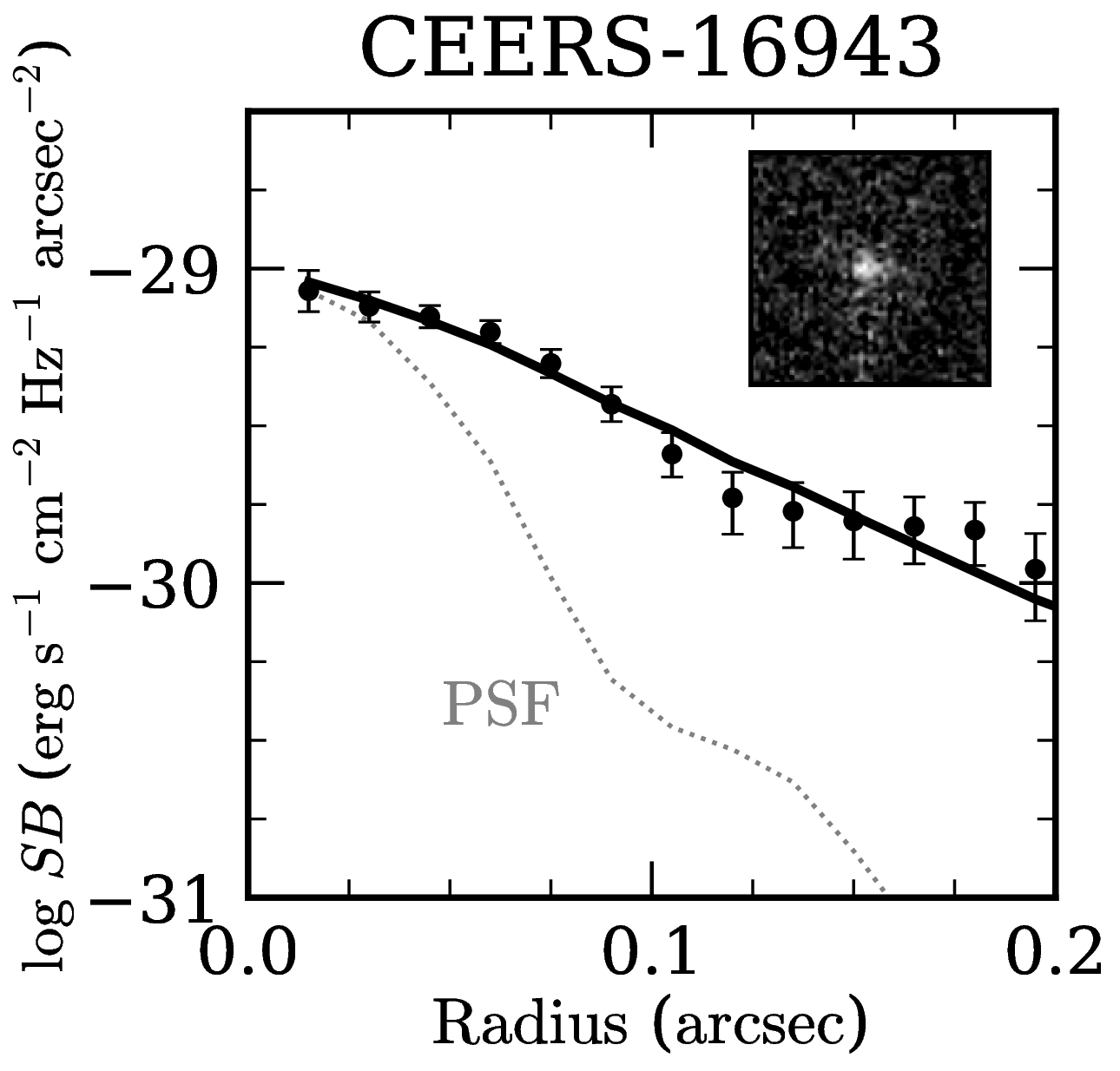}
   \includegraphics[width=0.35\textwidth]{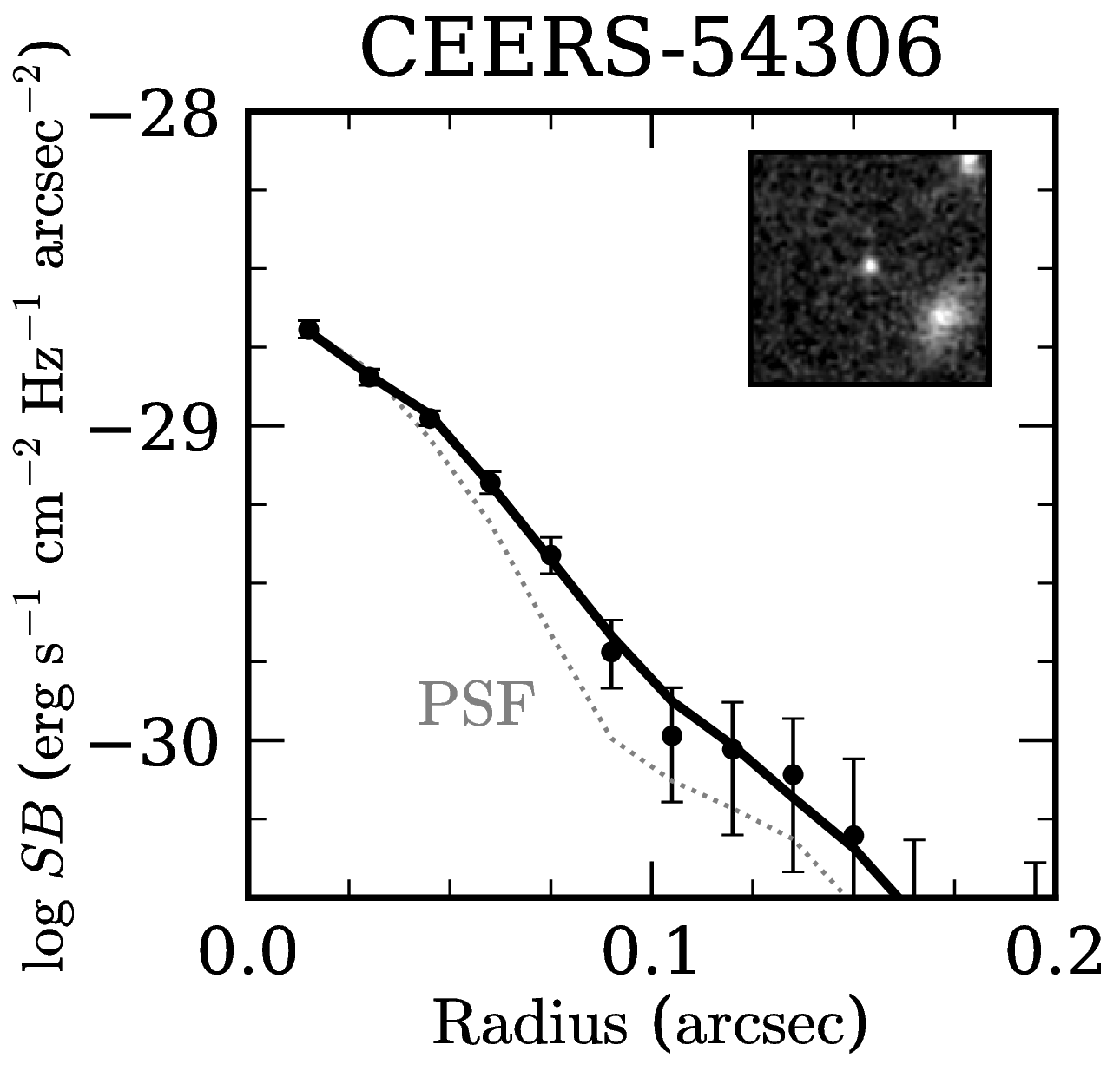}
\caption{Radial profiles of surface brightness profile fitting results 
for CEERS-16943 (left) and CEERS-54306 (right) 
shown as examples from our sample. 
In each panel, the black filled circles indicate the observed radial profile,
and the black solid line represents the best-fit S{\'e}rsic profile.
The gray dotted line shows the normalized PSF radial profile.
In the top-right corner of each panel, 
a $1\farcs5\times1\farcs5$ F200W image centered on the respective galaxy is displayed. 
}
\label{fig:fit_radial_SB_examples}
\end{center}
\end{figure*}

\subsection{Surface Brightness Profile Fitting Results} \label{subsec:SBfitting_results}

We perform surface brightness profile fitting using GALFIT for the $z\sim10$--$16$ galaxies 
in the CEERS and JADES fields summarized in Section \ref{sec:data}. 
For galaxies with an aperture magnitude $\mathrm{S/N} > 10$, we fit their profiles individually, 
while for fainter objects, we perform fitting on the stacked images to determine their average properties. 
Systematic uncertainties are corrected based on the results of the Monte Carlo simulations described 
in Section \ref{sec:SBprofile_fitting}.
Note that among the objects with $\mathrm{S/N} > 10$, 
the individual fitting processes do not converge successfully for five galaxies  in the JADES sample,\footnote{Their IDs are 
JADES-GS-53.06699-27.80884, 
JADES-GS-53.06708-27.80877,  
and 
JADES-GS-53.14528-27.82359 
in JADES Deep, and  
JADES-GS-53.06683-27.87294 
and 
JADES-GS-53.11763-27.88818 
in JADES Medium.} probably due to 
their very compact sizes and the slightly large pixel scale of the JADES images; 
they are excluded from this analysis for simplicity.

Figure \ref{fig:fit_results_CEERS}, Figure \ref{fig:fit_results_JADES}, 
Figure \ref{fig:fit_results_JADES_medium}, and Figure \ref{fig:fit_results_JADES_GN} 
present the surface brightness profile fitting results for individual objects with $\mathrm{S/N} > 10$ 
in CEERS, JADES Deep, JADES Medium, and JADES GOODS-N, respectively. 
From left to right, each panel shows the original image cutout at $1\farcs5\times1\farcs5$, 
the best-fit model image, the residual image, and the segmentation map. 
Their pseudo-color images are shown in Figure \ref{fig:pseudo_colors}.
In Figure \ref{fig:fit_radial_SB_examples}, 
the observed and best-fit radial profiles for two galaxies in the CEERS field are presented as examples.
In our fitting process, we prioritize fitting each galaxy individually. 
When a neighboring object is close by, we first attempt to mask it 
using a segmentation map to minimize its influence on the fit of the target galaxy. 
Alternatively, for galaxies composed of multiple components, 
we run the fitting on one component while masking the other using a segmentation map.
However, in cases where multiple components are too close to be effectively masked, 
we perform simultaneous fitting for all of them.
For example, 
CEERS-99715 and JADES-GS-53.07597-27.80654 are resolved into two components; 
we perform simultaneous fitting for both components
and report the obtained results separately.
For JADES-GS-z14-0, since a foreground object is located very close to it on the sky,
we perform simultaneous fitting for JADES-GS-z14-0 and the foreground object,
reporting only the results for JADES-GS-z14-0. 
The best-fit 
total magnitudes, circularized half-light radii, and axis ratios 
for all the individually analyzed galaxies with $\mathrm{S/N} > 10$ are listed in Table \ref{tab:GALFITresults}.
These values are corrected for systematic uncertainties and accompanied by statistical uncertainties
based on our Monte Carlo simulation results.

Among the objects analyzed in this study, 
some luminous galaxies have been investigated in detail including their sizes in the literature.
\cite{2024Natur.633..318C} have performed surface brightness profile fitting 
for JADES-GS-z14-0 and JADES-GS-z14-1
using ForcePho (\textcolor{blue}{B. D. Johnson in preparation})
and obtained results of $r_{\rm e} = 260 \pm 20$ pc and $r_{\rm e} < 160$ pc, respectively.
Our results are consistent with these findings.
\cite{2022ApJ...940L..55F} have analyzed CEERS-16943 (Maisie's Galaxy)
using GALFITM (\citealt{2013MNRAS.430..330H}), a modified version of GALFIT,
and STATMORPH (\citealt{2019MNRAS.483.4140R}) to perform S{\'e}rsic profile fitting, 
and reported $r_{\rm e} = 340 \pm 14$ pc.
Our results are in broad agreement with their result.
For GN-z11, \cite{2023ApJ...952...74T} have utilized ForcePho for surface brightness profile fitting,
providing a detailed analysis that has yielded $r_{\rm e} = 64 \pm 20$ pc.
Our results are consistent with this previous measurement within the $1\sigma$ uncertainty range.

\begin{figure}
\begin{center}
   \includegraphics[width=0.45\textwidth]{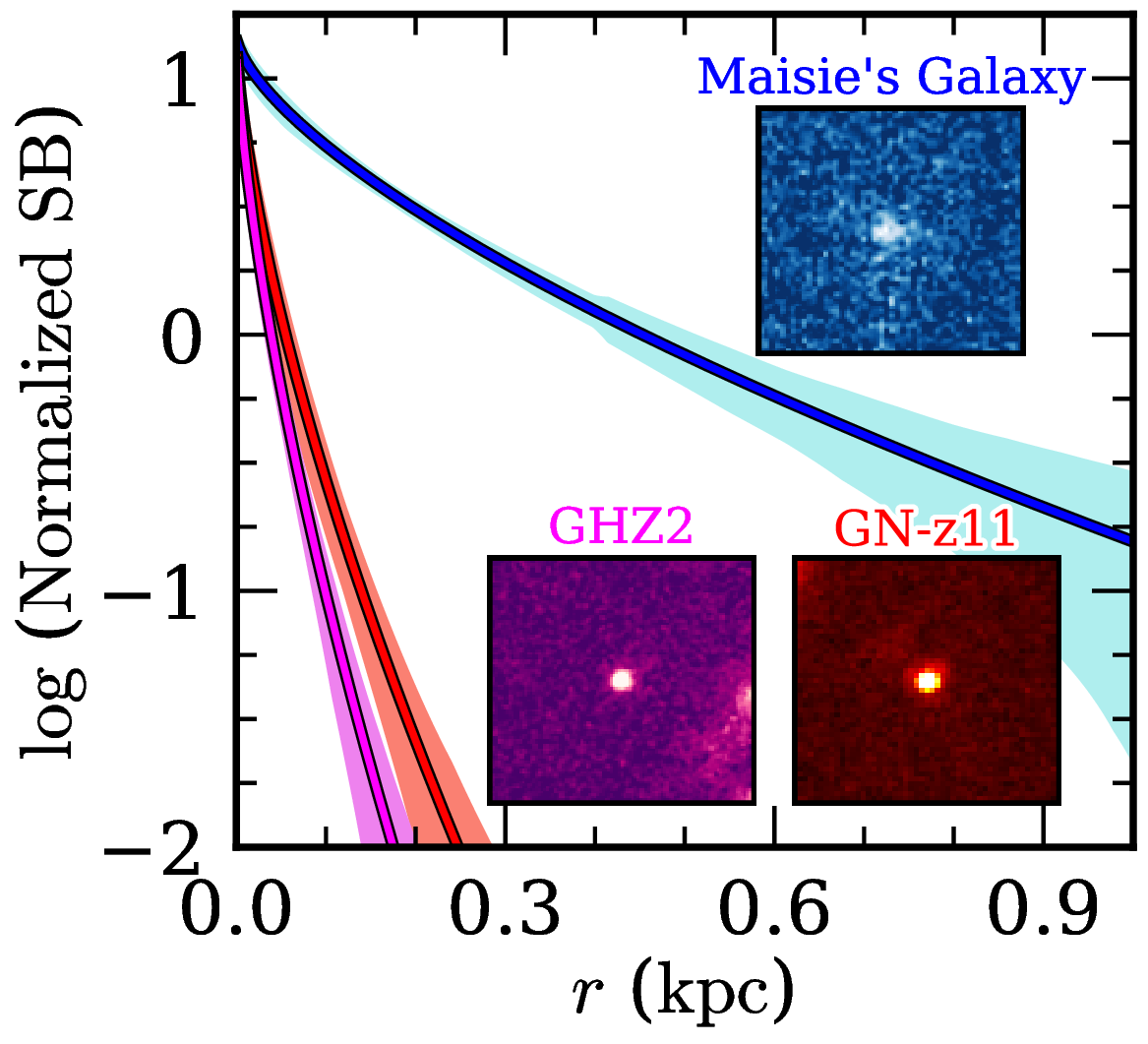}
\caption{
Normalized radial profiles of GN-z11 (red), GHZ2 (magenta), 
and Maisie's Galaxy (CEERS-16943, blue), 
selected as representative examples of a compact and an extended galaxy 
from the $z\sim10$--$16$ galaxies analyzed in this study and \cite{2023ApJ...951...72O}.
The solid curves are the best-fit S{\'e}rsic profiles 
and the shaded areas indicate the $1\sigma$ uncertainties,
computed from the 68th percentile ranges of radial profiles 
measured at randomly selected positions in the F200W images.
The inset panels display $1\farcs5 \times 1\farcs5$ cutouts of the original images.
}
\label{fig:Sersic_profiles}
\end{center}
\end{figure}

The sizes of the individual galaxies at $z\sim10$--$16$ obtained in this study  
range from compact values about $20$ pc to larger values around $500$ pc, 
with a median size of  $130$ pc.  
In Figure \ref{fig:Sersic_profiles},  
we present the normalized best-fit intrinsic S{\'e}rsic profiles of CEERS-16943 (Maisie's Galaxy) and GN-z11, 
selected as representative extended galaxy and compact galaxy examples from our sample, respectively.  
Additionally, as another example of compact galaxies,
we also present the normalized best-fit S{\'e}rsic profile of GHZ2, 
whose size is only $39 \pm 11$ pc as reported in \cite{2023ApJ...951...72O}.
These profiles represent the intrinsic light distributions of the galaxies before being convolved with the PSFs,
allowing for a direct comparison of their structural differences.
The origins of this diversity in galaxy sizes 
are discussed later by comparing with cosmological zoom-in simulation results 
in Section \ref{sec:discussion}.

In particular, for the extremely compact GN-z11 and GHZ2, 
two possible explanations have been suggested 
as summarized in \cite{2024arXiv240618352H} (see also, \citealt{2023ApJ...951...72O}).
One possibility is that they host AGNs. 
From deep NIRSpec spectroscopy, 
GN-z11 and GHZ2 exhibit high-ionization emission lines in their rest-frame UV spectra, 
with their emission line ratios and equivalent widths 
compared to photoionization models suggesting the possibility of AGN activity 
(for GN-z11, e.g., \citealt{2023A&A...677A..88B}; \citealt{2024ApJ...966...92S}; 
\citealt{2024Natur.627...59M}; \citealt{2024ApJ...976..142X}; \citealt{2024ApJ...976..122N}; 
\citealt{2024arXiv241212826A}; 
for GHZ2, e.g., \citealt{2024ApJ...972..143C}; \citealt{2024ApJ...975..245C}; 
\citealt{2024NatAs.tmp..258Z}; cf. \citealt{2024ApJ...977L...9Z}).
Alternatively, the compact morphologies could result from 
significantly enhanced star formation efficiency 
due to suppressed radiative feedback in high gas surface density regions, 
as indicated with recent 3D radiation hydrodynamic simulations 
(\citealt{2021MNRAS.506.5512F}; see also, \citealt{2018ApJ...859...68K}; \citealt{2020MNRAS.497.3830F}). 
\cite{2023MNRAS.523.3201D} have further discussed that 
in the feedback-free starburst scenario, 
compact galaxies with high star formation efficiency can form 
if the density of star-forming clouds is sufficiently high. 
Further investigations are required 
to elucidate the physical origins of these compact morphologies.

\begin{figure}[h]
\begin{center}
	\includegraphics[width=0.5\textwidth]{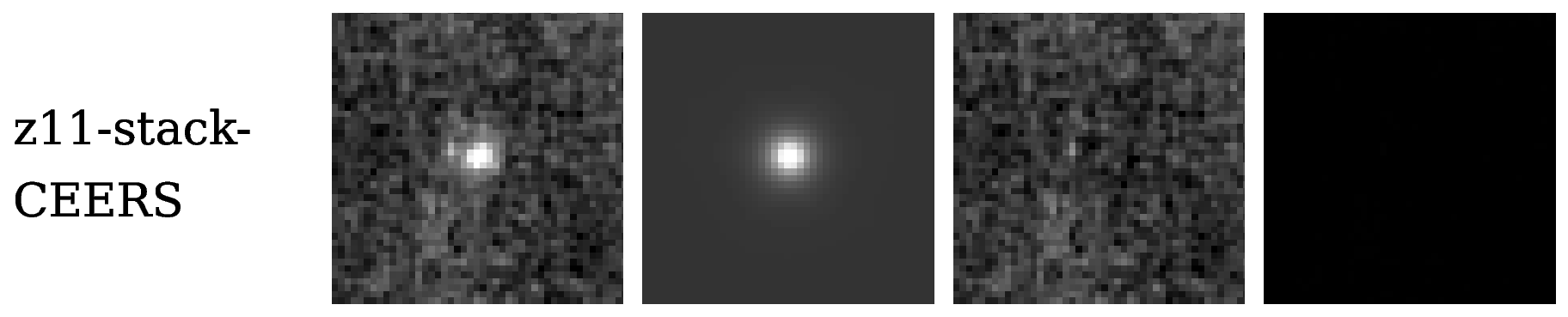}
	\includegraphics[width=0.5\textwidth]{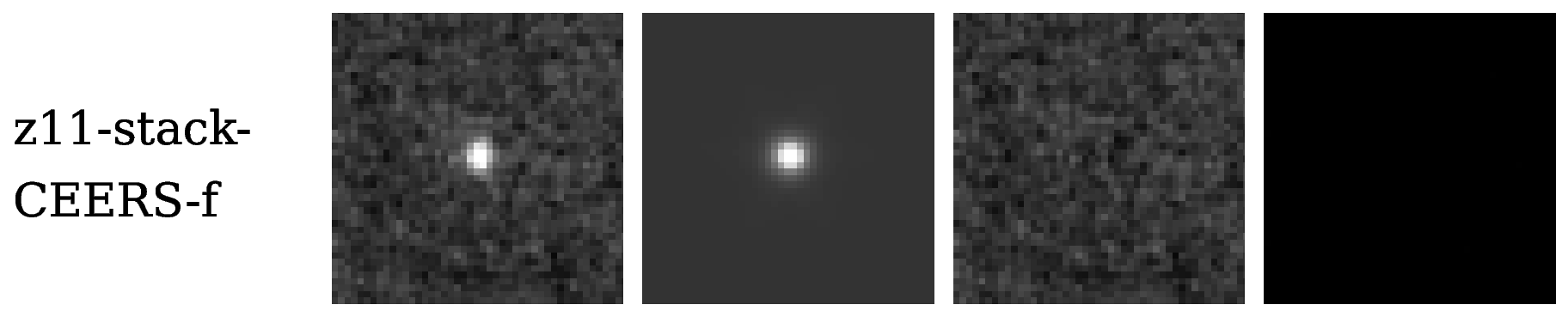}
	\includegraphics[width=0.5\textwidth]{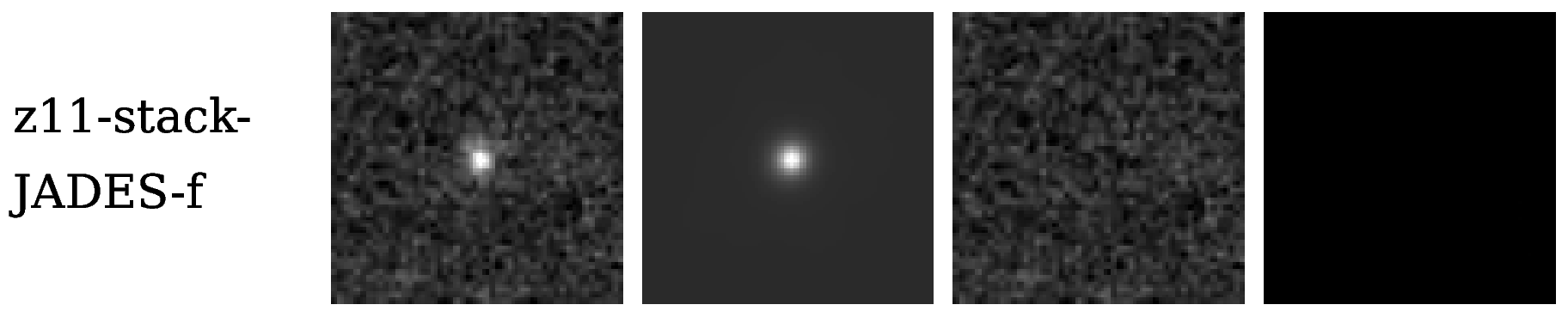}
	\includegraphics[width=0.5\textwidth]{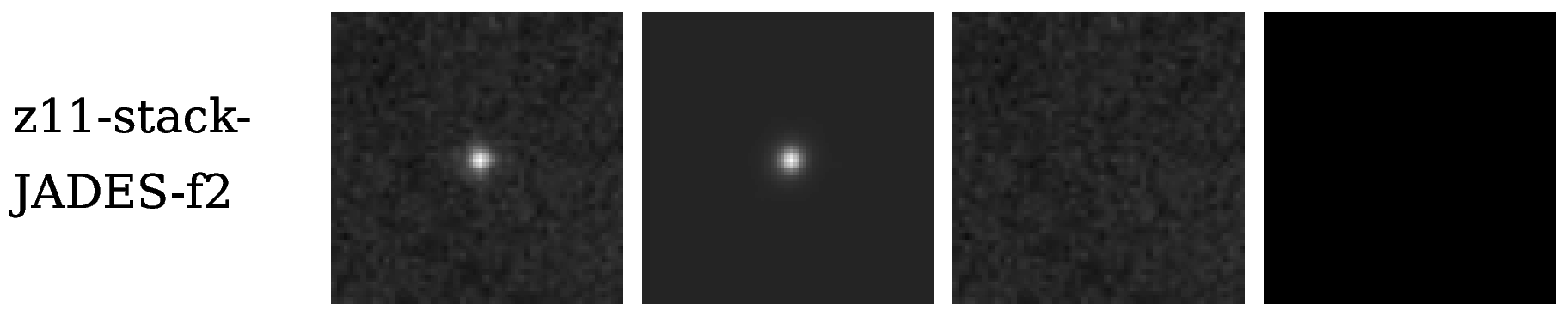}
	\includegraphics[width=0.5\textwidth]{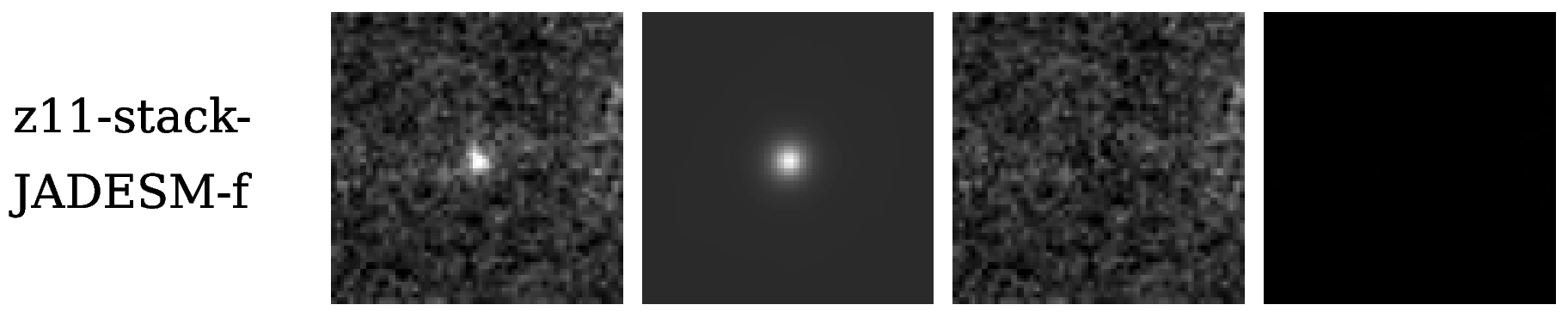}
	\includegraphics[width=0.5\textwidth]{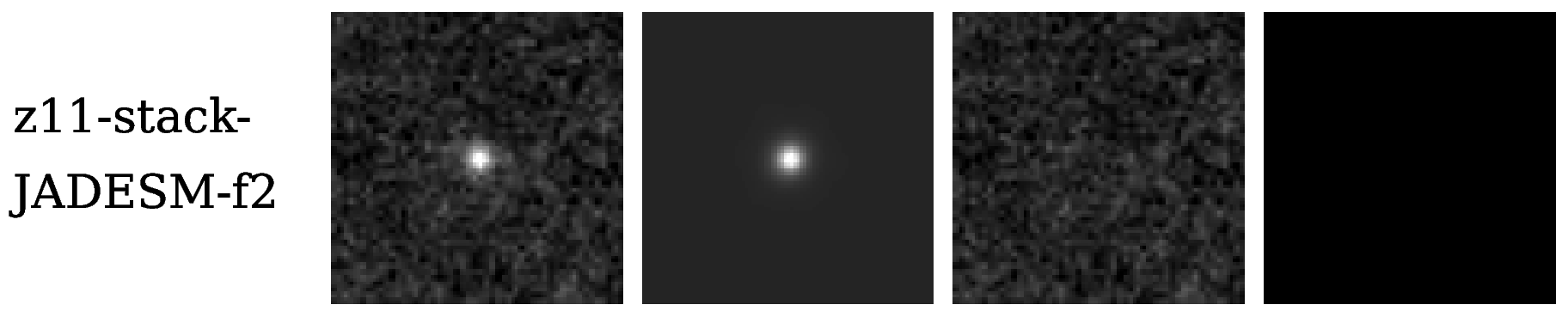}
	\includegraphics[width=0.5\textwidth]{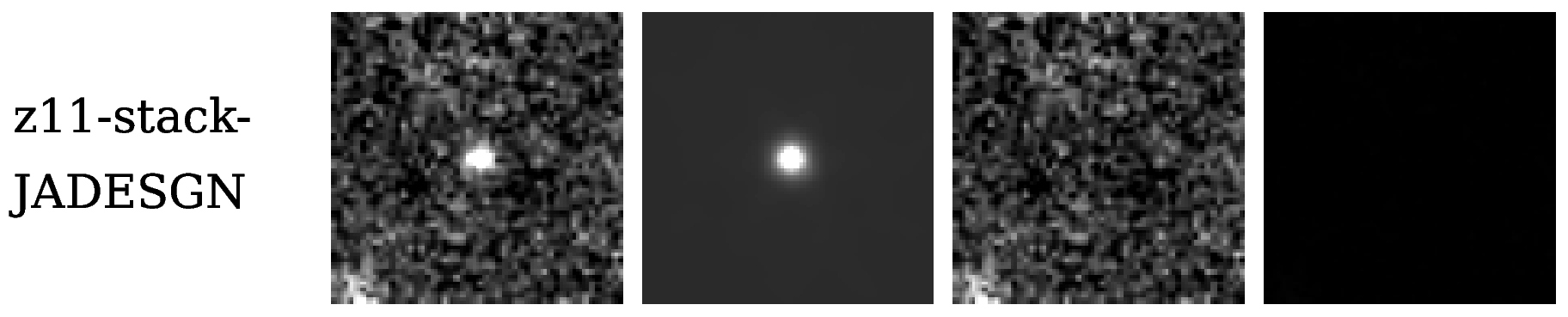}
	\includegraphics[width=0.5\textwidth]{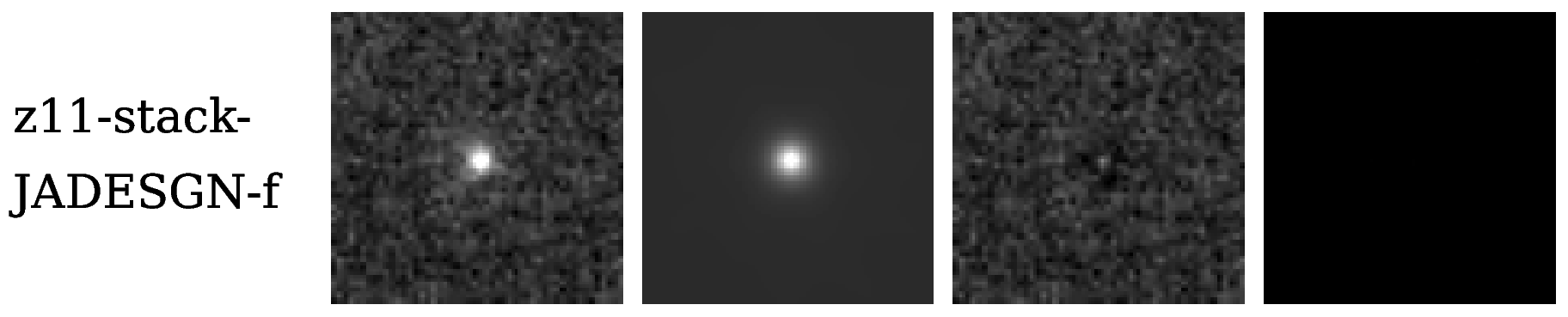}
	\includegraphics[width=0.5\textwidth]{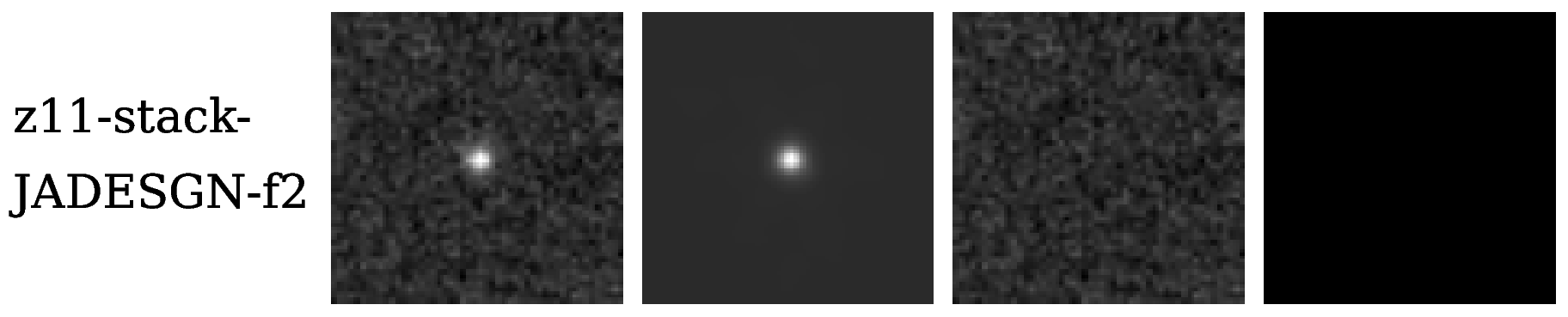}
	\includegraphics[width=0.5\textwidth]{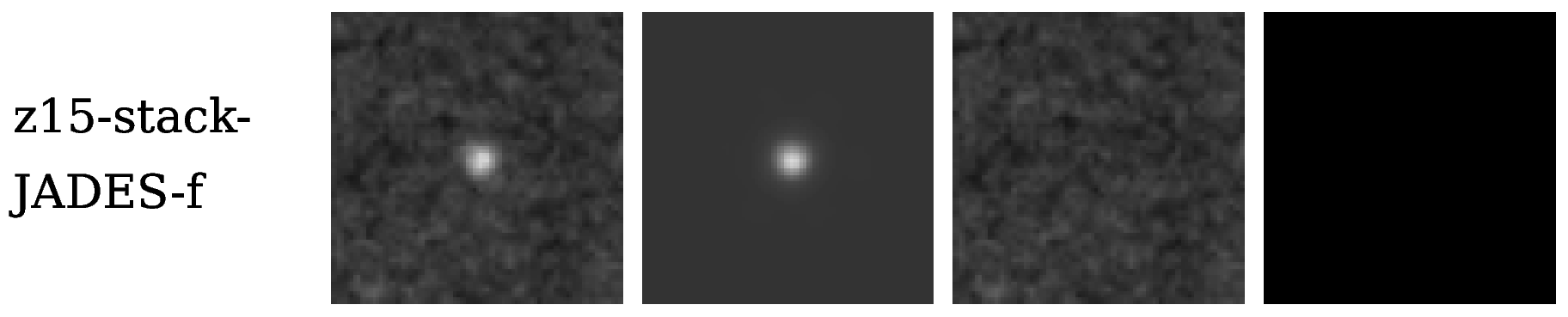}
\caption{
Same as Figure \ref{fig:fit_results_CEERS}, but for our stacked objects. 
The fields and luminosity ranges corresponding to each ID are summarized in 
Table \ref{tab:sample_stacked_numbers}.
}
\label{fig:fit_results_stack}
\end{center}
\end{figure}

Figure \ref{fig:fit_results_stack} shows the surface brightness profile fitting results 
for the stacked images of galaxies at $z\sim10$--$16$. 
Since the surface brightness profile of stacked galaxies is expected to be isotropic, 
the axis ratio is fixed to 1, and the position angle is set to 0 deg in these fits. 
The best-fit total magnitudes and circularized half-light radii for these stacked galaxies 
are summarized in Table \ref{tab:GALFITresults_stack}. 
As with the individual objects, the systematic uncertainties and statistical uncertainties 
in the size and total measurements are taken into account based on our Monte Carlo simulation results.
The measured sizes of the stacked images of $z\sim10$--$16$ galaxies 
are approximately in the range of $100$ pc to $300$ pc, 
with a median size of $160$ pc,  
comparable to the sizes of the individually analyzed objects. 
Our results suggest that the typical size of galaxies at $z\sim10$--$16$ 
is around $100$--$300$ pc.

\begin{deluxetable*}{lcccc} 
\tablecolumns{5} 
\tablewidth{0pt} 
\tablecaption{Surface Brightness Profile Fitting Results for Stacked Objects of $z\sim10$--$16$ Galaxies
\label{tab:GALFITresults_stack}}
\tablehead{
    \colhead{ID} 
    &  \colhead{$z_{\rm mean}$}
    &  \colhead{$m_{\rm UV}$}
    &  \colhead{$M_{\rm UV}$}
    &  \colhead{$r_{\rm e}$}
\\
    \colhead{ } 
    &  \colhead{ }
    &  \colhead{(mag)}
    &  \colhead{(mag)}
    &  \colhead{(kpc)}
\\
    \colhead{(1)} 
    &  \colhead{(2)}
    &  \colhead{(3)}
    &  \colhead{(4)}
    &  \colhead{(5)}
}
\startdata 
\multicolumn{5}{c}{CEERS} \\  
        z11-stack-CEERS &  11.55 &  27.80$^{+0.13}_{-0.13}$ &  $-19.89^{+0.13}_{-0.13}$ &  0.340$^{+0.024}_{-0.017}$ \\
        z11-stack-CEERS-f &  10.83 &  28.43$^{+0.06}_{-0.06}$ &  $-19.16^{+0.06}_{-0.06}$ &  0.188$^{+0.053}_{-0.046}$ \\
\multicolumn{5}{c}{JADES Deep} \\  
        z15-stack-JADES-f &  14.62 &  29.87$^{+0.12}_{-0.08}$ &  $-18.15^{+0.12}_{-0.08}$ &  0.059$^{+0.017}_{-0.019}$ \\
        z11-stack-JADES-f &  10.99 &  28.98$^{+0.08}_{-0.06}$ &  $-18.64^{+0.08}_{-0.06}$ &  0.174$^{+0.024}_{-0.025}$ \\
        z11-stack-JADES-f2 &  10.66 &  29.80$^{+0.08}_{-0.06}$ &  $-17.77^{+0.08}_{-0.06}$ &  0.101$^{+0.009}_{-0.010}$ \\
\multicolumn{5}{c}{JADES Medium} \\  
        z11-stack-JADESM-f &  10.25 &  28.81$^{+0.06}_{-0.04}$ &  $-18.71^{+0.06}_{-0.04}$ &  0.251$^{+0.012}_{-0.014}$ \\
        z11-stack-JADESM-f2 &  10.72 &  29.36$^{+0.08}_{-0.06}$ &  $-18.22^{+0.08}_{-0.06}$ &  0.142$^{+0.009}_{-0.010}$ \\
\multicolumn{5}{c}{JADES GOODS-N} \\  
        z11-stack-JADESGN &  11.08 &  28.55$^{+0.16}_{-0.11}$ &  $-19.08^{+0.16}_{-0.11}$ &  0.116$^{+0.010}_{-0.017}$ \\
        z11-stack-JADESGN-f &  10.53 &  28.48$^{+0.05}_{-0.04}$ &  $-19.07^{+0.05}_{-0.04}$ &  0.278$^{+0.018}_{-0.024}$ \\
        z11-stack-JADESGN-f2 &  10.55 &  29.40$^{+0.16}_{-0.11}$ &  $-18.16^{+0.16}_{-0.11}$ &  0.109$^{+0.010}_{-0.018}$ \\
\enddata 
\tablecomments{The systematic effects and statistical uncertainties 
in these obtained size and total magnitude measurements are considered 
based on our MC simulation results.  
(1) Object ID. The fields and luminosity ranges corresponding to each ID are summarized in 
Table \ref{tab:sample_stacked_numbers}.
(2) Mean redshift.  
(3) Total apparent UV magnitude measured by GALFIT. 
(4) Total absolute UV magnitude, calculated using $z_{\rm mean}$. 
(5) Circularized half-light radius $r_{\rm e} = a \sqrt{b/a}$. 
}
\end{deluxetable*} 

\begin{figure}[h]
\begin{center}
   \includegraphics[width=0.35\textwidth]{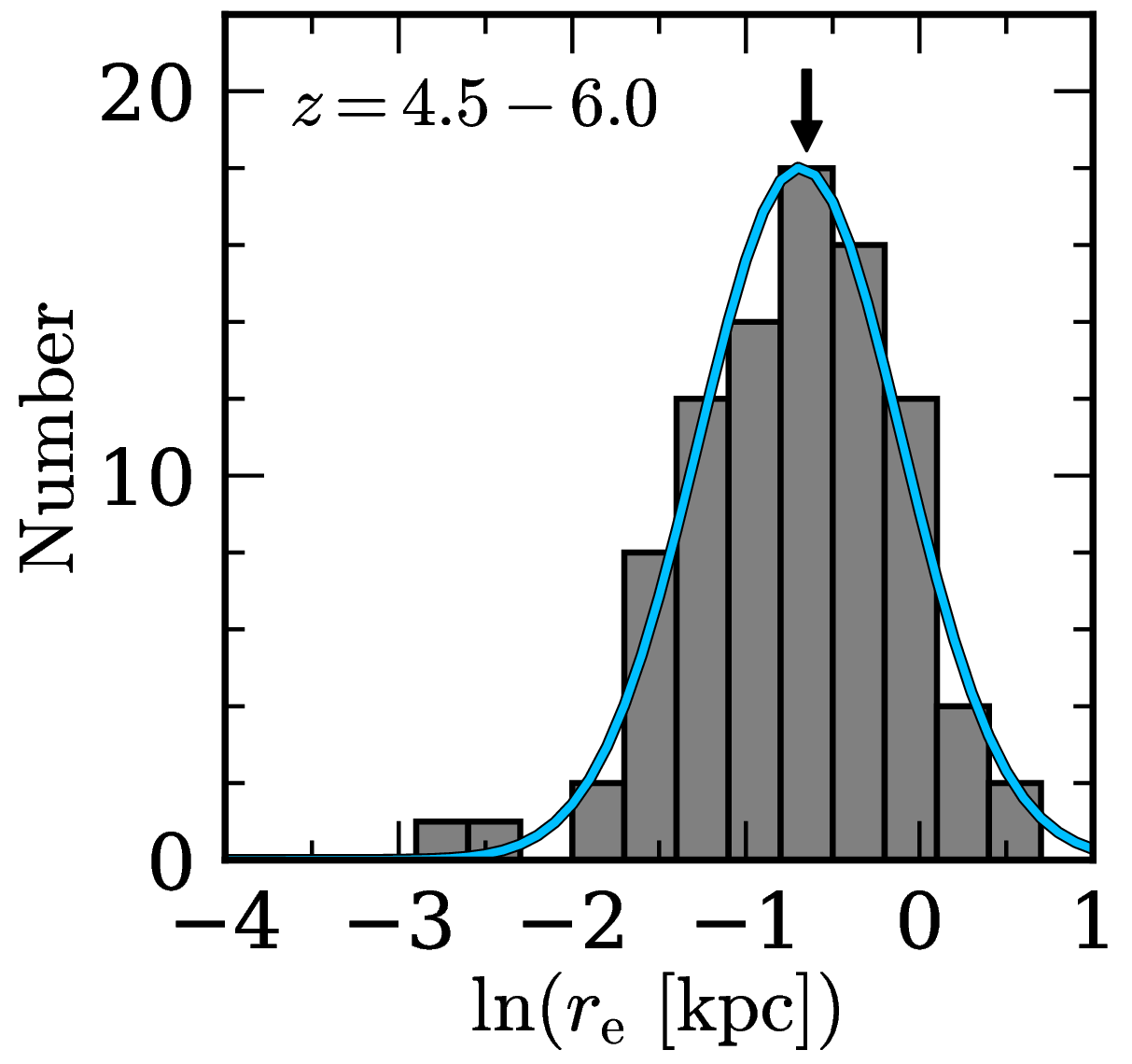}
   \includegraphics[width=0.35\textwidth]{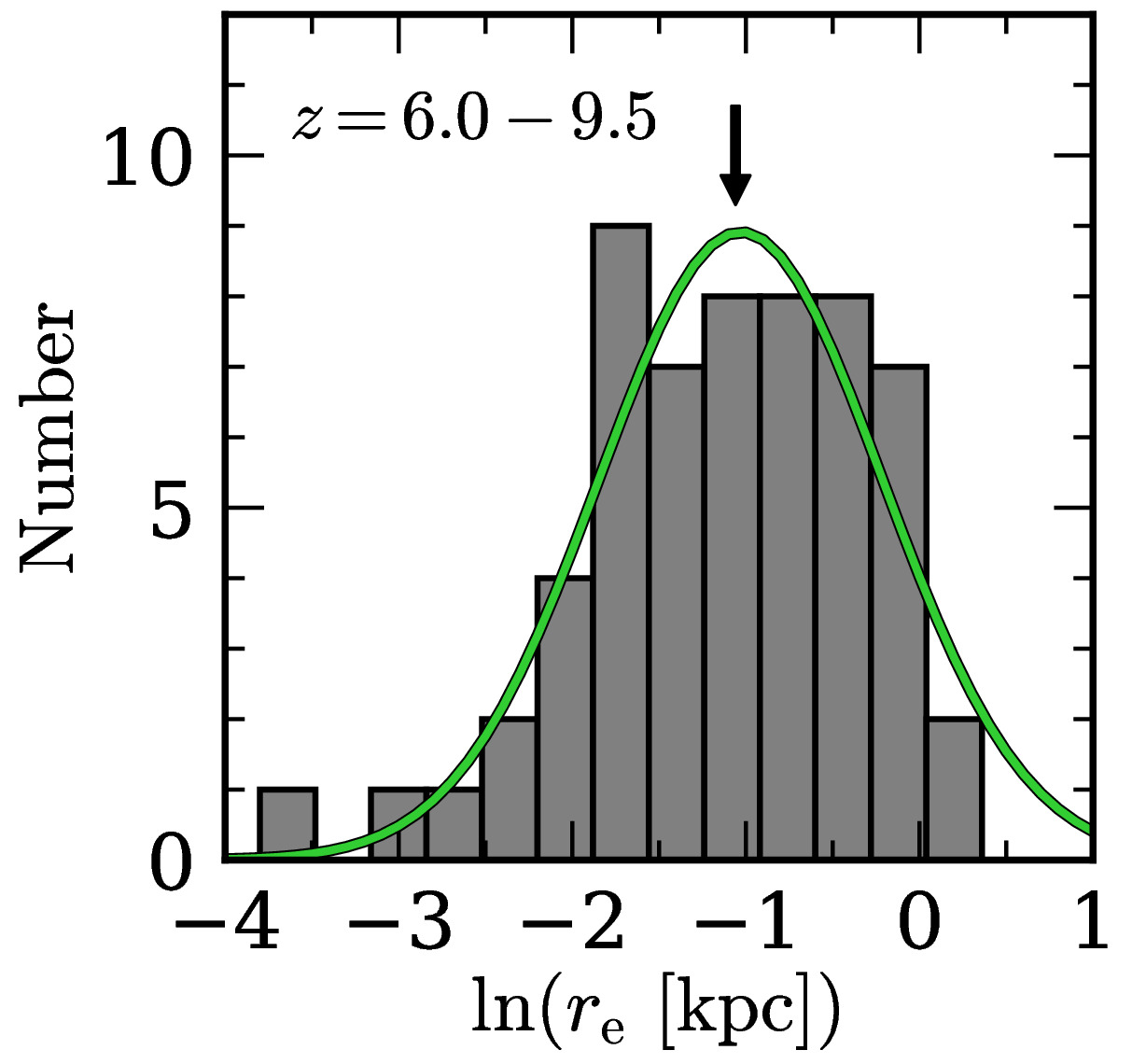}
   \includegraphics[width=0.35\textwidth]{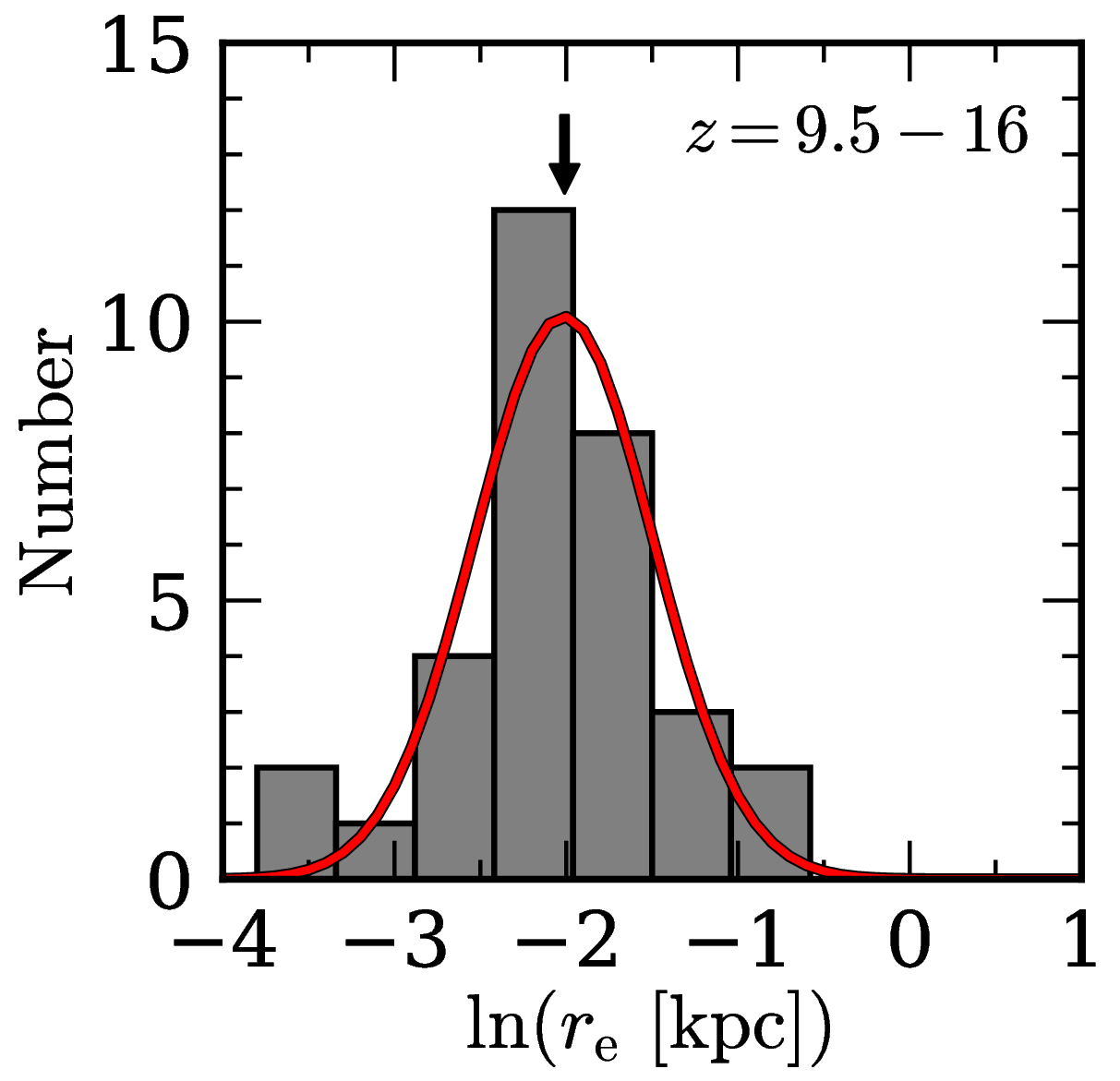}
\caption{
Distributions of $r_{\rm e}$ for high-$z$ galaxies. 
The top panel shows the distribution for galaxies at $z=4.5$--$6.0$, 
the middle panel for $z=6.0$--$9.5$, 
and the bottom panel for $z=9.5$--$16$. 
The downward black arrows indicate the median $r_{\rm e}$ values. 
The colored curves represent the best-fit log-normal distributions.
}
\label{fig:histo_lnre}
\end{center}
\end{figure}

To characterize the size measurement results for high-$z$ galaxies more quantitatively, 
Figure \ref{fig:histo_lnre} shows the histogram of $r_{\rm e}$, 
where $r_{\rm e}$ is displayed in natural logarithmic scale following the literature 
(\citealt{2015ApJS..219...15S}).\footnote{In this manuscript, 
{\lqq}ln{\rqq} refers to the natural logarithm (base $e$) 
and {\lqq}log{\rqq} refers to the common logarithm (base 10).}
Since the number of galaxies for which individual sizes are determined in this study is not large,
we combine them into a single histogram 
despite the wide redshift range.\footnote{Although our sample 
spans a wide range not only in redshift but also in luminosity, 
we present them in a single histogram due to the limited number of galaxies.
However, considering the galaxy size-luminosity relation as presented in Section \ref{subsec:sizeUV_relation}, 
this approach is suboptimal.
Future studies with sufficiently larger samples should allow for 
a more thorough characterization of galaxy size distributions including their luminosity dependence.} 
To extend the redshift range examined here, 
we incorporate the results of \cite{2024PASJ...76..219O}, 
who have individually measured the sizes of $149$ galaxies at lower redshifts 
of $z=4.5$--$9.5$ in the CEERS field using the same method as in this study. 
The top and middle panels are based on the results of \cite{2024PASJ...76..219O} for galaxies 
at $z=4.5$--$6.0$ and $z=6.0$--$9.5$, respectively, 
while the bottom panel represents the results of this study for galaxies at $z=9.5$--$16$. 
Their mean redshifts are $z=5.02$, $6.76$, and $10.94$, respectively. 
The median values of $\ln (r_{\rm e} \, \, [{\rm kpc}])$ are 
$-0.65$ for $z=4.5$--$6.0$, 
$-1.06$ for $z=6.0$--$9.5$, 
and $-2.01$ for $z=9.5$--$16$.\footnote{These values correspond to 
$0.52$ kpc, $0.35$ kpc, and $0.13$ kpc, respectively.}

\begin{figure}[h]
\begin{center}
   \includegraphics[width=0.5\textwidth]{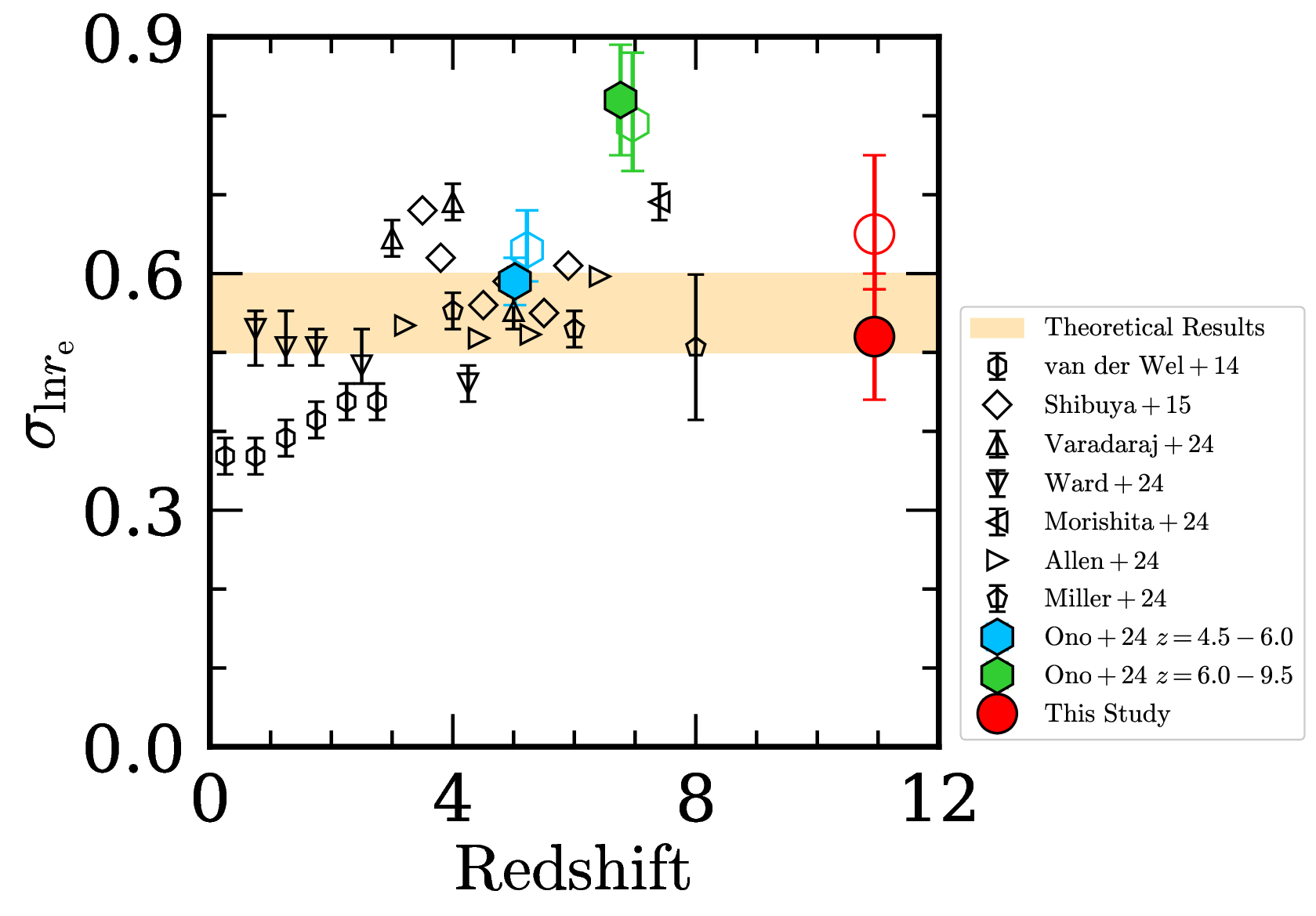}
\caption{
Standard deviation of the log-normal size distribution, $\sigma_{{\rm ln}r_{\rm e}}$, as a function of redshift. 
The red filled circle represents the results for $z\sim10$--$16$ galaxies in this study, 
while the green and cyan filled hexagons represent the results for $z=6.0$--$9.5$ and $z=4.5$--$6.0$ galaxies, respectively, 
based on the size measurement results of \cite{2024PASJ...76..219O}. 
The corresponding open symbols in these colors indicate the unbiased standard deviations, 
and for clarity, those at $z=6.0$--$9.5$ and $z=4.5$--$6.0$ are shifted by $+0.2$ in redshift. 
The black open symbols represent results from previous studies, 
with the symbol shapes and corresponding references shown in the legend 
(\citealt{2014ApJ...788...28V}; \citealt{2015ApJS..219...15S}; \citealt{2024MNRAS.533.3724V}; 
\citealt{2024ApJ...962..176W}; \citealt{2024ApJ...963....9M}; \citealt{2024arXiv241016354A}; 
\citealt{2024arXiv241206957M}).
The orange shaded region shows the width of the spin parameter $\lambda$ distribution, 
$\sigma_{{\rm ln}\lambda}$, predicted from theoretical studies 
(e.g., \citealt{1998MNRAS.295..319M}; \citealt{2001ApJ...555..240B}; \citealt{2001ApJ...557..616G}; 
\citealt{2002ApJ...576...21V}; \citealt{2002ApJ...581..799V}; \citealt{2005ApJ...627..647B}; 
\citealt{2009MNRAS.393.1498D}; \citealt{2017MNRAS.466.1625Z}). 
}
\label{fig:sigma_lnre_redshift}
\end{center}
\end{figure}

Following previous work (\citealt{2015ApJS..219...15S}), 
we fit these distributions with a log-normal function:
\begin{equation}
p(r_{\rm e}) d r_{\rm e} 
	= \dfrac{1}{ \sqrt{2 \pi} \sigma_{{\rm ln}r_{\rm e}} }
		\exp \left( - \dfrac{\left[ \ln (r_{\rm e} / \overline{r_{\rm e}}) \right]^2 }{2 \sigma^2_{{\rm ln}r_{\rm e}}} \right) 
		\dfrac{d r_{\rm e}}{r_{\rm e}},
\end{equation}
where $\overline{r_{\rm e}}$ corresponds to the half-light radius at the peak of the distribution, 
and $\sigma_{{\rm ln}r_{\rm e}}$ represents the standard deviation of the distribution.\footnote{Here  
we follow previous studies in using natural logarithms 
(e.g., \citealt{2015ApJS..219...15S}; see also, \citealt{1998MNRAS.295..319M}; \citealt{2001ApJ...555..240B}; 
\citealt{2001ApJ...557..616G}; \citealt{2002ApJ...576...21V}; \citealt{2002ApJ...581..799V}; \citealt{2005ApJ...627..647B}). 
However, some studies use common logarithms, which are based on 10 as the base;  
one should be careful when making comparisons.} 
The best-fit log-normal distributions are shown as solid curves in Figure \ref{fig:histo_lnre}. 
All the curves reasonably match the observed $r_{\rm e}$ histograms 
with reduced $\chi^2$ values of $\simeq 0.3$--$0.9$. 
Note that a small number of objects with very small sizes
appear to slightly deviate from the log-normal distribution.
In particular, at $z=9.5$--$16$, two objects are distributed around
$\ln (r_{\rm e} \, \, [{\rm kpc}]) \simeq -4$; 
these are JADES-GS-53.19051-27.74982 in JADES Deep
and JADES-GN-189.30296+62.21075 in JADES GOODS-N.
Neither of these objects has been spectroscopically confirmed,
and their nature remains unknown.
Although their size measurements have relatively large uncertainties,
they are interesting sources that could have sizes comparable to or even smaller than
those of GN-z11 and GHZ2,
which are suggested to host AGNs and/or compact star formation as described above.

The best-fit $\overline{r_{\rm e}}$ values are 
$\overline{r_{\rm e}} = 0.503^{+0.018}_{-0.017}$ kpc for $z=4.5$--$6.0$, 
$\overline{r_{\rm e}} = 0.357^{+0.028}_{-0.026}$ kpc for $z=6.0$--$9.5$, 
and $\overline{r_{\rm e}} = 0.133^{+0.013}_{-0.012}$ kpc for $z=9.5$--$16$.
The best-fit standard deviations are
$\sigma_{{\rm ln}r_{\rm e}} = 0.59 \pm 0.03$ for $z=4.5$--$6.0$, 
$\sigma_{{\rm ln}r_{\rm e}} = 0.82 \pm 0.07$ for $z=6.0$--$9.5$, 
and $\sigma_{{\rm ln}r_{\rm e}} = 0.52 \pm 0.08$ for $z=9.5$--$16$. 
For reference, the unbiased standard deviation of $\ln r_{\rm e}$ is 
calculated as $0.63^{+0.05}_{-0.04}$, $0.79^{+0.09}_{-0.06}$, and $0.65^{+0.10}_{-0.07}$ 
for these redshift ranges, respectively. 
For $z=4.5$--$6.0$ and $z=6.0$--$9.5$, 
the unbiased standard deviations are consistent with 
the best-fit log-normal standard deviations within the $1\sigma$ uncertainties. 
For $z=9.5$--$16$, likely due to the smaller sample size, 
the unbiased standard deviation slightly differs from the best-fit value. 
However, the difference, $0.65-0.52=0.13$, is still within $2\sigma$ of the combined uncertainty, 
which is calculated as $\sqrt{0.08^2 + 0.07^2} \simeq 0.11$.

Figure \ref{fig:sigma_lnre_redshift} presents the standard deviations 
of galaxy size distributions as a function of redshift, 
combining results from this study with those from previous studies
(\citealt{2014ApJ...788...28V}; \citealt{2015ApJS..219...15S}; \citealt{2024MNRAS.533.3724V};
\citealt{2024ApJ...962..176W}; \citealt{2024ApJ...963....9M}; \citealt{2024arXiv241016354A};
\citealt{2024arXiv241206957M}).
For the redshift value of the data point from \cite{2024ApJ...963....9M}, 
we use the median redshift of the galaxies in their catalog.
Previous observational studies have reported that 
the $\sigma_{{\ln}r_{\rm e}}$ values range from approximately $0.4$ to $0.7$ 
across a wide range of redshifts,  showing no significant evolution.
The standard deviation of the size distribution derived from the results of \cite{2024PASJ...76..219O} 
is broadly consistent with these earlier findings.
Moreover, the standard deviation of the size distribution for $z\sim10$--$16$ galaxies 
obtained in this study is also comparable to the values reported at lower redshifts 
within the $1\sigma$ uncertainty. 
This indicates that $\sigma_{{\ln}r_{\rm e}}$ shows no clear evolution toward $z\gtrsim10$.
The implication of this result for the disk formation scenario is discussed 
in Section \ref{sec:discussion}.

\subsection{Axis Ratio Distribution} \label{subsec:ba_distribution}

\begin{figure}[t]
\begin{center}
   \includegraphics[width=0.4\textwidth]{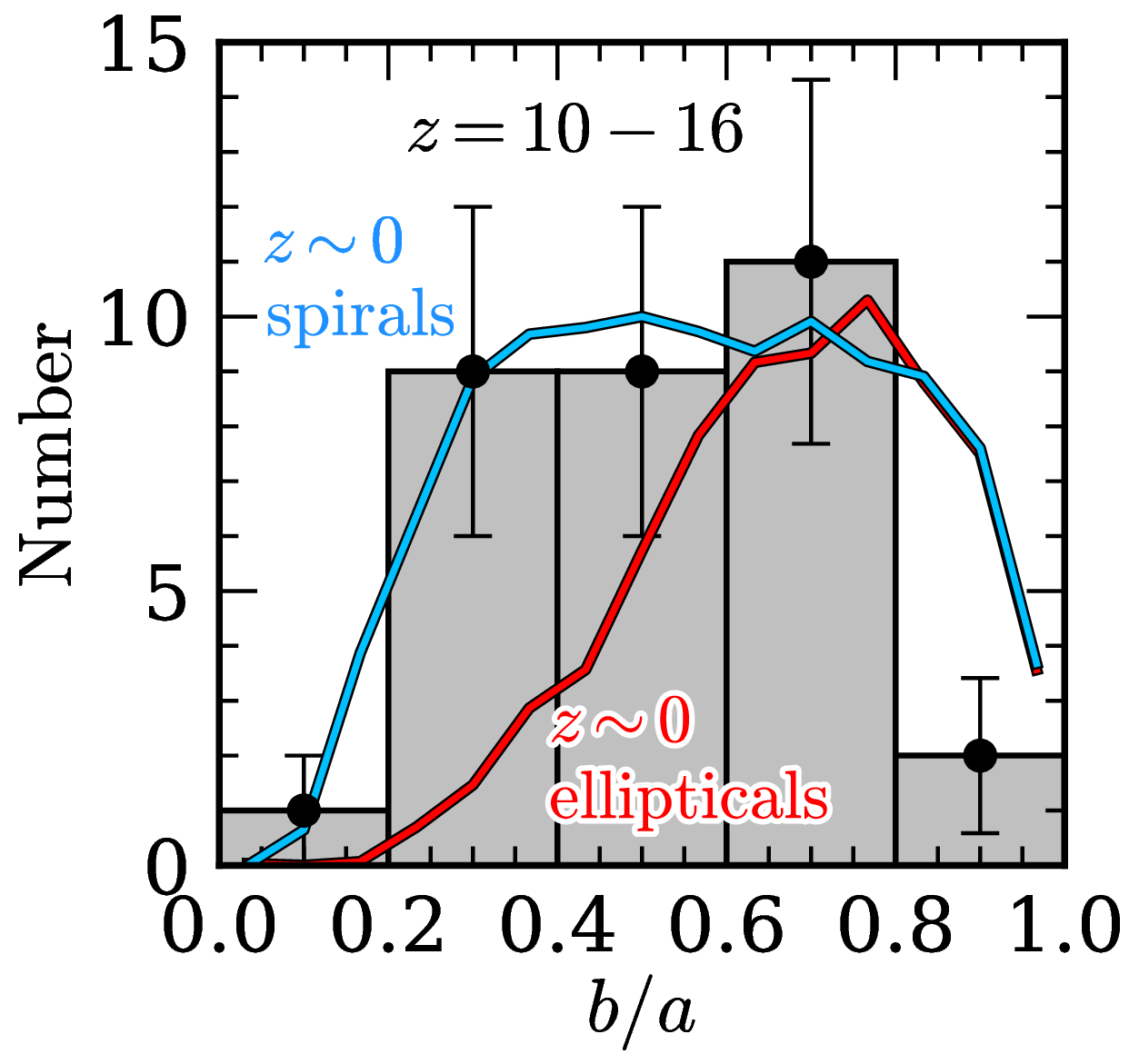}
\caption{Axis ratio $b/a$ distribution. 
The black circles and gray histogram represent 
the axis ratio distribution of the $z\sim10$--$16$ galaxies investigated in this study. 
The cyan and red solid lines denote the axis ratio distributions 
of local spiral and elliptical galaxies, respectively, 
obtained by \cite{1992MNRAS.258..404L} and scaled appropriately. 
}
\label{fig:histo_ba}
\end{center}
\end{figure}

We investigate the axis ratio distribution of $z\sim10$--$16$ galaxies.
Figure \ref{fig:histo_ba} shows the axis ratio distribution 
for the $z\sim10$--$16$ galaxies whose axis ratios are individually determined
through profile fitting in this study.
Although the number of the high-$z$ galaxies is limited,
resulting in a coarse frequency distribution,
the number of objects appears to be roughly constant across $b/a$ values from $0.2$ to $0.8$,
suggesting a distribution close to uniform.

As is well known from observations of local galaxies,
the axis ratio distribution for elliptical galaxies peaks around $0.8$
and decreases for both larger and smaller values.
In contrast, the axis ratio distribution for spiral galaxies is nearly uniform
(\citealt{1992MNRAS.258..404L}).
A similar uniform distribution has also been obtained for star-forming galaxies at $z\sim1$--$5$
(\citealt{2006ApJ...652..963R}; see also, \citealt{2004ApJ...600L.107F}).
For comparison, Figure \ref{fig:histo_ba} also displays
the axis ratio distributions for local elliptical and spiral galaxies, scaled appropriately, 
alongside the obtained distribution for our $z\sim10$--$16$ galaxies.
As can be seen from the figure,
the axis ratio distribution for the $z\sim10$--$16$ galaxies 
is closer to that of local spiral galaxies than to that of local elliptical galaxies.

To quantitatively evaluate this visual similarity,
we use the Kolmogorov-Smirnov (K-S) test to examine whether the axis ratio distribution
of our $z\sim10$--$16$ galaxies is statistically consistent with
those of local elliptical and spiral galaxies.
The K-S test is a non-parametric test that assesses the maximum vertical distance
(K-S statistic $D$) between the empirical cumulative distribution functions of two samples.

We perform the K-S test using our raw axis ratio data for the $z\sim10$--$16$ galaxies
and the axis ratio distribution data for local elliptical and spiral galaxies from \cite{1992MNRAS.258..404L}.
Here we adopt a statistical significance level of $\alpha=0.01$.
As a result, comparing our galaxy distribution with that of local elliptical galaxies,
we obtain $D=0.3590$ and a $p$-value of $0.0004$, 
which is much smaller than the significance level, 
providing strong evidence to reject the null hypothesis that both distributions are the same.
This indicates that the axis ratio distribution of our $z\sim10$--$16$ galaxies
is statistically significantly different from that of local elliptical galaxies. 
On the other hand, when comparing our galaxy distribution with that of local spiral galaxies,
we obtain $D=0.1876$ and a $p$-value of $0.1849$.
Since this $p$-value is larger than the significance level,
we cannot reject the null hypothesis, 
indicating that there is no statistically significant difference 
between the axis ratio distribution of our $z\sim10$--$16$ galaxies
and that of local spiral galaxies.
These quantitative results corroborate our visual inspection,
statistically suggesting that the axis ratio distribution of $z\sim10$--$16$ galaxies
differs from that of local elliptical galaxies and
is instead closer to the uniform distribution observed in local spiral galaxies.

\subsection{Size--UV Luminosity Relation} \label{subsec:sizeUV_relation}

\begin{figure}
\begin{center}
   \includegraphics[width=0.5\textwidth]{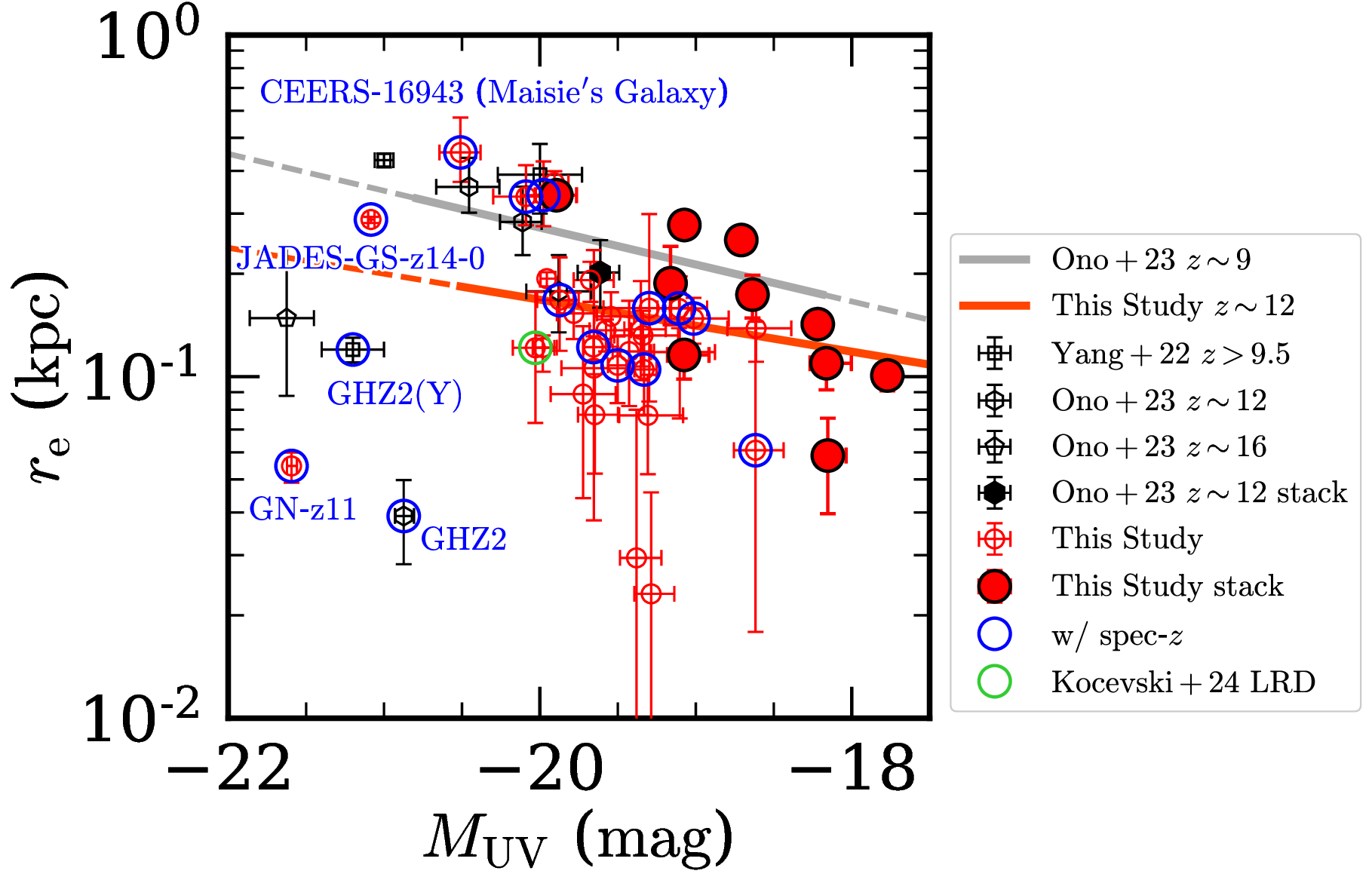}
\caption{
Relation between galaxy size $r_{\rm e}$ and $M_{\rm UV}$. 
The red open circles represent the results for relatively luminous $z\sim10$--$16$ galaxies 
analyzed individually in this study, 
while the red filled circles show the results for stacked images of relatively faint $z\sim10$--$16$ galaxies. 
The red solid line denotes the best-fit size--luminosity relation for our $z\sim10$--$13$ galaxies, 
and the red dashed line indicates its extrapolation. 
The black open symbols represent the results for relatively luminous $z\sim10$--$16$ galaxies 
analyzed individually in previous studies, 
and the black filled hexagon represents the results for stacked images from previous studies
(\citealt{2022ApJ...938L..17Y}; \citealt{2023ApJ...951...72O}). 
The blue large circles indicate spectroscopically confirmed galaxies 
(\citealt{2024ApJ...969L...2F}; \citealt{2024ApJ...964...71H}; \citealt{2024Natur.633..318C}; 
\citealt{2024ApJ...972..143C}; \citealt{2024arXiv240920549S}; \citealt{2024arXiv240920533C}).
The large green circle represents the object included in the catalog of little red dots 
(\citealt{2024arXiv240403576K}).
The names of the well-studied galaxies GN-z11, GHZ2, JADES-GS-z14-0, 
and CEERS-16943 (Maisie's Galaxy) are labeled in blue near their respective data points.
GHZ2(Y) indicates the measurement result for GHZ2 obtained by \cite{2022ApJ...938L..17Y}.
The gray solid line shows the best-fit size--luminosity relation for $z\sim9$ galaxies 
obtained in previous studies, and the gray dashed line indicates its extrapolation 
(\citealt{2023ApJ...951...72O}).
}
\label{fig:re_Muv}
\end{center}
\end{figure}

We investigate the relation between size and UV luminosity for $z\sim10$--$16$ galaxies. 
Figure \ref{fig:re_Muv} plots the size--luminosity relation for our $z\sim10$--$16$ galaxy sample, 
along with the galaxies at similar redshifts analyzed in the literature  
(\citealt{2022ApJ...938L..17Y}; \citealt{2023ApJ...951...72O}). 
As noted in Section \ref{subsec:SBfitting_results}, 
the sizes of galaxies with $M_{\rm UV} \lesssim -20$ mag show significant variation, 
ranging from relatively large sizes of up to around $500$ pc to very small sizes below $100$ pc  
such as GN-z11 and GHZ2. 
As described there, 
the compact morphologies are suggested to originate from AGN activity and/or compact star formation.
For the latter case, a comparison with cosmological zoom-in simulations is presented in Section \ref{sec:discussion}.

Note that, among the three objects overlapped with the little red dots selected by \cite{2024arXiv240403576K}, 
CEERS-98518 is the only object with an aperture magnitude $\mathrm{S/N} > 10$ 
and is marked with a large green circle in Figure \ref{fig:re_Muv}.
Its size is not remarkably small compared to the other galaxies with comparable UV luminosities at similar redshifts.  
Although little red dots are often selected as compact objects in the rest-frame optical 
(e.g., \citealt{2024ApJ...963..129M}; 
\citealt{2024ApJ...968....4P}; 
\citealt{2024ApJ...968...38K}; 
\citealt{2024arXiv240403576K}), 
our results may indicate that they are not necessarily as compact as point sources in the rest-frame UV 
probably because of the significant contribution from the stellar component 
(see also, \citealt{2024arXiv241114383R}).

Following previous studies 
(e.g., \citealt{2012A&A...547A..51G}; \citealt{2013ApJ...777..155O}; \citealt{2013ApJ...765...68H}; 
\citealt{2013ApJ...773..153J}; \citealt{2015ApJ...804..103K}; \citealt{2015ApJS..219...15S}; 
\citealt{2015ApJ...808....6H}; \citealt{2016MNRAS.457..440C}; \citealt{2017MNRAS.466.3612B}; 
\citealt{2017ApJ...843...41B}; \citealt{2018ApJ...855....4K}; \citealt{2020AJ....160..154H}; 
\citealt{2022ApJ...927...81B}; \citealt{2022ApJ...938L..17Y}; \citealt{2023ApJ...951...72O}; 
\citealt{2024ApJ...960..104S}; \citealt{2024ApJ...963....9M}; \citealt{2024MNRAS.533.3724V}), 
to characterize the size-luminosity relation, 
we fit a power-law function to the data points, 
\begin{equation}
r_{\rm e}
	= r_0 \left( \dfrac{L_{\rm UV}}{L_0} \right)^\alpha, 
\end{equation}
where $r_0$ is the effective radius at the luminosity of $L_0$, 
corresponding to $- 21.0$ mag 
(\citealt{2013ApJ...765...68H}; \citealt{2015ApJS..219...15S}),
and $\alpha$ is the slope of the size--luminosity relation. 
Since galaxy sizes evolve with redshift as presented in the next subsection,
we fit the power-law function to the data points for $z=9.5$--$13$ galaxies in our sample,
as well as the $z\sim12$ data points from \cite{2023ApJ...951...72O},
to limit the impact of redshift evolution by restricting the redshift range.
We do not perform the fit for the $z=13$--$16$ data points,
as their small sample size prevents a reliable constraint.
With this approach, we obtain best-fit values of
$r_0 = 0.20 \pm 0.01$ kpc and 
$\alpha = 0.19 \pm 0.03$. 
In this fitting, we exclude GN-z11 and GHZ2, 
for which the presence of AGN has been suggested (Section \ref{subsec:SBfitting_results}).
For $r_0$, \cite{2015ApJS..219...15S} fit their results with a functional form 
$r_0 \propto (1+z)^{\beta_z}$, as shown in their Figure 10. 
Compared to their extrapolated results, 
our value for $r_0$ is slightly smaller but broadly consistent. 
For $\alpha$, we find a slightly smaller value compared to \cite{2015ApJS..219...15S}.

For comparison, we also show the size--luminosity relation 
obtained in previous studies for galaxies at a slightly lower redshift of $z\sim9$ 
(\citealt{2023ApJ...951...72O}). 
At the same UV luminosity, our results suggest that 
the sizes of $z\sim10$--$16$ galaxies are slightly smaller on average.

Next, we compare the star formation rate (SFR) surface densities, 
$\Sigma_{\rm SFR}$, of the $z\sim10$--$16$ galaxies 
with previous results for lower-$z$ galaxies. 
The SFRs of the $z\sim10$--$16$ galaxies are derived 
from the UV luminosities using the following equation:
\begin{equation}
{\rm SFR} 
	= 1.4 \times 10^{-28} \alpha_{\rm SC} L_\nu, 
\label{eq:Kennicutt_SFR}
\end{equation}
where $L_\nu$ is the rest-frame UV luminosity density in units of erg s$^{-1}$ Hz$^{-1}$ 
(\citealt{1998ARA&A..36..189K}). 
We multiply by the factor $\alpha_{\rm SC} = 0.63$ (\citealt{2014ARA&A..52..415M}) 
to convert from the Salpeter initial mass function (IMF; \citealt{1955ApJ...121..161S}) 
to the Chabrier IMF (\citealt{2003PASP..115..763C}). 
Dust extinction is not corrected here, since high-$z$ galaxies tend to exhibit blue UV slopes 
(e.g., \citealt{2024MNRAS.529.4087T}; \citealt{2024MNRAS.531..997C}; \citealt{2024ApJ...976..193R}; 
\citealt{2024arXiv240410751A}; \citealt{2024arXiv241114532S}; \citealt{2024arXiv241119893Y}). 
In this conversion, 
we assume that the UV luminosities originate from star formation 
for all objects analyzed in this study including GN-z11 and GHZ2. 
We then calculate the SFR surface density in units of $M_\odot$ yr$^{-1}$ kpc$^{-2}$. 
This quantity represents the average SFR within a circular region enclosed by the half-light radius, 
given by
\begin{equation}
\Sigma_{\rm SFR}
	= \dfrac{\rm SFR}{2 \pi r_{\rm e}^2},
\end{equation}
where the factor $1/2$ is included because the SFR is estimated based on the total luminosity, 
while the area used for division is based on the half-light radius 
(e.g., \citealt{2008ApJ...673..686H}; \citealt{2013ApJ...768...74T}; \citealt{2016ApJ...833...70D}).

\begin{figure}
\begin{center}
   \includegraphics[width=0.5\textwidth]{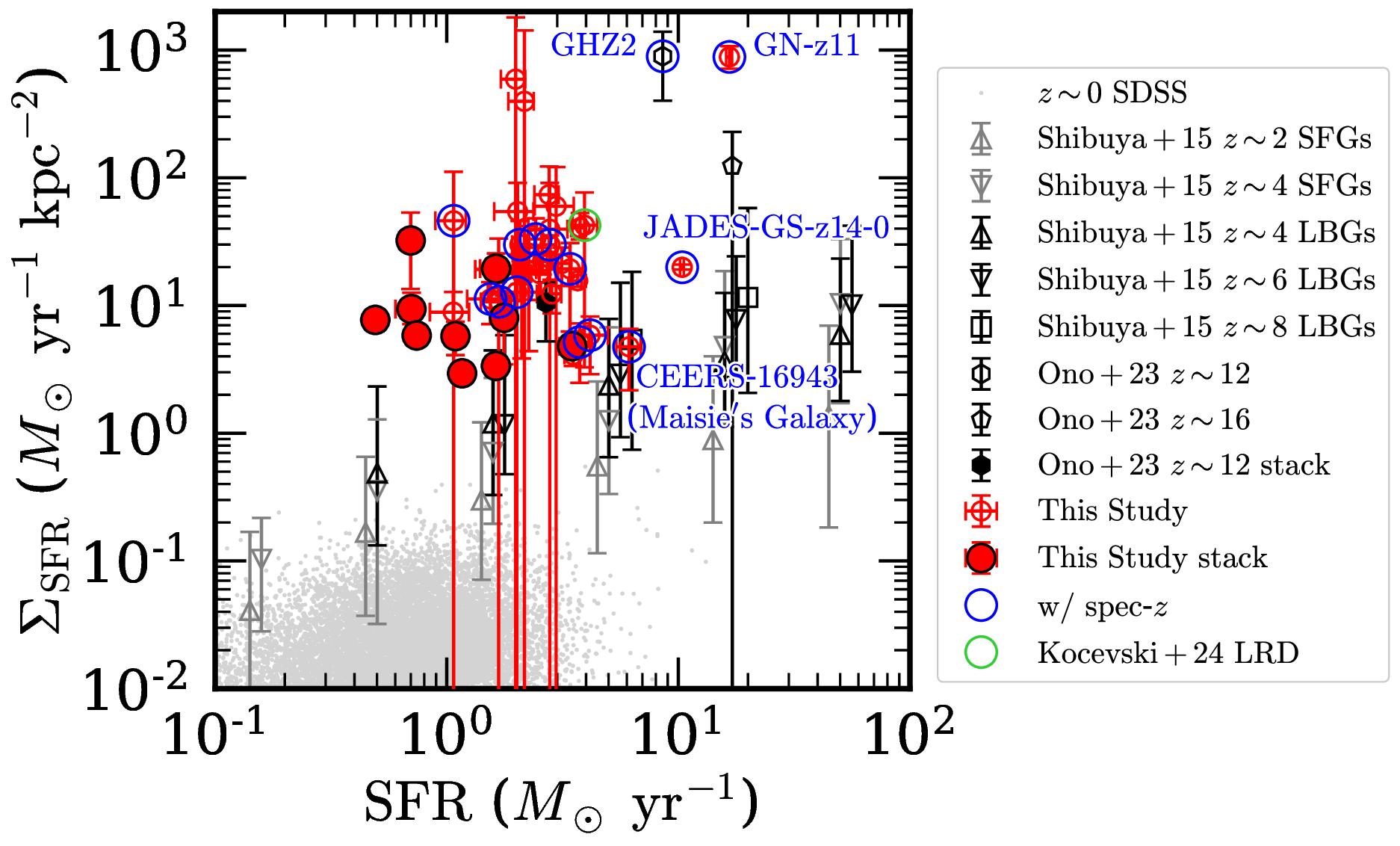}
\caption{
SFR surface density $\Sigma_{\rm SFR}$ vs. SFR.  
The red open circles show the results for relatively luminous $z\sim10$--$16$ galaxies individually analyzed in this study. 
The red filled circles indicate the results for stacked images of fainter $z\sim10$--$16$ galaxies.  
The black pentagons and hexagons represent previous results at similar redshifts (\citealt{2023ApJ...951...72O}).  
Among them, those marked with large blue circles 
correspond to spectroscopically confirmed galaxies 
(\citealt{2024ApJ...969L...2F}; \citealt{2024ApJ...964...71H}; \citealt{2024Natur.633..318C}; 
\citealt{2024ApJ...972..143C}; \citealt{2024arXiv240920549S}; \citealt{2024arXiv240920533C}). 
The large green circle denotes the object included in the catalog of little red dots 
(\citealt{2024arXiv240403576K}).
The names of well-studied objects, 
GN-z11, GHZ2, JADES-GS-z14-0, and CEERS-16943 (Maisie's Galaxy), 
are labeled near their respective data points.  
The gray dots represent $z\sim0$ galaxies compiled by \cite{2015ApJS..219...15S} 
from the Sloan Digital Sky Survey (SDSS), with $\Sigma_{\rm SFR}$ values 
calculated using the catalog of \cite{2012MNRAS.421.2277L}.  
The other black and gray data points 
are the results for $z\sim2$--$8$ galaxies reported by \cite{2015ApJS..219...15S}.
}
\label{fig:SigmaSFR_SFR}
\end{center}
\end{figure}

Figure \ref{fig:SigmaSFR_SFR} plots the SFR surface density on the vertical axis 
and the SFR on the horizontal axis, 
comparing the $z\sim10$--$16$ galaxies analyzed in this study 
with lower-$z$ galaxies investigated in previous work. 
As expected from the evolution of the size--UV luminosity relation, 
the $\Sigma_{\rm SFR}$ values increase on average with redshift even beyond $z=10$, 
which is consistent with previous results (\citealt{2023ApJ...951...72O}). 
The $\Sigma_{\rm SFR}$ values 
are on average comparable to those obtained 
within the framework of the L-GALAXIES2020 semi-analytical model by \cite{2025A&A...693L...6N}. 
Interestingly, the extremely compact luminous sources GN-z11 and GHZ2 
exhibit very high $\Sigma_{\rm SFR}$ values, around $1000 M_\odot$ yr$^{-1}$ kpc$^{-2}$, 
and occupy a distinct region in this plane compared to other galaxies. 
This origin can be attributed to the two possibilities described in Section \ref{subsec:SBfitting_results}.

\begin{figure}
\begin{center}
   \includegraphics[width=0.5\textwidth]{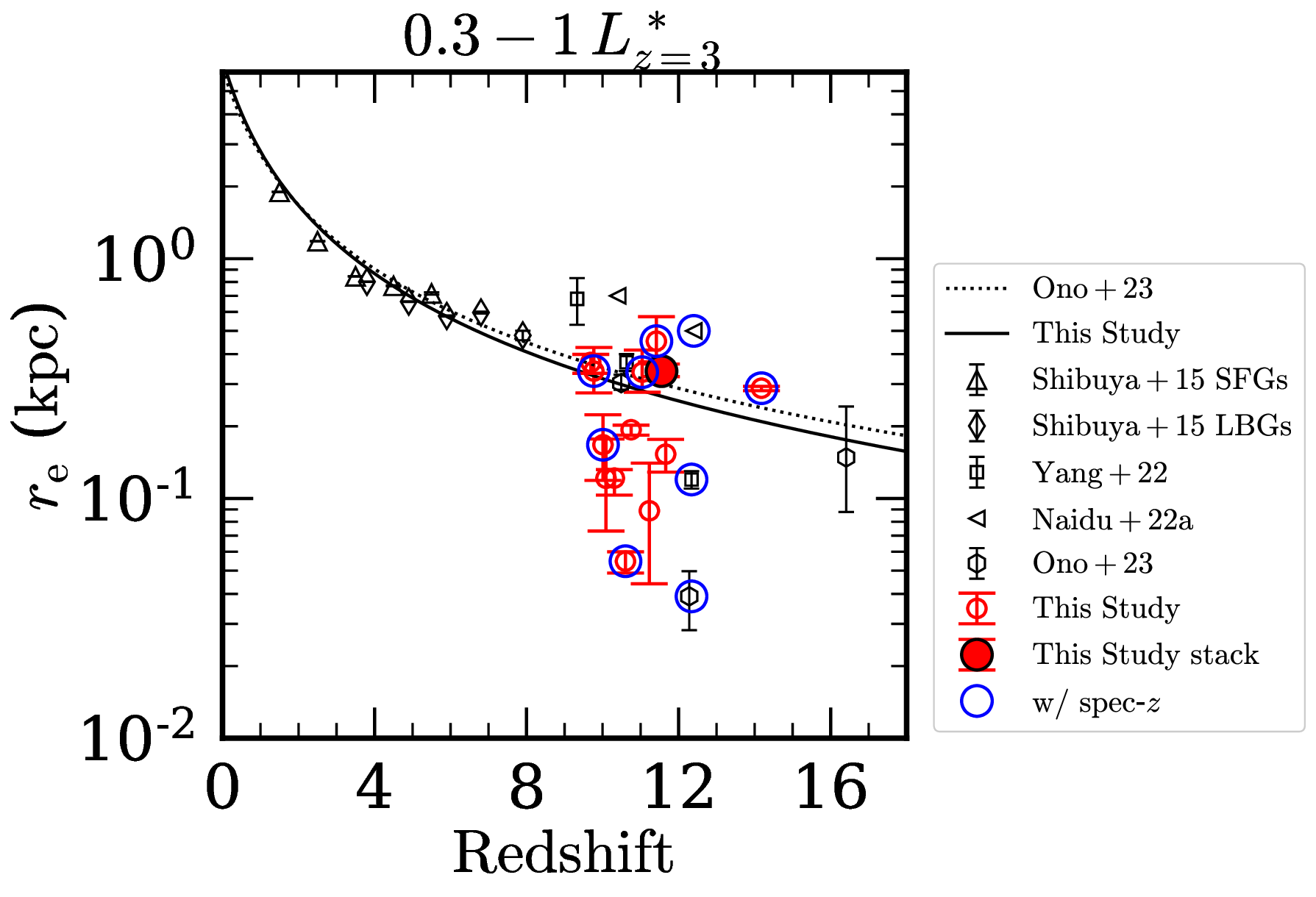}
   \includegraphics[width=0.5\textwidth]{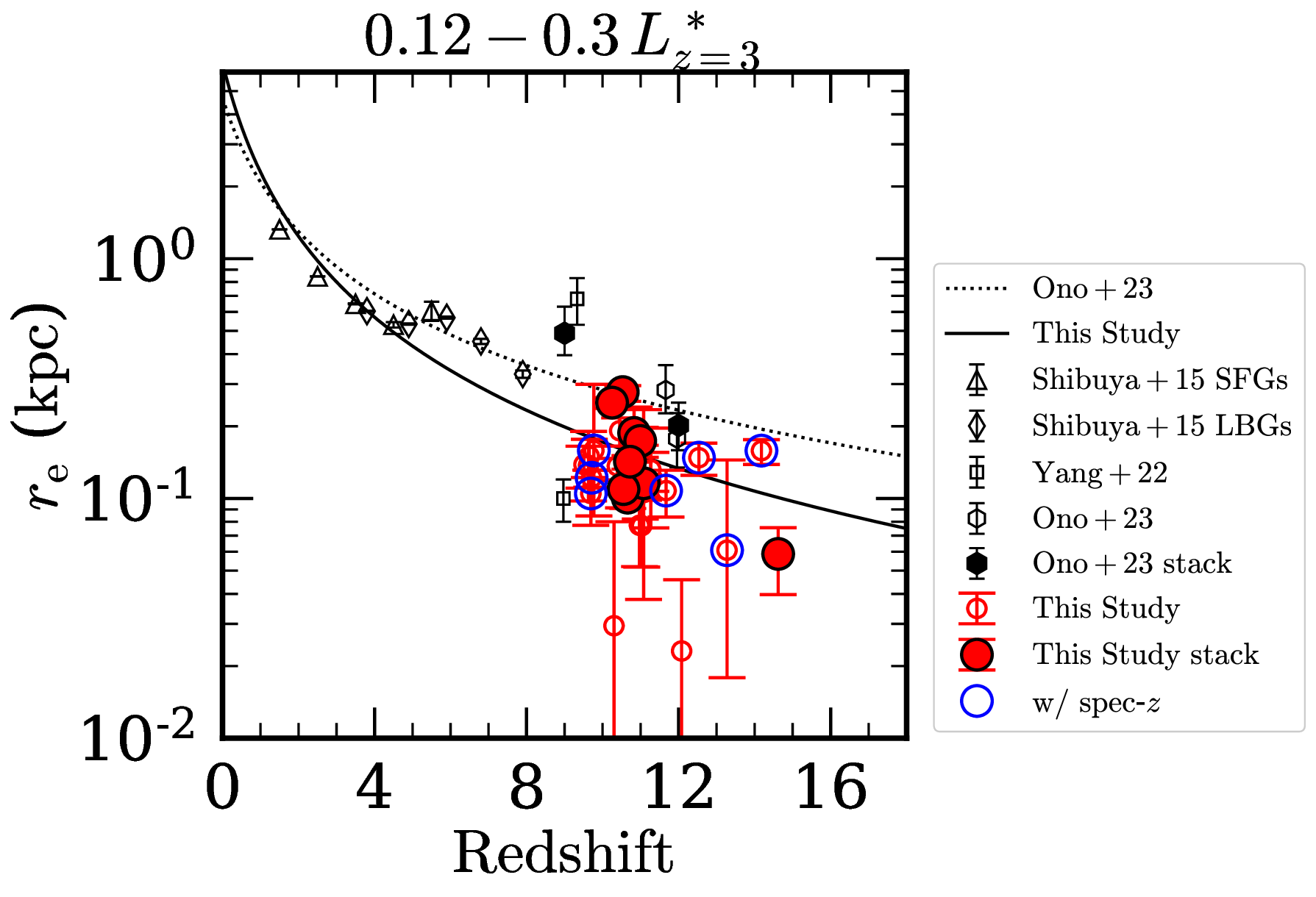}
\caption{
Evolution of the half-light radius of star-forming galaxies with UV luminosities 
in the ranges ($0.3$--$1$) $L_{z=3}^\ast$ (top) and ($0.12$--$0.3$) $L_{z=3}^\ast$ (bottom). 
The red open circles correspond to the sizes of relatively luminous $z\sim10$--$16$ galaxies 
measured individually in this study, while the red filled circles represent the results for 
stacked images of fainter $z\sim10$--$16$ galaxies. 
The black open symbols show the sizes of star-forming galaxies analyzed individually 
in previous studies, and the black filled hexagon indicates the results for stacked images 
in previous studies 
(\citealt{2015ApJS..219...15S}; \citealt{2022ApJ...938L..17Y}; \citealt{2023ApJ...951...72O}). 
The blue large circles mark spectroscopically confirmed galaxies 
(\citealt{2024ApJ...969L...2F}; \citealt{2024ApJ...964...71H}; \citealt{2024Natur.633..318C}; 
\citealt{2024ApJ...972..143C}; \citealt{2024arXiv240920549S}; \citealt{2024arXiv240920533C}). 
The solid curve shows the best-fit $r_{\rm e} \propto (1+z)^s$ function derived in this study, 
while the dotted curve represents the best-fit result from \cite{2023ApJ...951...72O}.
}
\label{fig:re_redshift}
\end{center}
\end{figure}

\begin{deluxetable*}{lcccccc} 
\tablecolumns{7} 
\tablewidth{0pt} 
\tablecaption{Summary of Galaxy Size Evolution Results 
\label{tab:SizeEvolution_results}}
\tablehead{
    \colhead{Reference} 
    &  \colhead{Telescope}
    &  \colhead{Software}
    &  \colhead{Redshift}
    &  \colhead{Rest-frame}
    &  \colhead{Luminosity or}
    &  \colhead{$s$ of $(1+z)^s$}
\\
    \colhead{ } 
    &  \colhead{ }
    &  \colhead{ }
    &  \colhead{Range}
    &  \colhead{Wavelength}
    &  \colhead{$M_{\rm star}$ Range}
    &  \colhead{ }
\\
    \colhead{(1)} 
    &  \colhead{(2)}
    &  \colhead{(3)}
    &  \colhead{(4)}
    &  \colhead{(5)}
    &  \colhead{(6)}
    &  \colhead{(7)}
}
\startdata 
\cite{2004ApJ...611L...1B}		&	HST		&	SExtractor			& 2--6		& UV			& ---						& $-1.05 \pm 0.21$ \\
\cite{2004ApJ...600L.107F}	&	HST		&	SExtractor			& 2--5		& UV			& ---						& $ \sim -1.5$$^{\dagger 1}$ \\
\cite{2006ApJ...653...53B}		&	HST		&	SExtractor			& $\sim 6$	& UV			& ---						& $-1.1 \pm 0.3 $ \\
\cite{2008ApJ...673..686H}	&	HST		&	SExtractor			& 4--6		& UV			& ---						& $\sim -1.5$$^{\dagger 1}$ \\
\cite{2010ApJ...709L..21O}	&	HST		&	SExtractor			& 7--8		& UV			& (0.3--1)$L_{z=3}^\ast$		& $-1.12 \pm 0.17$ \\
						&			&					&			& 			& (0.12--0.3)$L_{z=3}^\ast$	& $-1.32 \pm 0.52$ \\
\cite{2012ApJ...756L..12M}	&	HST		&	GALFIT			& 4--7		& UV			& $10^{9.5-10.4} M_\odot$	& $-1.20 \pm 0.11$ \\
						&			&					&			&			& $10^{8.6-9.5} M_\odot$		& $-1.18 \pm 0.10$ \\
\cite{2013ApJ...777..155O}	&	HST		&	GALFIT			& 7--8		& UV			& (0.12-1)$L_{z=3}^\ast$		& $-1.30^{+0.12}_{-0.14}$ \\
\cite{2015ApJ...804..103K}	&	HST		&	glafic				& 6--8		& UV			& (0.3-1)$L_{z=3}^\ast$		& $-1.24 \pm 0.1$ \\
\cite{2015ApJ...808....6H}		&	HST		&	GALFIT			& 9--10		& UV			& $>0.3 L_{z=3}^\ast$		& $-1.04 \pm 0.09$ \\
						&			&					&			&			& $<0.3 L_{z=3}^\ast$		& $-0.8 \pm 0.1$ \\
\cite{2015ApJS..219...15S}	&	HST		&	GALFIT			& 4--10		& UV			& (0.3--1)$L_{z=3}^\ast$		& $-1.10 \pm 0.06$ \\
						&			&					&			& 			& (0.12--0.3)$L_{z=3}^\ast$	& $-1.22 \pm 0.05$ \\
\cite{2016MNRAS.457..440C}	&	HST		&	SExtractor			& 4--8		& UV			& (0.3--1)$L_{z=3}^\ast$		& $-0.20 \pm 0.26$ \\
						&			&					&			& 			& (0.12--0.3)$L_{z=3}^\ast$	& $-0.47 \pm 0.62$ \\
\cite{2017ApJ...834L..11A}	&	HST		&	GALFIT			& 1--7		& UV			& $>10^{10} M_\odot$		& $-0.89 \pm 0.01$ \\
\cite{2018ApJ...855....4K}		&	HST		&	glafic				& 6--9		& UV			& (0.3--1)$L_{z=3}^\ast$		& $-1.28 \pm 0.11$ \\
\cite{2023ApJ...951...72O}		&	JWST	&	GALFIT			& 9--16		& UV			& (0.3--1)$L_{z=3}^\ast$		& $-1.22^{+0.17}_{-0.16}$ \\
						&			&					&			&			& (0.12--0.3)$L_{z=3}^\ast$	& $-1.17 \pm 0.16$ \\
\cite{2024ApJ...963....9M}		&	JWST	&	GALFIT			& 5--14		& UV			& ---$^{\dagger 2}$			& $-0.4 \pm 0.2$ \\
\cite{2024ApJ...960..104S}	&	JWST	&	GALFITM			& 4--10		& optical		& $10^{9.4-10.5} M_\odot$	& $-1.27 \pm 0.60$ \\	
						&			&					&			& 			& $10^{8-9.4} M_\odot$		& $-1.39 \pm 0.33$ \\
\cite{2024MNRAS.527.6110O}	& 	JWST	&	GALFIT			& 1--8		& optical		& $>10^{9.5} M_\odot$		& $-0.71 \pm 0.19$ \\
\cite{2024MNRAS.533.3724V}	& 	JWST	&	{\sc PyAutoGalaxy}	& 3--5		& optical		& $>10^9 M_\odot$			& $-0.60 \pm 0.22$ \\
\cite{2024ApJ...962..176W}	&	JWST	&	GALFITM			& 0.5--5.5		& optical		& $5 \times 10^{10} M_\odot$$^{\dagger 3}$	& $-0.63 \pm 0.07$ \\
\cite{2024arXiv241016354A}	& 	JWST	&	GALFITM			& 3--9		& optical		& $5 \times 10^{10} M_\odot$$^{\dagger 3}$	& $-0.807 \pm 0.026$ \\
\cite{2024arXiv241214970W}	&	JWST	&	GALFIT			& 7--13		& optical/UV	& ---						& $-0.67 \pm 0.06$ \\
\cite{2025arXiv250407185Y}	& 	JWST 	&	GaLight			& 2--10		& optical/UV	& $>10^9 M_\odot$			& $-1.21 \pm 0.05$ \\
This Study				& 	JWST	&	GALFIT			& 10--16		& UV			& (0.3--1)$L_{z=3}^\ast$		& $-1.28^{+0.26}_{-0.22}$ \\
						&			&					&			& 			& (0.12--0.3)$L_{z=3}^\ast$	& $-1.52^{+0.20}_{-0.1}$ \\
\enddata 
\tablecomments{(1) Reference. 
(2) Telescope used to obtain the imaging data.
(3) Software used for size measurements.
(4) Redshift range of the galaxy sample whose sizes are measured.
(5) Rest-frame wavelength in which size evolution is investigated.
(6) Luminosity or stellar mass range of the galaxies used for fitting the size evolution.
(7) Power-law index of $(1+z)$ fitted to the measured size evolution.} 
\vspace{-0.5em}
\tablenotetext{$^{\dagger 1}$}{Technically, reported to follow $\propto H(z)^{-1}$.} 
\vspace{-0.5em}
\tablenotetext{$^{\dagger 2}$}{Sizes corrected based on the size-stellar mass relation are used.}
\vspace{-0.5em}
\tablenotetext{$^{\dagger 3}$}{Sizes at a specific stellar mass based on the size-stellar mass relation are used.}
\end{deluxetable*} 

\subsection{Size Evolution} \label{subsec:size_evolution}

We investigate the size evolution of galaxies. 
Figure \ref{fig:re_redshift} shows the measured half-light radii of our $z\sim10$--$16$ galaxies 
plotted as a function of redshift, along with results from previous studies. 
Since galaxy size depends on UV luminosity, 
we follow previous studies and divide the data into two luminosity ranges: 
$L/L_{z=3}^\ast = 0.3$--$1$ and $0.12$--$0.3$.

The size evolution of galaxies has been characterized in previous studies 
using the functional form $r_{\rm e} \propto (1+z)^s$ 
(e.g., \citealt{2004ApJ...600L.107F}; 
\citealt{2004ApJ...611L...1B}; 
\citealt{2006ApJ...653...53B}; 
\citealt{2008ApJ...673..686H}; 
\citealt{2010ApJ...709L..21O}; 
\citealt{2013ApJ...777..155O}; 
\citealt{2015ApJ...804..103K}; 
\citealt{2015ApJ...808....6H}; 
\citealt{2015ApJS..219...15S}; 
\citealt{2016MNRAS.457..440C}; 
\citealt{2017ApJ...834L..11A}; 
\citealt{2018ApJ...855....4K}; 
\citealt{2019ApJ...882...42B}; 
\citealt{2020AJ....160..154H}), 
where $s$ represents the slope of the power-law relation. 
We fit this functional form to these data points. 
For the uncertainty in the \cite{2015ApJS..219...15S} data points, 
we follow their methodology and use the 68th percentile of the size distribution of individual galaxies. 
For the uncertainties in the other data points, 
we use the measurement uncertainties of the individual galaxy sizes. 
As a result, we obtain 
$s = -1.28^{+0.26}_{-0.22}$ for $L/L_{z=3}^\ast = 0.3$--$1$ 
and $s = -1.52^{+0.20}_{-0.19}$ for $L/L_{z=3}^\ast = 0.12$--$0.3$. 
The slope for the brighter luminosity sample is consistent with the previous results of \cite{2023ApJ...951...72O},  
while the slope for the fainter luminosity bin is slightly smaller than the previous results beyond the $1\sigma$ uncertainty. 
This may be due to the limited number of high-$z$ galaxies in the previous work.

Table \ref{tab:SizeEvolution_results} compiles 
a summary of previously reported power-law slopes from various studies. 
Observations primarily based on HST data generally show convergence towards slope values 
between approximately $-1.0$ and $-1.5$. 
Several recent studies utilizing JWST data have reported slopes with significantly smaller absolute values. 
While a detailed investigation into the exact reasons of these differences is beyond the scope of this paper, 
they may arise from variations in sample selection (e.g., stellar mass-selected versus UV luminosity-selected samples), 
the specific redshift ranges adopted for power-law fitting, 
and/or the rest-frame wavelength at which the sizes are measured.

\section{Discussion} \label{sec:discussion}

\subsection{Implications of the Log-Normal Size Distribution: 
Are $\lowercase{z}\sim10$--$16$ Galaxies Disk Galaxies?}

Previous studies have shown that 
the sizes of star-forming galaxies at $z\sim0$--$6$ follow a log-normal distribution 
with a standard deviation of $\sigma_{{\ln}r_{\rm e}} \sim 0.4$--$0.7$ independent of redshift 
(\citealt{2015ApJS..219...15S}; see also, \citealt{2015ApJ...804..103K}; \citealt{2018ApJ...855....4K}).
They have also reported that the S{\'e}rsic index is typically as low as $\simeq 1.5$ 
and that the stellar-to-halo size ratio $r_{\rm e}/r_{\rm vir}$, 
which is the ratio of galaxy size to host dark matter halo size, 
remains nearly constant at approximately $0.02$. 
According to the galaxy disk formation scenario, 
the rotationally supported gas that forms the basis of the galaxy disk 
acquires angular momentum from the host dark matter halo through tidal interactions 
(e.g., \citealt{1980MNRAS.193..189F}; \citealt{1984ApJ...286...38W}; \citealt{1987ApJ...319..575B}), 
and the galaxy size is expected to be proportional to $r_{\rm vir}$ and 
the dark matter halo spin parameter $\lambda$ 
(Equation 28 of \citealt{1998MNRAS.295..319M}; see also, \citealt{2013ApJ...764L..31K}). 
Moreover, theoretical studies suggest that $\lambda$ follows a log-normal distribution 
with a standard deviation of $\sigma_{{\ln}\lambda} \sim 0.5$--$0.6$ 
(e.g., \citealt{1998MNRAS.295..319M}; \citealt{2001ApJ...555..240B}; \citealt{2001ApJ...557..616G}; 
\citealt{2002ApJ...576...21V}; \citealt{2002ApJ...581..799V}; \citealt{2005ApJ...627..647B}; 
\citealt{2009MNRAS.393.1498D}; \citealt{2017MNRAS.466.1625Z}; 
see also, \citealt{1992ApJ...399..405W}; \citealt{1995MNRAS.272..570S}; \citealt{2007MNRAS.376..215B}), 
which is similar to those of the galaxy size distributions. 
Based on these results, it has been suggested that 
typical star-forming galaxies at $z\sim0$--$6$ have stellar components 
that can be explained by the galaxy disk formation scenario 
(\citealt{2015ApJS..219...15S}).

In Section \ref{subsec:SBfitting_results}, 
based on the deep JWST images and incorporating the results from our previous work, 
we have demonstrated that the size distribution of galaxies at $z\sim4$--$16$ also follows a log-normal distribution 
(Figure \ref{fig:histo_lnre}). 
Furthermore, by combining our results with those from previous studies
(\citealt{2014ApJ...788...28V}; \citealt{2015ApJS..219...15S}; \citealt{2024MNRAS.533.3724V};
\citealt{2024ApJ...962..176W}; \citealt{2024ApJ...963....9M}; \citealt{2024arXiv241016354A};
\citealt{2024arXiv241206957M}), 
we have reported that 
the standard deviation of the log-normal distribution of galaxy sizes 
shows no clear evolution toward $z\gtrsim10$ 
(Figure \ref{fig:sigma_lnre_redshift}). 
Additionally, for the $z\sim10$--$16$ galaxies investigated in this study,
we have derived their axis ratio distribution in Section \ref{subsec:ba_distribution},
finding it to be nearly uniform across $b/a \sim 0.2$--$0.8$,
similar to the distribution observed for local spiral galaxies.

In this section, we further estimate the stellar-to-halo size ratio, $r_{\rm e}/r_{\rm vir}$, 
for the $z\sim10$--$16$ galaxies analyzed in this study. 
The virial radii of the host dark matter halos for the $z\sim10$--$16$ galaxies 
are calculated based on their UV luminosities through the SFR and stellar mass, as follows, 
although the systematic uncertainties are not small.
First, the SFRs of the $z\sim10$--$16$ galaxies are derived 
from the UV luminosities using Equation \ref{eq:Kennicutt_SFR}. 
Next, based on the SFRs, we estimate the stellar masses of the $z\sim10$--$16$ galaxies 
using the star formation main sequence relation given as Equation 5 of 
\cite{2023arXiv231210152C} (see also their Table 2).\footnote{For galaxies at $z \gtrsim10$, 
the uncertainties of the stellar mass estimates would not be small, 
because NIRCam only probes their rest-frame UV. 
Additional data, such as those from MIRI, are required to probe the rest-frame optical 
and improve the accuracy of the stellar mass estimates.}
Subsequently, the virial masses of the host dark matter halos are calculated using 
the stellar mass -- halo mass relations from 
\cite{2019MNRAS.488.3143B} (see the right panel of their Figure 9). 
Specifically, we employ their Equations J1--J8 
and adopt the parameter values listed in the top row of their Table J. 
Note that their relations show a turnover at high redshifts,  
where stellar mass decreases with increasing halo mass,  
and the turnover mass decreases with redshift. 
For some of our galaxies, particularly the luminous ones at high redshifts,  
the derived halo mass exceeds the turnover mass, making it impossible to determine the halo mass. 
For such cases, we simply scale the $z=9$ stellar mass -- halo mass relation for our analysis. 
Finally, the virial radius $r_{\rm vir}$ is calculated from the halo mass using the following equation:
\begin{equation}
r_{\rm vir} 
	= \left( \dfrac{2 G M_{\rm vir}}{\Delta_{\rm vir} H(z)^2} \right)^{1/3}, 
\end{equation}
where 
$\Delta_{\rm vir} = 18 \pi^2 + 82 x - 39 x^2$ 
and 
$x = \Omega_{\rm m}(z) - 1$ 
(\citealt{1998ApJ...495...80B}; see also, \citealt{2015ApJ...804..103K}; 
\citealt{2015ApJS..219...15S}; \citealt{2018ApJ...855....4K}).

\begin{figure}
\begin{center}
   \includegraphics[width=0.5\textwidth]{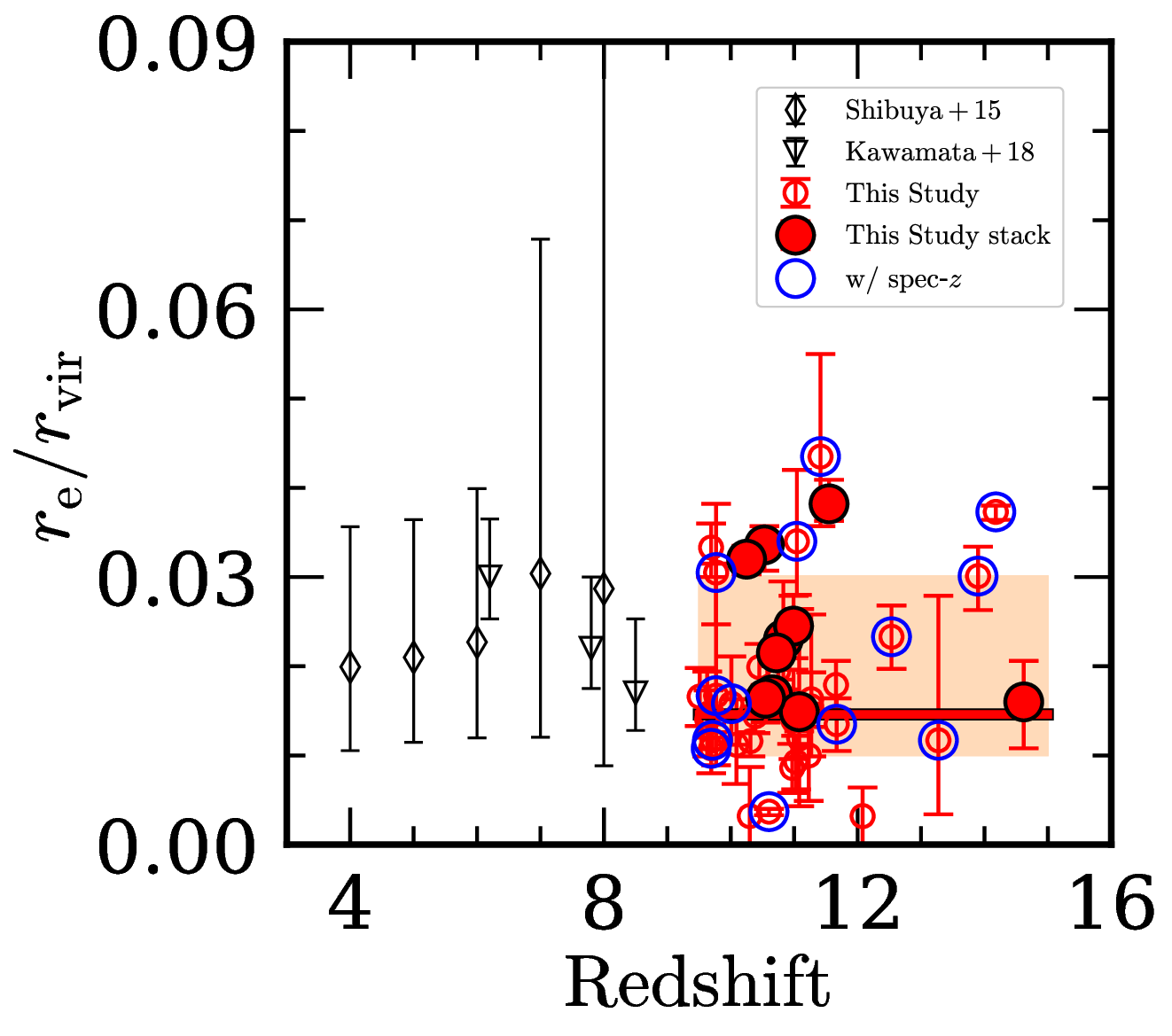}
\caption{
Ratio of galaxy UV continuum size to host dark matter halo size, 
$r_{\rm e}/r_{\rm vir}$, as a function of redshift. 
The red open circles represent the results of relatively luminous $z\sim10$--$16$ galaxies 
measured individually in this study, while the red filled circles show the results for 
stacked images of fainter $z\sim10$--$16$ galaxies. 
The blue large circles indicate spectroscopically confirmed galaxies 
(\citealt{2024ApJ...969L...2F}; \citealt{2024ApJ...964...71H}; \citealt{2024Natur.633..318C}; 
\citealt{2024ApJ...972..143C}; \citealt{2024arXiv240920549S}; \citealt{2024arXiv240920533C}). 
The red solid line represents the median of the results for the individually analyzed galaxies in this study, 
and the orange shaded region indicates the 68th percentile range. 
The black diamonds and downward triangles indicate the results from previous studies 
of $z\sim 4$--$8$ star-forming galaxies (\citealt{2015ApJS..219...15S}; \citealt{2018ApJ...855....4K}).
}
\label{fig:re_rvir_redshift}
\end{center}
\end{figure}

Figure \ref{fig:re_rvir_redshift} shows $r_{\rm e}/r_{\rm vir}$ as a function of redshift. 
The $z\sim10$--$16$ galaxies analyzed in this study exhibit $r_{\rm e}/r_{\rm vir}$ values of $\simeq 0$--$0.04$, 
with a median value and 68th percentile range of $r_{\rm e}/r_{\rm vir} = 0.015^{+0.015}_{-0.005}$. 
Previous studies have reported that the $r_{\rm e}/r_{\rm vir}$ values for galaxies at $z\sim 0$--$8$ 
remain approximately constant at $r_{\rm e}/r_{\rm vir} \simeq 0.02$ 
(\citealt{2015ApJ...804..103K}; \citealt{2015ApJS..219...15S}; \citealt{2018ApJ...855....4K}; 
see also, \citealt{2007MNRAS.378...41S}; \citealt{2017ApJ...838....6H}). 
Very recently, \cite{2025arXiv250407185Y} have estimated 
$r_{\rm e}/r_{\rm vir}$ values for $z\sim2$--$10$ galaxies 
based on JWST COSMOS-Web data, obtaining a median value of approximately $0.027$. 
Our results indicate that this trend continues to higher redshifts, up to $z\sim16$, with no significant evolution.

As described above, 
the size distribution of $z\sim10$--$16$ galaxies 
can be explained by a log-normal distribution, 
with its standard deviation comparable to that of the spin parameter of dark matter halos. 
Moreover, their axis ratio distribution is similar to that of local spiral galaxies. 
Additionally, the stellar-to-halo size ratio remains nearly constant regardless of redshift. 
Furthermore, as reported in Section \ref{sec:SBprofile_fitting}, 
the median S{\'e}rsic index for the luminous $z\sim10$--$16$ galaxies in our sample 
is consistent with the exponential disk profile. 
These findings suggest that $z\sim10$--$16$ galaxies, 
like the star-forming galaxies at lower redshifts studied in previous work, 
have stellar components that are consistent with the galaxy disk formation scenario.  
This may imply that stellar feedback is not very effective in these high-$z$ galaxies, 
which could be attributed to the compact nature of these high-$z$ galaxies, 
a consequence of the smaller virial radii of dark matter halos at higher redshifts for a given stellar mass.

Recent simulations have shown that at high redshifts, 
supernova feedback can significantly disrupt galaxy disks, 
making it challenging for them to sustain a stable disk structure over long periods, 
often resulting in clumpy morphologies instead 
(\citealt{2015MNRAS.451.2900K}; \citealt{2017ApJ...846...30Y}; \citealt{2023MNRAS.525.4832Y}). 
On the other hand, for compact galaxies at high redshifts, 
it has been suggested that feedback processes may be less effective, 
potentially allowing for more efficient star formation 
(\citealt{2023MNRAS.523.3201D}; \citealt{2024arXiv241020530A}).
Our results do not explicitly favor either scenario. 
However, the fact that the size distribution of $z\sim10$--$16$ galaxies is similar 
to that of the spin parameter distribution of dark matter halos suggests that the latter scenario might be more plausible. 
If feedback is indeed less effective in these galaxies, 
their star formation efficiency could be significantly enhanced. 
In this context, JWST may be preferentially observing such galaxies with high star formation efficiency. 
This potential observational bias warrants further investigation in future studies.

\begin{figure}
\begin{center}
   \includegraphics[width=0.45\textwidth]{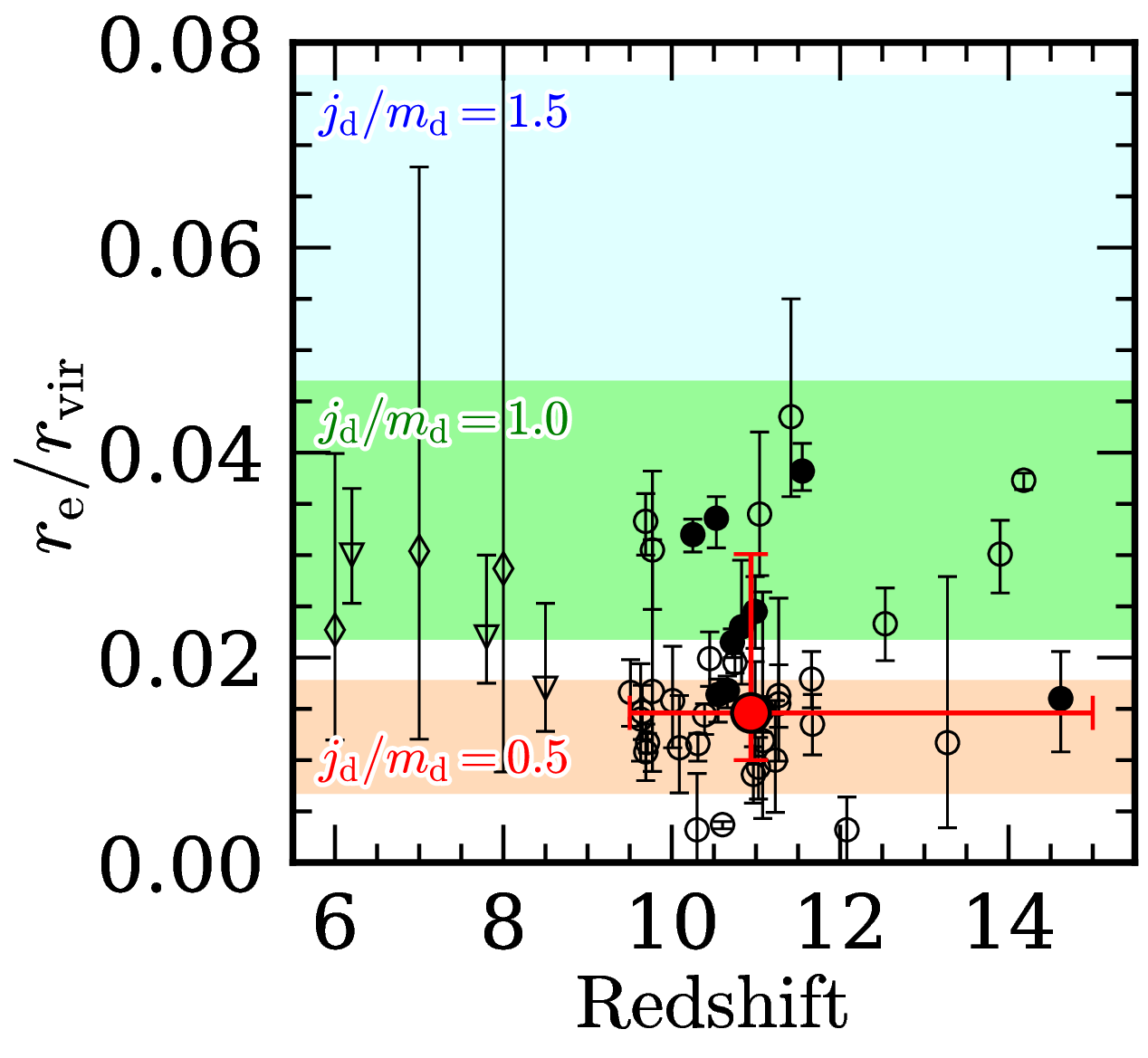}
\caption{Comparison between the observed galaxy-to-host dark matter halo size ratio  
and the expected values from the galaxy disk formation model.  
The black open circles represent the results for relatively luminous $z\sim10$--$16$ galaxies  
analyzed individually in this study,  
while the black filled circles show the results for stacked images of fainter $z\sim10$--$16$ galaxies.  
The red filled circle indicates the median value from this study, along with its 68th percentile range.  
The data points for lower-$z$ galaxies are the same as those in Figure \ref{fig:re_rvir_redshift}.  
The red, green, and blue shaded regions represent the ranges of values expected  
from Equation \ref{eq:re_rvir_theory} for specific disk angular momentum fractions of 
$j_{\rm d} / m_{\rm d} = 0.5$, $1.0$, and $1.5$, respectively,  
based on the model of \cite{1998MNRAS.295..319M}.
}
\label{fig:re_rvir_redshift_with_theory}
\end{center}
\end{figure}

In the case that $z\sim10$--$16$ galaxies host disks,  
a comparison with the galaxy disk formation scenario  
allows us to place rough constraints on their specific disk angular momentum fraction 
(e.g., \citealt{2013ApJ...764L..31K}; \citealt{2015ApJ...804..103K}; \citealt{2015ApJS..219...15S}).  
Based on the galaxy disk formation model of \cite{1998MNRAS.295..319M},  
the stellar-to-halo size ratio is given by  
\begin{equation}
\dfrac{r_{\rm e}}{r_{\rm vir}}
  =\dfrac{1.678}{\sqrt{2}} 
  \left(\dfrac{j_{\rm d}}{m_{\rm d}} \lambda \right)
  \dfrac{f_{\rm R} (\lambda, \, c_{\rm vir}, m_{\rm d}, j_{\rm d})}{\sqrt{f_{\rm c} (c_{\rm vir})}}. 
\label{eq:re_rvir_theory}
\end{equation}  
Here we adopt $r_{\rm e} = 1.678 R_{\rm d}$,  
where $R_{\rm d}$ is the scale length of an exponential disk.  
The parameters $m_{\rm d}$ and $j_{\rm d}$ represent  
the mass fraction and angular momentum fraction of the disk relative to the halo, respectively.  
The functions $f_{\rm c} (c_{\rm vir})$ and $f_{\rm R} (\lambda, \, c_{\rm vir}, m_{\rm d}, j_{\rm d})$  
depend on the halo concentration factor $c_{\rm vir}$,  
with their functional forms provided in Equation 23 and Equation 29 of \cite{1998MNRAS.295..319M}, 
respectively.\footnote{For the latter, the actual calculation follows Equation 32 of \cite{1998MNRAS.295..319M}.} 
By adopting the values of the spin parameter $\lambda$ and $c_{\rm vir}$ obtained from theoretical studies,  
we can use this comparison to obtain rough constraints 
on the specific disk angular momentum fraction, $j_{\rm d} / m_{\rm d}$.

\begin{figure*}[ht]
\begin{center}
   \includegraphics[width=0.7\textwidth]{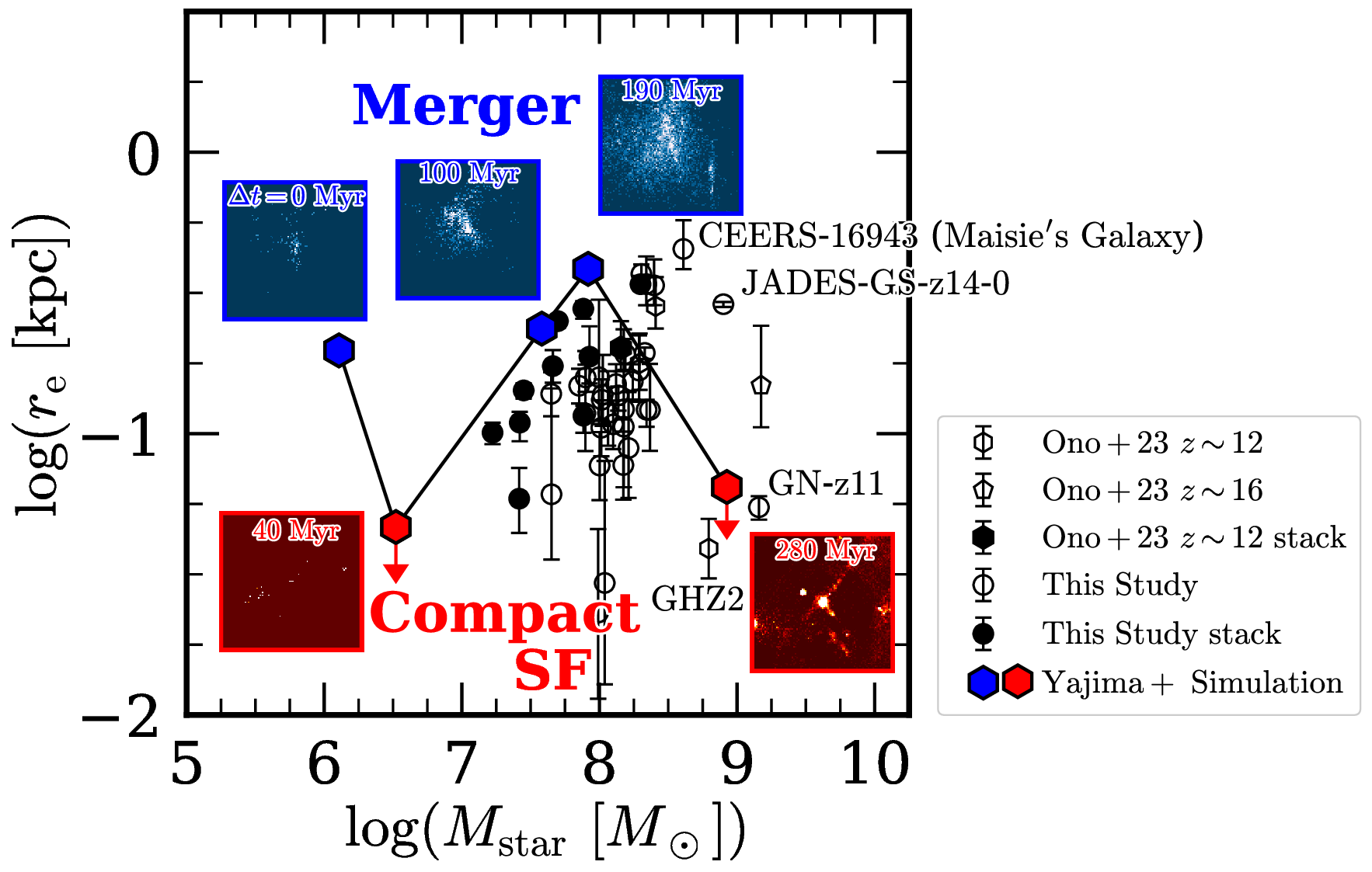}
\caption{Comparison with theoretical results on the half-light radius vs. stellar mass plane. 
The black open symbols represent the sizes of relatively luminous $z\sim10$--$16$ galaxies individually analyzed 
in this study and our previous work (\citealt{2023ApJ...951...72O}) 
and the black filled symbols indicate the results for stacked images of fainter $z\sim10$--$16$ galaxies. 
The well-studied objects, GN-z11, GHZ2, JADES-GS-z14-0, and CEERS-16943 (Maisie's Galaxy) 
are labeled near their corresponding data points.
The filled hexagons represent the evolutionary track of a galaxy with a similar stellar mass and redshift to our galaxies,  
obtained from cosmological hydrodynamic zoom-in simulations 
(\citealt{2022MNRAS.509.4037Y}; \citealt{2023MNRAS.525.4832Y}).  
For this simulated galaxy, data points with $\log r_{\rm e}$ (kpc) $> -1$ are marked in blue,  
whereas those with $\log r_{\rm e}$ (kpc) $< -1$ are in red.  
The inner images represent 2D projected stellar mass distribution snapshots of the corresponding simulated galaxy,  
each covering a $1\farcs5 \times 1\farcs5$ region, and are color-coded similarly:  
snapshots with $\log r_{\rm e} > -1$ are shaded in blue tones, while those with $\log r_{\rm e} \leq -1$ are in red tones.  
The time difference in Myr, relative to the starting data point at the top left, is indicated for each snapshot.  
}
\label{fig:re_Mstar_theory}
\end{center}
\end{figure*}

Figure \ref{fig:re_rvir_redshift_with_theory} compares 
the observed values of $r_{\rm e} / r_{\rm vir}$  
with the theoretical ranges corresponding to specific disk angular momentum fractions of  
$j_{\rm d} / m_{\rm d} = 0.5$, $1.0$, and $1.5$.  
The spin parameter $\lambda$ is varied within the range of the mean values  
obtained from theoretical studies, $\lambda = 0.035$--$0.045$  
(e.g., \citealt{2001ApJ...555..240B}; \citealt{2001ApJ...557..616G}; 
\citealt{2002ApJ...576...21V}; \citealt{2002ApJ...581..799V}; \citealt{2005ApJ...627..647B}; 
\citealt{2009MNRAS.393.1498D}; \citealt{2017MNRAS.466.1625Z}),  
and the halo concentration factor $c_{\rm vir}$ is varied within the range of $c_{\rm vir} = 3$--$5$   
following Figure 11 of \cite{2021MNRAS.506.4210I} (see also, \citealt{2012MNRAS.423.3018P}).  
Additionally, for the mass fraction, we adopt a conservative range of $m_{\rm d}=0.05$--$0.1$  
(\citealt{2015ApJS..219...15S}; see also, \citealt{1998MNRAS.295..319M}).  
This comparison suggests that the specific disk angular momentum fraction  
is roughly $j_{\rm d} / m_{\rm d} \sim 0.5$--$1$.  
This implies that disk galaxies acquire at least half of their specific angular momentum  
from their host dark matter halos.  
As shown in Figure \ref{fig:re_rvir_redshift_with_theory},  
this value is comparable to previous results for lower-$z$ galaxies
(\citealt{2015ApJ...804..103K}; \citealt{2015ApJS..219...15S}; 
\citealt{2018ApJ...854...22O}). 
Moreover, it is also in good agreement with values estimated from kinematic data  
for nearby disk galaxies ($j_{\rm d} / m_{\rm d} \sim 0.6$--$1.0$, 
e.g., \citealt{2012MNRAS.421..608D}; \citealt{2012ApJS..203...17R};  
\citealt{2013ApJ...769L..26F}; \citealt{2018ApJ...868..133F}; 
see also, \citealt{2023MNRAS.518.1002R}).  
These results suggest no significant evolution of $j_{\rm d} / m_{\rm d}$ even beyond $z=10$.

\subsection{Size Comparisons with Cosmological Zoom-in Simulation Results: 
Origins of Extended and Extremely Compact Morphologies}

As reported in Section \ref{subsec:SBfitting_results},
the sizes of $z\sim10$--$16$ galaxies exhibit significant diversity,
ranging from compact sources smaller than $100$ pc to extended ones reaching around $500$ pc.
As described there, two possible origins have been suggested for the compact galaxies:
one possibility is that they host AGNs, and the other is that they undergo compact star formation 
(e.g., \citealt{2024arXiv240618352H}).
Here, we focus on the latter case and explore the origin of this size diversity
by comparing our results with those from cosmological zoom-in simulations.
This discussion serves as a revised version of the analysis presented in Section 5.3 of \cite{2023ApJ...951...72O},
now incorporating the new observational results from this study.

We use the simulation results from the FOREVER22 project 
(\citealt{2022MNRAS.509.4037Y}),   
which has performed cosmological hydrodynamics zoom-in simulations 
to investigate the formation and evolution of galaxies within a large comoving volume of $(714 \, {\rm Mpc})^3$  
(see also, \citealt{2023MNRAS.525.4832Y}).  
Among their simulations, we focus on the highest-resolution run,  
which has a mass resolution of $\sim 8 \times 10^3 M_\odot$ (first run).  
We select a galaxy that has a stellar mass comparable to the galaxies analyzed in this study at $z\sim10$  
and extract its 2D projected stellar mass density distributions at four different epochs  
spanning from an initial redshift of $z=17$ to approximately $40$ Myr to $280$ Myr later.

Figure \ref{fig:re_Mstar_theory} shows the evolution of the half-light radius 
and stellar mass of the simulated galaxy.  
The stellar mass of the simulated galaxy increases over time,  
but its size exhibits both growth and shrinkage at different stages. 
As explained in \cite{2023MNRAS.525.4832Y},  
the simulated galaxy grows through mass accretion including minor mergers,  
leading to an increase in the size of its stellar components.  
However, when major mergers occur,  
the stellar components temporarily appear larger,  
but as angular momentum is lost, their size rapidly decreases.  
According to the Illustris simulation results,  
galaxies typically experience one to two major mergers by $z=10$ (\citealt{2015MNRAS.449...49R}),  
and at that time, the cosmic age corresponds to several times the dynamical timescale,  
suggesting that their morphologies are expected to be strongly affected by merger events.

We present our observational results alongside the simulated galaxy in Figure \ref{fig:re_Mstar_theory}.
Many of the $z\sim10$--$16$ galaxies exhibit sizes that fall between  
the compact phase and the relatively extended phase of the simulated galaxy.  
This suggests that a large fraction of $z\sim10$--$16$ galaxies  
are either in the process of increasing their size through mass accretion  
or in the stage of size reduction following a major merger.  
The particularly extended galaxies, CEERS-16943 (Maisie's Galaxy) and JADES-GS-z14-0,  
have sizes comparable to the largest size reached by the simulated galaxy  
at the snapshots taken in the simulation.  
This may indicate that CEERS-16943 and JADES-GS-z14-0  
are currently undergoing events such as major mergers and/or tidal interactions with surrounding objects.  
On the other hand, the compact objects GN-z11 and GHZ2  
have sizes consistent with those observed in the simulated galaxy  
when it undergoes significant size shrinkage after a major merger,  
leading to extremely compact star formation.  
Such compact phases naturally emerge in cosmological simulations,
where major mergers can trigger high-density gas concentrations,
resulting in intense star formation within a small region.
If GN-z11 and GHZ2 are indeed compact star-forming galaxies,  
this suggests that they may have just experienced size shrinkage due to a major merger.

These results indicate that mergers play an important role 
in shaping the diverse sizes of high-$z$ galaxies. 
While major mergers can significantly alter galaxy structures, 
either compacting them or temporarily making them appear more extended, 
gas accretion and feedback processes also contribute to their structural evolution. 
At high redshifts, the high baryon accretion (including minor mergers) rate  
ensures a continuous supply of gas 
with angular momentum comparable to those of the host dark matter halos. 
Additionally, feedback processes may not be highly effective in these early galaxies 
due to their high gas densities. 
This combination of mergers, high accretion rates, and weak feedback could explain 
why the observed size distribution follows a log-normal distribution 
with a standard deviation similar to that of the spin parameter of dark matter halos. 
The observed approximately constant stellar-to-halo size ratios further support the idea that, 
even in the presence of mergers, 
the morphological properties of high-$z$ galaxies are regulated 
by the angular momentum acquired from their dark matter halos.
However, these interpretations remain speculative, and further studies are required 
to fully understand the interplay between mergers, gas accretion, and feedback at high redshifts.

\section{Summary} \label{sec:summary}

In this study, we have performed morphological analyses 
for the statistical sample of $169$ galaxies at $z\sim10$–$16$ found in the CEERS and JADES surveys, 
$14$ of which have spectroscopic confirmations. 
Using the deep JWST NIRCam images, we have conducted surface brightness profile fitting for these galaxies. 
For the $30$ galaxies with aperture magnitudes S/N $> 10$, 
we have conducted individual analyses, 
while the remaining galaxies have been stacked to determine their average properties. 
When allowing the S{\'e}rsic index to vary as a free parameter in the profile fitting, 
the obtained value is $n = 1.3 \pm 0.6$ on average. 
However, due to the large uncertainty, 
we have fixed the S{\'e}rsic index for the profile fitting, following previous studies. 
Our main results are as follows:

\begin{enumerate}

\item The sizes of the individually analyzed $z\sim10$–$16$ galaxies 
range from compact sources smaller than $100$ pc 
to larger ones around $500$ pc, 
but are typically in the range of $100$–$300$ pc. 
Similarly, the sizes of stacked galaxies are also in the range of $100$–$300$ pc.
Thus, the typical size of $z\sim10$–$16$ galaxies is $100$–$300$ pc.

\item We have derived the size distribution of $z\sim10$--$16$ galaxies 
and found that it is well described by a log-normal distribution 
with a mean of $r_{\rm e} = 133^{+13}_{-12}$ pc 
and a standard deviation of $\sigma_{\ln r_{\rm e}} = 0.52 \pm 0.08$. 
Compared to previous results, 
we have found that the standard deviation of the log-normal distribution of galaxy sizes 
shows no significant evolution even at $z\gtrsim10$.

We have found that the axis ratio distribution of 
$z\sim10$--$16$ galaxies appears nearly uniform over $b/a\sim0.2$--$0.8$. 
A K-S test quantitatively confirms that this distribution is statistically similar 
to local spiral galaxies, unlike local ellipticals,  
suggesting that these high-$z$ galaxies may already exhibit disk-like morphologies.

\item Investigating the relation between galaxy size and UV luminosity 
has revealed that for galaxies brighter than $M_{\rm UV} \lesssim -20$ mag, 
the sizes span a wide range from large to small. 
On the other hand, for fainter galaxies, very large sizes are not obtained. 
Excluding the very compact luminous sources 
for which the presence of AGNs has been suggested based on previous spectroscopic studies, 
we have fitted the size--luminosity relation with a power-law function.  
We have found that the normalization broadly agrees with the trend of 
decreasing size with increasing redshift reported in previous studies, 
while the slope is slightly smaller compared to those studies.

\item Combining our results with previous findings, 
we have fitted the size evolution of galaxies with a power-law function of $r_{\rm e} \propto (1+z)^s$. 
For the sample with $L/L_{z=3}^\ast = 0.3$--$1$, the slope is consistent with previous results. 
However, for the sample with $L/L_{z=3}^\ast = 0.12$--$0.3$, 
the slope appears to be slightly steeper.

\item Using the UV luminosities, we have estimated the virial radii of the $z\sim10$--$16$ galaxies  
through the star formation main sequence relation and the stellar mass -- halo mass relation. 
We have then calculated their ratios of galaxy sizes to virial radii, $r_{\rm e} / r_{\rm vir}$, 
and obtained the median value of $r_{\rm e} / r_{\rm vir} = 0.015^{+0.015}_{-0.005}$. 
This value is consistent with those derived for lower-$z$ galaxies 
within the $1\sigma$ ranges, indicating no significant evolution toward $z\gtrsim10$.

\item Our findings that the size distribution of $z\sim10$--$16$ galaxies 
follows the log-normal distribution with the almost constant standard deviation, 
which is comparable to the standard deviation of the spin parameter distribution of dark matter halos, 
that their axis ratio distribution shows a uniform distribution similar to spiral galaxies,  
that the stellar-to-halo size ratio shows no significant evolution toward $z\gtrsim10$, 
and that the mean S{\'e}rsic index is around $1.3$ albeit with large uncertainties, 
are all consistent with the classical disk formation scenario. 
In summary, our results suggest that 
galaxies at $z\sim10$--$16$ possess stellar components 
that align with the galaxy disk formation scenario.

\item In the case that 
$z\sim10$--$16$ galaxies host disks, 
a comparison with the galaxy disk formation model suggests that 
their specific disk angular momentum fraction is approximately $j_{\rm d} / m_{\rm d} \sim 0.5$--$1$, 
implying that disk galaxies acquire at least half of their specific angular momentum from their host halos.  
This value is comparable to those for lower-$z$ galaxies including local disk galaxies.

\item Based on comparisons with cosmological simulations, 
in the case that these galaxies originate from star formation, 
their diverse sizes likely reflect different evolutionary stages. 
Extended galaxies like CEERS-16943 (Maisie's Galaxy) and JADES-GS-z14-0 
may be undergoing mergers or tidal interactions, 
while very compact objects such as GN-z11 and GHZ2 host compact star formation 
with sizes consistent with post-merger contraction phases.

\end{enumerate}

\section*{Acknowledgements}

We thank Kentaro Nagamine, Arjen van der Wel, and Kazuhiro Shimasaku 
for providing insightful comments on our results. 
We are also grateful to Takatoshi Shibuya for kindly sharing the data files from their previous work.
This work is based on observations
made with the NASA/ESA/CSA James Webb Space Telescope. 
The data 
(\citealt{https://doi.org/10.17909/z7p0-8481,https://doi.org/10.17909/8tdj-8n28})
were obtained from the Mikulski Archive for Space Telescopes 
at the Space Telescope Science Institute, 
which is operated by the Association of Universities for Research in Astronomy, Inc., 
under NASA contract NAS 5-03127 for JWST.
These observations are associated with programs 
ERS-1345 (CEERS), 
GTO-1180, GTO-1181, GTO-1210, GTO-1286, 
GO-1895, GO-1963, and GO-3215 (JADES). 
The authors acknowledge the CEERS team led by Steven L. Finkelstein 
and the JADES team led by Daniel Eisenstein and Nora Luetzgendorf 
for developing their observing programs. 
This work was partially performed using the computer facilities of
the Institute for Cosmic Ray Research, The University of Tokyo. 
This work was supported 
by the World Premier International
Research Center Initiative (WPI Initiative), MEXT, Japan, 
as well as 
KAKENHI Grant Numbers 
15K17602, 
15H02064, 
17H01110, 
17H01114, 
19K14752, 
20H00180, 
21H04467, 
21H04489, 
22K03670, 
and 24H00245 
through the Japan Society for the Promotion of Science (JSPS). 
This work was also supported by JST FOREST Program, Grant Number JP-MJFR202Z. 
This work was partially supported by the joint research program of 
the Institute for Cosmic Ray Research (ICRR), University of Tokyo. 

\software{GALFIT (\citealt{2002AJ....124..266P}; \citealt{2010AJ....139.2097P}), 
SExtractor (\citealt{1996A&AS..117..393B}), 
IRAF (\citealt{1986SPIE..627..733T,1993ASPC...52..173T}),\footnote{IRAF is distributed by the National Optical Astronomy Observatory, 
which is operated by the Association of Universities for Research in Astronomy (AURA) 
under a cooperative agreement with the National Science Foundation.} 
SAOImage DS9 \citep{2003ASPC..295..489J},
Numpy \citep{2020Natur.585..357H}, 
Matplotlib \citep{2007CSE.....9...90H}, 
Scipy \citep{2020NatMe..17..261V}, 
Astropy \citep{2013A&A...558A..33A,2018AJ....156..123A},\footnote{\url{http://www.astropy.org}}
Ned Wright's Javascript Cosmology Calculator \citep{2006PASP..118.1711W}.\footnote{\url{http://www.astro.ucla.edu/~wright/CosmoCalc.html}}
}

\bibliographystyle{aasjournal}
\bibliography{ref}

\appendix

\setcounter{table}{0}
\renewcommand{\thetable}{A.\arabic{table}}
\setcounter{figure}{0}
\renewcommand{\thefigure}{A.\arabic{figure}}

\section{$\lowercase{z} \sim10$--$16$ Galaxy Catalogs in the CEERS and JADES Fields Used in Our Size Measurements} \label{sec:galaxy_catalogs}

\begin{figure*}[h]
\begin{center}
   \includegraphics[width=0.18\textwidth]{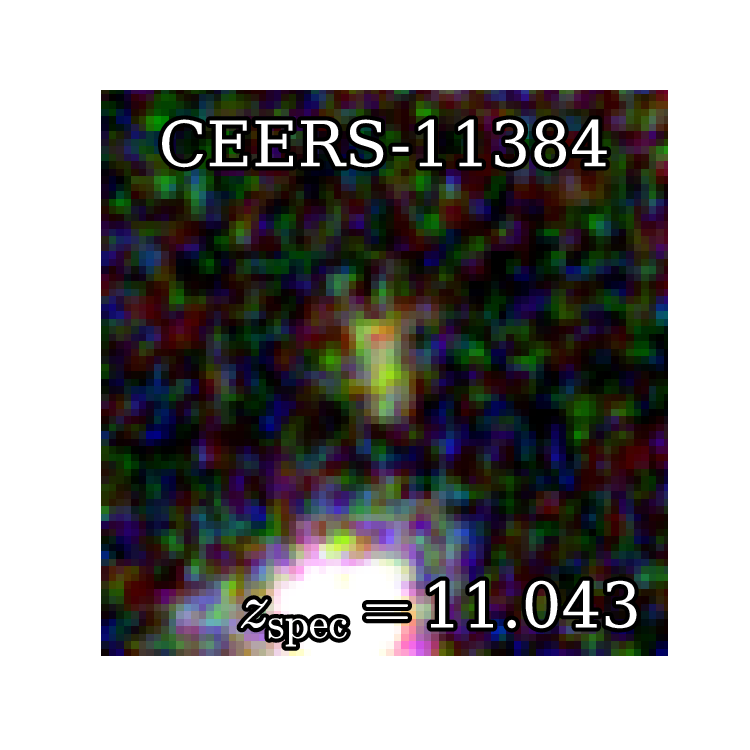}
   \includegraphics[width=0.18\textwidth]{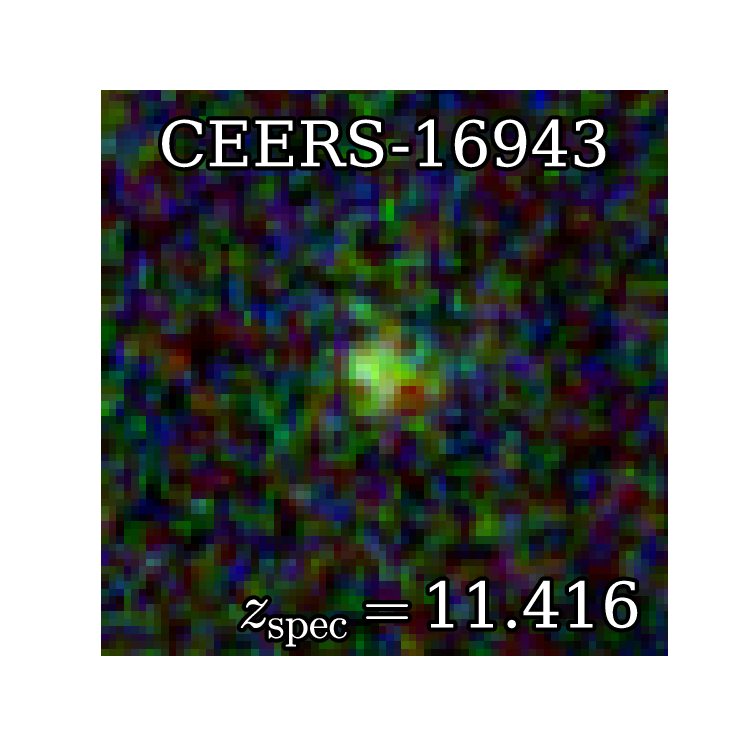}
   \includegraphics[width=0.18\textwidth]{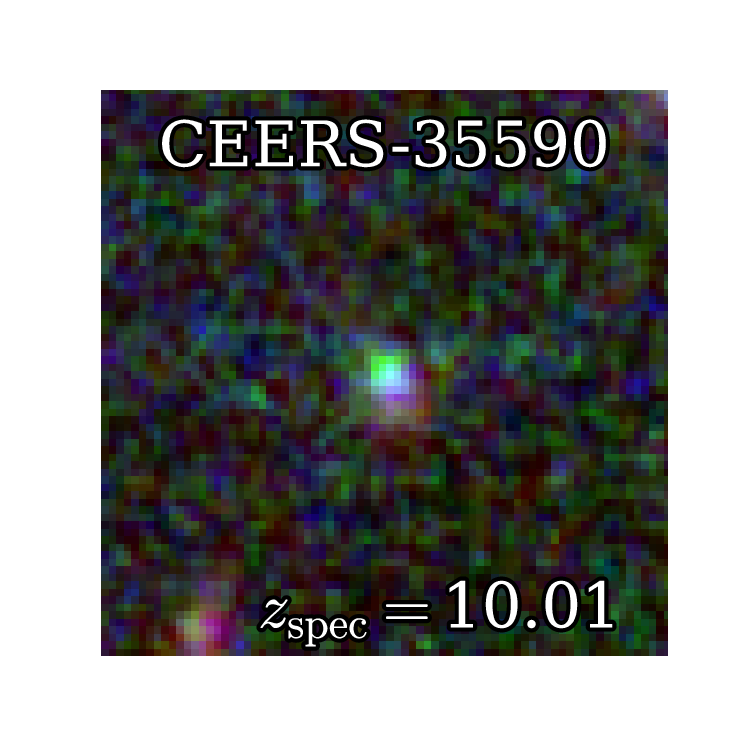}
   \includegraphics[width=0.18\textwidth]{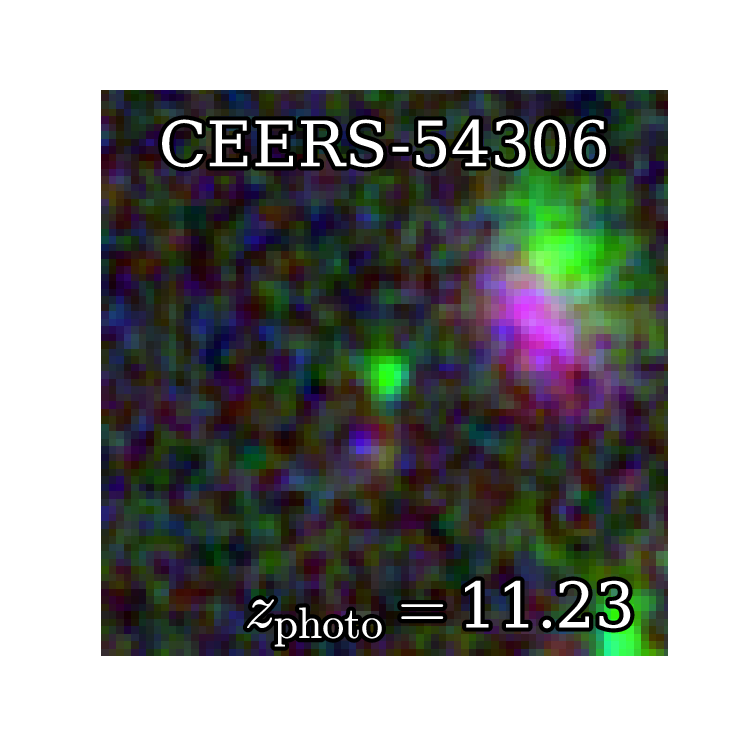}
   \includegraphics[width=0.18\textwidth]{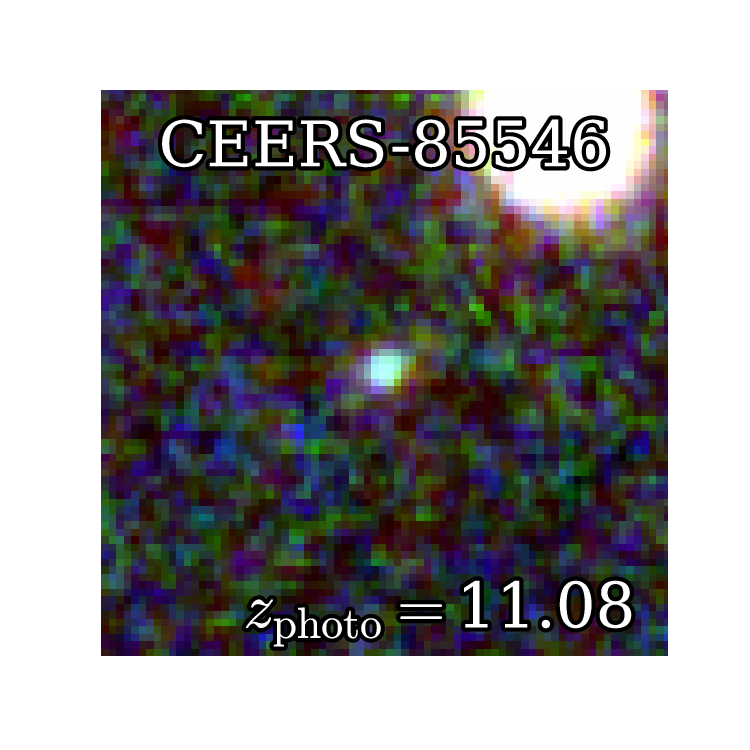}
   \includegraphics[width=0.18\textwidth]{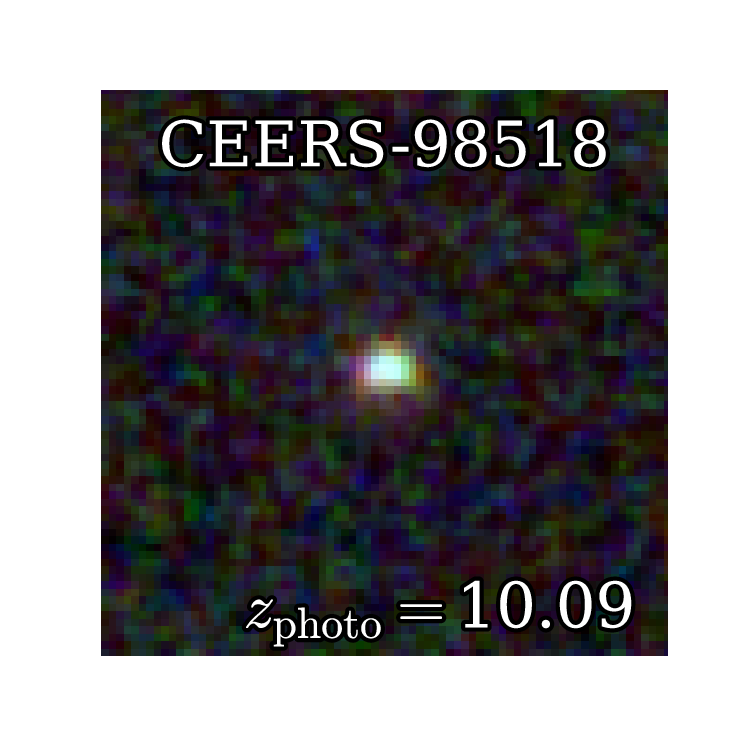}
   \includegraphics[width=0.18\textwidth]{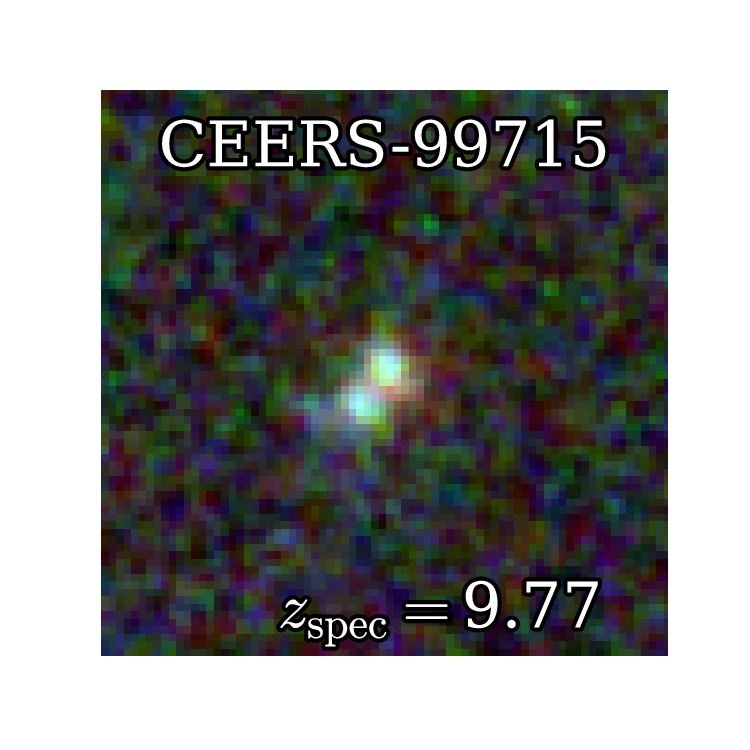}
   \includegraphics[width=0.18\textwidth]{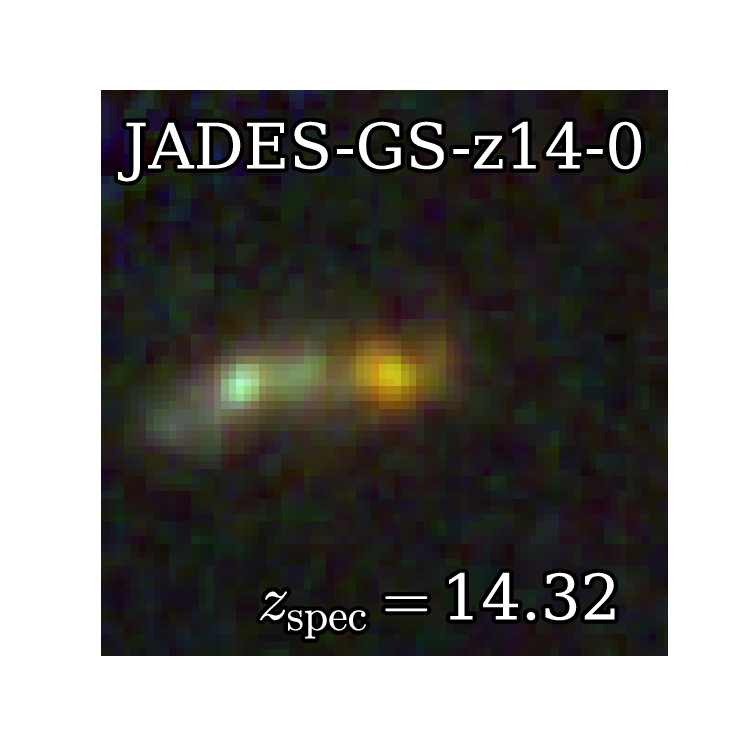}
   \includegraphics[width=0.18\textwidth]{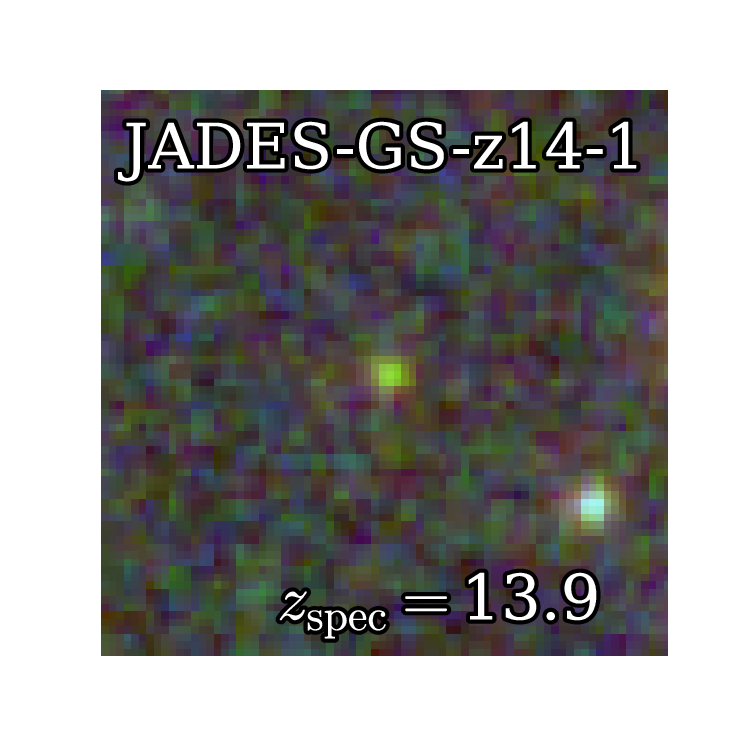}
   \includegraphics[width=0.18\textwidth]{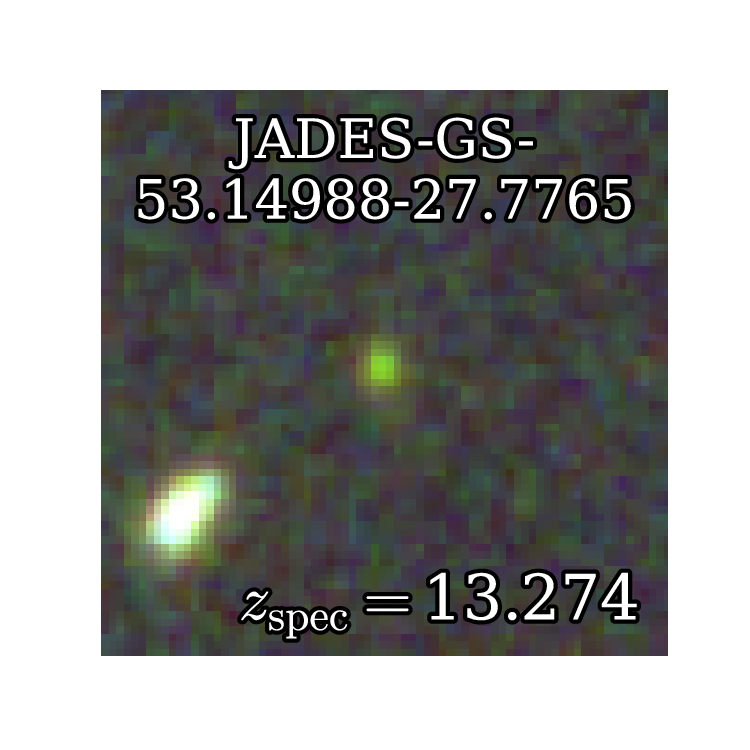}
   \includegraphics[width=0.18\textwidth]{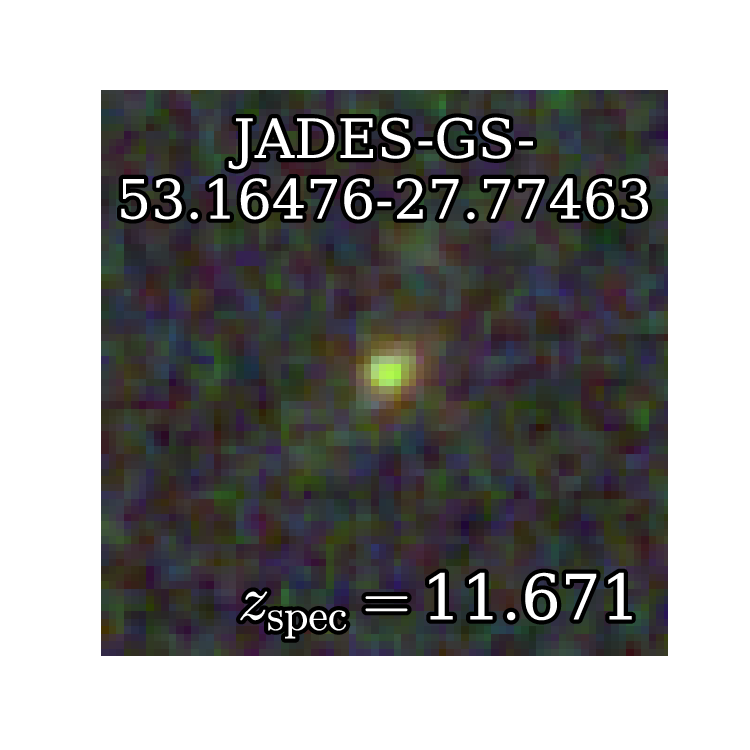}
   \includegraphics[width=0.18\textwidth]{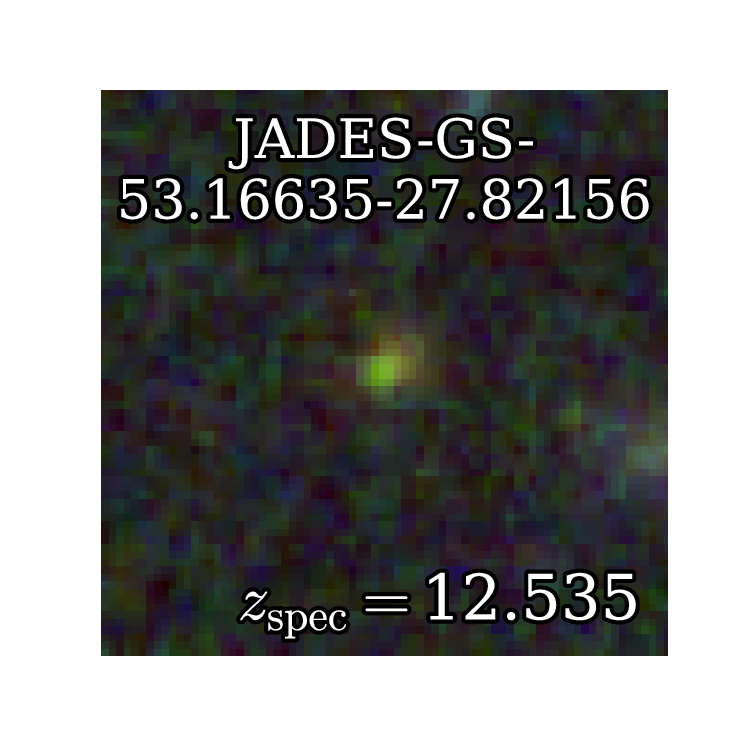}
   \includegraphics[width=0.18\textwidth]{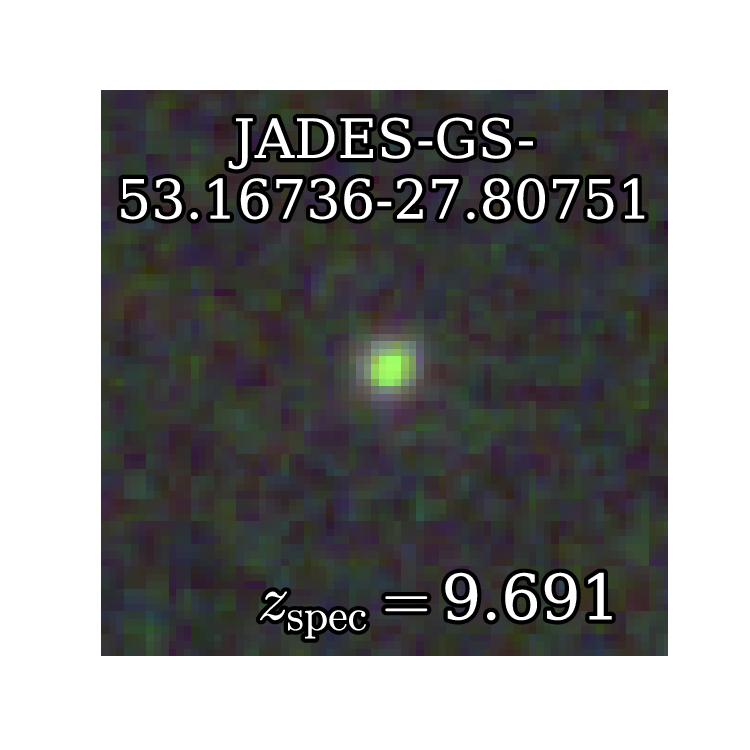}
   \includegraphics[width=0.18\textwidth]{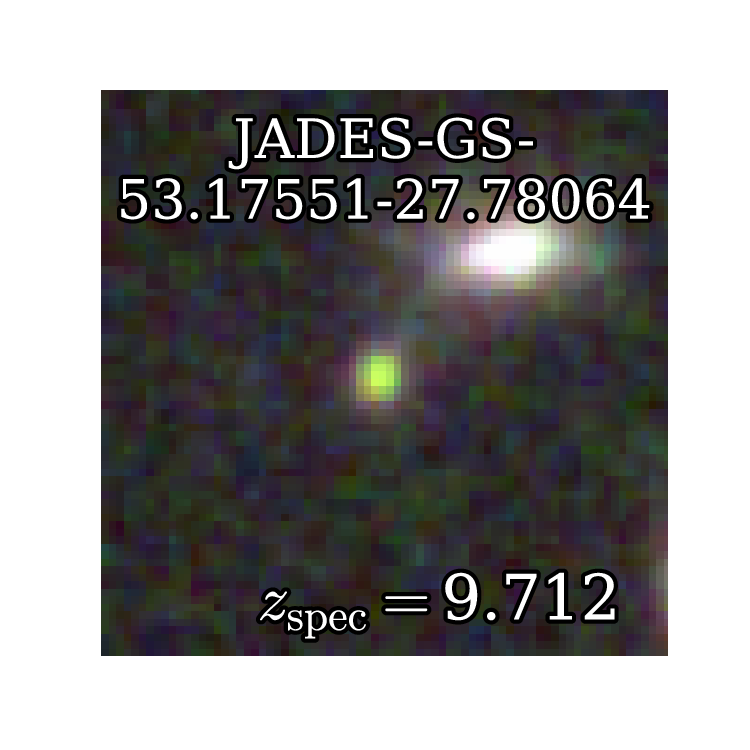}
   \includegraphics[width=0.18\textwidth]{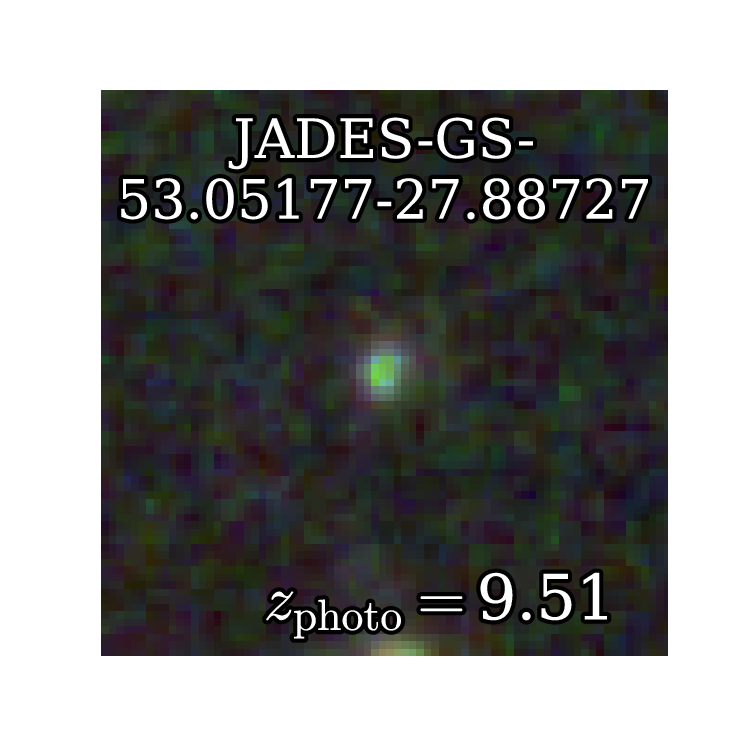}
   \includegraphics[width=0.18\textwidth]{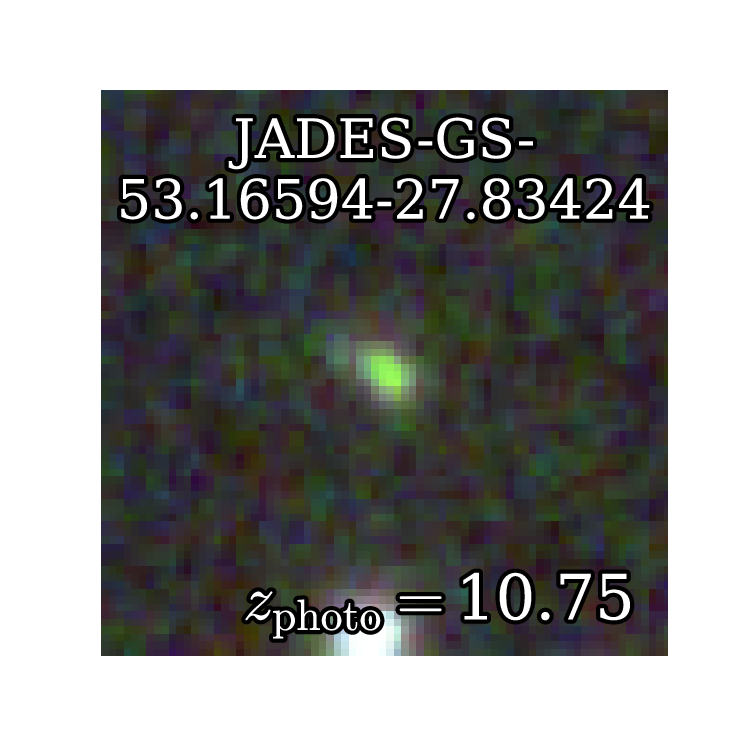}
   \includegraphics[width=0.18\textwidth]{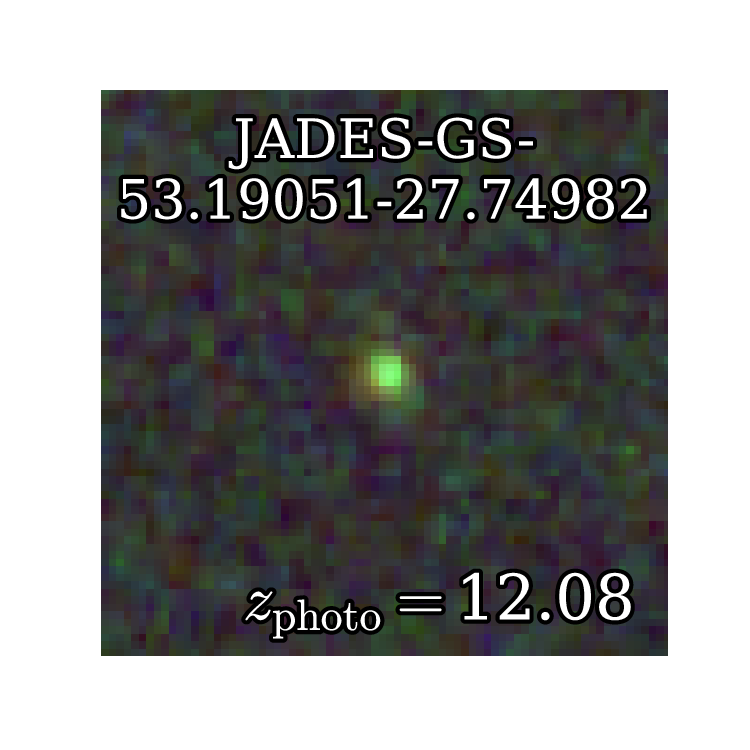}
   \includegraphics[width=0.18\textwidth]{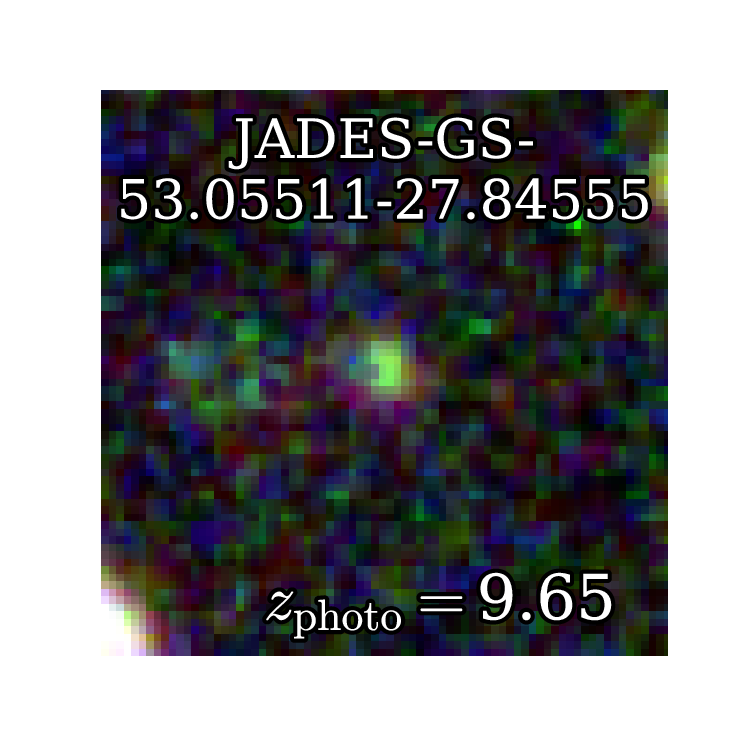}
   \includegraphics[width=0.18\textwidth]{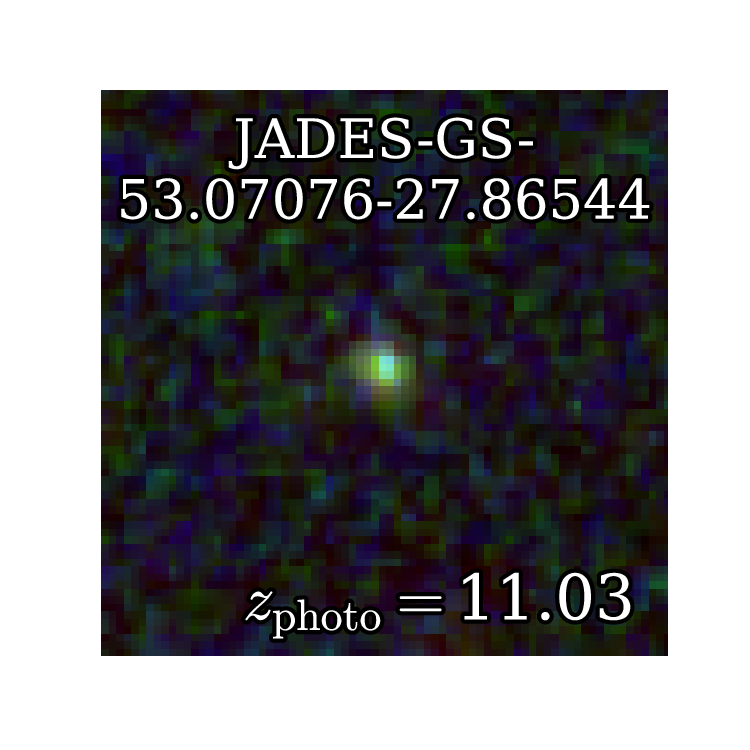}
   \includegraphics[width=0.18\textwidth]{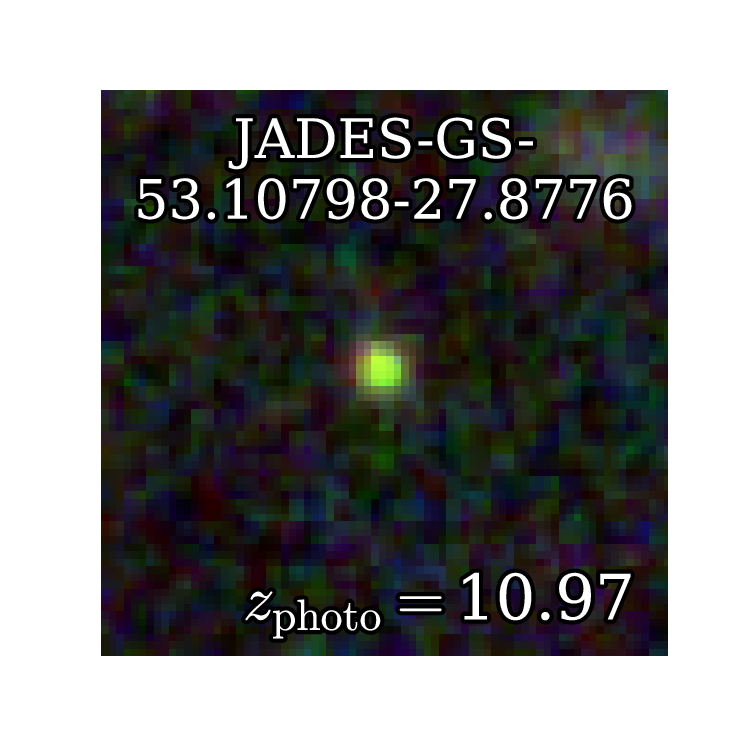}
   \includegraphics[width=0.18\textwidth]{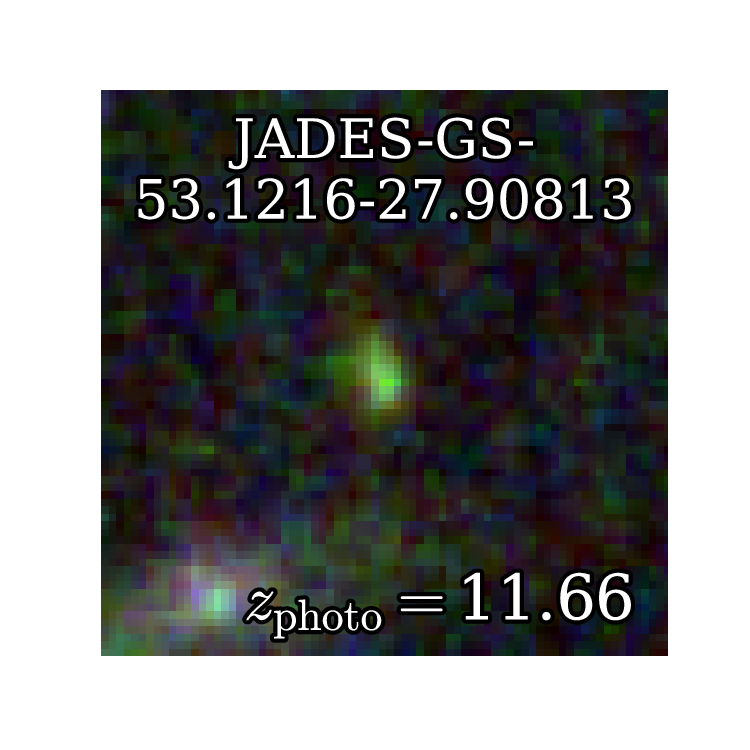}
   \includegraphics[width=0.18\textwidth]{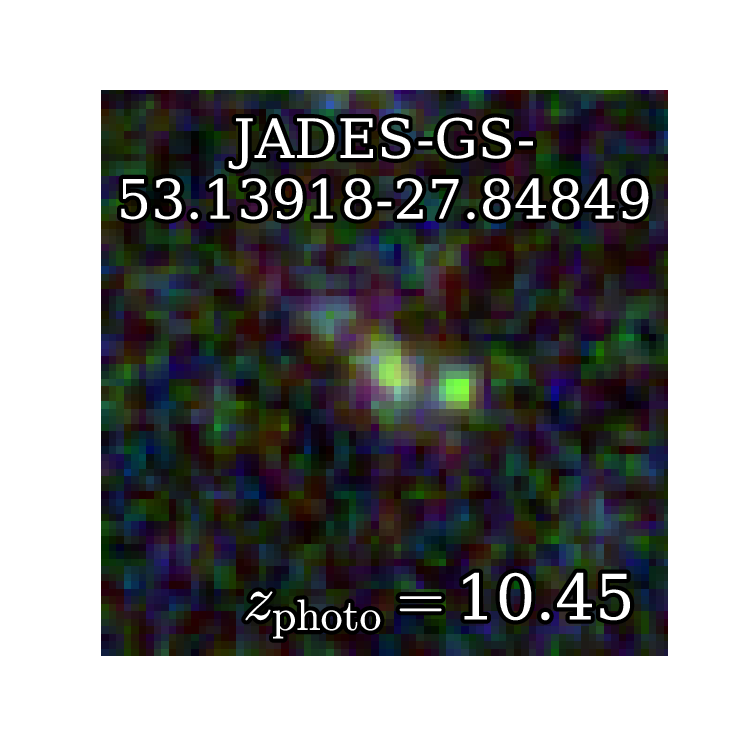}
   \includegraphics[width=0.18\textwidth]{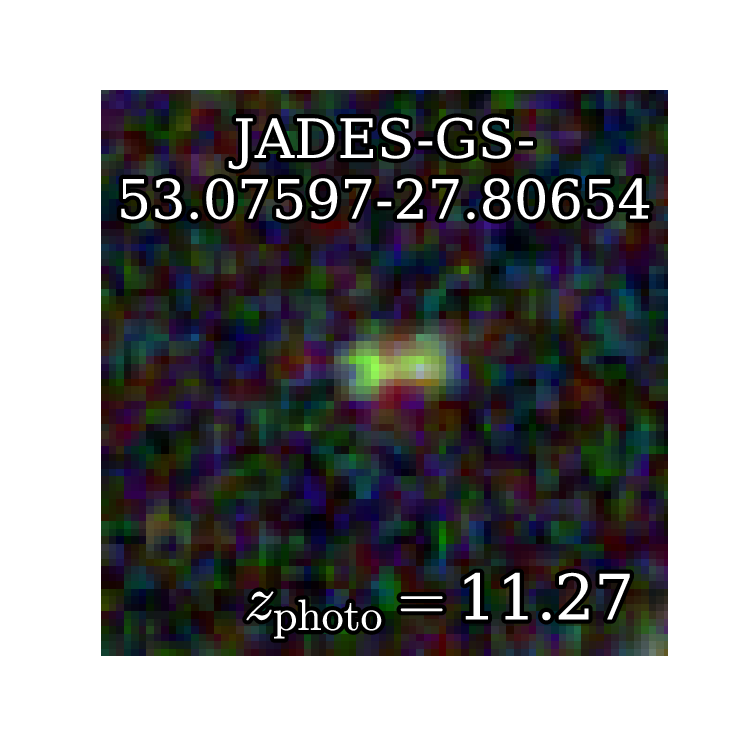}
   \includegraphics[width=0.18\textwidth]{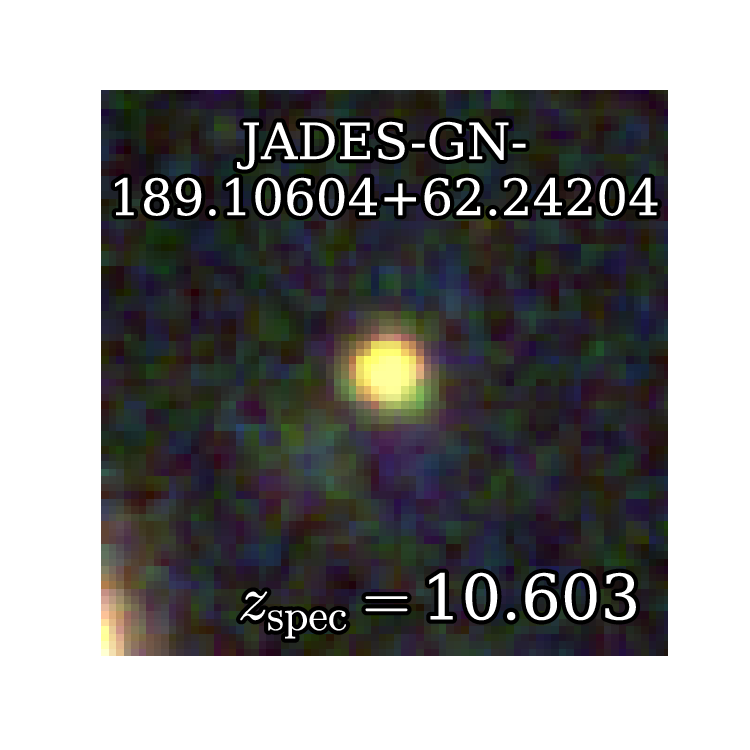}
   \includegraphics[width=0.18\textwidth]{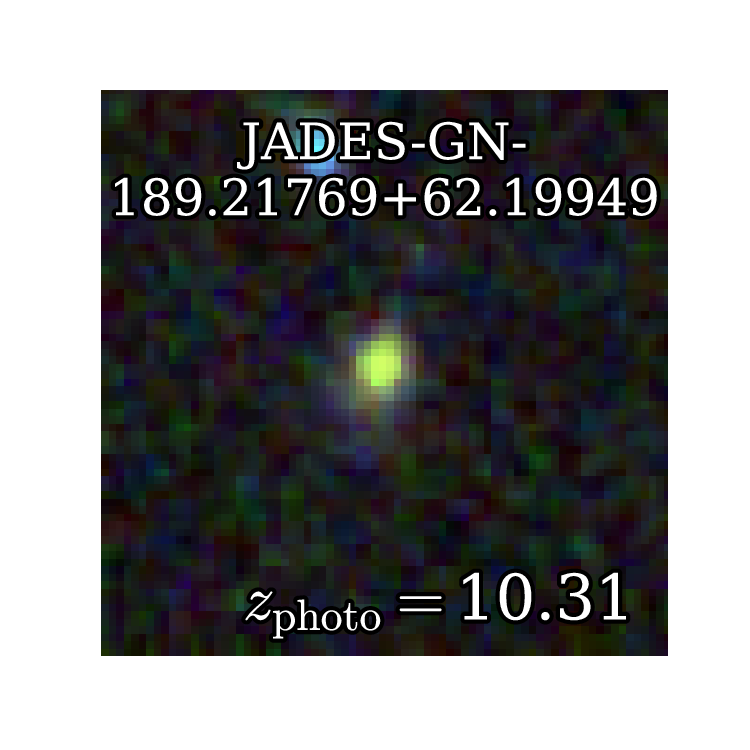}
   \includegraphics[width=0.18\textwidth]{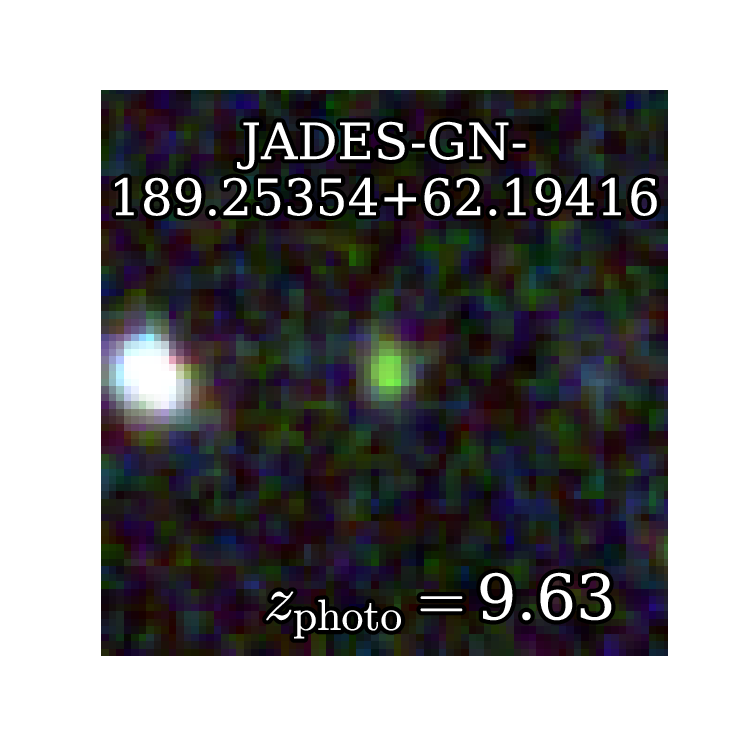}
   \includegraphics[width=0.18\textwidth]{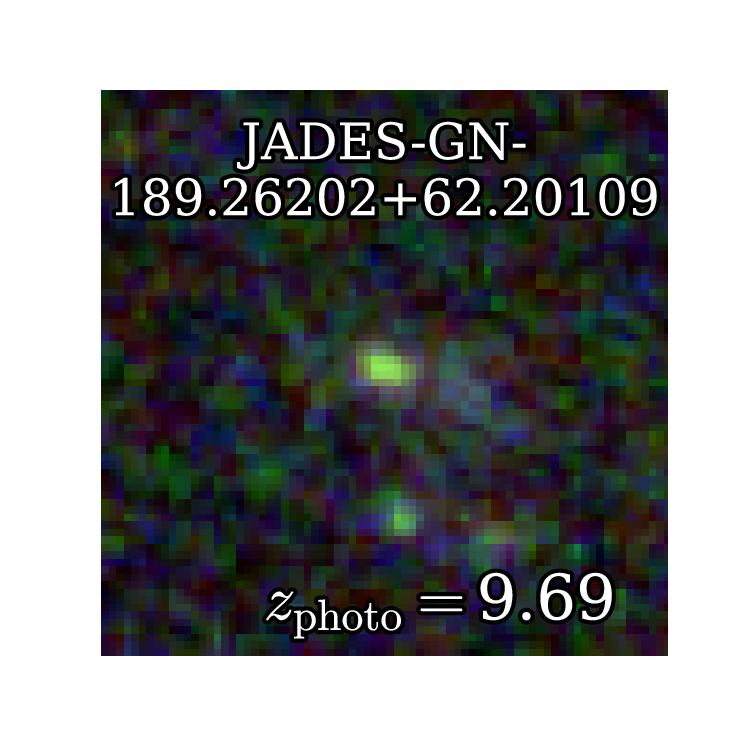}
   \includegraphics[width=0.18\textwidth]{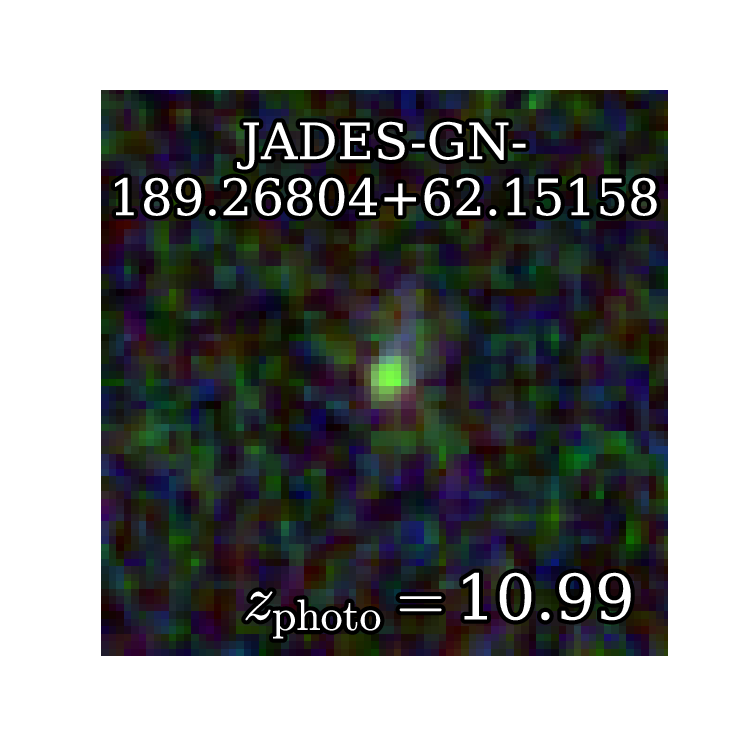}
   \includegraphics[width=0.18\textwidth]{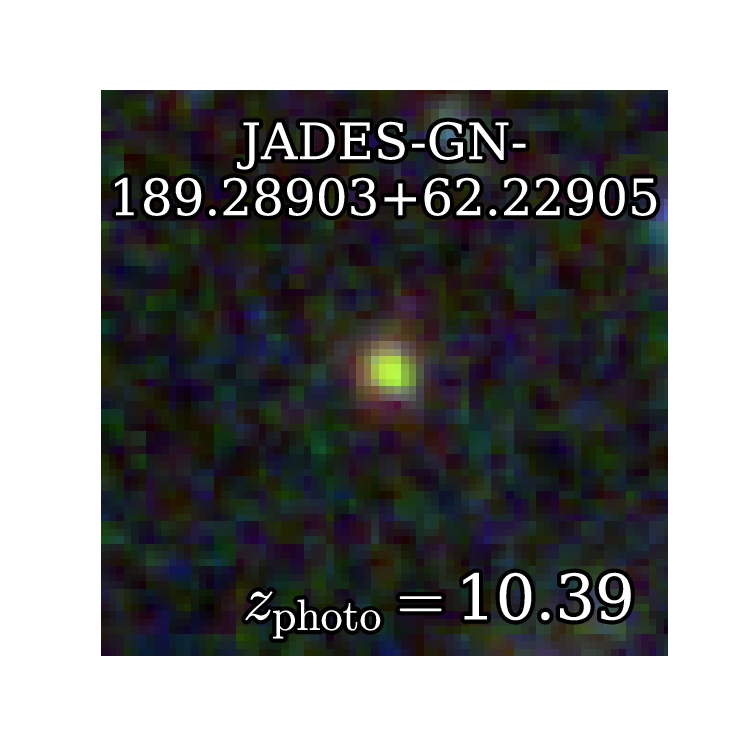}
   \includegraphics[width=0.18\textwidth]{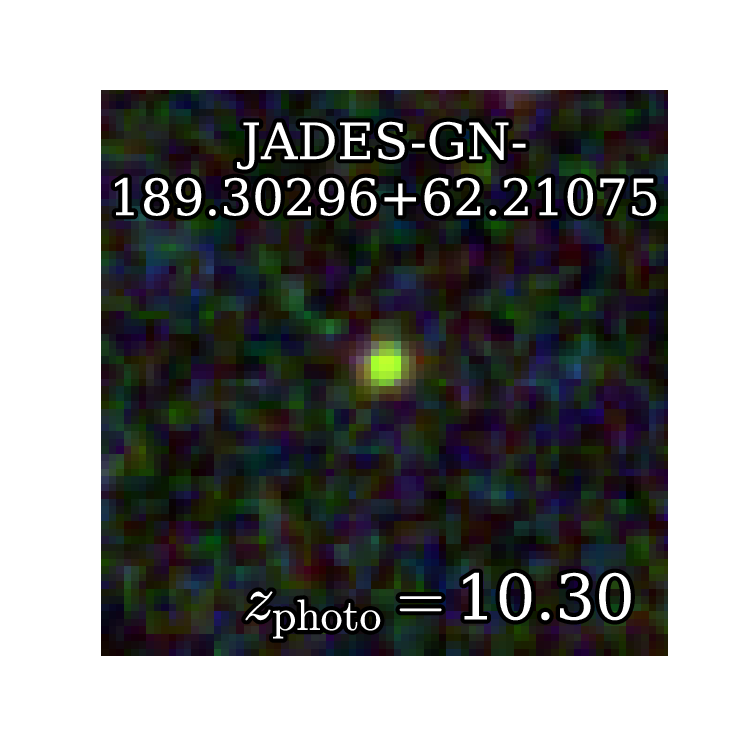}
\caption{Pseudo-color images of the $z\sim10$--$16$ galaxies analyzed in this study, 
for which the aperture magnitude $\mathrm{S/N} > 10$ 
and individual surface brightness profile fitting is successfully performed with GALFIT.
Blue, green, and red represent the F150W, F200W, and F277W images, respectively.
For spectroscopically confirmed galaxies, their spectroscopic redshifts are indicated, 
while photometric redshifts are provided for the others.
Each image has a size of $1\farcs5 \times 1\farcs5$.
}
\label{fig:pseudo_colors}
\end{center}
\end{figure*}

In Section \ref{sec:data}, we have compiled $z\sim10$--$16$ galaxies 
from the high-$z$ galaxy catalogs of CEERS and JADES 
(\citealt{2024ApJ...969L...2F}; \citealt{2024ApJ...964...71H}; \citealt{2024Natur.633..318C}; 
\citealt{2024arXiv240920549S}; \citealt{2024arXiv240920533C}). 
Table \ref{tab:sample_CEERS} summarizes the IDs, coordinates, redshifts, and UV magnitudes 
of objects in the CEERS field, while Table \ref{tab:sample_JADES} provides the same information for the JADES fields. 
Since the decision to perform size measurements individually is 
based on whether the S/N of the aperture magnitudes exceeded 10, 
the aperture magnitudes measured by SExtractor (\citealt{1996A&AS..117..393B}) are also included.
Among these, two objects in CEERS and one object in JADES are included 
in the little red dots catalog of \cite{2024arXiv240403576K}.
For these objects, we mark them with a dagger symbol and list their IDs in the table notes.
Figure \ref{fig:pseudo_colors} presents the pseudo-color images 
of the $z\sim10$--$16$ galaxies analyzed in this study,
for which the S/Ns of the aperture magnitude are larger than $10$ 
and the surface brightness profile fitting has been successfully performed with GALFIT.

\begin{deluxetable*}{lccccccc} 
\tablecolumns{8} 
\tablewidth{0pt} 
\tabletypesize{\small}
\tablecaption{$z>9.5$ Galaxies in the CEERS Fields Used in Our Size Analysis 
\label{tab:sample_CEERS}}
\tablehead{
    \colhead{ID} 
    &  \colhead{R.A.}
    &  \colhead{Decl.}
    &  \colhead{field}    
    &  \colhead{$z_{\rm photo}$}
    &  \colhead{$z_{\rm spec}$}
    &  \colhead{$m_{\rm UV}^{\rm (ap)}$}
    &  \colhead{$M_{\rm UV}$}
\\
    \colhead{ } 
    &  \colhead{(deg)}
    &  \colhead{(deg)}
    &  \colhead{ }
    &  \colhead{ }
    &  \colhead{ }
    &  \colhead{(mag)}
    &  \colhead{(mag)}
\\
    \colhead{(1)} 
    &  \colhead{(2)}
    &  \colhead{(3)}
    &  \colhead{(4)}
    &  \colhead{(5)}
    &  \colhead{(6)}
    &  \colhead{(7)}
    &  \colhead{(8)}
}
\startdata 
\multicolumn{8}{c}{$L/L^\ast_{z=3} = 0.3$--$1$} \\  
CEERS-87379 & 214.932064 & 52.841873 & C9 & $11.08^{+0.24}_{-0.48}$ & Nz & 28.56 & $-20.7$ \\
CEERS-99715 & 214.811852 & 52.737110 & C10 & $9.76^{+0.60}_{-0.09}$ & $9.77^{+0.37}_{-0.29}$ & 28.26 & $-20.5$ \\
CEERS-85546 & 214.885963 & 52.819060 & C9 & $11.08^{+0.24}_{-0.60}$ & --- & 28.41 & $-20.3$ \\
CEERS-35590 & 214.732525 & 52.758090 & C4 & $10.15^{+0.36}_{-0.42}$ & $10.01^{+0.14}_{-0.19}$ & 28.12 & $-20.2$ \\
CEERS-16943 & 214.943152 & 52.942442 & C2 & $11.08^{+0.39}_{-0.36}$ & $11.416^{+0.005}_{-0.005}$ & 28.41 & $-20.2$ \\
CEERS-2067 & 215.010026 & 53.013641 & C1 & $12.70^{+0.09}_{-0.72}$ & Nz & 28.87 & $-20.1$ \\
CEERS-77367 & 214.989018 & 52.879278 & C8 & $10.84^{+0.54}_{-0.51}$ & --- & 28.91 & $-20.0$ \\
CEERS-11384 & 214.906640 & 52.945504 & C2 & $11.53^{+0.30}_{-0.30}$ & $11.043^{+0.003}_{-0.003}$ & 28.67 & $-20.0$ \\
CEERS-98518$^{\dagger 1}$ & 214.817113 & 52.748343 & C10 & $10.09^{+0.18}_{-0.39}$ & --- & 27.83 & $-19.9$ \\
CEERS-36796 & 214.727248 & 52.748045 & C4 & $12.28^{+1.59}_{-0.24}$ & --- & 29.04 & $-19.9$ \\
CEERS-26112 & 214.818999 & 52.865299 & C3 & $11.38^{+0.30}_{-0.81}$ & --- & 29.57 & $-19.9$ \\
\multicolumn{8}{c}{$L/L^\ast_{z=3} = 0.12$--$0.3$} \\  
CEERS-70831 & 215.100921 & 52.936270 & C7 & $12.07^{+1.98}_{-1.08}$ & --- & 29.50 & $-19.6$ \\
CEERS-101746 & 214.881212 & 52.772963 & C10 & $10.33^{+0.24}_{-0.72}$ & --- & 28.83 & $-19.6$ \\
CEERS-10332 & 215.044001 & 52.994302 & C1 & $10.57^{+0.18}_{-1.05}$ & Nz & 28.58 & $-19.5$ \\
CEERS-34685 & 214.700083 & 52.752419 & C4 & $11.53^{+0.51}_{-0.72}$ & --- & 29.02 & $-19.4$ \\
CEERS-13452 & 214.861602 & 52.904604 & C2 & $9.55^{+0.78}_{-0.09}$ & Nz & 29.07 & $-19.4$ \\
CEERS-57400 & 214.869658 & 52.843646 & C6 & $10.60^{+0.60}_{-0.66}$ & Nz & 28.58 & $-19.4$ \\
CEERS-54306 & 214.858815 & 52.850712 & C6 & $11.23^{+0.36}_{-0.36}$ & --- & 28.40 & $-19.3$ \\
CEERS-74300 & 214.969256 & 52.882119 & C8 & $10.51^{+0.30}_{-0.57}$ & --- & 99.99 & $-19.3$ \\
CEERS-19996 & 214.922787 & 52.911529 & C2 & $11.32^{+0.30}_{-0.90}$ & $10.10^{+0.13}_{-0.26}$ & 29.08 & $-19.3$ \\
CEERS-76686 & 214.976311 & 52.873417 & C8 & $11.11^{+0.33}_{-0.36}$ & --- & 28.76 & $-19.2$ \\
CEERS-16984 & 214.866488 & 52.887854 & C2 & $10.63^{+0.39}_{-0.24}$ & --- & 29.06 & $-19.2$ \\
CEERS-76575 & 215.015299 & 52.913706 & C8 & $9.55^{+0.33}_{-0.06}$ & --- & 28.56 & $-19.1$ \\
CEERS-34925 & 214.738486 & 52.765665 & C4 & $11.89^{+1.92}_{-1.68}$ & --- & 29.67 & $-19.0$ \\
CEERS-77647 & 215.054220 & 52.923839 & C8 & $12.70^{+0.09}_{-0.66}$ & --- & 29.11 & $-18.9$ \\
CEERS-92463$^{\dagger 2}$ & 214.975831 & 52.841961 & C9 & $10.51^{+0.69}_{-0.60}$ & --- & 29.47 & $-18.8$ \\
\enddata 
\tablecomments{
The values presented in this table have been obtained in \cite{2024ApJ...969L...2F}, 
except for the 4th column and 7th column. 
(1) Object ID. 
(2) Right ascension. (3) Declination. 
(4) Field flag. C$i$ indicates that the object is located in the CEERS $i$ region. 
(5) Photometric redshift. 
(6) Spectroscopic redshift. Nz indicates that the object has been observed spectroscopically but its redshift is not determined. 
(7) Aperture magnitude in F200W measured in a $0\farcs2$ diameter circular aperture. 
(8) Total absolute UV magnitude.
}
\tablenotetext{$^{\dagger 1}$}{
CEERS 23931 (\citealt{2024arXiv240403576K}).} 
\tablenotetext{$^{\dagger 2}$}{
CEERS 36308 (\citealt{2024arXiv240403576K}).} 
\end{deluxetable*} 

\clearpage

\startlongtable
\begin{deluxetable*}{lccccccc} 
\tablecolumns{8} 
\tablewidth{0pt} 
\tabletypesize{\small}
\tablecaption{$z>9.5$ Galaxies in the JADES Fields Used in Our Size Analysis 
\label{tab:sample_JADES}}
\tablehead{
    \colhead{ID} 
    &  \colhead{R.A.}
    &  \colhead{Decl.}
    &  \colhead{field}    
    &  \colhead{$z_{\rm photo}$}
    &  \colhead{$z_{\rm spec}$}
    &  \colhead{$m_{\rm UV}^{\rm (ap)}$}
    &  \colhead{$M_{\rm UV}$}
\\
    \colhead{ } 
    &  \colhead{(deg)}
    &  \colhead{(deg)}
    &  \colhead{ }
    &  \colhead{ }
    &  \colhead{ }
    &  \colhead{(mag)}
    &  \colhead{(mag)}
\\
    \colhead{(1)} 
    &  \colhead{(2)}
    &  \colhead{(3)}
    &  \colhead{(4)}
    &  \colhead{(5)}
    &  \colhead{(6)}
    &  \colhead{(7)}
    &  \colhead{(8)}
}
\startdata 
\multicolumn{8}{c}{$L/L^\ast_{z=3} > 1$} \\  
JADES-GN-189.10604+62.24204 & 189.106043 & 62.242045 & JN & 11.0 & 10.603 & 26.45 & $-22.02$ \\
\multicolumn{8}{c}{$L/L^\ast_{z=3} = 0.3$--$1$} \\  
JADES-GS-z14-0 & 53.082937 & $-$27.855629 & JS & --- & $14.1793^{+0.0007}_{-0.0007}$ & $28.14$ & $-20.81$ \\
JADES-GS-53.13918-27.84849 & 53.139176 & $-27.84849$ & JSm & 10.45 & --- & 28.64 & $-20.40$ \\
JADES-GN-189.28497+62.22039 & 189.284974 & 62.220395 & JN & 11.4 & --- & 29.20 & $-20.38$ \\
JADES-GS-53.05511-27.84555 & 53.055107 & $-27.845547$ & JSm & 9.65 & --- & 28.41 & $-20.36$ \\
JADES-GN-189.21769+62.19949 & 189.217687 & 62.199492 & JN & 10.31 & --- & 28.11 & $-20.15$ \\
JADES-GS-53.06683-27.87294 & 53.066830 & $-27.872939$ & JSm & 10.64 & --- & 28.02 & $-20.13$ \\
JADES-GS-53.07597-27.80654 & 53.075972 & $-27.806544$ & JSm & 11.27 & --- & 28.70 & $-20.08$ \\
JADES-GN-189.26202+62.20109 & 189.262023 & 62.201085 & JN & 9.69 & --- & 28.81 & $-20.08$ \\
JADES-GS-53.17551-27.78064 & 53.175513 & $-27.780641$ & JS & 9.66 & 9.712 & 28.31 & $-20.00$ \\
JADES-GN-189.26284+62.23415 & 189.262835 & 62.234150 & JN & 9.65 & --- & 28.98 & $-19.99$ \\
JADES-GS-53.06131-27.90656 & 53.061314 & $-27.906563$ & JSm & 11.21 & --- & 29.12 & $-19.94$ \\
JADES-GS-53.1216-27.90813 & 53.121605 & $-27.908134$ & JSm & 11.66 & --- & 28.55 & $-19.88$ \\
JADES-GN-189.28903+62.22905 & 189.289034 & 62.229046 & JN & 10.39 & --- & 28.53 & $-19.82$ \\
JADES-GN-189.27641+62.20724 & 189.276408 & 62.207237 & JN & 12.19 & --- & 29.20 & $-19.82$ \\
JADES-GN-189.26804+62.15158 & 189.268037 & 62.151578 & JN & 10.99 & --- & 28.80 & $-19.75$ \\
JADES-GS-53.16476-27.77463 & 53.164762 & $-27.774626$ & JS & 12.31 & 11.671 & 28.79 & $-19.74$ \\
\multicolumn{8}{c}{$L/L^\ast_{z=3} = 0.12$--$0.3$} \\  
JADES-GS-53.16594-27.83424 & 53.165936 & $-27.834236$ & JS & 10.75 & --- & 28.46 & $-19.61$ \\
JADES-GS-53.10798-27.8776 & 53.107985 & $-27.877602$ & JSm & 10.97 & --- & 28.42 & $-19.55$ \\
JADES-GS-53.07076-27.86544 & 53.070763 & $-27.865438$ & JSm & 11.03 & --- & 28.71 & $-19.49$ \\
JADES-GN-189.2887+62.22923 & 189.288700 & 62.229233 & JN & 10.83 & --- & 29.58 & $-19.49$ \\
JADES-GS-53.16736-27.80751 & 53.167357 & $-27.807508$ & JS & 9.54 & 9.691 & 28.56 & $-19.48$ \\
JADES-GN-189.25354+62.19416 & 189.253536 & 62.194165 & JN & 9.63 & --- & 28.58 & $-19.48$ \\
JADES-GN-189.30296+62.21075 & 189.302960 & 62.210749 & JN & 10.3 & --- & 28.53 & $-19.47$ \\
JADES-GN-189.25361+62.18821 & 189.253611 & 62.188207 & JN & 9.96 & --- & 29.32 & $-19.45$ \\
JADES-GS-53.06422-27.8932 & 53.064224 & $-27.8932$ & JSm & 9.73 & --- & 29.45 & $-19.45$ \\
JADES-GN-189.18021+62.1804 & 189.180211 & 62.180405 & JN & 10.76 & --- & 99.99 & $-19.37$ \\
JADES-GS-53.14528-27.82359 & 53.145281 & $-27.823595$ & JS & 10.81 & --- & 29.42 & $-19.36$ \\
JADES-GN-189.2619+62.2012 & 189.261902 & 62.201201 & JN & 11.05 & --- & 29.66 & $-19.36$ \\
JADES-GN-189.25115+62.18742 & 189.251150 & 62.187420 & JN & 10.45 & --- & 29.32 & $-19.35$ \\
JADES-GS-53.13142-27.84713 & 53.131424 & $-27.847134$ & JSm & 10.25 & --- & 29.67 & $-19.32$ \\
JADES-GN-189.2234+62.19885 & 189.223400 & 62.198847 & JN & 10.64 & --- & 30.54 & $-19.29$ \\
JADES-GS-53.14297-27.84042 & 53.142965 & $-27.840424$ & JSm & 10.24 & --- & 29.75 & $-19.28$ \\
JADES-GS-53.06715-27.88317 & 53.067152 & $-27.883171$ & JSm & 10.95 & --- & 30.01 & $-19.27$ \\
JADES-GS-53.13269-27.84466 & 53.132690 & $-27.84466$ & JSm & 10.68 & --- & 99.99 & $-19.23$ \\
JADES-GN-189.24008+62.21415 & 189.240084 & 62.214146 & JN & 10.58 & --- & 29.24 & $-19.21$ \\
JADES-GN-189.28011+62.15677 & 189.280110 & 62.156770 & JN & 9.91 & --- & 29.29 & $-19.18$ \\
JADES-GS-53.11763-27.88818 & 53.117627 & $-27.888176$ & JSm & 11.47 & --- & 28.68 & $-19.17$ \\
JADES-GS-53.19051-27.74982 & 53.190505 & $-27.749824$ & JS & 12.08 & --- & 28.84 & $-19.15$ \\
JADES-GS-53.06708-27.80877 & 53.067076 & $-27.808773$ & JS & 11.53 & --- & 29.16 & $-19.14$ \\
JADES-GS-53.06699-27.80884 & 53.066989 & $-27.808839$ & JS & 11.1 & --- & 29.39 & $-19.11$ \\
JADES-GN-189.22085+62.22288 & 189.220850 & 62.222879 & JN & 10.22 & --- & 29.23 & $-19.04$ \\
JADES-GS-53.18993-27.77149 & 53.189927 & $-27.771494$ & JS & 11.48 & --- & 29.81 & $-19.03$ \\
JADES-GS-z14-1 & 53.074271 & $-$27.885927 & JS & --- & $13.90^{+0.17}_{-0.17}$ & $29.77$ & $-19.0$ \\
JADES-GS-53.16635-27.82156 & 53.166346 & $-27.821557$ & JS & 12.46 & 12.535 & 29.42 & $-18.99$ \\
JADES-GN-189.24382+62.24059 & 189.243820 & 62.240593 & JN & 10.51 & --- & 29.05 & $-18.99$ \\
JADES-GN-189.27993+62.21553 & 189.279929 & 62.215533 & JN & 11.99 & --- & 29.75 & $-18.98$ \\
JADES-GS-53.19592-27.7555 & 53.195917 & $-27.755498$ & JS & 15.32 & --- & 30.42 & $-18.98$ \\
JADES-GS-53.14988-27.7765 & 53.149881 & $-27.776503$ & JS & 13.41 & 13.274 & 29.83 & $-18.96$ \\
JADES-GS-53.16863-27.79276 & 53.168628 & $-27.792759$ & JS & 11.71 & --- & 29.49 & $-18.95$ \\
JADES-GN-189.2432+62.21009 & 189.243201 & 62.210094 & JN & 9.72 & --- & 29.36 & $-18.95$ \\
JADES-GS-53.05177-27.88727 & 53.051769 & $-27.887272$ & JS & 9.51 & --- & 29.35 & $-18.91$ \\
JADES-GS-53.12903-27.85492 & 53.129030 & $-27.854924$ & JS & 10.66 & --- & 29.59 & $-18.90$ \\
JADES-GS-53.12692-27.89778 & 53.126918 & $-27.897781$ & JSm & 10.76 & --- & 29.77 & $-18.86$ \\
JADES-GS-53.09748-27.85698 & 53.097475 & $-27.856984$ & JS & 9.64 & --- & 29.61 & $-18.84$ \\
JADES-GS-53.12025-27.88573 & 53.120253 & $-27.885728$ & JSm & 9.68 & --- & 29.28 & $-18.80$ \\
JADES-GS-53.18057-27.7455 & 53.180572 & $-27.745496$ & JS & 11.47 & --- & 29.68 & $-18.72$ \\
JADES-GS-53.1985-27.79675 & 53.198498 & $-27.796753$ & JSm & 9.72 & --- & 29.44 & $-18.71$ \\
\multicolumn{8}{c}{$L/L^\ast_{z=3} = 0.048$--$0.12$} \\  
JADES-GS-53.11375-27.88316 & 53.113749 & $-27.883159$ & JSm & 10.4 & --- & 29.28 & $-18.68$ \\
JADES-GN-189.27797+62.20034 & 189.277970 & 62.200342 & JN & 11.23 & --- & 29.45 & $-18.66$ \\
JADES-GS-53.10763-27.86014 & 53.107626 & $-27.860139$ & JS & 14.44 & --- & 30.48 & $-18.65$ \\
JADES-GS-53.15637-27.74844 & 53.156375 & $-27.748444$ & JS & 10.22 & --- & 29.65 & $-18.65$ \\
JADES-GS-53.02868-27.89301 & 53.028683 & $-27.893012$ & JS & 12.39 & --- & 29.84 & $-18.61$ \\
JADES-GS-53.12297-27.89089 & 53.122972 & $-27.890891$ & JSm & 11.69 & --- & 29.91 & $-18.61$ \\
JADES-GS-53.08138-27.8858 & 53.081379 & $-27.885803$ & JSm & 11.72 & --- & 30.14 & $-18.61$ \\
JADES-GS-53.06568-27.84119 & 53.065678 & $-27.841188$ & JSm & 11.73 & --- & 29.65 & $-18.60$ \\
JADES-GS-53.06671-27.87783 & 53.066706 & $-27.877826$ & JSm & 10.22 & --- & 29.32 & $-18.55$ \\
JADES-GS-53.09208-27.85278 & 53.092076 & $-27.852783$ & JS & 10.68 & --- & 29.84 & $-18.55$ \\
JADES-GS-53.06475-27.89024 & 53.064754 & $-27.890238$ & JS & 14.0 & --- & 30.16 & $-18.54$ \\
JADES-GS-53.16553-27.82303 & 53.165528 & $-27.823027$ & JS & 9.96 & --- & 29.88 & $-18.51$ \\
JADES-GN-189.32099+62.18479 & 189.320992 & 62.184791 & JN & 10.34 & --- & 29.54 & $-18.50$ \\
JADES-GN-189.34241+62.15816 & 189.342411 & 62.158156 & JN & 10.56 & --- & 29.73 & $-18.49$ \\
JADES-GN-189.30304+62.16551 & 189.303038 & 62.165511 & JN & 11.68 & --- & 29.84 & $-18.47$ \\
JADES-GN-189.25292+62.2362 & 189.252917 & 62.236195 & JN & 10.74 & --- & 29.43 & $-18.46$ \\
JADES-GS-53.06131-27.81951 & 53.061314 & $-27.819506$ & JS & 9.86 & --- & 29.78 & $-18.46$ \\
JADES-GS-53.14159-27.87173 & 53.141588 & $-27.87173$ & JSm & 9.92 & --- & 29.59 & $-18.45$ \\
JADES-GS-53.04813-27.86053 & 53.048134 & $-27.860527$ & JSm & 10.73 & --- & 29.83 & $-18.45$ \\
JADES-GN-189.22825+62.22154 & 189.228245 & 62.221540 & JN & 10.42 & --- & 29.54 & $-18.43$ \\
JADES-GS-53.15883-27.7735 & 53.158829 & $-27.773498$ & JS & 10.84 & 10.420 & 29.59 & $-18.42$ \\
JADES-GS-53.05715-27.81617 & 53.057155 & $-27.816173$ & JSm & 10.25 & --- & 29.35 & $-18.39$ \\
JADES-GS-53.07607-27.85602 & 53.076073 & $-27.856023$ & JS & 9.62 & --- & 29.78 & $-18.34$ \\
JADES-GS-53.20065-27.79802 & 53.200650 & $-27.798022$ & JSm & 11.71 & --- & 30.43 & $-18.32$ \\
JADES-GN-189.20016+62.18314 & 189.200159 & 62.183138 & JN & 9.94 & --- & 30.11 & $-18.29$ \\
JADES-GN-189.32172+62.19397 & 189.321719 & 62.193966 & JN & 10.5 & --- & 99.99 & $-18.29$ \\
JADES-GS-53.14673-27.77901 & 53.146726 & $-27.779009$ & JS & 13.68 & --- & 30.42 & $-18.28$ \\
JADES-GN-189.33638+62.16733 & 189.336380 & 62.167326 & JN & 12.53 & --- & 29.82 & $-18.27$ \\
JADES-GN-189.30384+62.15493 & 189.303836 & 62.154925 & JN & 10.19 & --- & 29.82 & $-18.27$ \\
JADES-GS-53.12239-27.78814 & 53.122385 & $-27.788141$ & JS & 11.5 & --- & 30.94 & $-18.24$ \\
JADES-GN-189.22712+62.14806 & 189.227122 & 62.148063 & JN & 9.85 & --- & 29.64 & $-18.24$ \\
JADES-GS-53.21339-27.76951 & 53.213386 & $-27.76951$ & JS & 11.13 & --- & 30.70 & $-18.21$ \\
JADES-GN-189.26099+62.20112 & 189.260990 & 62.201124 & JN & 9.85 & --- & 29.48 & $-18.21$ \\
JADES-GS-53.079-27.86359 & 53.079003 & $-27.86359$ & JS & 10.5 & --- & 30.34 & $-18.20$ \\
JADES-GS-53.18602-27.81748 & 53.186023 & $-27.81748$ & JS & 10.03 & --- & 30.00 & $-18.20$ \\
JADES-GS-53.1485-27.74159 & 53.148504 & $-27.741589$ & JS & 11.2 & --- & 30.20 & $-18.20$ \\
JADES-GS-53.10469-27.86187 & 53.104686 & $-27.861872$ & JSm & 12.27 & --- & 30.37 & $-18.19$ \\
JADES-GN-189.26695+62.20538 & 189.266955 & 62.205384 & JN & 10.33 & --- & 29.64 & $-18.18$ \\
JADES-GS-53.09789-27.90578 & 53.097895 & $-27.905785$ & JSm & 9.68 & --- & 30.10 & $-18.16$ \\
JADES-GS-53.07477-27.84542 & 53.074768 & $-27.845421$ & JS & 9.92 & --- & 30.18 & $-18.15$ \\
JADES-GS-53.14501-27.87926 & 53.145007 & $-27.879264$ & JSm & 10.76 & --- & 29.82 & $-18.14$ \\
JADES-GS-53.17847-27.75591 & 53.178469 & $-27.755909$ & JS & 15.13 & --- & 30.85 & $-18.13$ \\
JADES-GS-53.09872-27.8602 & 53.098718 & $-27.860204$ & JS & 10.69 & --- & 30.49 & $-18.12$ \\
JADES-GS-53.13506-27.79161 & 53.135065 & $-27.791608$ & JS & 11.24 & --- & 30.16 & $-18.12$ \\
JADES-GS-53.17301-27.80085 & 53.173013 & $-27.800852$ & JS & 11.28 & --- & 30.64 & $-18.12$ \\
JADES-GS-53.08468-27.86666 & 53.084682 & $-27.866663$ & JS & 12.48 & --- & 30.35 & $-18.12$ \\
JADES-GS-53.03138-27.87219 & 53.031381 & $-27.872188$ & JS & 10.65 & --- & 30.33 & $-18.11$ \\
JADES-GS-53.09967-27.87844 & 53.099670 & $-27.878442$ & JSm & 10.68 & --- & 30.12 & $-18.08$ \\
JADES-GS-53.13871-27.819 & 53.138705 & $-27.819003$ & JS & 10.05 & --- & 30.12 & $-18.07$ \\
JADES-GS-53.17345-27.81217 & 53.173448 & $-27.812166$ & JS & 9.58 & --- & 29.64 & $-18.07$ \\
JADES-GS-53.08188-27.84431 & 53.081885 & $-27.844315$ & JS & 10.46 & --- & 30.73 & $-18.06$ \\
JADES-GS-53.07901-27.87154 & 53.079012 & $-27.871536$ & JS & 11.97 & --- & 30.31 & $-18.05$ \\
JADES-GS-53.14124-27.80343 & 53.141244 & $-27.803431$ & JS & 10.38 & --- & 30.51 & $-18.04$ \\
JADES-GN-189.31881+62.18663 & 189.318808 & 62.186627 & JN & 10.76 & --- & 30.27 & $-18.04$ \\
JADES-GS-53.06006-27.89143 & 53.060058 & $-27.891429$ & JS & 10.92 & --- & 29.88 & $-18.04$ \\
JADES-GN-189.28836+62.22878 & 189.288362 & 62.228776 & JN & 9.59 & --- & 29.86 & $-18.03$ \\
JADES-GS-53.02619-27.88716 & 53.026185 & $-27.887162$ & JS & 11.35 & --- & 30.26 & $-18.03$ \\
JADES-GS-53.07557-27.87268 & 53.075568 & $-27.872677$ & JS & 15.31 & --- & 31.53 & $-18.02$ \\
JADES-GN-189.24505+62.21991 & 189.245048 & 62.219905 & JN & 10.09 & --- & 30.01 & $-17.98$ \\
JADES-GS-53.1551-27.75395 & 53.155100 & $-27.753951$ & JS & 10.29 & --- & 30.39 & $-17.97$ \\
JADES-GS-53.16874-27.81698 & 53.168738 & $-27.816975$ & JS & 9.92 & --- & 30.36 & $-17.95$ \\
JADES-GS-53.06235-27.84774 & 53.062349 & $-27.847743$ & JSm & 10.18 & --- & 29.85 & $-17.92$ \\
JADES-GS-53.17153-27.84392 & 53.171530 & $-27.843922$ & JS & 9.92 & --- & 30.35 & $-17.92$ \\
JADES-GN-189.15382+62.21145 & 189.153820 & 62.211454 & JN & 9.71 & --- & 29.77 & $-17.91$ \\
JADES-GS-53.04017-27.87602 & 53.040167 & $-27.876022$ & JS & 11.92 & --- & 30.33 & $-17.90$ \\
JADES-GS-53.08537-27.85047 & 53.085365 & $-27.850474$ & JS & 9.69 & --- & 29.92 & $-17.90$ \\
JADES-GS-53.02804-27.86158 & 53.028043 & $-27.861585$ & JSm & 9.74 & --- & 99.99 & $-17.87$ \\
JADES-GS-53.13918-27.78273 & 53.139184 & $-27.782726$ & JS & 10.49 & --- & 30.45 & $-17.85$ \\
JADES-GS-53.07277-27.85209 & 53.072767 & $-27.852093$ & JS & 9.88 & --- & 30.81 & $-17.84$ \\
JADES-GS-53.04887-27.89045 & 53.048871 & $-27.890447$ & JS & 9.61 & --- & 30.11 & $-17.83$ \\
JADES-GS-53.07248-27.85535$^{\dagger 1}$ & 53.072480 & $-27.855352$ & JS & 10.6 & --- & 30.24 & $-17.81$ \\
JADES-GS-53.13018-27.89975 & 53.130183 & $-27.899752$ & JSm & 10.62 & --- & 99.99 & $-17.80$ \\
JADES-GS-53.12737-27.81805 & 53.127367 & $-27.818052$ & JS & 10.71 & --- & 30.41 & $-17.80$ \\
JADES-GS-53.13639-27.81306 & 53.136386 & $-27.813064$ & JS & 11.01 & --- & 30.33 & $-17.79$ \\
JADES-GS-53.16627-27.81377 & 53.166270 & $-27.813766$ & JS & 11.46 & --- & 30.34 & $-17.79$ \\
JADES-GS-53.17016-27.82454 & 53.170158 & $-27.824542$ & JS & 10.7 & --- & 30.50 & $-17.79$ \\
JADES-GN-189.31828+62.20169 & 189.318282 & 62.201687 & JN & 11.53 & --- & 30.46 & $-17.79$ \\
JADES-GS-53.09871-27.86016 & 53.098714 & $-27.860157$ & JS & 9.59 & --- & 30.73 & $-17.78$ \\
JADES-GS-53.10484-27.86807 & 53.104839 & $-27.868072$ & JSm & 9.94 & --- & 29.84 & $-17.78$ \\
JADES-GS-53.18129-27.81043 & 53.181291 & $-27.810427$ & JS & 12.52 & --- & 30.20 & $-17.75$ \\
JADES-GS-53.13274-27.7967 & 53.132735 & $-27.796699$ & JS & 10.47 & --- & 30.04 & $-17.71$ \\
JADES-GN-189.27007+62.21786 & 189.270070 & 62.217863 & JN & 10.53 & --- & 30.37 & $-17.70$ \\
\enddata 
\tablecomments{
The values presented in this table have been obtained in \cite{2024ApJ...964...71H} and \cite{2024Natur.633..318C}, 
except for the 4th column and 7th column. 
(1) Object ID. 
(2) Right ascension. (3) Declination. 
(4) Field flag. JS, JSm, and JN indicate that the object is located in JADES Deep, JADES Medium, and JADES GOODS-N, respectively. 
(5) Photometric redshift. 
(6) Spectroscopic redshift. 
(7) Aperture magnitude measured in a $0\farcs2$ diameter circular aperture. 
For $z<13$, F200W images are used, and for $z>13$, F277W images are used. 
(8) Total absolute UV magnitude.
}
\tablenotetext{$^{\dagger 1}$}{
JADES 67592 (\citealt{2024arXiv240403576K}).} 
\end{deluxetable*} 

\end{document}